\documentclass[11pt,draftclsnofoot,journal,onecolumn]{IEEEtran}

\usepackage{graphicx}
\usepackage{amsmath,amssymb, amsthm, bbm}
\usepackage{epsfig}
\usepackage{varioref}
\usepackage{subfigure}
\usepackage[sort&compress,square,numbers]{natbib}
\usepackage{arydshln}
\usepackage{paralist}
\usepackage[colorlinks]{hyperref}

\begin{document}
\title{On Measure Transformed Canonical Correlation Analysis}
\author{{Koby Todros and Alfred O. Hero}\\
{Dept. of Electrical Engineering and Computer Science}\\
{University of Michigan}, 
{Ann-Arbor 48105, MI, U.S.A}\\
{Email: ktodros@umich.edu, hero@eecs.umich.edu}}

\maketitle

\newtheorem{Theorem}{Theorem}
\newtheorem{Lemma}{Lemma}
\newtheorem{Corollary}{Corollary}
\newtheorem{Conclusion}{Conclusion}
\newtheorem{Proposition}{Proposition}
\newtheorem{Definition}{Definition}
\newtheorem{Remark}{Remark}
\newtheorem{Identity}{Identity}

\def\Csp{{\mathbb{C}}}

\def\Rsp{{\mathbb{R}}}

\def\Nsp{{\mathbb{N}}}

\def\Thetasp{{\mbox{\boldmath $\Theta$}}}

\def\Thetaspsc{{\mbox{\boldmath $\Thetasc$}}}

\def\Lambdasp{{\mbox{\boldmath $\Lambda$}}}

\def\Lambdaspsc{{\mbox{\boldmath $\Lambdasc$}}}

\def\Vmatsc{{\mbox{\boldmath $\Vsc$}}}

\def\VTau{\Vmat\left(\btau\right)}

\def\VHTau{\Vmat^{H}\left(\btau\right)}

\def\VHTauPrime{\Vmat^{H}\left(\btau^{\prime}\right)}

\def\LTwo{{\mathcal{L}}_{2}\left(\mathcal{X},\mathcal{F},{\mathcal{P}}_{\thetavecsc_t}\right)}

\def\LTwoBrev{{\mathcal{L}}_{2}\left(\pxy\right)}

\def\LTwoBrevMod{{\mathcal{L}}_{2}\left(\Tuv\right)}

\def\LTwoBrevEtag{{\mathcal{L}}_{2}\left(\eta_{g}\right)}

\def\LTwoBrevn{{\mathcal{L}}_{2}\left(\mathcal{X},p_{\theta_{n}}\right)}
\def\LTwoBrevZero{{\mathcal{L}}_{2}\left(\mathcal{X},p_{\theta_{0}}\right)}

\def\LTwoBrevAlpha{{\mathcal{L}}_{2}\left(\mathcal{X},p_{\alpha}\right)}

\def\LTwoBrevTrue{{\mathcal{L}}_{2}\left(\mathcal{X},p_{\theta_{t}}\right)}
\def\LTwoBrevPrime{{\mathcal{L}}_{2}\left(\mathcal{X},p_{{\theta^{\prime}}}\right)}
\def\LTwoBrevVar{{\mathcal{L}}_{2}\left(\mathcal{X},p_{\vartheta}\right)}

\def\LTwoBrevTwo{{\mathcal{L}}^{2}_{2}\left(\mathcal{X}\times\Thetasp\right)}

\def\LTwoBrevsc{{\mathcal{L}}_{2}\left(\XCalsc\times\Thetaspsc_{\rm{r}}\right)}

\def\LTwoBrevL{{\mathcal{L}}_{2}^{L}\left(\XCal\right)}

\def\LTwoBrevLd{{\mathcal{L}}_{2}^{L_{\rm{d}}}\left(\XCal\times\Thetasp_{\rm{r}}\right)}

\def\LTwoBrevLr{{\mathcal{L}}_{2}^{L_{\rm{r}}}\left(\XCal\times\Thetasp_{\rm{r}}\right)}

\def\Tuv{{\rm{T}}_{u,v}\left[\mu_{xy}\right]}

\def\errvec{\evec_{\hat{\gvec}}\left(\xvec,\thetavec\right)}

\def\errvecMMSE{\evec_{\hat{\gvec}_{\rm{MMSE}}}\left(\xvec,\thetavec\right)}

\def\errvecH{\evec^{H}_{\hat{\gvec}}\left(\xvec,\thetavec\right)}

\def\ProjHvarphi{\pvec_{\mathcal{J}}\left(\gtheta|{{\mathcal{H}_{\varphi}}}\right)}

\def\ProjHphi{\pvec_{\mathcal{J}}\left(\gtheta|{{\mathcal{H}_{\phi}}}\right)}

\def\ProjHphiH{\pvec^{H}_{\mathcal{J}}\left(\gtheta|{{\mathcal{H}_{\phi}}}\right)}

\def\ProjHvarphiPsi{\pvec_{\mathcal{J}}\left(\gtheta|{{\mathcal{H}_{\varphi}^{\left(\psivecsc\right)}}}\right)}

\def\ProjHvarphiPsiH{\pvec^{H}_{\mathcal{J}}\left(\gtheta|{{\mathcal{H}_{\varphi}^{\left(\psivecsc\right)}}}\right)}

\def\muxy{\mu_{\xvec\yvec}}
\def\mux{\mu_{\xvec}}
\def\muy{\mu_{\yvec}}

\def\pxy{P_{\Xmatsc\Ymatsc}}
\def\px{P_{\Xmatsc}}
\def\py{P_{\Ymatsc}}

\def\etaxy{\eta^{\left(u,v\right)}_{\xvec\yvec}}
\def\etax{\eta^{\left(u,v\right)}_{\xvec}}
\def\etay{\eta^{\left(u,v\right)}_{\yvec}}

\def\qxy{Q^{\left(u,v\right)}_{\Xmatsc\Ymatsc}}
\def\qx{Q^{\left(u,v\right)}_{\Xmatsc}}
\def\qy{Q^{\left(u,v\right)}_{\Ymatsc}}

\def\uexp{u_{\expsc}}
\def\vexp{v_{\expsc}}

\def\uGauss{u^{(\sigma)}_{\Gausssc}}
\def\vGauss{v^{(\tau)}_{\Gausssc}}

\def\uGausss{u_{\Gausssc}}
\def\vGausss{v_{\Gausssc}}

\newcommand{\expsc}{{\mbox{\tiny${\rm{E}}$}}}

\newcommand{\Gausssc}{{\mbox{\tiny${\rm{G}}$}}}

\newcommand{\Gsc}{{\mbox{\tiny${\rm{G}}$}}}

\newcommand{\svecsc}{{\mbox{\tiny\svec}}}

\newcommand{\tvecsc}{{\mbox{\tiny\tvec}}}

\def\LPlusXTheta{{\mathcal{L}}^{+}\left(\mathcal{X}\times\Thetasp_{\rm{r}}\times\Lambdasp\right)}

\def\LPlusTheta{{\mathcal{L}}^{+}\left(\Thetasp\right)}

\def\LPlusThetaXThetaPrime{{\mathcal{L}}^{+}\left(\Lambdasp\times\Lambdasp\times\mathcal{X}\times\Thetasp_{\rm{r}}\right)}

\def\LPlusLambda{{\mathcal{L}}^{+}\left(\Lambdasp\right)}

\def\Ptheta{p_{\theta}}
\def\Pvartheta{p_{\vartheta}}
\def\PvarthetaX{p_{\vartheta}\left(\xvec\right)}

\def\PthetatrueX{p_{\thetavecsc_{t}}\left(\xvec\right)}
\def\PthetaX{p_{\theta}\left(\xvec\right)}
\def\PvarthetaX{p_{\vartheta}\left(\xvec\right)}

\def\PthetanX{p_{\thetavecsc_{n}}\left(\xvec\right)}

\def\PthetaXPrime{p_{\thetavecsc^{\prime}}\left(\xvec\right)}

\def\Pthetatrue{p_{\thetavecsc_{t}}}

\def\MSEMatDot{{\rm{{MSE}}}\left(\cdot\right)}

\def\MSEMat{{\rm{{MSE}}}\left(\hat{g}\left(\xvec\right)\right)}

\def\MSEMatTilde{{\rm{{MSE}}}\left(\tilde{\gvec}\right)}

\def\MSEMatLBU{{\rm{\textbf{MSE}}}\left(\hat{\gvec}_{\rm{LBU}}\left(\xvec\right)\right)}

\def\MSEMatMMSE{{\rm{\textbf{MSE}}}\left(\hat{\gvec}_{\rm{MMSE}}\left(\xvec\right)\right)}

\def\MSEMatTilde{{\rm{\textbf{MSE}}}\left(\tilde{\gvec}\left(\xvec\right)\right)}

\def\Hphisub{\mathcal{H}_{\varphi}^{\left(\hvec\right)}}
\def\Hv{\mathcal{H}_{\varphi}^{\left(\vvec\right)}}
\def\Hvk{\mathcal{H}_{\varphi}^{\left({\vvec_{k}}\right)}}

\def\Hphisubk{\mathcal{H}_{\varphi}^{\left(\hvec{k}\right)}}

\def\Hvarphi{\mathcal{H}_{\varphi}}

\def\HphiH{\mathcal{H}_{\varphi}^{\left(\hvec\right)}}

\def\PhiaH{\varphi_{\avec, \hvec}\left(\xvec,\thetavec\right)}

\def\Hphisubn{\mathcal{H}_{\varphi}^{\left(\hvec_{n}\right)}}

\def\Hphi{\mathcal{H}_{\phi}}

\def\HphiN{\mathcal{H}_{\phi}^{N}}

\def\HphiL{\mathcal{H}_{\phi}^{L}}

\def\HphiTwo{\mathcal{H}_{\phi}^{2}}

\def\PhiahX{\varphi_{\avec,\hvec}\left(\xvec\right)}

\def\PhiTildealhX{\varphi_{\tilde{\avec}_{l},\hvec}\left(\xvec\right)}

\def\PhiTildeakhX{\varphi_{\tilde{\avec}_{k},\hvec}\left(\xvec\right)}

\def\LOneTheta{\mathcal{L}_{1}\left(\Thetasp,\lambda\right)}

\def\LOneXTheta{\mathcal{L}_{1}\left(\XCal\times\Thetasp\right)}

\def\LOneLambdaTheta{\mathcal{L}_{1}\left(\Lambdasp\times\Thetasp\right)}

\def\LOneLambda{\mathcal{L}_{1}\left(\Lambdasp\right)}

\def\LOnePLambda{\mathcal{L}_{1}^{P}\left(\Lambdasp\right)}

\def\LOneThetaL{\mathcal{L}_{1}^{L}\left(\Thetasp\right)}

\def\ghatLBU{\hat{\gvec}_{\rm{LBU}}\left(\xvec\right)}

\def\ghatMMSE{\hat{\gvec}_{\rm{MMSE}}\left(\xvec\right)}

\def\gTilde{\tilde{\gvec}\left(\xvec\right)}

\def\LTwoBayes{{\mathcal{L}}_{2}\left(\mathcal{X}\times\Thetasp,\mathcal{F}\times\mathcal{M},{\mathcal{P}}\right)}

\def\thetavec{{\bf{\theta}}}

\def\gtheta{g\left(\theta\right)}
\def\btheta{b\left(\theta\right)}

\def\gthetad{\gvec_{\rm{d}}\left(\thetavec_{\rm{d}}\right)}
\def\gthetadT{\gvec^{T}_{\rm{d}}\left(\thetavec_{\rm{d}}\right)}

\def\gthetar{\gvec_{\rm{r}}\left(\thetavec_{\rm{r}}\right)}
\def\gthetarT{\gvec^{T}_{\rm{r}}\left(\thetavec_{\rm{r}}\right)}

\def\gthetadPrime{\gvec_{\rm{d}}\left(\thetavec^{\prime}_{\rm{d}}\right)}

\def\gthetaT{\gvec^{T}\left(\thetavec\right)}

\def\gthetasc{\gvec\left(\thetavecsc\right)}

\def\gTtheta{\gvec^{T}\left(\thetavec\right)}

\def\gthetaTrue{g\left(\theta_{t}\right)}

\def\gthetam{\gvec\left(\thetavec_{m}\right)}

\def\gthetan{g\left(\theta_{n}\right)}
\def\bthetan{b\left(\theta_{n}\right)}

\def\gthetazero{\gvec\left(\thetavec_{t}\right)}

\def\gthetaFin{\gvec\left(\thetavec_{N-1}\right)}

\def\gthetaSc{\gvec\left(\thetavecsc\right)}

\def\gthetaPrime{\gvec\left(\thetavec^{\prime}\right)}

\def\gthetaPlusTau{\gvec\left(\thetavec+\btau\right)}

\def\gthetaPlusTaun{\gvec\left(\thetavec+\btau_{n}\right)}

\def\gthetaH{\gvec^{T}\left(\thetavec\right)}

\def\gTau{\gvec\left(\btau\right)}

\def\gTauSc{\gvec\left(\btausc\right)}

\def\gThetaSc{\gvec\left(\thetavecsc\right)}

\def\gTauPrimeSc{\gvec\left(\btausc^{\prime}\right)}

\def\gthetatrue{g\left(\theta_{t}\right)}

\def\ghatX{\hat{g}\left(\xvec\right)}

\def\ghatXd{\hat\gvec_{\rm{d}}\left(\xvec\right)}
\def\ghatXdT{\hat\gvec^{T}_{\rm{d}}\left(\xvec\right)}

\def\ghatXr{\hat\gvec_{\rm{r}}\left(\xvec\right)}
\def\ghatXrT{\hat\gvec^{T}_{\rm{r}}\left(\xvec\right)}

\def\gTildeX{\tilde\gvec\left(\xvec\right)}

\def\gthetaScal{g\left(\theta\right)}

\def\ghatXScal{\hat{g}\left(\xvec\right)}

\def\PthetaXScal{{p_{\theta}\left(\xvec\right)}}

\def\eXTheta{\evec\left(\Xmat,\thetavec\right)}

\def\eXThetatrue{\evec\left(\Xmat,\thetavec_{t}\right)}

\def\eXThetatrueStar{\evec^{*}\left(\Xmat,\thetavec_{t}\right)}

\def\ghatMVUX{\hat\gvec_{MVU}\left(\Xmat\right)}

\def\ghatMVU{\hat\gvec_{MVU}}

\def\ghat{\hat\gvec}

\def\LRSym{r\left(\xvec,\vartheta,\theta\right)}
\def\LRSymTrue{r\left(\xvec,\vartheta,\theta\right)}
\def\LRSymScal{r_{\theta}\left(\xvec,\vartheta\right)}

\def\LRSymTau{\nu\left(\xvec,\btau\right)}

\def\LRSymTauSc{\nu\left(\xvec,\btausc\right)}

\def\LRSymThetaSc{\nu\left(\xvec,\thetavecsc,\btau_{sc}\right)}

\def\LRSymTauPrime{\nu\left(\xvec,\btau^{\prime}\right)}

\def\LRSymTauPrimeSc{\nu\left(\xvec,\btausc^{\prime}\right)}

\def\LRSymn{\nu\left(\xvec,\thetavec,\btau_{n}\right)}

\def\dNidTau{\frac{\partial\nu\left(\xvec,\btau\right)}{\partial\btau}}

\def\dNidTauAtThetaTrue{\left.\frac{\partial\nu\left(\xvec,\btau\right)}{\partial\btau}

\right|_{\thetavecsc=\thetavecsc_{t}}}

\def\LRSymZero{\nu\left(\xvec,\thetavec,\btau_{0}\right)}
\def\LRSymN{\nu\left(\xvec,\thetavec,\btau_{N-1}\right)}

\def\dNidTauPrime{\frac{\partial\nu\left(\xvec,\btau^{\prime}\right)}{\partial\btau^{\prime}}}

\def\dLogFxdThetaAtThetaTrue{\left.\frac{\partial{\log{f\left(\xvec;\thetavecsc\right)}}}{\partial{\thetavecsc}}

\right|_{\thetavecsc=\thetavecsc_{t}}}

\def\LRSymPrime{\nu\left(\xvec,\thetavec,\btau^{\prime}\right)}

\def\LR{\frac{f\left(\xvec;\vartheta\right)}{f\left(\xvec;\theta\right)}}
\def\LRTrue{\frac{f\left(\xvec;\theta\right)}{f\left(\xvec;\theta_{t}\right)}}

\def\LRBig{\frac{f\left(\xvec;\thetavec\right)}{f\left(\xvec;\theta_{t}\right)}}

\def\LRn{\frac{f\left(\xvec;\thetavecsc_{n}\right)}{f\left(\xvec;\theta_{t}\right)}}

\def\LROp{\frac{f\left(\xvec;\thetavecnsc\right)}{f\left(\xvec;\thetavecsc\right)}}

\def\LROpTau{\frac{f\left(\xvec;\thetavecnsc\right)}{f\left(\xvec;\btausc\right)}}

\def\IFIM{\Imat_{\text{FIM}}}

\def\IBFIM{\Imat_{\rm{BFIM}}}

\def\LFunc{f\left(\xvecsc,\thetavecsc\right)}

\def\LRPosSym{\rho\left(\xvec;\thetavec+\btau,\thetavec\right)}

\def\LRPosNegSym{\rho\left(\xvec;\thetavec\pm\btau,\thetavec\right)}
\def\LRXiSym{\rho\left(\xvec;\bxi,\thetavec\right)}

\def\LRNegSym{\rho\left(\xvec;\thetavec-\btau,\thetavec\right)}

\def\LRPosSymWW{\rho^{\beta\left(\btausc\right)}{\left(\xvec;\thetavec+\btau,\thetavec\right)}}

\def\LRPosSymWWn{\rho^{\beta\left(\btausc_{n}\right)}{\left(\xvec;\thetavec+\btau_{n},\thetavec\right)}}

\def\LRPosSymWWHalf{\rho^{\frac{1}{2}}{\left(\xvec;\thetavec+\btau_{n},\thetavec\right)}}

\def\LRNegSymWW{\rho^{\left(1-\beta\left(\btausc\right)\right)}{\left(\xvec;\thetavec-\btau,\thetavec\right)}}

\def\LRPos{\frac{f\left(\xvec,\thetavecsc+\btausc\right)}{f\left(\xvec,\thetavecsc\right)}}

\def\LRPosPres{\frac{f\left(\xvec,\thetavec+\btau\right)}{f\left(\xvec,\thetavec\right)}}

\def\LRPosNegPres{\frac{f\left(\xvec,\thetavec\pm\btau\right)}{f\left(\xvec,\thetavec\right)}}

\def\LRPosNegPresSc{\frac{f\left(\xvecsc,\thetavecsc\pm\btausc\right)}{f\left(\xvecsc,\thetavecsc\right)}}
\def\LRXiPresSc{\frac{f\left(\xvecsc,\xivecsc\right)}{f\left(\xvecsc,\thetavecsc\right)}}

\def\ATau{\Amat\left(\btau\right)}

\def\APrimeTau{\Amat^{\prime}\left(\btau\right)}

\def\ATauPrime{\Amat\left(\btau^{\prime}\right)}

\def\APrimeTauPrime{\Amat^{\prime}\left(\btau^{\prime}\right)}

\def\APrimeTauPrimeH{\Amat^{\prime H}\left(\btau^{\prime}\right)}

\def\ATauDoublePrime{\Amat\left(\btau^{\prime\prime}\right)}

\def\ATauH{\Amat^{H}\left(\btau\right)}

\def\ATauPrimeH{\Amat^{H}\left(\btau^{\prime}\right)}

\def\AAlpha{\Amat\left(\alphavec\right)}

\def\AAlphaH{\Amat^{H}\left(\alphavec\right)}

\def\AAlphaPrime{\Amat\left(\alphavec^{\prime}\right)}

\def\AAlphaPrimeH{\Amat^{H}\left(\alphavec^{\prime}\right)}

\def\DgDthetaAtTheta{\frac{d\gvec\left(\thetavecsc\right)}{d\thetavecsc}}

\def\DgDthetaAtThetaBig{\frac{d\gvec\left(\thetavec\right)}{d\thetavec}}

\def\DgDthetaAtThetaTrue{\left.\frac{d\gvec\left(\thetavecsc\right)}{d\thetavecsc}\right|_{\thetavecsc=\thetavecsc_t}}

\def\DgDTauAtThetaTrue{\left.\frac{\partial\gvec\left(\btausc\right)}{\partial\btausc}\right|_{\btausc=\thetavecsc_t}}

\def\DgHDthetaAtThetaTrue{\left.\frac{\partial\gvec^{T}\left(\thetavec\right)}{\partial\thetavec}\right|_{\thetavec=\thetavec_t}}

\def\GammaMu{\bGamma_{\bxisc,\muvecsc}}

\def\GammaMuH{\bGamma^{H}_{\bxisc,\muvecsc}}

\def\GammaTau{\bGamma_{\hvec}\left(\btau\right)}

\def\GammaTauH{\bGamma_{\hvec}^{H}\left(\btau\right)}

\def\GammaTauPrimeH{\bGamma_{\hvec}^{H}\left(\btau^{\prime}\right)}

\def\GammaTauPrime{\bGamma_{\hvec}\left(\btau^{H}\right)}

\def\GammaTauDoublePrime{\bGamma_{\hvec}\left(\btau^{\prime\prime}\right)}

\def\GammaTauDoublePrimeH{\bGamma_{\hvec}^{H}\left(\btau^{\prime\prime}\right)}

\def\GammaTauTriplePrime{\bGamma_{\hvec}\left(\btau^{\prime\prime\prime}\right)}

\def\GammaTauH{\bGamma_{\hvec}^{H}\left(\btau\right)}

\def\GammaAlpha{\bGamma_{\Hmat}\left(\alphavec\right)}

\def\GammaAlphaPrime{\bGamma_{\Hmat}\left(\alphavec^{\prime}\right)}

\def\GammaAlphaH{\bGamma_{\Hmat}^{H}\left(\alphavec\right)}

\def\GammaAlphaPrimeH{\bGamma_{\Hmat,\psivecTausc}^{H}\left(\alphavec^{\prime}\right)}

\def\KerTauTauPrime{\Kmat_{\hvec}\left(\alphavec,\alphavec^{\prime}\right)}

\def\KerMu{\Kmat_{\muvecsc}}

\def\KerMuInv{\Kmat^{-1}_{\muvecsc}}

\def\KerhkTauTauPrime{\Kmat_{\hvec_{k}}\left(\btau,\btau^{\prime}\right)}

\def\KerTauTauPrimeFourier{\Kmat_{\hvec_{\rm{Fourier}}}\left(\btau,\btau^{\prime}\right)}

\def\KerTauPrimeTauDoublePrime{\Kmat_{\hvec}\left(\btau^{\prime},\btau^{\prime\prime}\right)}

\def\KerTauDoublePrimeTauPrime{\Kmat_{\hvec}\left(\btau^{\prime\prime},\btau^{\prime}\right)}

\def\KerHTauDoublePrimeTauPrime{\Kmat_{\hvec}^{H}\left(\btau^{\prime\prime},\btau^{\prime}\right)}

\def\KeraaTauTauPrime{\Kmat_{\hvec{1,1}}\left(\btau,\btau^{\prime}\right)}

\def\KerabTauTauPrime{\Kmat_{\hvec{1,2}}\left(\btau,\btau^{\prime}\right)}

\def\KerbaTauTauPrime{\Kmat_{\hvec{2,1}}\left(\btau,\btau^{\prime}\right)}

\def\KerbbTauTauPrime{\Kmat_{\hvec{2,2}}\left(\btau,\btau^{\prime}\right)}

\def\KerAlphaAlphaPrime{\Kmat_{\Hmatsc,\psivecTausc}\left(\alphavec,\alphavec^{\prime}\right)}

\def\hTauTheta{\hvec\left(\alphavec,\btau\right)}

\def\hkTauTheta{\hvec_{k}\left(\btau,\thetavec\right)}

\def\hOnekTauTheta{\hvec_{1,k}\left(\btau,\thetavec\right)}

\def\hTwokTauTheta{h_{2,k}\left(\btau,\thetavec\right)}

\def\hTauPrimeThetaPrime{\hvec\left(\btau^{\prime},\thetavec^{\prime}\right)}

\def\hTauDoublePrimeTheta{\hvec\left(\btau^{\prime\prime},\thetavec\right)}

\def\hHTauPrimeThetaPrime{\hvec^{H}\left(\alphavec^{\prime},\btau^{\prime}\right)}

\def\hHTauTheta{\hvec^{H}\left(\btau,\thetavec\right)}

\def\hPrimeTheta{\hvec^{\prime}\left(\thetavec\right)}

\def\hPrimeThetaPrime{\hvec^{\prime}\left(\thetavec^{\prime}\right)}

\def\HAlphaTau{\Hmat\left(\alphavec,\btau\right)}

\def\HHAlphaPrimeTauPrime{\Hmat^{H}\left(\alphavec^{\prime},\btau^{\prime}\right)}

\def\ZTau{\Zmat\left(\btau\right)}

\def\ZHTau{\Zmat^{T}\left(\btau\right)}

\def\ZTauPrime{\Zmat\left(\btau^{\prime}\right)}

\def\ZHTauPrime{\Zmat^{T}\left(\btau^{\prime}\right)}

\def\PsiXThetaTau{\psivec\left(\xvec,\thetavec;\btau\right)}

\def\PsiBBHBXThetaTau{{\psivec}_{\rm{BBH}}^{\left(P^{\prime}\right)}{\left(\xvec,\thetavec;\btau\right)}}

\def\PsiBBHBXThetaTauPone{{\psivec}_{\rm{BBH}}^{\left(1\right)}{\left(\xvec,\thetavec;\btau\right)}}

\def\PsiBBHBTXThetaTau{{\psivec}_{\rm{BBH}}^{{T}\left(P^{\prime}\right)}{\left(\xvec,\thetavec;\btau\right)}}

\def\PsiWWXThetaTau{\psi_{\rm{WW}}{\left(\xvec,\thetavec;\btau\right)}}

\def\PsiWWXThetaTaun{\psi_{\rm{WW}}{\left(\xvec,\thetavec;\btau_{n}\right)}}

\def\PsiCWWCRBXThetaTau{\psivec_{\rm{CCRWW}}{\left(\xvec,\thetavec;\btau\right)}}

\def\PsiRMXThetaTau{\psi_{\rm{RM}}{\left(\xvec,\thetavec;\btau\right)}}

\def\PsiAbelXThetaTau{\psivec_{{\rm{BA}}}^{\left(P^{\prime}\right)}{\left(\xvec,\thetavec;\btau\right)}}

\def\PsiHXThetaTauPrime{\psivec^{T}\left(\xvec,\thetavec;\btau^{\prime}\right)}

\def\PsiHXThetaTau{\psivec^{T}{\left(\xvec,\thetavec;\btau\right)}}

\def\PsiPrimeXThetaTau{\psivec^{{\prime}\left(P^{\prime}\right)}{\left(\xvec,\thetavec;\btau\right)}}

\def\hTauThetaPrime{\hvec\left(\btau,\thetavec^{\prime}\right)}

\def\hTauThetaDoublePrime{\hvec\left(\btau,\thetavec^{\prime\prime}\right)}

\def\hHTauThetaPrime{\hvec^{H}\left(\btau,\thetavec^{\prime}\right)}

\def\hTauPrimeThetaPrime{\hvec\left(\btau^{\prime},\thetavec^{\prime}\right)}

\def\hTauDoublePrimeThetaDoublePrime{\hvec\left(\btau^{\prime\prime},\thetavec^{\prime\prime}\right)}

\def\hHTauTheta{\hvec^{H}\left(\btau,\thetavec\right)}

\def\hPrimeHTheta{\hvec^{\prime H}\left(\thetavec\right)}

\def\hPrimeHThetaPrime{\hvec^{\prime H}\left(\thetavec^{\prime}\right)}

\def\pThetaTau{\pvec\left(\thetavec,\btau\right)}

\def\pThetaTauPrime{\pvec\left(\thetavec,\btau^{\prime}\right)}

\def\pThetaPrimeTau{\pvec\left(\thetavec^{\prime},\btau\right)}

\def\pThetaPrimeTauPrime{\pvec\left(\thetavec^{\prime},\btau^{\prime}\right)}

\def\pThetaDoublePrimeTauDoublePrime{\thetavec^{\prime\prime},\pvec\left(\btau^{\prime\prime}\right)}

\def\pHThetaTau{\pvec^{H}\left(\thetavec,\btau\right)}

\def\pHThetaPrimeTau{\pvec^{H}\left(\thetavec^{\prime},\btau\right)}

\def\pHThetaPrimeTauPrime{\pvec^{H}\left(\thetavec^{\prime},\btau^{\prime}\right)}

\def\pHThetaTriplePrimeTau{\pvec^{H}\left(\thetavec^{\prime\prime\prime},\btau\right)}

\def\GTauTauPrime{\Gmat_{\hvec}\left(\btau,\btau^{\prime}\right)}

\def\GPrimeTau{\Gmat_{\Hmat,\psivecTausc}\left(\btau^{\prime},\btau\right)}

\def\TildeGTauPrimeTau{\tilde{\Gmat}_{\Hmat,\psivecTausc}\left(\btau^{\prime},\btau\right)}

\def\GThetamThetan{\Gmat_{\hvec}\left(\thetavec_{m},\thetavec_{n}\right)}

\def\GAlphaPrimeAlpha{\Gmat_{\Hmatsc,\psivecTausc}\left(\alphavec^{\prime},\alphavec\right)}

\def\GAlphaPrimeAlphaDoublePrime{\Gmat_{\Hmatsc,\psivecTausc}\left(\alphavec^{\prime},\alphavec^{\prime\prime}\right)}

\def\GMatTauPrimeTau{\Gmat\left(\btau^{\prime},\btau\right)}

\def\GMatTauPrimeTauDoublePrime{\Gmat\left(\btau^{\prime},\btau^{\prime\prime}\right)}

\def\GaaTauPrimeTau{\Gmat_{\hvec_{1,1}}\left(\btau^{\prime},\btau^{\prime}\right)}

\def\GabTauPrimeTau{\Gmat_{\hvec_{1,2}}\left(\btau^{\prime},\btau\right)}

\def\GbaTauPrimeTau{\Gmat_{\hvec_{2,1}}\left(\btau^{\prime},\btau\right)}

\def\GbbTauPrimeTau{\Gmat_{\hvec_{2,2}}\left(\btau^{\prime},\btau\right)}

\def\GaaTauPrimeTauDoublePrime{\Gmat_{\hvec_{1,1}}\left(\btau^{\prime},\btau^{\prime\prime}\right)}

\def\GabTauPrimeTauDoublePrime{\Gmat_{\hvec_{1,2}}\left(\btau^{\prime},\btau^{\prime\prime}\right)}

\def\GbaTauPrimeTauDoublePrime{\Gmat_{\hvec_{2,1}}\left(\btau^{\prime},\btau^{\prime\prime}\right)}

\def\GbbTauPrimeTauDoublePrime{\Gmat_{\hvec_{2,2}}\left(\btau^{\prime},\btau^{\prime\prime}\right)}

\def\Ghaa{\Gmat_{\hvec_{1,1}}\left(\cdot,\cdot\right)}

\def\Ghab{\Gmat_{\hvec_{1,2}}\left(\cdot,\cdot\right)}

\def\Ghba{\Gmat_{\hvec_{2,1}}\left(\cdot,\cdot\right)}

\def\Ghbb{\Gmat_{\hvec_{2,2}}\left(\cdot,\cdot\right)}

\def\GaaThetaTrueThetaTrue{\Gmat_{\hvec_{1,1}}\left(\thetavec_{t},\thetavec_{t}\right)}

\def\GabThetaTrueOmegaj{\Gmat_{\hvec_{1,2}}\left(\thetavec_{t},\bOmega_{j}\right)}

\def\GabThetaTrueOmegal{\Gmat_{\hvec_{1,2}}\left(\thetavec_{t},\bOmega_{l}\right)}

\def\GabThetaTrueOmegaOne{\Gmat_{\hvec_{1,2}}\left(\thetavec_{t},\bOmega_{1}\right)}

\def\GabThetaTrueOmegaJ{\Gmat_{\hvec_{2,1}}\left(\bOmega_{J},\thetavec_{t}\right)}

\def\GbbOmegamOmegaj{\Gmat_{\hvec_{2,2}}\left(\bOmega_{j},\bOmega_{k}\right)}

\def\GbbOmegamOmegal{\Gmat_{\hvec_{2,2}}\left(\bOmega_{j^{\prime}},\bOmega_{l}\right)}

\def\GbbOmegaOneOmegaOne{\Gmat_{\hvec_{2,2}}\left(\bOmega_{1},\bOmega_{1}\right)}

\def\GbbOmegaOneOmegaJ{\Gmat_{\hvec_{2,2}}\left(\bOmega_{1},\bOmega_{J}\right)}

\def\GbbOmegaJOmegaOne{\Gmat_{\hvec_{2,2}}\left(\bOmega_{J},\bOmega_{1}\right)}

\def\GbbOmegaJOmegaJ{\Gmat_{\hvec_{2,2}}\left(\bOmega_{J},\bOmega_{J}\right)}

\def\GbaOmegamThetaTrue{\Gmat_{\hvec_{2,1}}\left(\bOmega_{j},\thetavec_{t}\right)}

\def\GbaOmegaOneThetaTrue{\Gmat_{\hvec_{2,1}}\left(\bOmega_{1},\thetavec_{t}\right)}

\def\GbaOmegaJThetaTrue{\Gmat_{\hvec_{2,1}}\left(\bOmega_{J},\thetavec_{t}\right)}

\def\TildeGaa{\left[{\begin{array}{*{20}c}

\Gmat_{\hvec_{1,1}}\left(\alphavec_{1},\alphavec_{1}\right) & \cdots &

\Gmat_{\hvec_{1,1}}\left(\alphavec_{1},\alphavec_{J}\right) \\ \vdots & \ddots & \vdots \\

\Gmat_{\hvec_{1,1}}\left(\alphavec_{J},\alphavec_{1}\right) & \cdots &

\Gmat_{\hvec_{1,1}}\left(\alphavec_{J},\alphavec_{J}\right) \end{array}}\right]}

\def\TildeGab{\left[{\begin{array}{*{20}c}

\Gmat_{\hvec_{1,2}}\left(\alphavec_{1},\betavec_{1}\right) & \cdots &

\Gmat_{\hvec_{1,2}}\left(\alphavec_{1},\betavec_{N}\right) \\ \vdots & \ddots & \vdots \\

\Gmat_{\hvec_{1,2}}\left(\alphavec_{J},\betavec_{1}\right) & \cdots &

\Gmat_{\hvec_{1,2}}\left(\alphavec_{J},\betavec_{N}\right) \end{array}}\right]}

\def\TildeGba{\left[{\begin{array}{*{20}c}

\Gmat_{\hvec_{2,1}}\left(\betavec_{1},\alphavec_{1}\right) & \cdots &

\Gmat_{\hvec_{2,1}}\left(\betavec_{1},\alphavec_{J}\right) \\ \vdots & \ddots & \vdots \\

\Gmat_{\hvec_{2,1}}\left(\betavec_{N},\alphavec_{1}\right) & \cdots &

\Gmat_{\hvec_{2,1}}\left(\betavec_{N},\alphavec_{J}\right) \end{array}}\right]}

\def\TildeGbb{\left[{\begin{array}{*{20}c}

\Gmat_{\hvec_{2,2}}\left(\betavec_{1},\betavec_{1}\right) & \cdots &

\Gmat_{\hvec_{2,2}}\left(\betavec_{1},\betavec_{N}\right) \\ \vdots & \ddots & \vdots \\

\Gmat_{\hvec_{2,2}}\left(\betavec_{N},\betavec_{1}\right) & \cdots &

\Gmat_{\hvec_{2,2}}\left(\betavec_{N},\betavec_{N}\right) \end{array}}\right]}

\def\GTauDoublePrimeTauTriplePrime{\Gmat_{\hvec}\left(\btau^{\prime\prime},\btau^{\prime\prime\prime}\right)}

\def\GTauPrimeTauDoublePrime{\Gmat_{\hvec}\left(\btau^{\prime},\btau^{\prime\prime}\right)}

\def\GTauPrimeTauTriplePrime{\Gmat_{\hvec}\left(\btau^{\prime},\btau^{\prime\prime\prime}\right)}

\def\GTauDoublePrimeTau{\Gmat_{\hvec}\left(\btau^{\prime\prime},\btau\right)}

\def\GThetaTrue{\Gmat_{\hvec}\left(\thetavec_{t},\thetavec_{t}\right)}

\def\GThetaThetaPrime{{G}\left(\thetavec,\thetavec^{\prime}\right)}

\def\GThetaThetaDoublePrime{{G}\left(\thetavec,\thetavec^{\prime\prime}\right)}

\def\TildeGThetaThetaPrime{{\tilde{G}_{\hvec}}\left(\thetavec,\thetavec^{\prime}\right)}

\def\GThetaPrimeTheta{{G}\left(\thetavec^{\prime},\thetavec\right)}

\def\TildeGThetaPrimeTheta{{\tilde{G}_{\hvec}}\left(\thetavec^{\prime},\thetavec\right)}

\def\TildeGThetaPrimeThetaDoublePrime{{\tilde{G}_{\hvec}}\left(\thetavec^{\prime},\thetavec^{\prime\prime}\right)}

\def\GTauPrimeTau{{{\Gmat_{\hvec}}}\left(\btau^{\prime},\btau\right)}

\def\KThetaThetaPrime{{K}\left(\btau,\btau^{\prime}\right)}

\def\GThetaPrimeThetaDoublePrime{{G}\left(\thetavec^{\prime},\thetavec^{\prime\prime}\right)}

\def\GThetaPrimeThetaTriplePrime{{G}\left(\thetavec^{\prime},\thetavec^{\prime\prime\prime}\right)}

\def\GThetaDoublePrimeThetaTriplePrime{{G}\left(\thetavec^{\prime\prime},\thetavec^{\prime\prime\prime}\right)}

\def\GThetaDoublePrimeTheta{{G}\left(\thetavec^{\prime\prime},\thetavec\right)}

\def\KThetaThetaPrime{{K}\left(\btau,\btau^{\prime}\right)}

\def\KThetaDoublePrimeThetaPrime{{K}\left(\thetavec^{\prime\prime},\thetavec^{\prime}\right)}

\def\KThetaTriplePrimeThetaPrime{{K}\left(\thetavec^{\prime\prime\prime},\thetavec^{\prime}\right)}

\def\KThetaPrimeThetaDoublePrime{{K}\left(\thetavec^{\prime},\thetavec^{\prime\prime}\right)}

\def\KTauTauPrime{{\Kmat_{\psivecTausc}}\left(\btau,\btau^{\prime}\right)}

\def\TildeKaa{\left[{\begin{array}{*{20}c}

\Smat\left(\alphavec_{1},\alphavec_{1}\right) & \cdots &

\Smat\left(\alphavec_{1},\alphavec_{J}\right) \\ \vdots & \ddots & \vdots \\

\Smat\left(\alphavec_{J},\alphavec_{1}\right) & \cdots &

\Smat\left(\alphavec_{J},\alphavec_{J}\right) \end{array}}\right]}

\def\TildeKab{\left[{\begin{array}{*{20}c}

\Bmat\left(\alphavec_{1},\betavec_{1}\right) & \cdots &

\Bmat\left(\alphavec_{1},\betavec_{N}\right) \\ \vdots & \ddots & \vdots \\

\Bmat\left(\alphavec_{J},\betavec_{1}\right) & \cdots &

\Bmat\left(\alphavec_{J},\betavec_{N}\right) \end{array}}\right]}

\def\TildeKba{\left[{\begin{array}{*{20}c}

\Bmat^{H}\left(\betavec_{1},\alphavec_{1}\right) & \cdots &

\Bmat^{H}\left(\betavec_{1},\alphavec_{J}\right) \\ \vdots & \ddots & \vdots \\

\Bmat^{H}\left(\betavec_{N},\alphavec_{1}\right) & \cdots &

\Bmat^{H}\left(\betavec_{N},\alphavec_{J}\right) \end{array}}\right]}

\def\TildeKbb{\left[{\begin{array}{*{20}c}

\Fmat\left(\betavec_{1},\betavec_{1}\right) & \cdots &

\Fmat\left(\betavec_{1},\betavec_{N}\right) \\ \vdots & \ddots & \vdots \\

\Fmat\left(\betavec_{N},\betavec_{1}\right) & \cdots &

\Fmat\left(\betavec_{N},\betavec_{N}\right) \end{array}}\right]}

\def\TildeKaaFB{\left[{\begin{array}{*{20}c}

\Smat\left(\thetavec_{t},\thetavec_{t}\right)\end{array}}\right]}

\def\TildeGaaFB{\left[{\begin{array}{*{20}c}

\Gmat_{{\hvec_{1,1}}}\left(\thetavec_{t},\thetavec_{t}\right)\end{array}}\right]}

\def\TildeKabFB{\left[{\begin{array}{*{20}c}

\Bmat\left(\thetavec_{t},\bOmega_{0}\right) & \cdots &

\Bmat\left(\thetavec_{t},\bOmega_{J-1}\right)\end{array}}\right]}

\def\TildeGabFB{\left[{\begin{array}{*{20}c}

\Gmat_{{\hvec_{1,2}}}\left(\thetavec_{t},\bOmega_{0}\right) & \cdots &

\Gmat_{{\hvec_{1,2}}}\left(\thetavec_{t},\bOmega_{J-1}\right)\end{array}}\right]}

\def\TildeKbaFB{\left[{\begin{array}{*{20}c}

\Bmat^{H}\left(\thetavec_{t},\bOmega_{0}\right) \\ \vdots \\

\Bmat^{H}\left(\thetavec_{t},\bOmega_{J-1}\right) \end{array}}\right]}

\def\TildeGbaFB{\left[{\begin{array}{*{20}c}

\Gmat_{{\hvec_{2,1}}}\left(\bOmega_{0},\thetavec_{t}\right) \\ \vdots \\

\Gmat_{{\hvec_{2,1}}}\left(\bOmega_{J-1},\thetavec_{t}\right) \end{array}}\right]}

\def\TildeKbbFB{\left[{\begin{array}{*{20}c}

\Fmat\left(\bOmega_{0},\bOmega_{0}\right) & \cdots &

\Fmat\left(\bOmega_{0},\bOmega_{J-1}\right) \\ \vdots & \ddots & \vdots \\

\Fmat\left(\bOmega_{J-1},\bOmega_{0}\right) & \cdots &

\Fmat\left(\bOmega_{J-1},\bOmega_{J-1}\right) \end{array}}\right]}

\def\TildeGbbFB{\left[{\begin{array}{*{20}c}

\Gmat_{{\hvec_{2,1}}}\left(\bOmega_{0},\bOmega_{0}\right) & \cdots &

\Gmat_{{\hvec_{2,1}}}\left(\bOmega_{0},\bOmega_{J-1}\right) \\ \vdots & \ddots & \vdots \\

\Gmat_{{\hvec_{2,1}}}\left(\bOmega_{J-1},\bOmega_{0}\right) & \cdots &

\Gmat_{{\hvec_{2,1}}}\left(\bOmega_{J-1},\bOmega_{J-1}\right) \end{array}}\right]}

\def\XsiTheta{\bxi\left(\btau\right)}

\def\XsiThetasc{\bxi\left(\btau\right)}

\def\XsiHTheta{\bxi^{T}\left(\btau\right)}

\def\XsiThetaPrime{\bxi\left(\btau ^{\prime}\right)}

\def\XsiPrimeTheta{\bxi^{\prime}\left(\thetavec\right)}

\def\XsiPrimeThetaPrime{\bxi^{\prime}\left(\thetavec^{\prime}\right)}

\def\XsiThetaDoublePrime{\bxi\left(\thetavec^{\prime\prime}\right)}

\def\XsiHTheta{\bxi^{T}\left(\thetavec\right)}

\def\XsiPrimeHTheta{\bxi^{{\prime}T}\left(\thetavec\right)}

\def\XsiHThetaPrime{\bxi^{T}\left(\thetavec^{\prime}\right)}

\def\XsiHThetaDoublePrime{\bxi^{T}\left(\thetavec^{\prime\prime}\right)}

\def\XsiHThetaTriplePrime{\bxi^{T}\left(\thetavec^{\prime\prime\prime}\right)}

\def\XsiTau{\bxi\left(\btau\right)}

\def\XsiHTau{\bxi^{T}\left(\btau\right)}

\def\XsiTauPrime{\bxi\left(\btau^{\prime}\right)}

\def\XsiThetan{{\bxi}\left({\btau}_{n}\right)}

\def\XsiHThetan{{\bxi}^{T}\left({\btau}_{n}\right)}

\def\XsiThetam{{\bxi}\left({\btau}_{m}\right)}

\def\XsiHThetam{{\bxi}^{T}\left({\btau}_{m}\right)}

\def\XsiThetaOne{\bxi\left({\btau}_{0}\right)}

\def\XsiThetaK{\bxi\left({\btau}_{K}\right)}

\def\XsiThetaN{{\bxi}\left({\btau_{N-1}}\right)}

\def\XsiThetaZero{{\bxi}\left({\btau_{0}}\right)}

\def\phiTaun{\bphi\left({\btau}_{n}\right)}

\def\phiTaunscal{\bphi\left({\tau}_{n}\right)}

\def\phiTauOne{\bphi\left({\btau}_{1}\right)}

\def\phiTauK{\bphi\left({\btau}_{K}\right)}

\def\phiTauN{\bphi\left({\btau_N}\right)}

\def\dXsidTauPrimeatThetaTrue{\left.\frac{\partial\XsiTauPrime}{\partial\btau^{\prime}}\right|_{\btausc^{\prime}=\thetavecsc_{t}}}

\def\dXsiHdTauatThetaTrue{\left.\frac{\partial\XsiHTau}{\partial\btau}\right|_{\btau=\thetavec_{t}}}

\def\dXsiDthetaAtThetaTrue{\left.\frac{\partial\xsivec\left(\thetavec\right)}{\partial\thetavec}\right|_{\thetavec=\thetavec_t}}

\def\XsiDotThetaTrue{\dot{\xsivec}\left(\thetavec_{t}\right)}

\def\XsiDotTThetaTrue{\dot{\xsivec}^{T}\left(\thetavec_{t}\right)}

\def\XsiDotThetaTrue{\dot{\gvec}\left(\thetavec_{t}\right)}

\def\XsiDotTThetaTrue{\dot{\gvec}^{T}\left(\thetavec_{t}\right)}

\newcommand{\dThetaOne}{{\dvec\left(\thetavec_{1}\right)}}

\newcommand{\dThetan}{{\dvec\left(\thetavec_{n}\right)}}

\newcommand{\dthetan}{{\dvec\left(\theta_{n}\right)}}

\newcommand{\dThetaN}{{\dvec\left(\thetavec_{N}\right)}}

\newcommand{\dTauOne}{{\dvec\left(\btau_{1}\right)}}

\newcommand{\dTaun}{{\dvec\left(\btau_{n}\right)}}

\newcommand{\dtaun}{{\dvec\left(\tau_{n}\right)}}

\newcommand{\dTauN}{{\dvec\left(\btau_{N}\right)}}

\def\phimTheta{\phi_{m}\left(\thetavec\right)}

\def\phimThetaPrime{\phi_{m}\left(\thetavec^{\prime}\right)}

\def\phinTheta{\phi_{n}\left(\thetavec\right)}

\def\phinThetaPrime{\phi_{n}\left(\thetavec^{\prime}\right)}

\def\phiOmegam{\bphi\left(\bOmega_{m}\right)}

\def\phiOmegaOne{\bphi\left(\bOmega_{1}\right)}

\def\phiOmegaJ{\bphi\left(\bOmega_{J}\right)}

\def\phiHOmegam{\bphi^{T}\left(\bOmega_{m}\right)}

\def\phiHOmegaOne{\bphi^{T}\left(\bOmega_{1}\right)}

\def\phiHOmegaJ{\bphi^{T}\left(\bOmega_{J}\right)}

\def\hCramer{{\left(\frac{\partial\delta\left(\btausc-\thetavecsc\right)}{\partial\btausc}\right)^{T}

\delta\left(\btau-\thetavec_{t}\right)}}

\def\hCramerPrime{{\frac{\partial\delta\left(\btausc^{\prime}-\thetavecsc^{\prime}\right)}{\partial\btausc^{\prime}}

\delta\left(\btau^{\prime}-\thetavec^{\prime}_{0}\right)}}

\def\Th{\mathcal{T}_{\hvec}}

\def\ThH{\mathcal{T}_{\hvec ^{H}}}

\def\Wp{W_{\pvec}}

\def\WpH{W_{\pvec^{T}}}

\def\ThPhiTau{\left(T_{\hvec}\phi_{m}\right)\left(\btau\right)}

\def\ThHPhiTauPrime{\left(T_{\hvec^{T}}\phi_{m}\right)\left(\btau^{\prime}\right)}

\def\WpHPhiTau{\left(W_{\pvec^{T}}\phi_{n}\right)\left(\btau\right)}

\def\WpPhiTauPrime{\left(W_{\pvec}\phi_{n}\right)\left(\btau^{\prime}\right)}

\def\HTauPrimeTau{\Hmat\left(\btau^{\prime},\btau\right)}

\def\HTauPrimeTauDoublePrime{\Hmat\left(\btau^{\prime},\btau^{\prime\prime}\right)}

\newcommand{\Phimat}{{\bf{\Phi}}}

\newcommand{\niXTau}{\vartheta\left(\Xmat,\btau\right)}

\newcommand{\Upsilonmat}{{\bf{\Upsilon}}}

\newcommand{\atheta}{\bvec\left(\omega\right)}

\newcommand{\athetaZero}{\bvec\left(\theta_t\right)}

\newcommand{\athetan}{\bvec\left(\theta_n\right)}

\newcommand{\athetam}{\bvec\left(\theta_{m}\right)}

\newcommand{\aDotthetaZero}{\dot{\bvec}\left(\theta_{t}\right)}

\newcommand{\aDotHthetaZero}{\dot{\bvec}^{H}\left(\theta_{t}\right)}

\newcommand{\avec}{{\bf{a}}}

\newcommand{\bvec}{{\bf{b}}}

\newcommand{\cvec}{{\bf{c}}}

\newcommand{\dvec}{{\bf{d}}}

\newcommand{\evec}{{\bf{e}}}

\newcommand{\fvec}{{\bf{f}}}

\newcommand{\epsvec}{{\bf{\epsilon}}}

\newcommand{\pvec}{{\bf{p}}}

\newcommand{\qvec}{{\bf{q}}}

\newcommand{\Yvec}{{\bf{Y}}}

\newcommand{\yvec}{{\bf{y}}}

\newcommand{\uvec}{{\bf{u}}}

\newcommand{\wvec}{{\bf{w}}}

\newcommand{\xvec}{{\bf{x}}}

\newcommand{\xvecsc}{{\mbox{\boldmath \tiny $\xvec$}}}

\newcommand{\bSnSc}{{\mbox{\tiny $\bvec_{{\mathcal{S}}_{N}}$}}}

\newcommand{\zvec}{{\bf{z}}}

\newcommand{\mvec}{{\bf{m}}}

\newcommand{\nvec}{{\bf{n}}}

\newcommand{\rvec}{{\bf{r}}}

\newcommand{\Svec}{{\bf{S}}}

\newcommand{\Tvec}{{\bf{T}}}

\newcommand{\svec}{{\bf{s}}}

\newcommand{\vvec}{{\bf{v}}}

\newcommand{\gvec}{{\bf{g}}}

\newcommand{\varsigmavec}{{\bf{\varsigma}}}

\newcommand{\WWsc}{{\mbox{\boldmath \tiny {WW}}}}

\newcommand{\gveca}{\gvec_{\alphavec}}

\newcommand{\uveca}{\uvec_{\alphavec}}

\newcommand{\hvec}{{\bf{h}}}

\newcommand{\ivec}{{\bf{i}}}

\newcommand{\kvec}{{\bf{k}}}

\newcommand{\etavec}{{\bf{\eta}}}

\newcommand{\onevec}{{\bf{1}}}

\newcommand{\zerovec}{{\bf{0}}}

\newcommand{\bnu}{{\bf{\nu}}}

\newcommand{\bnusc}{{\mbox{\tiny $\nu$}}}

\newcommand{\xisc}{{\mbox{\tiny $\xi$}}}

\newcommand{\musc}{{\mbox{\tiny $\mu$}}}

\newcommand{\alphavec}{{\bf{\alpha}}}

\newcommand{\Phivec}{{\bf{\Phi}}}

\newcommand{\phivec}{{\bf{\phi}}}

\newcommand{\deltavec}{{\bf{\Delta}}}

\newcommand{\Lambdamat}{{\bf{\Lambda}}}

\newcommand{\invLambdamat}{\Lambdamat^{-1}}

\newcommand{\Gammamat}{{\bf{\Gamma}}}

\newcommand{\Amat}{{\bf{A}}}

\newcommand{\Bmat}{{\bf{B}}}

\newcommand{\Cmat}{{\bf{C}}}

\newcommand{\Dmat}{{\bf{D}}}

\newcommand{\Emat}{{\bf{E}}}

\newcommand{\Fmat}{{\bf{F}}}

\newcommand{\Gmat}{{\bf{G}}}

\newcommand{\Hmat}{{\bf{H}}}

\newcommand{\Hmatsc}{{\mbox{\boldmath \tiny $\Hmat$}}}

\newcommand{\jvec}{{\bf{j}}}

\newcommand{\Jmat}{{\bf{J}}}

\newcommand{\Kmat}{{\bf{K}}}

\newcommand{\Imat}{{\bf{I}}}

\newcommand{\Lmat}{{\bf{L}}}

\newcommand{\Pmat}{{\bf{P}}}

\newcommand{\Pmatperp}{{\bf{P^{\bot}}}}

\newcommand{\Ptmatperp}{{\bf{P_2^{\bot}}}}

\newcommand{\Qmat}{{\bf{Q}}}

\newcommand{\invQmat}{\Qmat^{-1}}

\newcommand{\Smat}{{\bf{S}}}

\newcommand{\Tmat}{{\bf{T}}}

\newcommand{\Tmattilde}{\tilde{\bf{T}}}

\newcommand{\Tmatcheck}{\check{\bf{T}}}

\newcommand{\Tmatbar}{\bar{\bf{T}}}

\newcommand{\Rmat}{{\bf{R}}}

\newcommand{\Umat}{{\bf{U}}}

\newcommand{\Vmat}{{\bf{V}}}

\newcommand{\Wmat}{{\bf{W}}}

\newcommand{\Xmat}{{\bf{X}}}

\newcommand{\Xmatsc}{{\mbox{\boldmath \tiny $\Xmat$}}}

\newcommand{\Ymatsc}{{\mbox{\boldmath \tiny $\Ymat$}}}

\newcommand{\Ymat}{{\bf{Y}}}

\newcommand{\Zmat}{{\bf{Z}}}

\newcommand{\Ry}{\Rmat_{\yvec}}

\newcommand{\Rz}{\Rmat_{\zvec}}

\newcommand{\RyInv}{\Rmat_{\yvec}^{-1}}

\newcommand{\Ryhat}{\hat{\Rmat}_{\yvec}}

\newcommand{\Rs}{\Rmat_{\svec}}

\newcommand{\Rn}{\Rmat_{\nvec}}

\newcommand{\Rninv}{\Rmat_{\nvec}^{-1}}

\newcommand{\Reta}{\Rmat_{\etavec}}

\newcommand{\Ralpha}{\Rmat_{\alphavec}}

\newcommand{\Ck}{\Cmat_{\kvec}}

\newcommand{\Cn}{\Cmat_{\nvec}}

\newcommand{\Cg}{\Cmat_{\gvec}}

\newcommand{\invRn}{\Rmat_{\nvec}^{-1}}

\newcommand{\w}{{\rm{w}}}

\newcommand{\dbsdalpha}{\Tmat_1 \frac{\partial b\svec_1(\alphavec )}{\partial \alphavec}}

\newcommand{\dbsHdalpha}{\frac{\partial b\svec_1^H(\alphavec )}{\partial \alphavec} \Tmat_1^H}

\newcommand{\dbstdalpha}{\Tmat_2 \frac{\partial b\svec_2(\alphavec )}{\partial \alphavec}}

\newcommand{\dbstHdalpha}{\frac{\partial b\svec_2^H(\alphavec )}{\partial \alphavec} \Tmat_2^H}

\newcommand{\dtdw}{\left[ \frac{d\hat{\theta} (\hat{\wvec})}{d \wvec} \right]_{\wvec = \wvec_o}}

\newcommand{\dtdu}{\left[ \frac{d\hat{\theta} (\hat{\uvec})}{d \uvec} \right]_{\uvec = \uvec_o}}

\newcommand{\Jww}{J_{\wvec \wvec}}

\newcommand{\Jbb}{J_{\bvec \bvec}}

\newcommand{\Jaa}{J_{\alphavec \alphavec}}

\newcommand{\Jtb}{J_{\thetavec \bvec}}

\newcommand{\Jbt}{J_{\bvec \thetavec}}

\newcommand{\Jwt}{J_{\wvec \thetavec}}

\newcommand{\Jtw}{J_{\thetavec \wvec}}

\newcommand{\Jtu}{J_{\thetavec \uvec}}

\newcommand{\Jtt}{J_{\thetavec \thetavec}}

\newcommand{\Jee}{J_{\etavec \etavec}}

\newcommand{\Jae}{J_{\alphavec \etavec}}

\newcommand{\Jea}{J_{\etavec \alphavec}}

\newcommand{\fww}{f_{\wvec \wvec}}

\newcommand{\fwt}{f_{\wvec \thetavec}}

\newcommand{\ftw}{f_{\thetavec \wvec}}

\newcommand{\ftt}{f_{\thetavec \thetavec}}

\newcommand{\Juu}{J_{\uvec \uvec}}

\newcommand{\Jub}{J_{\uvec \bvec}}

\newcommand{\Jbu}{J_{\bvec \uvec}}

\newcommand{\Jaatilde}{\tilde{J}_{\alphavec \alphavec}}

\newcommand{\Jqq}{J_{\qvec \qvec}}

\newcommand{\Jaq}{J_{\alphavec \qvec}}

\newcommand{\Jqa}{J_{\qvec \alphavec}}

\newcommand{\JTT}{J_{\Thetavec \Thetavec}}

\newcommand{\Tr}{{\rm Tr}}

\newcommand{\vecc}{{\rm vec}}

\newcommand{\Imm}{{\rm Im}}

\newcommand{\Ree}{{\rm Re}}

\newcommand{\define}{\stackrel{\triangle}{=}}

\newcommand{\Psimat}{\mbox{\boldmath $\Psi$}}

\newcommand{\bzeta}{\mbox{\boldmath $\zeta$}}

\def\bzeta{{\mbox{\boldmath $\zeta$}}}

\def\bgamma{{\mbox{\boldmath $\gamma$}}}

\def\Beta{{\mbox{\boldmath $\eta$}}}

\def\lam{{\mbox{\boldmath $\Gamma$}}}

\def\bomega{{\mbox{\boldmath $\omega$}}}

\def\bOmega{{\mbox{\boldmath $\Omega$}}}

\def\bxi{{\mbox{\boldmath $\xi$}}}

\def\bxisc{{\mbox{\boldmath \tiny $\bxi$}}}

\def\bOmegasc{{\mbox{\boldmath \tiny $\bOmega$}}}

\def\brho{{\mbox{\boldmath $\rho$}}}

\def\brhosc{{\mbox{\boldmath \tiny $\rho$}}}

\def\bmu{{\mbox{\boldmath $\mu$}}}

\def\bvartheta{{\mbox{\boldmath $\vartheta$}}}

\def\bvarthetasc{{\mbox{\boldmath \tiny $\vartheta$}}}

\def\bnu{{\mbox{\boldmath $\nu$}}}

\def\btau{{\mbox{\boldmath $\tau$}}}

\def\bphi{{\mbox{\boldmath $\phi$}}}

\def\bSigma{{\mbox{\boldmath $\Sigma$}}}

\def\bLambda{{\mbox{\boldmath $\Lambda$}}}

\def\Thetavec{{\mbox{\boldmath $\Theta$}}}

\def\bomega{{\mbox{\boldmath $\omega$}}}

\def\brho{{\mbox{\boldmath $\rho$}}}

\def\bmu{{\mbox{\boldmath $\mu$}}}

\def\bGamma{{\mbox{\boldmath $\Gamma$}}}

\def\bnu{{\mbox{\boldmath $\nu$}}}

\def\btau{{\mbox{\boldmath $\tau$}}}

\def\btausc{{\mbox{\boldmath \tiny $\btau$}}}

\def\xivecsc{{\mbox{\boldmath \tiny $\xi$}}}

\def\bomegasc{{\mbox{\boldmath \tiny $\omega$}}}

\def\btausm{{\mbox{\boldmath \tiny $\btau$}}}

\def\bphi{{\mbox{\boldmath $\phi$}}}

\def\blambda{{\mbox{\boldmath $\lambda$}}}

\def\bPhi{{\mbox{\boldmath $\Phi$}}}

\def\bxi{{\mbox{\boldmath $\xi$}}}

\def\bxsisc{{\mbox{\boldmath \tiny $\bxsi$}}}

\def\bvarphi{{\mbox{\boldmath $\varphi$}}}

\def\bvarphisc{{\mbox{\boldmath \tiny $\varphi$}}}

\def\varphisc{{\mbox{\tiny $\varphi$}}}

\def\bvarphism{{\mbox{\boldmath \tiny $\bvarphi$}}}

\def\bepsilon{{\mbox{\boldmath $\epsilon$}}}

\def\balpha{{\mbox{\boldmath $\alpha$}}}

\def\bvarepsilon{{\mbox{\boldmath $\varepsilon$}}}

\def\bXsi{{\mbox{\boldmath $\Xi$}}}

\def\betavec{{\mbox{\boldmath $\beta$}}}

\def\betavecsc{{\mbox{\boldmath \tiny $\beta$}}}

\def\psivec{{\mbox{\boldmath $\psi$}}}

\def\psivecsc{{\mbox{\boldmath \tiny{$\psivec$}}}}

\def\psivecTau{{\psivec}}

\def\psivecTausc{{\mbox{\boldmath \tiny{$\psivecTau$}}}}

\def\xsivec{{\mbox{\boldmath $\xi$}}}

\def\xivecsc{{\mbox{\boldmath \tiny $\xsivec$}}}

\def\Bmatsc{{\mbox{\boldmath \tiny $\Bmat$}}}

\def\bvecsc{{\mbox{\boldmath \tiny $\bvec$}}}

\def\alphavec{{\mbox{\boldmath $\alpha$}}}

\def\alphavecsc{{\mbox{\boldmath \tiny $\alpha$}}}

\def\alphasc{{\mbox{\tiny $\alpha$}}}

\def\gammavec{{\mbox{\boldmath $\gamma$}}}

\def\gammavecsc{{\mbox{\boldmath \tiny{$\gamma$}}}}

\def\svec{{\mbox{\boldmath $s$}}}

\def\tvec{{\mbox{\boldmath $t$}}}

\def\etavec{{\mbox{\boldmath $\eta$}}}

\def\thetavec{{\mbox{\boldmath $\theta$}}}

\def\thetavecsc{{\mbox{\boldmath \tiny {$\theta$}}}}

\def\thetavecnsc{{\mbox{\boldmath \tiny {$\theta_{n}$}}}}

\def\etavecsc{{\mbox{\boldmath \tiny {$\eta$}}}}

\def\etavecsch{{\mbox{\boldmath \tiny {$\eta_{\hvec}$}}}}

\def\KCalsc{{\mbox{\tiny {$\mathcal{K}$}}}}

\def\thetasc{{\mbox{\tiny {$\theta$}}}}

\def\Thetasc{{\mbox{\tiny {$\Theta$}}}}

\def\Lambdasc{{\mbox{\tiny {$\Lambda$}}}}

\def\Vsc{{\mbox{\tiny {$V$}}}}

\def\Ssc{{\mbox{\tiny $S$}}}

\def\Esc{{\mbox{\tiny $E$}}}

\def\XCalsc{{\mbox{\tiny $\mathcal{X}$}}}

\def\XCal{{\mbox{$\mathcal{X}$}}}

\def\YCal{{\mbox{$\mathcal{Y}$}}}

\def\YCalsc{{\mbox{\tiny $\mathcal{Y}$}}}

\def\PCal{{\mbox{$\mathcal{P}$}}}

\def\ACal{{\mbox{$\mathcal{A}$}}}

\def\BCal{{\mbox{$\mathcal{B}$}}}

\def\FCal{{\mbox{$\mathcal{F}$}}}

\def\SCal{{\mbox{$\mathcal{S}$}}}

\def\SCalsc{{\mbox{ \tiny $\mathcal{S}$}}}

\def\ACalsc{{\mbox{\tiny $\mathcal{A}$}}}

\def\BCalsc{{\mbox{\tiny $\mathcal{B}$}}}

\def\FCalsc{{\mbox{\tiny $\mathcal{F}$}}}

\def\muvec{{\mbox{\boldmath $\mu$}}}

\def\muvecsc{{\mbox{\boldmath \tiny{$\mu$}}}}

\def\Ximat{{\mbox{\boldmath $\Xi$}}}

\newcommand{\be}{\begin{equation}}

\newcommand{\ee}{\end{equation}}

\newcommand{\beqna}{\begin{eqnarray}}

\newcommand{\eeqna}{\end{eqnarray}}

\begin{abstract}
In this paper linear canonical correlation analysis (LCCA) is generalized by applying a structured transform to the joint probability distribution of the considered pair of random vectors, i.e., a transformation of the joint probability measure defined on their joint observation space. This framework, called measure transformed canonical correlation analysis (MTCCA), applies LCCA to the data after transformation of the joint probability measure. We show that judicious choice of the transform leads to a modified canonical correlation analysis, which, in contrast to  LCCA, is capable of detecting non-linear relationships between the considered pair of random vectors. Unlike kernel canonical correlation analysis, where the transformation is applied to the random vectors, in MTCCA the transformation is applied to their joint probability distribution. This results in performance advantages and reduced implementation complexity. The proposed approach is illustrated for graphical model selection in simulated data having non-linear dependencies, and for measuring long-term associations between companies traded in the NASDAQ and NYSE stock markets. 
\end{abstract}
\begin{IEEEkeywords}
Association analysis, canonical correlation analysis, graphical model selection, multivariate data analysis, probability measure transform.
\end{IEEEkeywords}
\IEEEpeerreviewmaketitle
\section{Introduction}
Linear canonical correlation analysis (LCCA) \cite{Hotelling} is a technique for multivariate data analysis and dimensionality reduction, which quantifies the linear associations between a pair of random vectors. In particular, LCCA generates a sequence of pairwise unit variance linear combinations of the considered random vectors, such that the Pearson correlation coefficient between the elements of each pair is maximal, and each pair is uncorrelated with its predecessors. The coefficients of these linear combinations, called the linear canonical directions, give insight into the underlying relationships between the random vectors. They are easily obtained by solving a simple generalized eigenvalue decomposition (GEVD) problem, which only involves the covariance and cross-covariance matrices of the considered random vectors. LCCA has been applied to blind source separation \cite{BSSCCA}, image set matching \cite{ImProcCCA}, direction-of-arrival estimation  \cite{DOA_CCA_1}, \cite{DOA_CCA_2}, data fusion and group inference in medical imaging data \cite{CCA_Med_Im}, localization of visual events associated with sound sources \cite{AV_SYNCH_1}, audio-video synchronization \cite{AV_SYNCH_2}, undersea target classification \cite{CCA_Sharf} among others.

The Pearson correlation coefficient is only sensitive to linear associations between random variables. Therefore, in cases where the considered random vectors are statistically dependent yet uncorrelated, LCCA is not an informative tool.

In order to overcome the linear dependence limitation several generalizations of LCCA have been proposed in the literature. In \cite{MICCA} an information-theoretic approach to canonical correlation analysis, called ICCA, was proposed. This method generates a sequence pairwise unit variance linear combinations of the considered random vectors, such that the mutual-information (MI) \cite{InfTheory} between the elements of each pair is maximal, and each pair is uncorrelated with its predecessors. Since the MI is a general measure of statistical dependence, which is sensitive to non-linear relationships, the ICCA \cite{MICCA} is capable of capturing pairs of linear combinations  exhibiting non-linear dependence. However, in contrast to LCCA, the ICCA does not reduce to a simple GEVD problem. Indeed, in \cite{MICCA} each pair of linear combinations must be obtained separately via an iterative Newton-Raphson \cite{ConvOpt} algorithm, which may converge to undesired local maxima. Moreover each step of the Newton-Raphson algorithm involves re-estimation of the MI in a non-parametric manner at a potentially high computational cost. 

Another approach to non-linear generalization of LCCA is kernel canonical correlation analysis (KCCA) \cite{Akaho}-\cite{Bach}. KCCA applies LCCA to high-dimensional non-linear transformations of the considered random vectors that map them into some reproducing kernel Hilbert spaces. Although the KCCA approach can be successful in extracting non-linear relations \cite{Bach}, \cite{Yoshihiro}-\cite{Suetani}, it suffers from the following drawbacks. First, the high-dimensional mappings may have high computational complexity. Second, the method is highly prone to over-fitting errors, and requires regularization of the covariance matrices of the transformed random vectors to increase numerical stability. Finally, the non-linear mappings of the random vectors may mask the dependencies between their original coordinates.

In this paper we propose a different non-linear generalization of LCCA called measure transformed canonical correlation analysis (MTCCA). We apply a structured transform to the joint probability distribution of the considered pair of random vectors, i.e., a transformation of the joint probability measure defined on their joint observation space. The proposed transform is structured by a pair of non-negative functions called the MT-functions. It preserves statistical independence and maps the joint probability distribution into a set of probability measures on the joint observation space. By modifying the MT-functions classes of measure transformations can be obtained that have different properties. Two types of MT-functions, the exponential and the Gaussian,  are developed in this paper. The former has a translation invariance property while the latter has a localization property. 

MTCCA applies LCCA to the considered pair of random vectors under the proposed probability measure transform. By modifying the MT-functions the correlation coefficient under the transformed probability measure, called the MT-correlation coefficient, is modified, resulting in a new general framework for canonical correlation analysis. In MTCCA, the MT-correlation coefficients between the elements of each generated pair of linear combinations are called the MT-canonical correlation coefficients.

The MT-functions are selected from exponential and Gaussian families of functions parameterized by scale and location parameters. Under these function classes it is shown that pairs of linear combinations with non-linear dependence can be detected by MTCCA. The parameters of the MT-functions are selected via maximization of a lower bound on the largest MT-canonical correlation coefficient. We show that, for these selected parameters, the corresponding largest MT-canonical correlation coefficient constitutes a measure for statistical independence under the original probability distribution. In this case it is also shown that the considered random vectors are statistically independent under both transformed and original probability distributions if and only if they are uncorrelated under the transformed probability distribution. 

In the paper an empirical implementation of MTCCA is proposed that uses strongly consistent estimators of the measure transformed covariance and cross-covariance matrices of the considered random vectors. 

The MTCCA approach has the following advantages over LCCA, ICCA, and the KCCA discussed above:
\begin{inparaenum}
\item
In contrast to LCCA, MTCCA is capable of detecting non-linear dependencies. Moreover, under appropriate selection of the MT-functions, the largest MT-canonical correlation coefficient is a measure of statistical independence between the considered random vectors. 
\item
In comparison to the ICCA, MTCCA is easier to implement from the following reasons. First, it reduces to a simple GEVD problem, which only involves the measure transformed covariance and cross-covariance matrices of the considered random vectors. Second, while MTCCA with exponential and Gaussian MT-functions involves a {\it{single}} maximization for choosing the MT-functions parameters, the ICCA involves a {\it{sequence}} of maximization problems, each having the same dimensionality as in MTCCA.
\item
In the paper we show that unlike the empirical ICCA and KCCA, the computational complexity of the empirical MTCCA is linear in the sample size which makes it favorable in large sample size scenarios.
\item
Unlike KCCA, MTCCA does not expand the dimensions of the random vectors, nor does it require regularization of their measure transformed covariance matrices. 
\item
Finally, unlike KCCA, in MTCCA the original coordinates of the observation vectors are retained after the probability measure transform. 
Therefore, MTCCA can be easily applied to variable selection \cite{StatDict} by discarding a subset of the variables for which the corresponding entries of the measure transformed canonical directions are practically zero.  
\end{inparaenum}

The proposed approach is illustrated for two applications. The first is a simulation of graphical models with known dependency structure. In this simulated example we show 
that in similar to ICCA, the MTCCA outperforms the LCCA in selecting valid linear/non-linear graphical model topology. The second application is construction of networks that analyze long-term associations between companies traded in the NASDAQ and NYSE stock markets. We show that MTCCA and KCCA better associate companies in the same sector (technology, pharmaceutical, financial) than does LCCA and ICCA. Furthermore, MTCCA is able to achieve this by finding strong non-linear dependencies between the daily log-returns of these companies.

The paper is organized as follows. In Section \ref{CCA_REVIEW}, LCCA is reviewed. In Section \ref{PMT_CCA}, LCCA is generalized by applying a transform to the joint probability distribution. Selection of the MT-functions associated with the transform is discussed in Section \ref{ChoiceOfKernels}. In Section \ref{Estimation}, empirical implementation of MTCCA is obtained. In Section \ref{Examp}, the proposed approach is illustrated via simulation experiment. In Section \ref{Disc}, the main points of this contribution are summarized. The propositions and theorems stated throughout the paper are proved in the Appendix.
\section{Linear canonical correlation analysis: Review}
\label{CCA_REVIEW}
\subsection{Preliminaries}
\label{Perlim}
Let $\Xmat$ and $\Ymat$ denote two random vectors, whose observation spaces are given by $\XCal\subseteq\Rsp^{p}$ and $\YCal\subseteq\Rsp^{q}$, respectively. We define the measure space $\left(\XCal\times\YCal,\mathcal{S}_{\XCalsc\times\YCalsc},\pxy\right)$, where $\mathcal{S}_{\XCalsc\times\YCalsc}$ is a $\sigma$-algebra over $\XCal\times\YCal$, and $\pxy$ is the joint probability measure on $\mathcal{S}_{\XCalsc\times\YCalsc}$. The marginal probability measures of $\pxy$ on $\mathcal{S}_{\XCalsc}$ and $\mathcal{S}_{\YCalsc}$ are denoted by $\px$ and $\py$, where $\mathcal{S}_{\XCalsc}$ and $\mathcal{S}_{\YCalsc}$ are the $\sigma$-algebras over $\XCal$ and $\YCal$, respectively. Let $g\left(\cdot,\cdot\right)$ denote an integrable scalar function on $\XCal\times\YCal$. The expectation of $g\left(\Xmat,\Ymat\right)$ under $\pxy$ is defined as
\begin{equation}
\label{ExpDef}
{\rm{E}}\left[g\left(\Xmat,\Ymat\right);\pxy\right]\triangleq\int\limits_{\XCalsc\times\YCalsc}g\left(\xvec,\yvec\right)d\pxy\left(\xvec,\yvec\right),
\end{equation}
where $\xvec\in\XCal$ and $\yvec\in\YCal$. The random vectors $\Xmat$ and $\Ymat$ will be said to be statistically independent under $\pxy$ if 
\begin{equation}
\label{IndDef}
{\rm{E}}\left[g_{1}\left(\Xmat\right)g_{2}\left(\Ymat\right);\pxy\right]={\rm{E}}\left[g_{1}\left(\Xmat\right);\px\right]
{\rm{E}}\left[g_{2}\left(\Ymat\right);\py\right]
\end{equation}
for all integrable scalar functions $g_{1}\left(\cdot\right)$, $g_{2}\left(\cdot\right)$ on $\XCal$ and $\YCal$, respectively. The random vectors $\Xmat$ and $\Ymat$ will be said to be uncorrelated under $\pxy$ if
\begin{equation}
\label{IndDef2}
{\rm{E}}\left[\Xmat\Ymat^{T};\pxy\right]={\rm{E}}\left[\Xmat;\px\right]{\rm{E}}\left[\Ymat^{T};\py\right],
\end{equation}
where $\left(\cdot\right)^{T}$ denotes the transpose operator.
\subsection{The LCCA procedure}
\label{ClassCCA}
LCCA generates a sequence of pairwise unit-variance linear combinations $\left(\avec^{T}_{k}\Xmat,\bvec^{T}_{k}\Ymat\right)$, $k=1,\ldots,r=\min\left(p,q\right)$ in the following manner. The first pair $\left(\avec^{T}_{1}\Xmat,\bvec^{T}_{1}\Ymat\right)$ is determined by maximizing the Pearson correlation coefficient between $\avec^{T}\Xmat$ and $\bvec^{T}\Ymat$ over $\avec\in\Rsp^{p}$ and $\bvec\in\Rsp^{q}$ with the constraint that both $\avec^{T}\Xmat$ and $\bvec^{T}\Ymat$ have unit variance. Similarly, the $k$-th pair $\left(\avec^{T}_{k}\Xmat,\bvec^{T}_{k}\Ymat\right)$ $(1<k\leq{r})$ is determined by maximizing the Pearson correlation coefficient between $\avec^{T}\Xmat$ and $\bvec^{T}\Ymat$ over $\avec\in\Rsp^{p}$ and $\bvec\in\Rsp^{q}$ with the constraints that both $\avec^{T}\Xmat$ and $\bvec^{T}\Ymat$ have unit variance and $\left(\avec^{T}\Xmat,\bvec^{T}\Ymat\right)$ are uncorrelated with all the previously obtained pairs $\left(\avec^{T}_{l}\Xmat,\bvec^{T}_{l}\Ymat\right)$, $l=1,\ldots,k-1$. The pairs $\left(\avec_{k},\bvec_{k}\right)$ and $\left(\avec^{T}_{k}\Xmat,\bvec^{T}_{k}\Ymat\right)$ are called the $k$-th order {\it{linear canonical directions}} and the $k$-th order {\it{linear canonical variates}}, respectively. The Pearson correlation coefficient between $\avec^{T}_{k}\Xmat$ and  $\bvec^{T}_{k}\Ymat$ is called the $k$-th order {\it{linear canonical correlation coefficient}}.

The Pearson correlation coefficient between $\avec^{T}\Xmat$ and $\bvec^{T}\Ymat$ under $\pxy$ is given by 
\begin{eqnarray}
\label{RhoDef}
{\rm{Corr}}\left[\avec^{T}\Xmat,\bvec^{T}\Ymat;{\pxy}\right]\triangleq\frac{{\rm{Cov}}\left[\avec^{T}\Xmat,\bvec^{T}\Ymat;{\pxy}\right]}{\sqrt{{\rm{Var}}\left[\avec^{T}\Xmat;{\px}\right]}\sqrt{{\rm{Var}}\left[\bvec^{T}\Ymat;{\py}\right]}}=\frac{\avec^{T}\bSigma_{\Xmatsc\Ymatsc}\bvec}{\sqrt{\avec^{T}\bSigma_{\Xmatsc}\avec}\sqrt{\bvec^{T}\bSigma_{\Ymatsc}\bvec}},
\end{eqnarray}
where ${\rm{Var}}\left[\cdot;\px\right]$ and ${\rm{Cov}}\left[\cdot,\cdot;\pxy\right]$ denote the variance and covariance under $\px$ and $\pxy$, respectively. The last equality in (\ref{RhoDef}) can be easily verified using the basic definitions of variance and covariance, where $\bSigma_{\Xmatsc}\in\Rsp^{p\times{p}}$, $\bSigma_{\Ymatsc}\in\Rsp^{q\times{q}}$ and $\bSigma_{\Xmatsc\Ymatsc}\in\Rsp^{p\times{q}}$ denote the covariance matrix of $\Xmat$ under $\px$, the covariance matrix of $\Ymat$ under $\py$, and their cross-covariance matrix under $\pxy$, respectively, and it is assumed that $\bSigma_{\Xmatsc}$ and $\bSigma_{\Ymatsc}$ are non-singular. 

Hence, LCCA solves the following constraint maximization sequentially over $k=1,\ldots,{r}$.
\begin{eqnarray}
\label{CCAOptProb}
&&\rho_{k}\left(\bSigma_{\Xmatsc},\bSigma_{\Ymatsc},\bSigma_{\Xmatsc\Ymatsc}\right)=\max\limits_{\avec,\bvec}{\avec^{T}\bSigma_{\Xmatsc\Ymatsc}\bvec},\\\nonumber
&&{\rm{s.t.}}\hspace{0.4cm}\avec^{T}\bSigma_{\Xmatsc}\avec=\bvec^{T}\bSigma_{\Ymatsc}\bvec=1,\\\nonumber
&&
{\rm{and}}\hspace{0.4cm}\avec^{T}\bSigma_{\Xmatsc\Ymatsc}\bvec_{l}=
\bvec^{T}\bSigma^{T}_{\Xmatsc\Ymatsc}\avec_{l}=
\avec^{T}\bSigma_{\Xmatsc}\avec_{l}=\bvec^{T}\bSigma_{\Ymatsc}\bvec_{l}=0\hspace{0.4cm}
\forall\hspace{0.1cm}{1\leq{l}<k},
\end{eqnarray}
where $\rho_{k}\left(\bSigma_{\Xmatsc},\bSigma_{\Ymatsc},\bSigma_{\Xmatsc\Ymatsc}\right)$ denotes the $k$-th order linear canonical correlation coefficient. Since the number of constraints in (\ref{CCAOptProb}) increases with $k$, it is implied that the linear canonical correlation coefficients satisfy the following order relation
$1\geq\rho_{1}\left(\bSigma_{\Xmatsc},\bSigma_{\Ymatsc},\bSigma_{\Xmatsc\Ymatsc}\right)\geq\ldots\geq\rho_{r}\left(\bSigma_{\Xmatsc},\bSigma_{\Ymatsc},\bSigma_{\Xmatsc\Ymatsc}\right)\geq0$.

It is well known that the constrained maximization problem in (\ref{CCAOptProb}) reduces to the set of $r$ distinct solutions of the following generalized eigenvalue problem \cite{Anderson}
\begin{equation}
\label{GEVD}
\left[\begin{array}{cc}{\zerovec} & {\bSigma_{\Xmatsc\Ymatsc}} \\ {\bSigma^{T}_{\Xmatsc\Ymatsc}} & {\zerovec}\end{array}\right]\left[\begin{array}{c}{\avec} \\{\bvec}\end{array}\right]=\rho\left[\begin{array}{cc}{\bSigma_{\Xmatsc}} & {\zerovec} \\ {\zerovec} & {\bSigma_{\Ymatsc}}\end{array}\right]\left[\begin{array}{c}{\avec} \\{\bvec}\end{array}\right],
\end{equation}
where $\rho=\rho_{k}\left(\bSigma_{\Xmatsc},\bSigma_{\Ymatsc},\bSigma_{\Xmatsc\Ymatsc}\right)$ is the $k$-th largest generalized eigenvalue of the pencil in (\ref{GEVD}), and $\left[\avec^{T},\bvec^{T}\right]^{T}=\left[\avec^{T}_{k},\bvec^{T}_{k}\right]^{T}$ is its corresponding generalized eigenvector.
\section{Measure transformed canonical correlation analysis}
\label{PMT_CCA}
In this section LCCA is generalized by applying a transform to the joint probability measure $\pxy$. First, a transform which maps $\pxy$ into a set of joint probability measures $\left\{\qxy\right\}$ on $\mathcal{S}_{\XCalsc\times\YCalsc}$ is derived that have the property that they preserve statistical independence of $\Xmat$ and $\Ymat$ under $\pxy$. The MTCCA method is obtained by applying LCCA to $\Xmat$ and $\Ymat$ under the transformed probability measure $\qxy$. 
\subsection{Transformation of the joint probability measure $\pxy$}
\begin{Definition}
\label{Def1}
Given two non-negative functions $u:\Rsp^{p}\rightarrow\Rsp$ and $v:\Rsp^{q}\rightarrow\Rsp$ satisfying 
\begin{equation}
\label{Assumption2}  
0<{{{\rm{E}}}\left[u\left(\Xmat\right)v\left(\Ymat\right);\pxy\right]}<\infty,
\end{equation}
a transform on the joint probability measure $\pxy$ is defined via the following relation
\begin{equation}
\label{MeasureTransform} 
\qxy\left(A\right)\triangleq{\rm{T}}_{u,v}\left[\pxy\right]\left(A\right)=\int\limits_{A}\varphi_{u,v}\left(\xvec,\yvec\right)d\pxy\left(\xvec,\yvec\right),
\end{equation}
where $A\in\mathcal{S}_{\XCalsc\times\YCalsc}$, $\xvec\in\XCal$, $\yvec\in\YCal$, and
\begin{equation}
\label{VarPhiDef} 
\varphi_{u,v}\left(\xvec,\yvec\right)\triangleq\frac{u\left(\xvec\right)v\left(\yvec\right)}{{{\rm{E}}}\left[u\left(\Xmat\right)v\left(\Ymat\right);\pxy\right]}.
\end{equation}
The functions $u\left(\cdot\right)$ and $v\left(\cdot\right)$, associated with the transform ${\rm{T}}_{u,v}\left[\cdot\right]$, are called the MT-functions.
\end{Definition}
In the following Proposition, some properties of the measure transform (\ref{MeasureTransform}) are given.   
\begin{Proposition}
\label{Prop1}
Let $\qxy$ be defined by relation (\ref{MeasureTransform}). 
Then
\begin{enumerate}
\item
\label{P1}
$\qxy$ is a probability measure on $\mathcal{S}_{\XCalsc\times\YCalsc}$.
\item
\label{P2}
$\qxy$ is absolutely continuous w.r.t. $\pxy$, with Radon-Nikodym derivative \cite{Folland} given by
\begin{equation}
\label{MeasureTransformRadNik} 
\frac{d\qxy\left(\xvec,\yvec\right)}{d\pxy\left(\xvec,\yvec\right)}=\varphi_{u,v}\left(\xvec,\yvec\right).
\end{equation}
\item
\label{P3} 
If $\Xmat$ and $\Ymat$ are statistically independent under $\pxy$, then they are statistically independent under $\qxy$.
\item
\label{P4} 
Assume that the MT-functions $u\left(\cdot\right)$ and $v\left(\cdot\right)$ are strictly positive. If $\Xmat$ and $\Ymat$ are statistically independent under $\qxy$, then they are statistically independent under $\pxy$.
\end{enumerate} 
[A proof is given in Appendix \ref{Prop1Proof}]
\end{Proposition}
By modifying the MT-functions $u\left(\cdot\right)$ and $v\left(\cdot\right)$, such that the conditions in Definition \ref{Def1} are satisfied, an infinite set of joint probability measures on $\mathcal{S}_{\XCalsc\times\YCalsc}$ can be obtained. 
\subsection{The MTCCA procedure}
\label{GeneralizedCCA}
MTCCA generates a sequence of pairwise linear combinations $\left(\avec^{T}_{k}\Xmat,\bvec^{T}_{k}\Ymat\right)$, $k=1,\ldots,r=\min\left(p,q\right)$ that have the following properties under the transformed probability measure $\qxy$: $\avec^{T}_{k}\Xmat$ and $\bvec_{k}^{T}\Ymat$ have unit variance, the correlation coefficient between $\avec_{k}^{T}\Xmat$ and $\bvec_{k}^{T}\Ymat$ is maximal, and $\left(\avec^{T}_{k}\Xmat,\bvec^{T}_{k}\Ymat\right)$ are uncorrelated with $\left(\avec^{T}_{l}\Xmat,\bvec^{T}_{l}\Ymat\right)$ for all $1\leq{l}<k$. In MTCCA, the pairs $\left(\avec_{k},\bvec_{k}\right)$ and $\left(\avec^{T}_{k}\Xmat,\bvec^{T}_{k}\Ymat\right)$ are called the $k$-th order {\it{MT-canonical directions}} and the $k$-th order {\it{MT-canonical variates}}, respectively. The correlation coefficient between $\avec_{k}^{T}\Xmat$ and $\bvec_{k}^{T}\Ymat$ under $\qxy$ is called the $k$-th order {\it{MT-canonical correlation coefficient}}.

The correlation coefficient between $\avec^{T}\Xmat$ and $\bvec^{T}\Ymat$ under $\qxy$ is given by
\begin{equation}
\label{MTRhoDef}
{{\rm{Corr}}}\left[\avec^{T}\Xmat,\bvec^{T}\Ymat;{\qxy}\right]\triangleq\frac{{\rm{Cov}}\left[\avec^{T}\Xmat,\bvec^{T}\Ymat;{\qxy}\right]}{\sqrt{{\rm{Var}}\left[\avec^{T}\Xmat;{\qx}\right]}\sqrt{{\rm{Var}}\left[\bvec^{T}\Ymat;{\qy}\right]}}=\frac{\avec^{T}\bSigma^{\left(u,v\right)}_{\Xmatsc\Ymatsc}\bvec}{\sqrt{\avec^{T}
\bSigma^{\left(u,v\right)}_{\Xmatsc}\avec}\sqrt{\bvec^{T}\bSigma^{\left(u,v\right)}_{\Ymatsc}\bvec}},
\end{equation}
where ${{\rm{Corr}}}\left[\cdot,\cdot;{\qxy}\right]$ is called the {\it{MT-correlation coefficient}}, and the measures $\qx$ and $\qy$ are the marginal probability measures of $\qxy$ on $\mathcal{S}_{\XCalsc}$ and $\mathcal{S}_{\YCalsc}$, respectively. The matrices $\bSigma^{\left(u,v\right)}_{\Xmatsc}$, $\bSigma^{\left(u,v\right)}_{\Ymatsc}$ and $\bSigma^{\left(u,v\right)}_{\Xmatsc\Ymatsc}$ denote the covariance matrix of $\Xmat$ under $\qx$, the covariance matrix of $\Ymat$ under $\qy$, and their cross-covariance matrix under $\qxy$, respectively, where it is assumed that $\bSigma^{\left(u,v\right)}_{\Xmatsc}$ and $\bSigma^{\left(u,v\right)}_{\Ymatsc}$ are non-singular.

Using (\ref{ExpDef}) and (\ref{MeasureTransformRadNik}) it can be shown that ${\rm{E}}\left[\Gmat\left(\Xmat,\Ymat\right);\qxy\right]={\rm{E}}\left[\Gmat\left(\Xmat,\Ymat\right)\varphi_{u,v}\left(\Xmat,\Ymat\right);\pxy\right]$, where $\Gmat\left(\Xmat,\Ymat\right)$ is some arbitrary matrix function of $\Xmat$ and $\Ymat$. Therefore, one can easily verify that 
\begin{equation} 
\label{RxMod_pxy}
\bSigma^{\left(u,v\right)}_{\Xmatsc}={\rm{E}}\left[\Xmat\Xmat^{T}\varphi_{u,v}\left(\Xmat,\Ymat\right);\pxy\right]-{\rm{E}}\left[\Xmat\varphi_{u,v}\left(\Xmat,\Ymat\right);\pxy\right]{\rm{E}}\left[\Xmat^{T}\varphi_{u,v}\left(\Xmat,\Ymat\right);\pxy\right],
\end{equation}
\begin{equation}
\label{RyMod_pxy}
\bSigma^{\left(u,v\right)}_{\Ymatsc}={\rm{E}}\left[\Ymat\Ymat^{T}\varphi_{u,v}\left(\Xmat,\Ymat\right);\pxy\right]-{\rm{E}}\left[\Ymat\varphi_{u,v}\left(\Xmat,\Ymat\right);\pxy\right]{\rm{E}}\left[\Ymat^{T}\varphi_{u,v}\left(\Xmat,\Ymat\right);\pxy\right],
\end{equation}
and
\begin{equation}
\label{RxyMod_pxy}  
\bSigma^{\left(u,v\right)}_{\Xmatsc\Ymatsc}={\rm{E}}\left[\Xmat\Ymat^{T}\varphi_{u,v}\left(\Xmat,\Ymat\right);\pxy\right]-{\rm{E}}\left[\Xmat\varphi_{u,v}\left(\Xmat,\Ymat\right);\pxy\right]{\rm{E}}\left[\Ymat^{T}\varphi_{u,v}\left(\Xmat,\Ymat\right);\pxy\right].
\end{equation}
Equations (\ref{RxMod_pxy})-(\ref{RxyMod_pxy}) imply that $\bSigma^{\left(u,v\right)}_{\Xmatsc}$, $\bSigma^{\left(u,v\right)}_{\Ymatsc}$ and $\bSigma^{\left(u,v\right)}_{\Xmatsc\Ymatsc}$ are weighted covariance and cross-covariance matrices of $\Xmat$ and $\Ymat$ under $\pxy$, with weighting function $\varphi_{u,v}\left(\cdot,\cdot\right)$.

MTCCA solves the following constrained maximization sequentially over $k=1,\ldots,{r}$.
\begin{eqnarray}
\label{CCAOptProbMod}
&&\rho_{k}\left(\bSigma^{\left(u,v\right)}_{\Xmatsc},\bSigma^{\left(u,v\right)}_{\Ymatsc},\bSigma^{\left(u,v\right)}_{\Xmatsc\Ymatsc}\right)=\max\limits_{\avec,\bvec}{\avec^{T}\bSigma^{\left(u,v\right)}_{\Xmatsc\Ymatsc}\bvec},\\\nonumber
&&{\rm{s.t.}}\hspace{0.4cm}\avec^{T}\bSigma^{\left(u,v\right)}_{\Xmatsc}\avec=\bvec^{T}\bSigma^{\left(u,v\right)}_{\Ymatsc}\bvec=1,
\\\nonumber
&&{\rm{and}}\hspace{0.4cm}
\avec^{T}\bSigma^{\left(u,v\right)}_{\Xmatsc\Ymatsc}\bvec_{l}=
\bvec^{T}\bSigma^{\left(u,v\right)T}_{\Xmatsc\Ymatsc}\avec_{l}=
\avec^{T}\bSigma^{\left(u,v\right)}_{\Xmatsc}\avec_{l}=
\bvec^{T}\bSigma^{\left(u,v\right)}_{\Ymatsc}\bvec_{l}=0\hspace{0.4cm}
\forall\hspace{0.1cm}{1\leq{l}<k},
\end{eqnarray}
where $\rho_{k}\left(\bSigma^{\left(u,v\right)}_{\Xmatsc},\bSigma^{\left(u,v\right)}_{\Ymatsc},\bSigma^{\left(u,v\right)}_{\Xmatsc\Ymatsc}\right)$ denotes the $k$-th order MT-canonical correlation coefficient. 
Since the number of constraints in (\ref{CCAOptProbMod}) increases with $k$, the MT-canonical correlation coefficients satisfy the following order relation
$1\geq\rho_{1}\left(\bSigma^{\left(u,v\right)}_{\Xmatsc},\bSigma^{\left(u,v\right)}_{\Ymatsc},\bSigma^{\left(u,v\right)}_{\Xmatsc\Ymatsc}\right)\geq\ldots\geq\rho_{r}\left(\bSigma^{\left(u,v\right)}_{\Xmatsc},\bSigma^{\left(u,v\right)}_{\Ymatsc},\bSigma^{\left(u,v\right)}_{\Xmatsc\Ymatsc}\right)\geq0.$

Similarly to (\ref{CCAOptProb}) the constrained maximization problem in (\ref{CCAOptProbMod}) reduces to the following generalized eigenvalue problem
\begin{equation}
\label{GEVD_PMTCCA} 
\left[\begin{array}{cc}{\zerovec} & {\bSigma^{\left(u,v\right)}_{\Xmatsc\Ymatsc}} \\ {\bSigma^{\left(u,v\right)T}_{\Xmatsc\Ymatsc}} & {\zerovec}\end{array}\right]\left[\begin{array}{c}{\avec} \\{\bvec}\end{array}\right]=\rho\left[\begin{array}{cc}{\bSigma^{\left(u,v\right)}_{\Xmatsc}} & {\zerovec} \\ {\zerovec} & {\bSigma^{\left(u,v\right)}_{\Ymatsc}}\end{array}\right]\left[\begin{array}{c}{\avec} \\{\bvec}\end{array}\right],
\end{equation}
where $\rho=\rho_{k}\left(\bSigma_{\Xmatsc},\bSigma_{\Ymatsc},\bSigma_{\Xmatsc\Ymatsc}\right)$ is the $k$-th largest generalized eigenvalue of the pencil in (\ref{GEVD_PMTCCA}), and $\left[\avec^{T},\bvec^{T}\right]^{T}=\left[\avec^{T}_{k},\bvec^{T}_{k}\right]^{T}$ is its corresponding generalized eigenvector.

By modifying the MT-functions $u\left(\cdot\right)$ and $v\left(\cdot\right)$, such that the condition in (\ref{Assumption2}) is satisfied,  the MT-correlation coefficient under $\qxy$ is modified, resulting in a family of canonical correlation analyses, generalizing LCCA described in Subsection \ref{ClassCCA}. In particular, by choosing $u\left(\xvec\right)\equiv{1}$ and $v\left(\yvec\right)\equiv{1}$, then $\qxy=\pxy$, ${{\rm{Corr}}}\left[\avec^{T}\Xmat,\bvec^{T}\Ymat;{\qxy}\right]={{\rm{Corr}}}\left[\avec^{T}\Xmat,\bvec^{T}\Ymat;{\pxy}\right]$, and the LCCA is obtained. Other choices of $u\left(\cdot\right)$ and $v\left(\cdot\right)$ are discussed below.
\section{Selection of the MT-functions}
\label{ChoiceOfKernels}
In this section we parameterize the MT-functions $u\left(\xvec;\svec\right)$ and $v\left(\yvec;\tvec\right)$ with parameters $\svec\in\Rsp^{p}$ and $\tvec\in\Rsp^{q}$ under the exponential and Gaussian families of functions. This will result in the corresponding cross-covariance matrix $\bSigma^{\left(u,v\right)}_{\Xmatsc\Ymatsc}\left(\tvec,\svec\right)$ gaining sensitivity to non-linear relationships between the entries of $\Xmat$ and $\Ymat$. Optimal choice of the parameters $\svec$ and $\tvec$ is also discussed.
\subsection{Exponential MT-functions}
\label{ExpMT}
Let $u\left(\cdot;\cdot\right)$ and $v\left(\cdot;\cdot\right)$ be defined as the parameterized functions
\begin{eqnarray}
\label{ExpKernel}   
\uexp\left(\xvec;\svec\right)\triangleq\exp\left(\svec^{T}\xvec\right) &{\rm{and}}& \vexp\left(\yvec;\tvec\right)\triangleq\exp\left(\tvec^{T}\yvec\right),
\end{eqnarray}
where $\svec\in\Rsp^{p}$ and $\tvec\in\Rsp^{q}$. Using (\ref{VarPhiDef}), (\ref{RxyMod_pxy}) and (\ref{ExpKernel}) one can easily verify that the cross-covariance matrix of $\Xmat$ and $\Ymat$ under $Q^{\left(\uexp,\vexp\right)}_{\Xmatsc\Ymatsc}$ takes the form 
\begin{equation}
\label{OffHessxy}
\bSigma^{\left(\uexp,\vexp\right)}_{\Xmatsc\Ymatsc}\left(\svec,\tvec\right)
=\frac{\partial^{2}\log{M_{\Xmatsc\Ymatsc}}\left(\svec,\tvec\right)}{\partial\svec\partial\tvec^{T}},
\end{equation}
where
\begin{equation}
\label{MonGenFunc} 
{M}_{\Xmatsc\Ymatsc}\left(\svec,\tvec\right)\triangleq{\rm{E}}\left[\exp\left(\svec^{T}\Xmat+\tvec^{T}\Ymat\right);\pxy\right]
\end{equation}
is the joint moment generating function of $\Xmat$ and $\Ymat$, and it is assumed that ${M}_{\Xmatsc\Ymatsc}\left(\svec,\tvec\right)$ is finite in some open region in $\Rsp^{p}\times\Rsp^{q}$ containing the origin. Note that the cross-covariance matrix in (\ref{OffHessxy}) involves higher-order statistics of $\Xmat$ and $\Ymat$. Additionally, observe that $\bSigma^{\left(\uexp,\vexp\right)}_{\Xmatsc\Ymatsc}\left(\svec,\tvec\right)$ reduces to the standard cross-covariance matrix $\bSigma_{\Xmatsc\Ymatsc}$ for $\svec=\zerovec$ and $\tvec=\zerovec$. Finally, note that the quantity in (\ref{OffHessxy}) has been proposed in \cite{Yeredor1}-\cite{Yeredor8} for blind source separation, blind channel estimation, blind channel equalization, and auto-regression parameter estimation. To the best of our knowledge this paper is the first to propose this quantity for generalizing LCCA.

In the following Theorem, which follows directly from (\ref{OffHessxy}) and the properties of ${M}_{\Xmatsc\Ymatsc}\left(\svec,\tvec\right)$ \cite{Severini}, \cite{DasGupta}, one sees that $\bSigma^{\left(\uexp,\vexp\right)}_{\Xmatsc\Ymatsc}\left(\svec,\tvec\right)$ preserves statistical independence and can capture non-linear dependencies when they exist.
\begin{Theorem}
\label{OffOriginIndepProp}
Let $U$ denote an arbitrary open region in $\Rsp^{p}\times\Rsp^{q}$ containing the origin, and assume that ${M}_{\Xmatsc\Ymatsc}\left(\svec,\tvec\right)$ is finite on $U$. The random vectors $\Xmat$ and $\Ymat$ are statistically independent under the joint probability measure $\pxy$ if and only if 
\begin{equation}
\label{IndCond}
\bSigma^{\left(\uexp,\vexp\right)}_{\Xmatsc\Ymatsc}\left(\svec,\tvec\right)=\zerovec\hspace{0.2cm}\forall\left(\svec,\tvec\right)\in{U}.
\end{equation}
[A proof is given in Appendix \ref{ExpKerTh}].
\end{Theorem}
The ``if'' is the interesting part of the theorem since the ``only if'' part follows directly from Property \ref{P3} of Proposition \ref{Prop1}. In particular, if $\Xmat$ and $\Ymat$ are statistically dependent under $\pxy$, then there exist $\avec\in\Rsp^{p}$, $\bvec\in\Rsp^{q}$, $\svec\in\Rsp^{p}$ and $\tvec\in\Rsp^{q}$, such that $\avec^{T}\bSigma^{\left(\uexp,\vexp\right)}_{\Xmatsc\Ymatsc}\left(\svec,\tvec\right)\bvec\neq{0}$. Thus, (\ref{MTRhoDef}) implies that if $\Xmat$ and $\Ymat$ are statistically dependent under $\pxy$ then there exist linear combinations of the form $\avec^{T}\Xmat$ and $\bvec^{T}\Ymat$ whose MT-correlation coefficient under $Q^{\left(\uexp,\vexp\right)}_{\Xmatsc\Ymatsc}$ is non-zero.

Finally, we show that MTCCA with the exponential MT-functions in (\ref{ExpKernel}) is translation-invariant. Let $\Xmat^{\prime}\triangleq\Xmat+\alphavec$ and $\Ymat^{\prime}\triangleq\Ymat+\betavec$, where $\alphavec$ and $\betavec$ are deterministic vectors in $\Rsp^{p}$ and $\Rsp^{q}$, respectively. According to (\ref{VarPhiDef}) and (\ref{ExpKernel}) $\varphi_{u,v}\left(\Xmat,\Ymat\right)=\varphi_{u,v}\left(\Xmat^{\prime},\Ymat^{\prime}\right)$. Therefore, by (\ref{RxMod_pxy})-(\ref{RxyMod_pxy}): $\bSigma^{\left(\uexp,\vexp\right)}_{\Xmatsc}\left(\svec,\tvec\right)=\bSigma^{\left(\uexp,\vexp\right)}_{\Xmatsc^{\prime}}\left(\svec,\tvec\right)$, 
$\bSigma^{\left(\uexp,\vexp\right)}_{\Ymatsc}\left(\svec,\tvec\right)=\bSigma^{\left(\uexp,\vexp\right)}_{\Ymatsc^{\prime}}\left(\svec,\tvec\right)$, and 
$\bSigma^{\left(\uexp,\vexp\right)}_{\Xmatsc\Ymatsc}\left(\svec,\tvec\right)=\bSigma^{\left(\uexp,\vexp\right)}_{\Xmatsc^{\prime}\Ymatsc^{\prime}}\left(\svec,\tvec\right)$.
Thus, by (\ref{CCAOptProbMod}), the MT-canonical correlation coefficients are invariant to translation, i.e. $$\rho_{k}\left(\bSigma^{\left(\uexp,\vexp\right)}_{\Xmatsc}\left(\svec,\tvec\right),\bSigma^{\left(\uexp,\vexp\right)}_{\Ymatsc}\left(\svec,\tvec\right),\bSigma^{\left(\uexp,\vexp\right)}_{\Xmatsc\Ymatsc}\left(\svec,\tvec\right)\right)=\rho_{k}\left(\bSigma^{\left(\uexp,\vexp\right)}_{\Xmatsc^{\prime}}\left(\svec,\tvec\right),\bSigma^{\left(\uexp,\vexp\right)}_{\Ymatsc^{\prime}}\left(\svec,\tvec\right),\bSigma^{\left(\uexp,\vexp\right)}_{\Xmatsc^{\prime}\Ymatsc^{\prime}}\left(\svec,\tvec\right)\right)$$ for $k=1,\ldots,r$.
\subsection{Gaussian MT-functions}
\label{GaussMT}
Next we define the MT-functions $u\left(\cdot;\cdot,\cdot\right)$ and $v\left(\cdot;\cdot,\cdot\right)$ by 
\begin{eqnarray}
\label{GaussKernel} 
\uGausss\left(\xvec;\svec,\sigma\right)\triangleq\frac{1}{\left(2\pi\sigma^{2}\right)^{\frac{p}{2}}}\exp\left(-\frac{\left\|\xvec-\svec\right\|^{2}_{2}}{2\sigma^{2}}\right) &{\rm{and}}& \vGausss\left(\yvec;\tvec,\tau\right)\triangleq\frac{1}{\left(2\pi\tau^{2}\right)^{\frac{q}{2}}}\exp\left(-\frac{\left\|\yvec-\tvec\right\|^{2}_{2}}{2\tau^{2}}\right),
\end{eqnarray}
where $\svec\in\Rsp^{p}$, $\tvec\in\Rsp^{q}$, $\sigma\in\Rsp^{+}$, $\tau\in\Rsp^{+}$, and $\left\|\cdot\right\|_{2}$ denotes the $l_{2}$-norm. 
Since $\uGausss\left(\cdot;\cdot,\cdot\right)$ and $\vGausss\left(\cdot;\cdot,\cdot\right)$ are strictly positive and bounded, one can easily verify that the condition in (\ref{Assumption2}) is satisfied. Relations (\ref{VarPhiDef}) and (\ref{RxyMod_pxy}) imply that the MT-functions (\ref{GaussKernel}) produce a weighted cross-covariance matrix, for which the observations are weighted in an inverse proportion to the distances $\left\|\xvec-\svec\right\|^{2}_{2}$ and $\left\|\yvec-\tvec\right\|^{2}_{2}$. Hence, the resulting MT-correlation coefficient is a measure of local linear dependence in the vicinity of $\left(\svec,\tvec\right)$. We note that local linear dependence exists whenever there are global non-linear dependencies.

Sensitivity of $\bSigma^{\left(\uGausss,\vGausss\right)}_{\Xmatsc\Ymatsc}\left(\svec,\tvec\right)$ to non-linear relationships between $\Xmat$ and $\Ymat$ is shown via the following Theorem.
\begin{Theorem}
\label{GaussIndepProp}  
Let $\sigma$, $\tau$ be fixed and positive. Additionally, let $U$ denote an arbitrary open region in $\Rsp^{p}\times\Rsp^{q}$ containing the origin. The random vectors $\Xmat$ and $\Ymat$ are statistically independent under the joint probability measure $\pxy$ if and only if 
\begin{equation}
\label{IndCond2}
\bSigma^{\left(\uGausss,\uGausss\right)}_{\Xmatsc\Ymatsc}\left(\svec,\tvec\right)=\zerovec\hspace{0.2cm}\forall\left(\svec,\tvec\right)\in{U}.
\end{equation}
[A proof is given in Appendix \ref{GaussKerTh}].
\end{Theorem}
Hence, if $\Xmat$ and $\Ymat$ are statistically dependent under $\pxy$, then there exist $\avec\in\Rsp^{p}$, $\bvec\in\Rsp^{q}$, $\svec\in\Rsp^{p}$ and $\tvec\in\Rsp^{q}$, such that $\avec^{T}\bSigma^{\left(\uGausss,\vGausss\right)}_{\Xmatsc\Ymatsc}\left(\svec,\tvec\right)\bvec\neq{0}$. Therefore, again, non-linear dependencies can be detected using MTCCA.
\subsection{Comparison between the exponential and Gaussian MT-functions:}
\label{MTFuncComp}
Unlike MTCCA with Gaussian MT-fucntions (\ref{GaussKernel}), MTCCA with exponential MT-functions (\ref{ExpKernel}) is translation invariant. 
Moreover, in MTCCA with Gaussian MT-functions, in addition to the location parameters $\svec$, $\tvec$, which share the same dimensionality of the scaling parameters of the exponential MT-functions, one has to set two width parameters $\sigma$ and $\tau$. On the other hand, unlike the exponential MT-functions, the Gaussian MT-functions are bounded in the joint observation space $\XCal\times\YCal$. Hence, MTCCA with Gaussian MT-functions is more robust to outliers. Additionally, the Gaussian MT-functions has the property that they localize linear dependence over the observation space. This property is illustrated in Subsection \ref{SimExamp1}. Additional common properties of the exponential and Gaussian MT-functions are given in the following remarks:
\begin{Remark}
Since the exponential and Gaussian MT-functions are strictly positive, by Property \ref{P4} of Proposition \ref{Prop1} we conclude that $Q^{\left(u_{\rm{E}},v_{\rm{E}}\right)}_{\Xmatsc\Ymatsc}$ and $Q^{\left(u_{\rm{G}},v_{\rm{G}}\right)}_{\Xmatsc\Ymatsc}$ preserve statistical dependence under $\pxy$.    
\end{Remark}
\begin{Remark}
The exponential and Gaussian MT-functions preserve Gaussianity in the sense that if $\Xmat$ and $\Ymat$ are jointly Gaussian under $\pxy$, then they are jointly Gaussian under  $Q^{\left(u_{\rm{E}},v_{\rm{E}}\right)}_{\Xmatsc\Ymatsc}$ and $Q^{\left(u_{\rm{G}},v_{\rm{G}}\right)}_{\Xmatsc\Ymatsc}$.
\end{Remark}
\subsection{Selection of the MT-functions parameters}
\label{KernelParamChoice}
A natural choice of the parameters $\svec$ and $\tvec$, would be those that maximize the first-order MT-canonical correlation coefficient 
$\rho_{1}\left(\bSigma^{\left(u,v\right)}_{\Xmatsc}\left(\svec,\tvec\right),\bSigma^{\left(u,v\right)}_{\Ymatsc}\left(\svec,\tvec\right),\bSigma^{\left(u,v\right)}_{\Xmatsc\Ymatsc}\left(\svec,\tvec\right)\right)$ in (\ref{CCAOptProbMod}). However, this maximization is analytically cumbersome. Therefore, as an alternative, we propose maximizing a lower bound on $\rho_{1}\left(\bSigma^{\left(u,v\right)}_{\Xmatsc}\left(\svec,\tvec\right),\bSigma^{\left(u,v\right)}_{\Ymatsc}\left(\svec,\tvec\right),\bSigma^{\left(u,v\right)}_{\Xmatsc\Ymatsc}\left(\svec,\tvec\right)\right)$. We show that the resultant first-order MT-canonical correlation coefficient will be sensitive to dependence between $\Xmat$ and $\Ymat$.
\begin{Proposition}
\label{LowerBoundTh}
Define the following element-by-element average:
\begin{equation}  
\label{PsiDef}
\psi\left(\bSigma^{\left(u,v\right)}_{\Xmatsc}\left(\svec,\tvec\right),\bSigma^{\left(u,v\right)}_{\Ymatsc}\left(\svec,\tvec\right),\bSigma^{\left(u,v\right)}_{\Xmatsc\Ymatsc}\left(\svec,\tvec\right)\right)\triangleq
\left(\frac{1}{pq}\sum\limits_{i=1}^{p}\sum\limits_{j=1}^{q}\frac{\left[\bSigma^{\left(u,v\right)}_{\Xmatsc\Ymatsc}\left(\svec,\tvec\right)\right]^{2}_{i,j}}
{\left[\bSigma^{\left(u,v\right)}_{\Xmatsc}\left(\svec,\tvec\right)\right]_{i,i}\left[\bSigma^{\left(u,v\right)}_{\Ymatsc}\left(\svec,\tvec\right)\right]_{j,j}}\right)^{1/2},
\end{equation}
where $\left[\Amat\right]_{i,j}$ denotes the $i,j$-th  entry of $\Amat$.
\begin{equation}
\label{LowerBoundEq}
\psi\left(\bSigma^{\left(u,v\right)}_{\Xmatsc}\left(\svec,\tvec\right),\bSigma^{\left(u,v\right)}_{\Ymatsc}\left(\svec,\tvec\right),\bSigma^{\left(u,v\right)}_{\Xmatsc\Ymatsc}\left(\svec,\tvec\right)\right)
\leq
\rho_{1}\left(\bSigma^{\left(u,v\right)}_{\Xmatsc}\left(\svec,\tvec\right),\bSigma^{\left(u,v\right)}_{\Ymatsc}\left(\svec,\tvec\right),\bSigma^{\left(u,v\right)}_{\Xmatsc\Ymatsc}\left(\svec,\tvec\right)\right).
\end{equation}
[A proof is given in Appendix \ref{LowerBoundProof}]
\end{Proposition}
Proposition \ref{LowerBoundTh} suggests choosing the optimal MT-functions parameters by maximizing the lower bound in (\ref{LowerBoundEq}):
\begin{equation}
\label{MaxProb}
\left(\svec^{*},\tvec^{*}\right)=\arg\max\limits_{\left(\svecsc,\tvecsc\right)\in{V}}\psi\left(\bSigma^{\left(u,v\right)}_{\Xmatsc}\left(\svec,\tvec\right),\bSigma^{\left(u,v\right)}_{\Ymatsc}\left(\svec,\tvec\right),\bSigma^{\left(u,v\right)}_{\Xmatsc\Ymatsc}\left(\svec,\tvec\right)\right),
\end{equation}
where $V$ a closed region in $\Rsp^{p}\times\Rsp^{q}$ containing the origin. Under the MT-functions pairs in (\ref{ExpKernel}) and (\ref{GaussKernel}) one can verify that $\psi\left(\bSigma^{\left(u,v\right)}_{\Xmatsc}\left(\svec,\tvec\right),\bSigma^{\left(u,v\right)}_{\Ymatsc}\left(\svec,\tvec\right),\bSigma^{\left(u,v\right)}_{\Xmatsc\Ymatsc}\left(\svec,\tvec\right)\right)$ is continuous in $\Rsp^{p}\times\Rsp^{q}$. Therefore, by the extreme value theorem \cite{AdCal} it has a maximum in $V$.
The maximization problem in (\ref{MaxProb}) can be solved numerically, e.g., using gradient ascent \cite{ConvOpt} or greedy search over the region $V$. 

The following theorem justifies the use of the first-order MT-canonical correlation coefficient as a measure of statistical independence.
\begin{Theorem}
\label{IndProp}
The random vectors $\Xmat$ and $\Ymat$ are statistically independent under $\pxy$ if and only if $$\rho_{1}\left(\bSigma^{\left(u,v\right)}_{\Xmatsc}\left(\svec^{*},\tvec^{*}\right),\bSigma^{\left(u,v\right)}_{\Ymatsc}\left(\svec^{*},\tvec^{*}\right),\bSigma^{\left(u,v\right)}_{\Xmatsc\Ymatsc}\left(\svec^{*},\tvec^{*}\right)\right)=0,$$
where $\left(u,v\right)$ are the MT-functions in (\ref{ExpKernel}) or (\ref{GaussKernel}), and $\left(\svec^{*},\tvec^{*}\right)$ are selected according to (\ref{MaxProb}).
[A proof is given in Appendix \ref{IndPropProof}]
\end{Theorem}
Therefore, if the MT-functions and their parameters are selected as in Theorem \ref{IndProp}, we conclude that $\Xmat$ and $\Ymat$ are statistically independent under 
$\pxy$ if and only if they are uncorrelated under $\qxy$. Hence, since by Property \ref{P3} of Proposition \ref{Prop1} $\qxy$ preserves statistical independence under $\pxy$, we also conclude that $\Xmat$ and $\Ymat$ are statistically independent under $\qxy$ if and only if they are uncorrelated under $\qxy$.
\section{Empirical implementation of MTCCA}
\label{Estimation}
Given $N$ i.i.d. samples of $\left(\Xmat,\Ymat\right)$ an empirical version of MTCCA (\ref{CCAOptProbMod}) can be implemented by replacing 
$\bSigma^{\left(u,v\right)}_{\Xmatsc}$, $\bSigma^{\left(u,v\right)}_{\Ymatsc}$ and $\bSigma^{\left(u,v\right)}_{\Xmatsc\Ymatsc}$ in (\ref{CCAOptProbMod}), (\ref{GEVD_PMTCCA}) and (\ref{MaxProb}) with their sample covariance estimates. Hence, strongly consistent estimators of $\bSigma^{\left(u,v\right)}_{\Xmatsc}$, $\bSigma^{\left(u,v\right)}_{\Ymatsc}$ and $\bSigma^{\left(u,v\right)}_{\Xmatsc\Ymatsc}$ are constructed, based on $N$ i.i.d. samples of $\left(\Xmat,\Ymat\right)$. 
\begin{Proposition}
\label{ConsistentEst}
Let $\left(\Xmat_{n},\Ymat_{n}\right)$, $n=1,\ldots,N$ denote a sequence of i.i.d. samples from the joint distribution $\pxy$, and define the empirical covariance estimates
\begin{equation}
\label{Rx_uv_Est}
\hat{\bSigma}^{\left(u,v\right)}_{\Xmatsc}\triangleq\frac{1}{N-1}\sum\limits_{n=1}^{N}\Xmat_{n}\Xmat^{T}_{n}\hat{\varphi}_{u,v}\left(\Xmat_{n},\Ymat_{n}\right)
-\frac{N}{N-1}\hat{\muvec}^{\left(u,v\right)}_{\xvec}\hat{\muvec}^{\left(u,v\right)T}_{\xvec},
\end{equation}
\begin{equation} 
\label{Ry_uv_Est} 
\hat{\bSigma}^{\left(u,v\right)}_{\Ymatsc}\triangleq
\frac{1}{N-1}\sum\limits_{n=1}^{N}\Ymat_{n}\Ymat^{T}_{n}\hat{\varphi}_{u,v}\left(\Xmat_{n},\Ymat_{n}\right)
-\frac{N}{N-1}\hat{\muvec}^{\left(u,v\right)}_{\Ymatsc}\hat{\muvec}^{\left(u,v\right)T}_{\Ymatsc},
\end{equation}
and
\begin{equation}
\label{Rxy_uv_Est}  
\hat{\bSigma}^{\left(u,v\right)}_{\Xmatsc\Ymatsc}\triangleq
\frac{1}{N-1}\sum\limits_{n=1}^{N}\Xmat_{n}\Ymat^{T}_{n}\hat{\varphi}_{u,v}\left(\Xmat_{n},\Ymat_{n}\right)
-\frac{N}{N-1}\hat{\muvec}^{\left(u,v\right)}_{\xvec}\hat{\muvec}^{\left(u,v\right)T}_{\Ymatsc},
\end{equation}
where
\begin{eqnarray}
\label{Mu_x_Mu_y}
\hat{\muvec}^{\left(u,v\right)}_{\Xmatsc}\triangleq\frac{1}{N}\sum\limits_{n=1}^{N}\Xmat_{n}\hat{\varphi}_{u,v}\left(\Xmat_{n},\Ymat_{n}\right),&&
\hat{\muvec}^{\left(u,v\right)}_{\Ymatsc}\triangleq\frac{1}{N}\sum\limits_{n=1}^{N}\Ymat_{n}\hat{\varphi}_{u,v}\left(\Xmat_{n},\Ymat_{n}\right),
\end{eqnarray}
and
\begin{equation}
\label{varphi_hat}
\hat{\varphi}_{u,v}\left(\Xmat_{n},\Ymat_{n}\right)\triangleq\frac{u\left(\Xmat_{n}\right)v\left(\Ymat_{n}\right)}{\frac{1}{N}\sum\limits_{n=1}^{N}u\left(\Xmat_{n}\right)v\left(\Ymat_{n}\right)}.
\end{equation}
Assume 
\begin{eqnarray}
\label{Cond1}
{\rm{E}}\left[u^{4}\left(\Xmat\right);\px\right]<\infty,&&{\rm{E}}\left[v^{4}\left(\Ymat\right);\py\right]<\infty,
\end{eqnarray}
\begin{eqnarray}
\label{Cond2} 
{\rm{E}}\left[X^{4}_{k};\px\right]<\infty\hspace{0.2cm}\forall{k=1,\ldots,p}&\rm{and}&{\rm{E}}\left[Y^{4}_{l};\py\right]<\infty\hspace{0.2cm}\forall{l=1,\ldots,q},
\end{eqnarray}
where $X_{k}$ and $Y_{l}$ denote the $k$-th and the $l$-th entries of $\Xmat$ and $\Ymat$, respectively. Then $\hat{\bSigma}^{\left(u,v\right)}_{\Xmatsc}\rightarrow{\bSigma}^{\left(u,v\right)}_{\Xmatsc}$, $\hat{\bSigma}^{\left(u,v\right)}_{\Ymatsc}\rightarrow{\bSigma}^{\left(u,v\right)}_{\Ymatsc}$ and $\hat{\bSigma}^{\left(u,v\right)}_{\Xmatsc\Ymatsc}\rightarrow{\bSigma}^{\left(u,v\right)}_{\Xmatsc\Ymatsc}$ almost surely as $N\rightarrow\infty$. [A proof is given in Appendix \ref{ConsistEstProof}]
\end{Proposition}
Note that for $u\left(\Xmat\right)\equiv{1}$ and $v\left(\Ymat\right)\equiv{1}$, the estimators $\hat{\bSigma}^{\left(u,v\right)}_{\Xmatsc}$, $\hat{\bSigma}^{\left(u,v\right)}_{\Ymatsc}$, and $\hat{\bSigma}^{\left(u,v\right)}_{\Xmatsc\Ymatsc}$ reduce to the standard unbiased estimators of the covariance and cross-covariance matrices $\bSigma_{\Xmatsc}$, $\bSigma_{\Ymatsc}$ and $\bSigma_{\Xmatsc\Ymatsc}$, respectively.

The empirical MTCCA procedure with the exponential and Gaussian MT-functions is given in Appendix \ref{EmpMTCCA}. In the first stage of the procedure, 
the parameters of the MT-functions are selected by solving a {\it{single}} $(p+q)$-dimensional maximization problem (\ref{MaxProb2}) using gradient ascent. It can be shown that each iteration of the gradient ascent algorithm, which only involves the empirical measure transformed covariance and cross-covariance matrices, has asymptotic computational load (ACL) of $O((p+q)^{2}N)$ flops per iteration. In the second stage, the empirical MT-canonical correlation coefficients and directions are obtained {\it{simultaneously}} by solving the GEVD problem (\ref{GEVD_PMTCCA_2}) with ACL of $O((p+q)^{3})$ flops. Unlike the empirical MTCCA, the empirical ICCA \cite{MICCA} involves a {\it{sequence}} of $\left(p+q\right)$-dimensional numerical maximizations, one for each pair of canonical directions, using an iterative Newton-Raphson algorithm. It can be shown that each iteration of the Newton-Rafson algorithm, which involves re-estimation of the mutual-information in a non-parametric manner and inversion of a Hessian matrix, has ACL of $O((p+q)N^{2} + (p+q+2k)^{3})$ flops, where $k$ denotes a canonical directions pair index. The empirical KCCA procedure \cite{Akaho}-\cite{Bach}, which involves computation of two $N\times{N}$ Gram matrices followed by solving a GEVD problem, has ACL of $O((p+q)N^{2}+N^{3})$ flops. Hence, one sees that unlike the empirical ICCA and KCCA, the computational complexity of the empirical MTCCA is linear in $N$, which makes it favorable in large samples size scenarios.
\section{Numerical examples}
\label{Examp}
In this section, we illustrate the use of empirical LCCA, ICCA, KCCA and MTCCA for graphical model selection. In every example below the empirical MTCCA was performed 
with the exponential and Gaussian MT-functions via the procedure in Appendix \ref{EmpMTCCA}. In ICCA, the empirical mutual-information, $\hat{I}_{k}$, between each pair of canonical variates was mapped to the interval $[0,1]$ via the formula $\hat{\rho}_{k}=\sqrt{1-\exp(-2\hat{I}_{k})}$ which produce the empirical informational canonical correlation coefficients. The empirical KCCA was performed using Gaussian radial basis function kernels. Since KCCA masks the original coordinates of $\Xmat$ and $\Ymat$, it is not illustrated for the graphical model selection tasks in simulation examples 1 and 2, which involve variable selection. In simulation examples 1 and 2, the canonical correlation coefficients and canonical directions were estimated using $N=1000$ i.i.d. samples of $\Xmat$ and $\Ymat$. The statistical significance of the empirical canonical correlation coefficients was tested using empirical estimates of $p$-values associated with rejecting the null-hypothesis of no statistical dependence between $\Xmat$ and $\Ymat$ (see Appendix \ref{PVal}). 
\subsection{Simulation example 1: Selection of graphical model with non-linear connections}
\label{SimExamp1}
\begin{figure}[htbp!]
\centerline{\psfig{figure=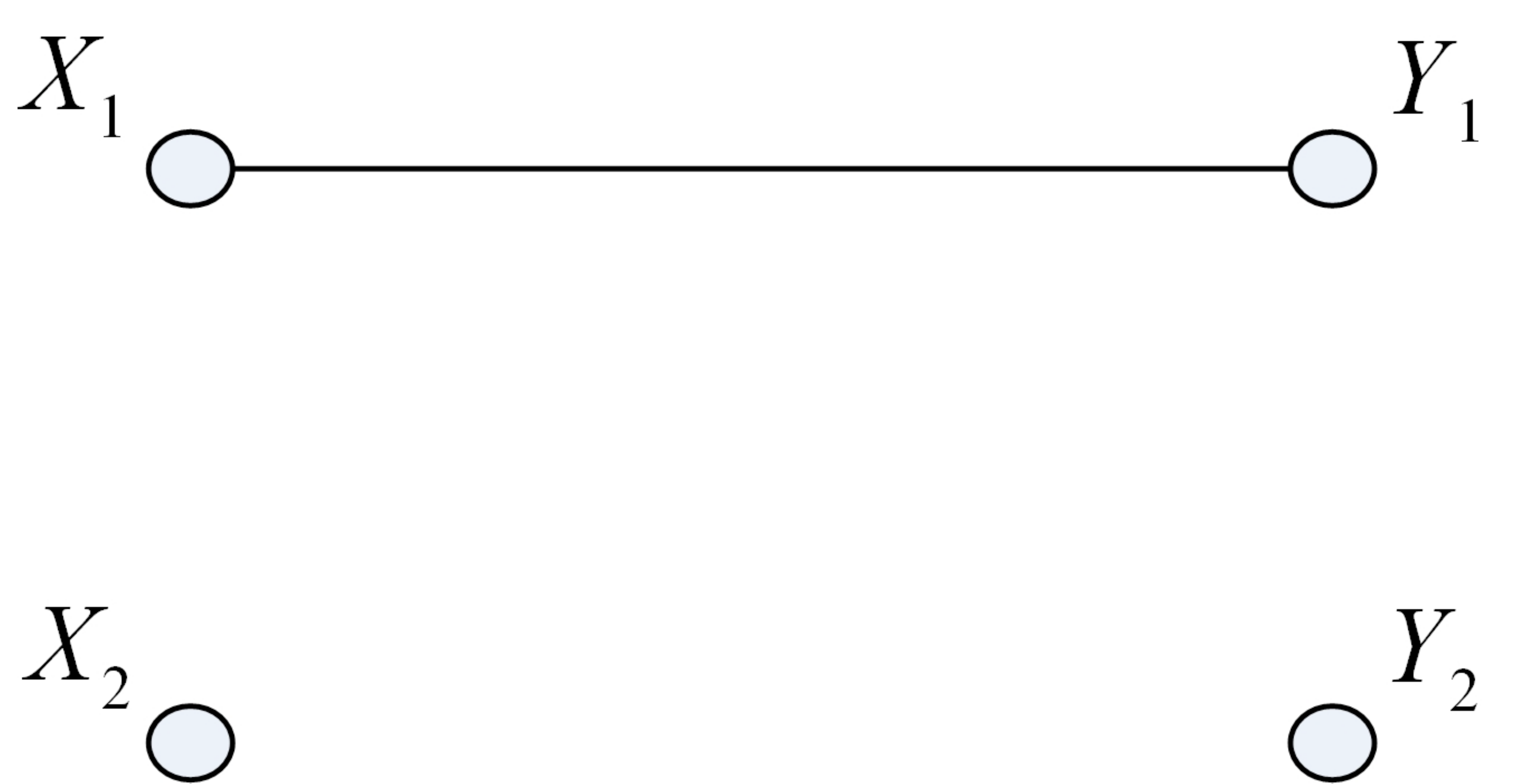,scale=0.13}}
\caption{The graphical model of dependencies in simulation example 1. A single edge exists between $X_{1}$ and $Y_{1}$ due to the non-linear relation model $Y_{1}=\cos\left(X_{1}\right)+0.1W$. The correlation between $X_1$ and $Y_1$ under $\pxy$ is equal to zero even though they are dependent.}
\label{GraphMod1}
\end{figure}

In this example, we consider the random vectors $\Xmat=\left[X_{1},X_{2}\right]^{T}$ and $\Ymat=\left[Y_{1},Y_{2}\right]^{T}$, where 
$$Y_{1}=\cos\left(X_{1}\right)+0.1W,$$ and $X_{1}$, $X_{2}$, $Y_{2}$, and $W$ are mutually independent standard normal random variables. For this example, the pair of linear combinations of the form $\left(\avec^{T}\Xmat,\bvec^{T}\Ymat\right)$ having maximal dependency is obtained for the vector pair $(\avec_{1}=\left[1,0\right]^{T},\bvec_{1}=\left[1,0\right]^{T})$ which are identical to the true first-order MT-canonical directions. In this example, all pairs of linear combinations of the form $\avec^{T}\Xmat$ and $\bvec^{T}\Ymat$ have zero linear correlation even though they are not statistically independent. The dependencies between $\Xmat$ and $\Ymat$ are depicted by the bipartite graphical model in Fig. \ref{GraphMod1}.

The averaged estimates of the MT, linear, and informational canonical correlation coefficients and their corresponding averaged $p$-values, based on 1000 Monte-Carlo simulations, are given in Table \ref{Tab1}. The sample means and standard deviations of the absolute dot products of $(\avec_{1}/{\|\avec_{1}\|}_{2},{\hat{\avec}_{1}}/{\|\hat{\avec}_{1}\|}_{2})$ and $(\bvec_{1}/{\|\bvec_{1}\|}_{2},{\hat{\bvec}_{1}}/{\|\hat{\bvec}_{1}\|}_{2})$, based on $1000$ Monte-Carlo simulations, are given in Table {\ref{Tab2}}. The absolute dot products should be equal to 1 when the estimated canonical directions $\hat{\avec}$, $\hat{\bvec}$ are equal to $\avec_{1}=\left[1,0\right]^{T}$, $\bvec_{1}=\left[1,0\right]^{T}$, respectively. One can notice that in contrast to LCCA, the MTCCA and ICCA detect the true dependencies between $\Xmat$ and $\Ymat$, depicted by the bipartite graphical model in Figs. \ref{GraphMod1}.   
\begin{table}[htdp]
\caption{Simulation example 1: The averaged estimates of the MT, linear, and informational canonical correlation coefficients and their corresponding averaged $p$-values (in parentheses).}
\begin{center}
\begin{tabular}{| c | c | c | c | c |}
\hline
&\hspace{0.3cm}\textbf{Exponential MT-functions}      & \hspace{0.3cm}\textbf{Gaussian MT-functions} & \hspace{0.3cm}\textbf{LCCA} & \hspace{0.3cm}\textbf{ICCA}\\
\hline
$\hat{\rho}_{1}$                &\hspace{0.3cm}  0.83 (0)         & \hspace{0.3cm} 0.88 (0)            & \hspace{0.3cm} 0.06 (0.37)  & \hspace{0.3cm} 0.85 (0)        \\
\hline
$\hat{\rho}_{2}$                 &\hspace{0.3cm} 0.04 (0.38)          & \hspace{0.3cm} 0.03 (0.36)           & \hspace{0.3cm} 0.01 (0.45)  & \hspace{0.3cm} 0.23 (0.42)    \\       
\hline
\end{tabular}
\end{center}
\label{Tab1}
\end{table}
\begin{table}[htdp]
\caption{Simulation example 1: The sample means and standard deviations (in parenthesis) of $c(\avec_{1},\hat{\avec}_{1})$ and $c(\bvec_{1},\hat{\bvec}_{1})$, where $c(\uvec,\vvec)\triangleq|\frac{\uvec^{T}\vvec}{{\|\uvec\|}_{2}{\|\vvec\|}_{2}}|$.}
\begin{center}
\begin{tabular}{| c | c | c | c | c |}
\hline
&\hspace{0.3cm}\textbf{Exponential MT-functions}      & \hspace{0.3cm}\textbf{Gaussian MT-function}  & \hspace{0.3cm}\textbf{LCCA} & \hspace{0.3cm}\textbf{ICCA}\\
\hline
$c(\avec_{1},{\hat{\avec}_{1}})$                &\hspace{0.3cm}  0.99 ($7\cdot{10}^{-4}$)         & \hspace{0.3cm} 0.99 ($3\cdot{10}^{-4}$)            & \hspace{0.3cm} 0.73 (0.27)    & \hspace{0.3cm} 0.99 ($2\cdot{10}^{-5}$)        \\
\hline
$c(\bvec_{1},{\hat{\bvec}_{1}})$                 &\hspace{0.3cm} 0.99 ($4\cdot{10}^{-4}$)          & \hspace{0.3cm} 0.99 ($1\cdot{10}^{-4}$)           & \hspace{0.3cm} 0.75 (0.22)    & \hspace{0.3cm} 0.99 ($1\cdot{10}^{-5}$)       \\
\hline 
\end{tabular}
\end{center}
\label{Tab2}
\end{table}

\begin{figure}[htb]
  \begin{center}
    {\subfigure[]{\label{EXP_CANN_VAR_1_NL}\includegraphics[scale=0.5]{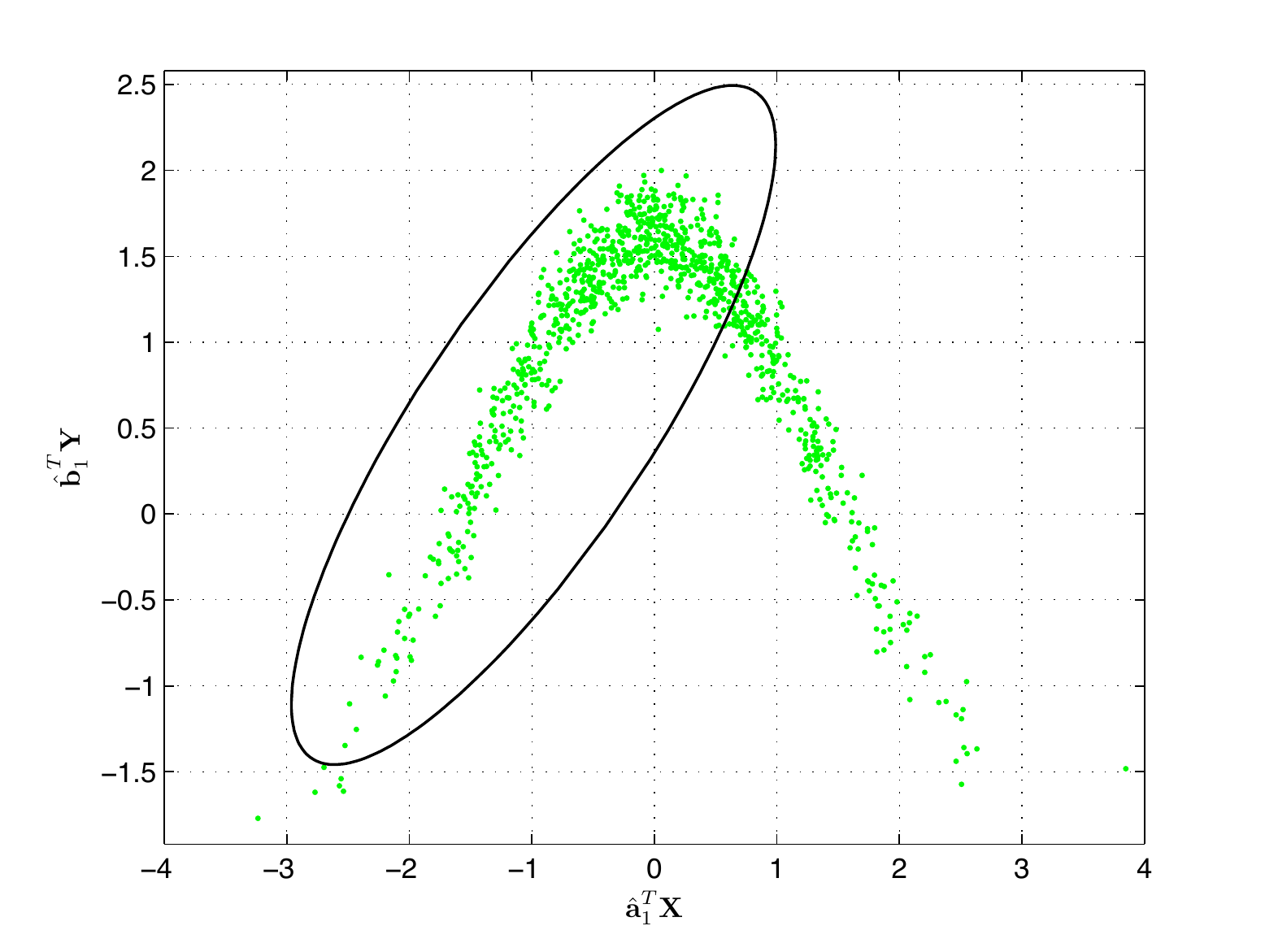}}}
    {\subfigure[]{\label{GAUSS_CANN_VAR_1_NL}\includegraphics[scale=0.5]{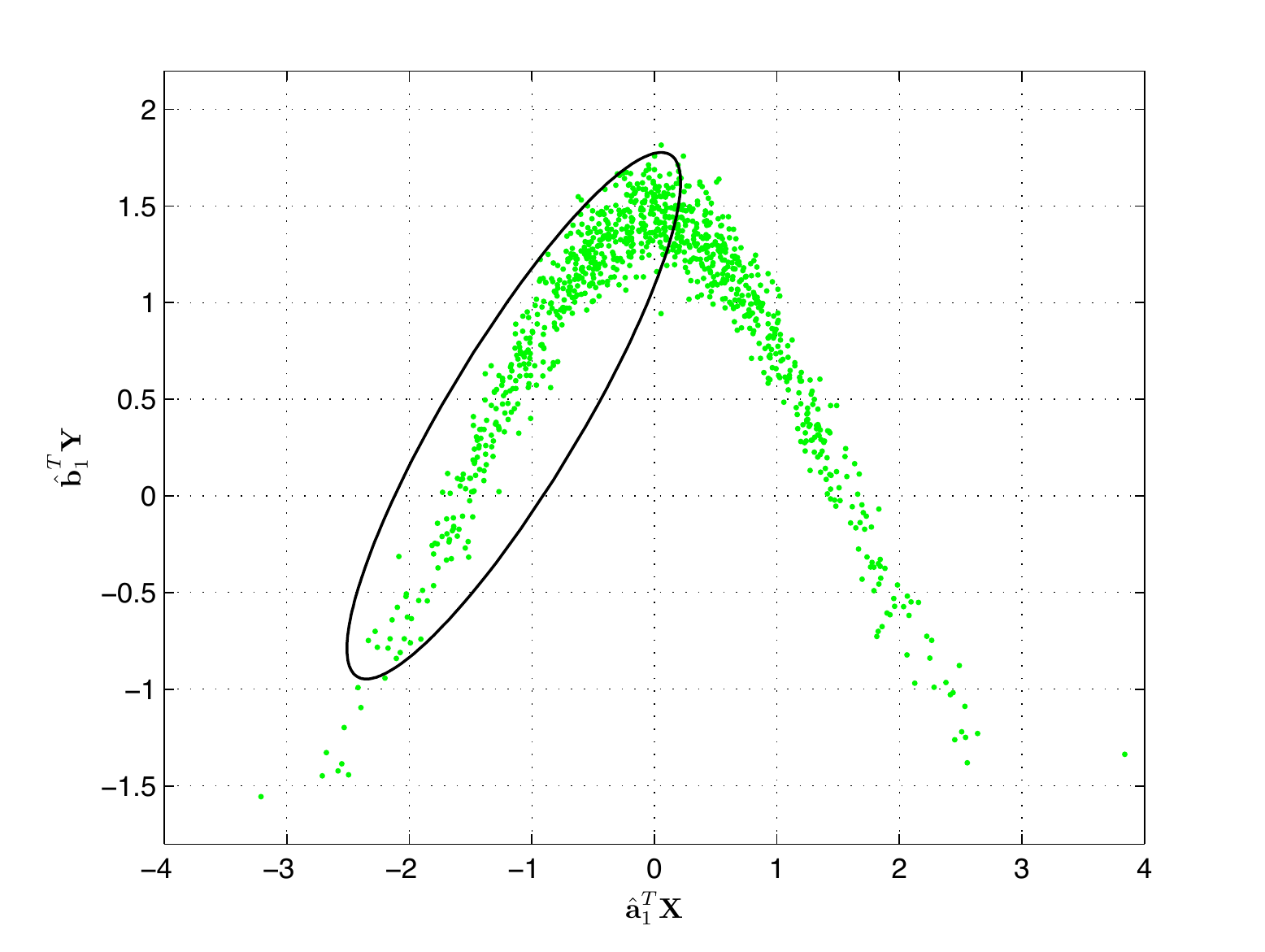}}}
    {\subfigure[]{\label{LIN_CANN_VAR_1_NL}\includegraphics[scale=0.5]{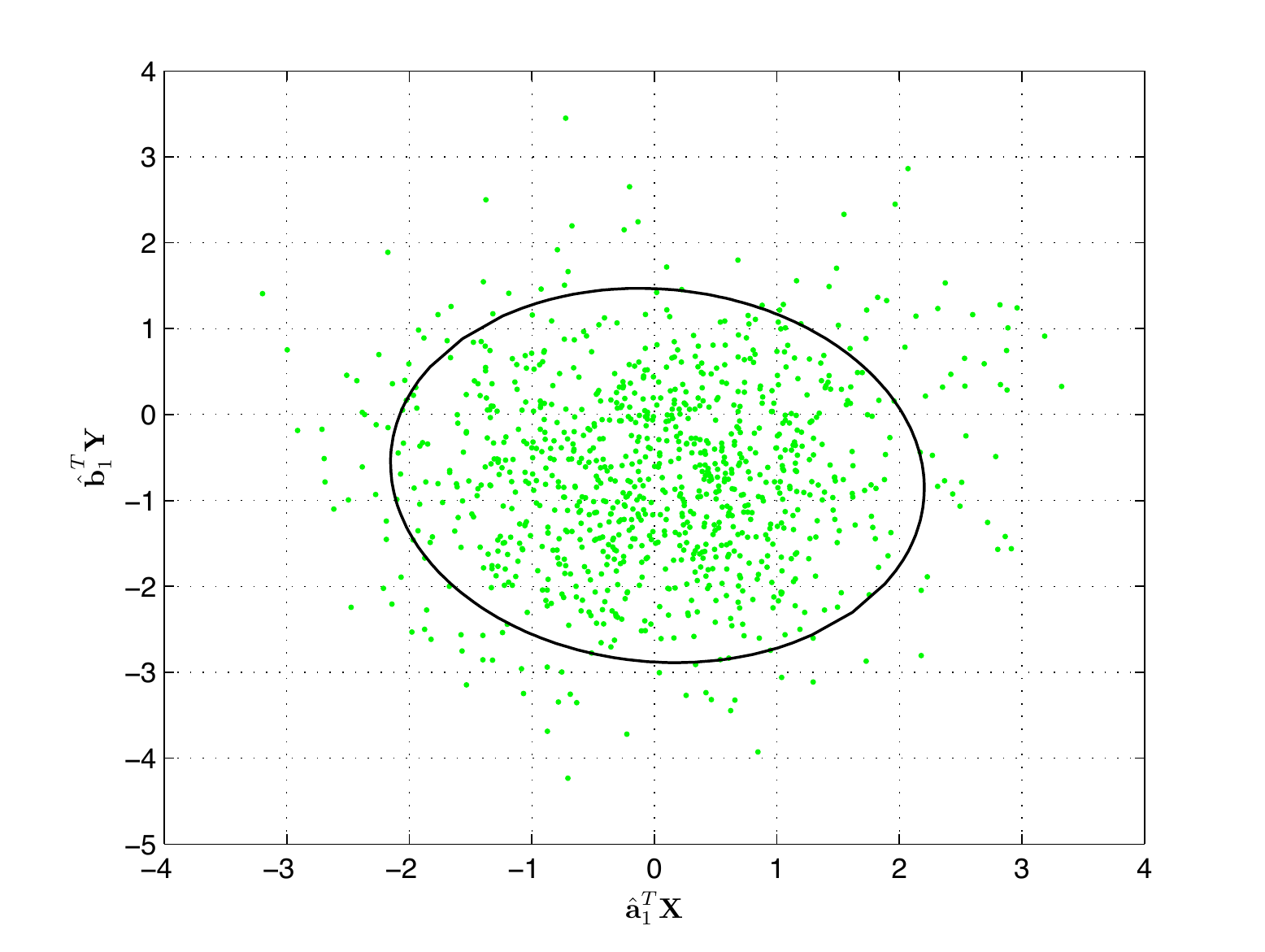}}}
    {\subfigure[]{\label{I_CANN_VAR_1_NL}\includegraphics[scale=0.5]{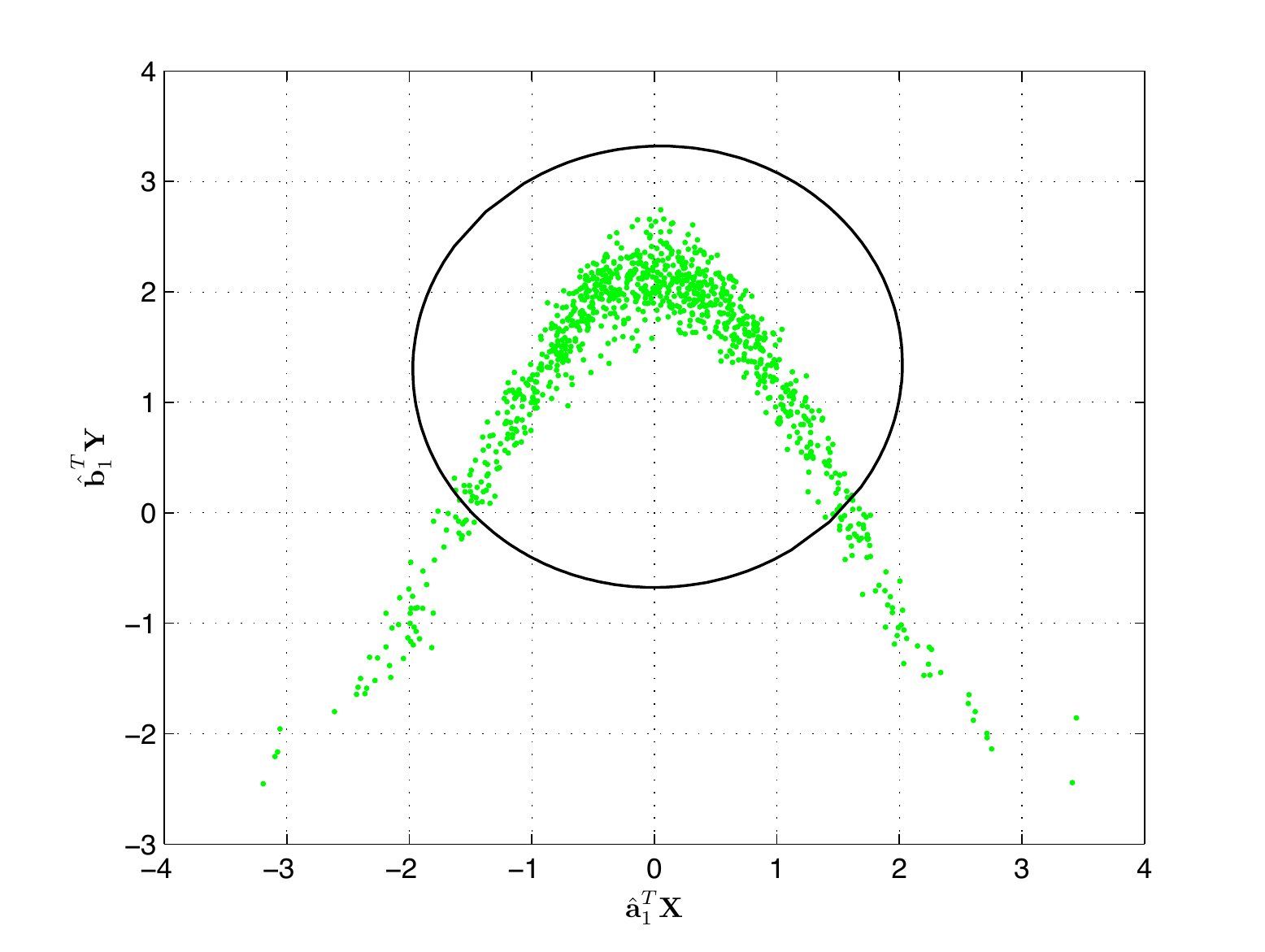}}}
  \end{center}
  \caption{Simulation example 1: Scatter plots of the empirical first-order canonical variates obtained by: (a) MTCCA with exponential MT-functions, (b) MTCCA with Gaussian MT-functions, (c) LCCA, and (d) ICCA. Note that, while the linear canonical variates are uninformative (circular Gaussian distributed), the MT and informational canonical variates have captured the non-linear structure (raised cosine shape) of the non-linear model. This occurs since all variables in example 1 have zero correlation but some variables are non-linearly dependent. The ellipses represent the associated covariance matrices under the probability measures $Q^{\left(u_{{\rm{E}}},v_{{\rm{E}}}\right)}_{\Xmatsc\Ymatsc}$, $Q^{\left(u_{{\rm{G}}},v_{{\rm{G}}}\right)}_{\Xmatsc\Ymatsc}$, and $\pxy$, respectively.}
\label{CANVAR}
\end{figure}

Scatter plots of the empirical first-order MT, linear, and informational canonical variates $(\hat{\avec}^{T}_{1}\Xmat,\hat{\bvec}^{T}_{1}\Ymat)$ are shown in Figs. \ref{EXP_CANN_VAR_1_NL}-\ref{I_CANN_VAR_1_NL}. Observe that unlike LCCA, MTCCA and ICCA recover the true non-linear relation between $\Xmat$ and $\Ymat$, which has a raised cosine shape. In these figures, we have also plotted the ellipses associate with the empirical covariance matrices of $[\hat{\avec}_{1}^{T}\Xmat,\hat{\bvec}_{1}^{T}\Ymat]^{T}$ under the probability measures $Q^{\left(u_{{\rm{E}}},v_{{\rm{E}}}\right)}_{\Xmatsc\Ymatsc}$,  $Q^{\left(u_{{\rm{G}}},v_{{\rm{G}}}\right)}_{\Xmatsc\Ymatsc}$, and $\pxy$, respectively. Observing Figs. \ref{EXP_CANN_VAR_1_NL} and \ref{GAUSS_CANN_VAR_1_NL} one can notice that the local linear trend is better captured by MTCCA with Gaussian MT-functions due to their localization property, discussed in Subsection \ref{GaussMT}.
\subsection{Simulation example 2: Selection of graphical model with linear and non-linear connections}
\begin{figure}[htbp!]
\centerline{\psfig{figure=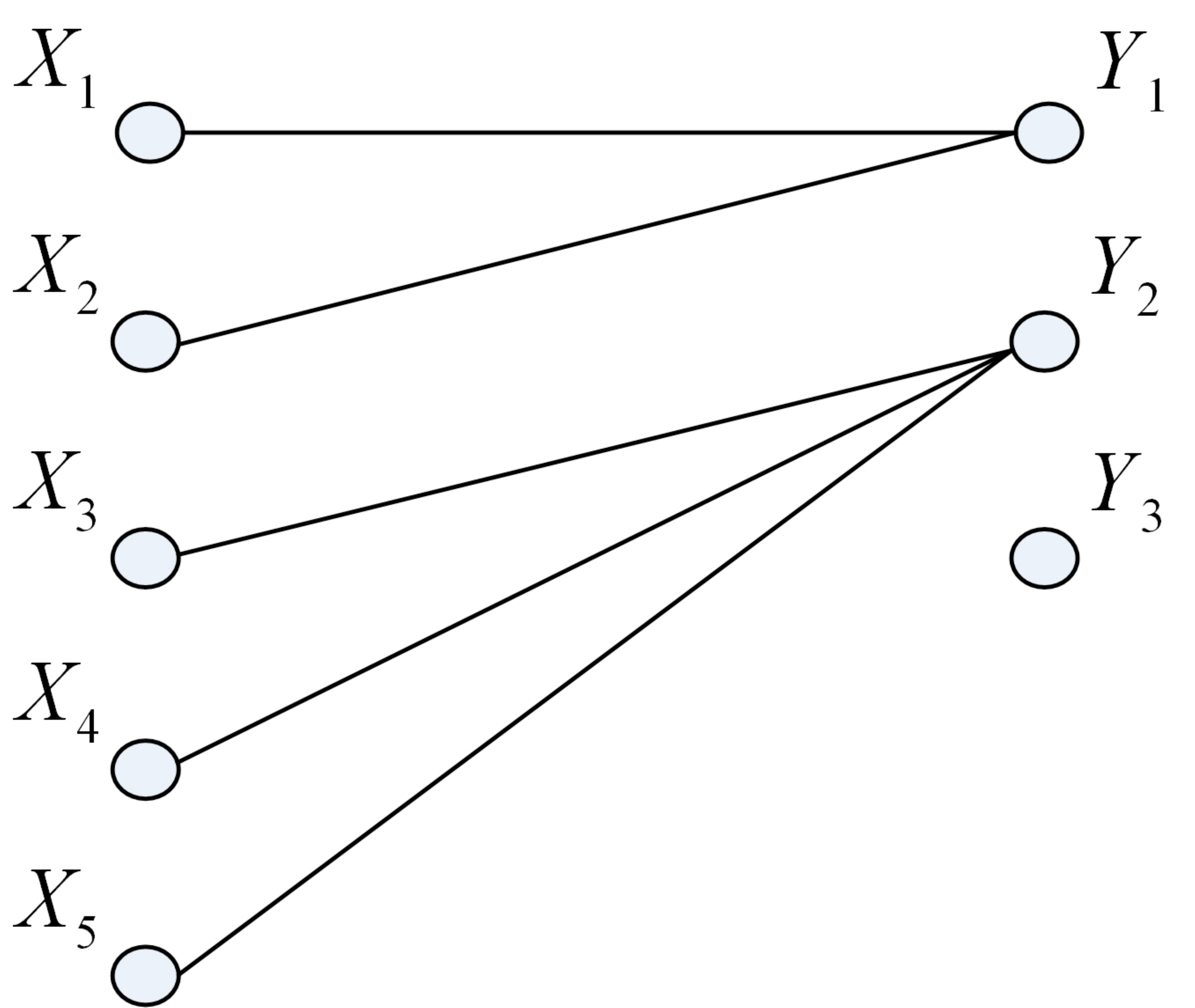,scale=0.13}}
\caption{The dependency graphical model corresponding to simulation example 2. There are two connected components $\{(X_{1},Y_{1}),(X_{2},Y_{1})\}$ and $\{(X_{3},Y_{2}),(X_{4},Y_{2}),(X_{5},Y_{2})\}$.}
\label{GraphMod2}
\end{figure}

In this example, we consider a more complex model. Let the random vectors $\Xmat=\left[X_{1},X_{2},X_{3},X_{4},X_{5}\right]^{T}$ and $\Ymat=\left[Y_{1},Y_{2},Y_{3}\right]^{T}$ satisfy
$$Y_{1}=X_{1}+0.5X_{2}+0.1W_{1},$$ $$Y_{2}=\cos\left(X_{3}+0.75X_{4}+0.5X_{5}\right)+0.1W_{2},$$ where $X_{i}$, $i=1,\ldots,5$, $W_{i}$, $i=1,2$, and $Y_{3}$ are mutually independent standard normal random variables. In this example there exist two independent pairs of linear combinations $(\avec_{k}^{T}\Xmat,\bvec_{k}^{T}\Ymat)$, $k=1,2,$ with maximal inter-dependencies. These maximally dependent canonical variates are obtained for the vector pairs $(\avec_{1}=\left[1,0.5,0,0,0\right]^{T},\bvec_{1}=\left[1,0,0\right]^{T})$ and $(\avec_{2}=\left[0,0,1,0.75,0.5\right]^{T},\bvec_{2}=\left[0,1,0\right]^{T})$, which are also the first-order and second-order MT-canonical directions. The dependencies between $\Xmat$ and $\Ymat$ are depicted by the bipartite graphical model in Fig. \ref{GraphMod2}.

The averaged estimates of the MT, linear, and informational canonical correlation coefficients and their corresponding averaged $p$-values, based on 1000 Monte-Carlo simulations, are given in Table \ref{Tab3}. The sample means and standard deviations of the absolute dot products of the pairs  $(\avec_{k}/{\|\avec_{k}\|}_{2},{\hat{\avec}_{k}}/{\|\hat{\avec}_{k}\|}_{2})$ and $(\bvec_{k}/{\|\bvec_{k}\|}_{2},{\hat{\bvec}_{k}}/{\|\hat{\bvec}_{k}\|}_{2})$, $k=1,2$, based on 1000 Monte-Carlo simulations, are given in Table {\ref{Tab4}}. Observe that both MTCCA and ICCA detect the true dependencies between $\Xmat$ and $\Ymat$, depicted by the bipartite graphical model in Fig. \ref{GraphMod2}. As expected, the  LCCA detects only the linearly dependent combinations.
\begin{table}[htdp]
\caption{Simulation example 2: The averaged estimates of the MT, linear, and informational canonical correlation coefficients and their corresponding averaged $p$-values (in parenthesis).}
\begin{center}
\begin{tabular}{| c | c | c | c | c|}
\hline
              &\hspace{0.3cm}\textbf{Exponential MT-functions}      & \hspace{0.3cm}\textbf{Gaussian MT-functions} & \hspace{0.3cm}\textbf{LCCA} & \hspace{0.3cm}\textbf{ICCA}\\
\hline
$\hat{\rho}_{1}$   &\hspace{0.3cm}  1 ($0$)    & \hspace{0.3cm} 1 ($0$)  & \hspace{0.3cm} 1 ($0$)    & \hspace{0.3cm} 0.93 ($0$)         \\
\hline
$\hat{\rho}_{2}$   &\hspace{0.3cm} 0.75 ($0$)               & \hspace{0.3cm} 0.9 (0)           & \hspace{0.3cm} 0.08 (0.22)   & \hspace{0.3cm} 0.89 (0)\\
\hline
$\hat{\rho}_{3}$   &\hspace{0.3cm} 0.08 ($0.2$)     & \hspace{0.3cm} 0.1 ($0.18$)  & \hspace{0.3cm} 0.04 ($0.35$)  & \hspace{0.3cm} 0.24 ($0.27$)      \\
\hline    
\end{tabular}
\end{center}
\label{Tab3}
\end{table}
\begin{table}[htdp]
\caption{Simulation example 2: The sample means and standard deviations (in parenthesis) of $c(\avec_{k},\hat{\avec}_{k})$ and $c(\bvec_{k},\hat{\bvec}_{k})$, $k=1,2$, where $c(\uvec,\vvec)\triangleq|\frac{\uvec^{T}\vvec}{{\|\uvec\|}_{2}{\|\vvec\|}_{2}}|$.}
\begin{center}
\begin{tabular}{| c | c | c | c | c |}
\hline
&\hspace{0.3cm}\textbf{Exponential MT-functions}      & \hspace{0.3cm}\textbf{Gaussian MT-functions} & \hspace{0.3cm}\textbf{LCCA} & \hspace{0.3cm}\textbf{ICCA}\\
\hline
$c(\avec_{1},{\hat{\avec}}_{1})$   &\hspace{0.3cm}  1 ($5\cdot{10}^{-5}$)    & \hspace{0.3cm} 1 ($5\cdot{10}^{-4}$)  & \hspace{0.3cm} 1 (${10}^{-5}$)   & \hspace{0.3cm} 0.99 ($7\cdot{10}^{-4}$)            \\
\hline
$c(\avec_{2},{\hat{\avec}}_{2})$   &\hspace{0.3cm} 0.99 ($6\cdot{10}^{-3}$)               & \hspace{0.3cm} 0.99 ($8\cdot{10}^{-3}$)           & \hspace{0.3cm} 0.5 (0.28)  & \hspace{0.3cm} 0.99 ($1\cdot{10}^{-3}$)                \\
\hline
$c(\bvec_{1},{\hat{\bvec}}_{1})$   &\hspace{0.3cm} 1 ($8\cdot{10}^{-5}$)     & \hspace{0.3cm} 1 ($9\cdot{10}^{-4}$)  & \hspace{0.3cm} 1 ($2\cdot{10}^{-5}$)  & \hspace{0.3cm} 0.99 ($2\cdot{10}^{-3}$)            \\
\hline
$c(\bvec_{2},{\hat{\bvec}}_{2})$   &\hspace{0.3cm} 0.99 ($2\cdot{10}^{-3}$)               & \hspace{0.3cm} 0.99 ($6\cdot{10}^{-3}$)            & \hspace{0.3cm} 0.7 (0.26) 
& \hspace{0.3cm} 0.99 ($3\cdot{10}^{-3}$)     \\
\hline      
\end{tabular}
\end{center}
\label{Tab4}
\end{table}
\subsection{Measuring long-term associations between NASDAQ/NYSE traded companies}
Here, MTCCA is applied to a real world example of capturing long-term associations between pairs of companies traded on the NASDAQ and NYSE stock markets. The compared companies were Microsoft (MSFT), Intel (INTC), Apple (AAPL), Merck (MRK), Pfizer (PFE), Johnson and Johnson (JNJ), American express (AXP), JP Morgan (JPM), and Bank of America (BAC). For each pair of companies, we considered the random vectors $\Xmat=[X_{1},X_{2}]^{T}$ and $\Ymat=[Y_{1},Y_{2}]^{T}$. The variables $X_{1}$ and $Y_{1}$ are the log-ratios of two consecutive daily closing prices of a stock, called log-returns. The variables $X_{2}$ and $Y_{2}$ are the log-ratios of two consecutive daily trading volumes of a stock, called log-volume ratios. Consecutive daily measurements of $\Xmat$ and $\Ymat$ from January 2, 2001 to December 31, 2010, comprising 2514 samples, were obtained from the WRDS database \cite{WRDS}.

\begin{figure}[htbp!]
  \begin{center}  
    \subfigure[]{\label{Fig_2_a}\includegraphics[scale=0.5]{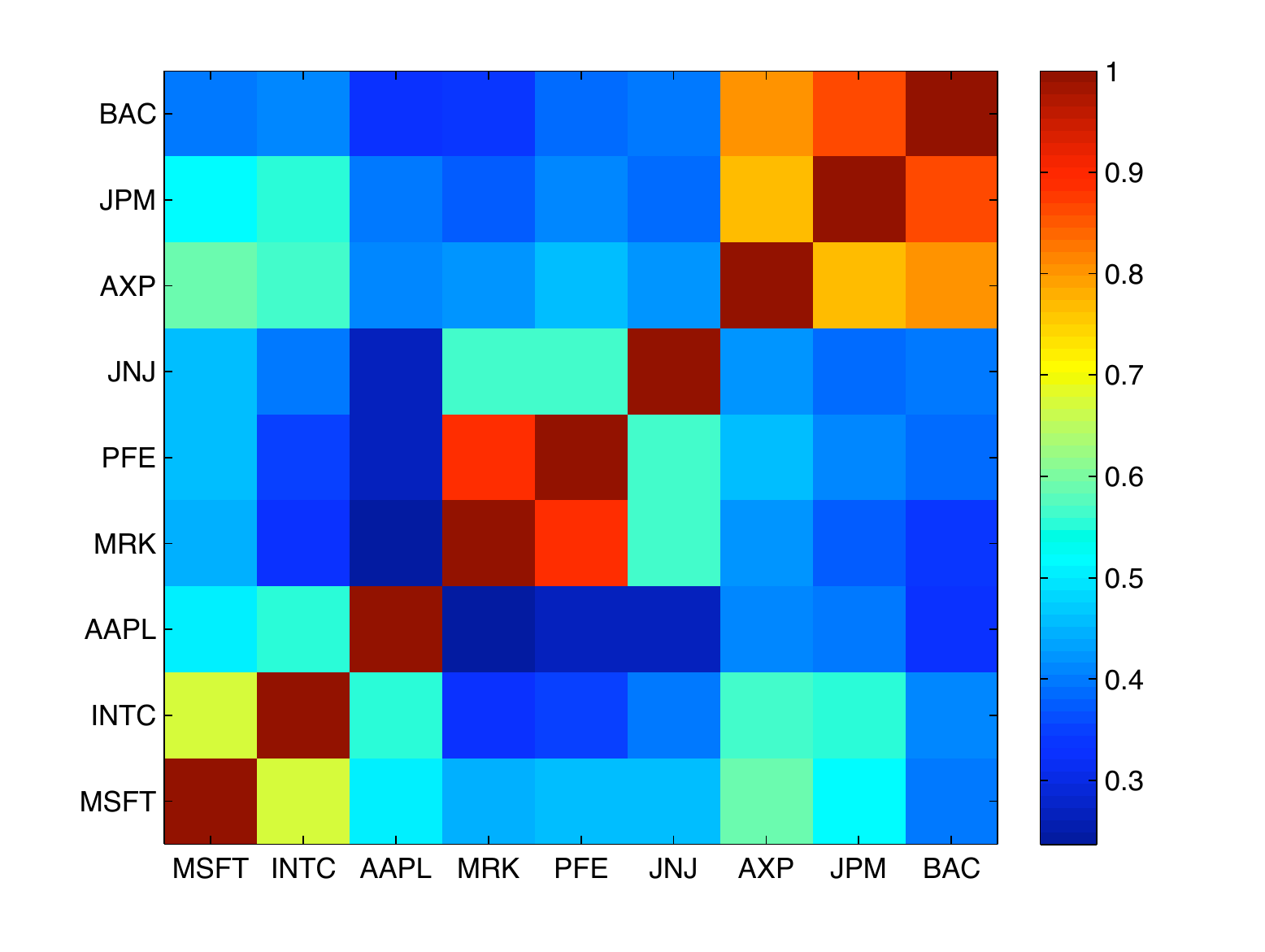}}
    \subfigure[]{\label{Fig_2_b}\includegraphics[scale=0.5]{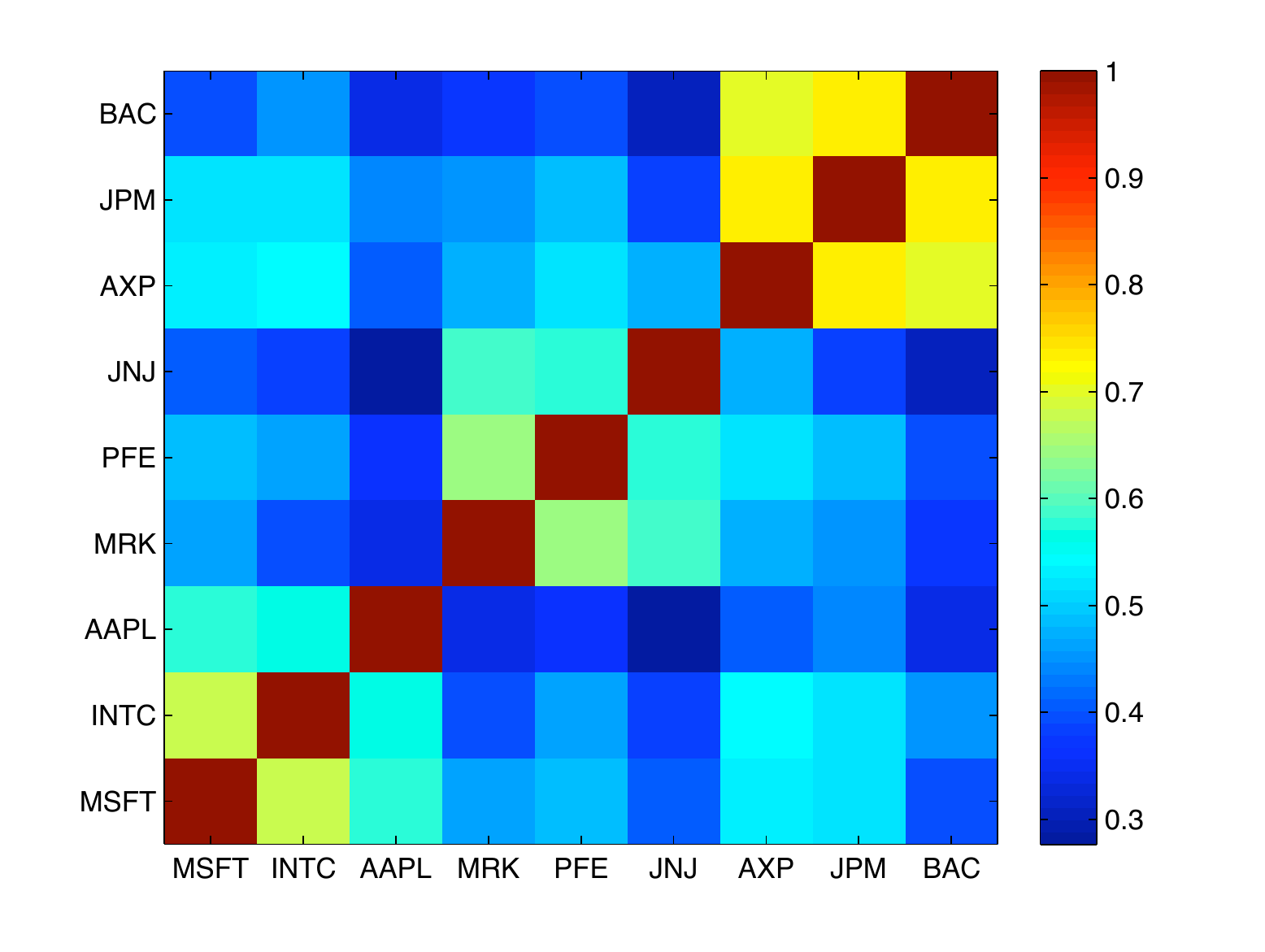}}
    \subfigure[]{\label{Fig_2_c}\includegraphics[scale=0.5]{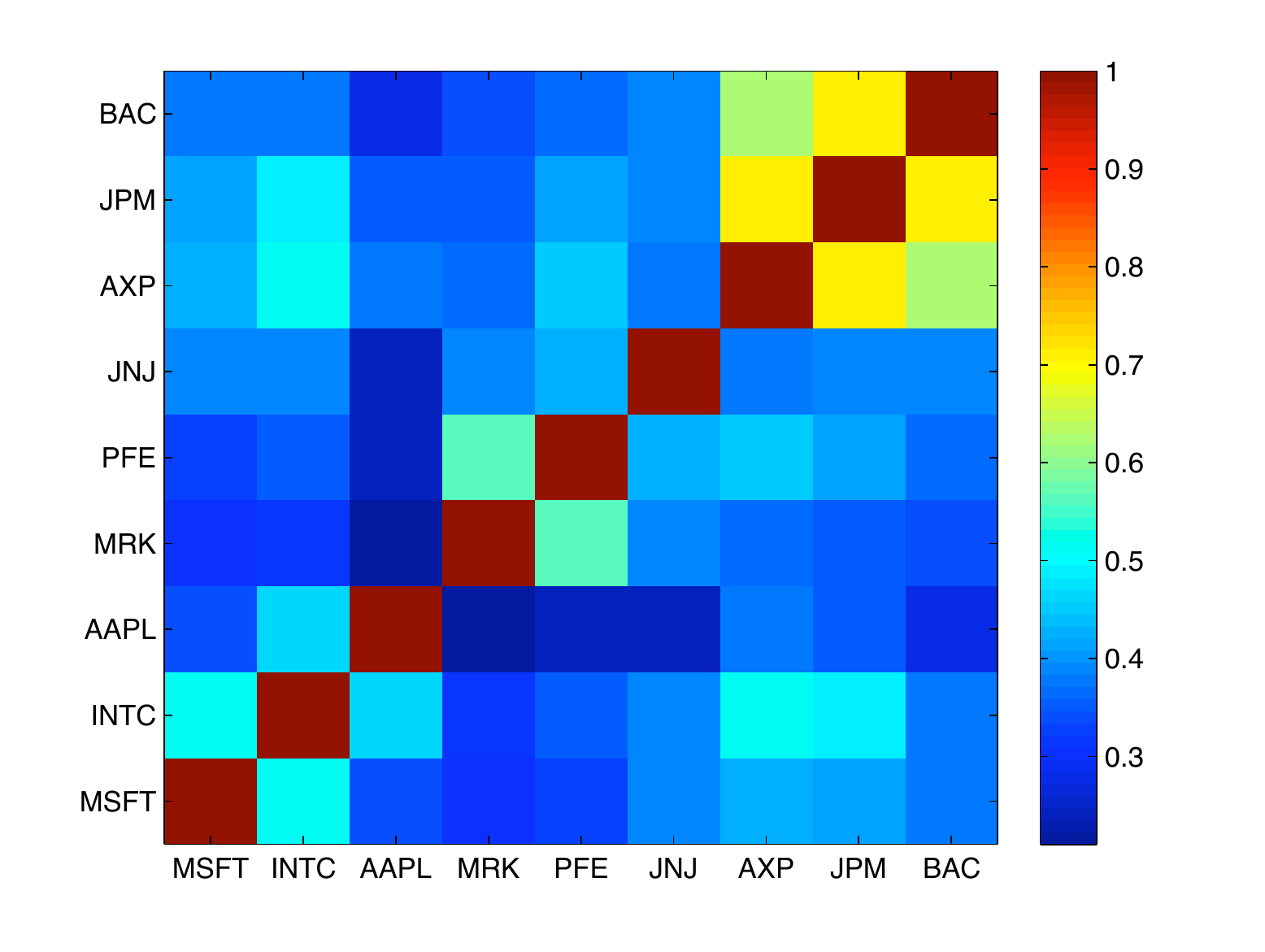}}
    \subfigure[]{\label{Fig_2_d}\includegraphics[scale=0.5]{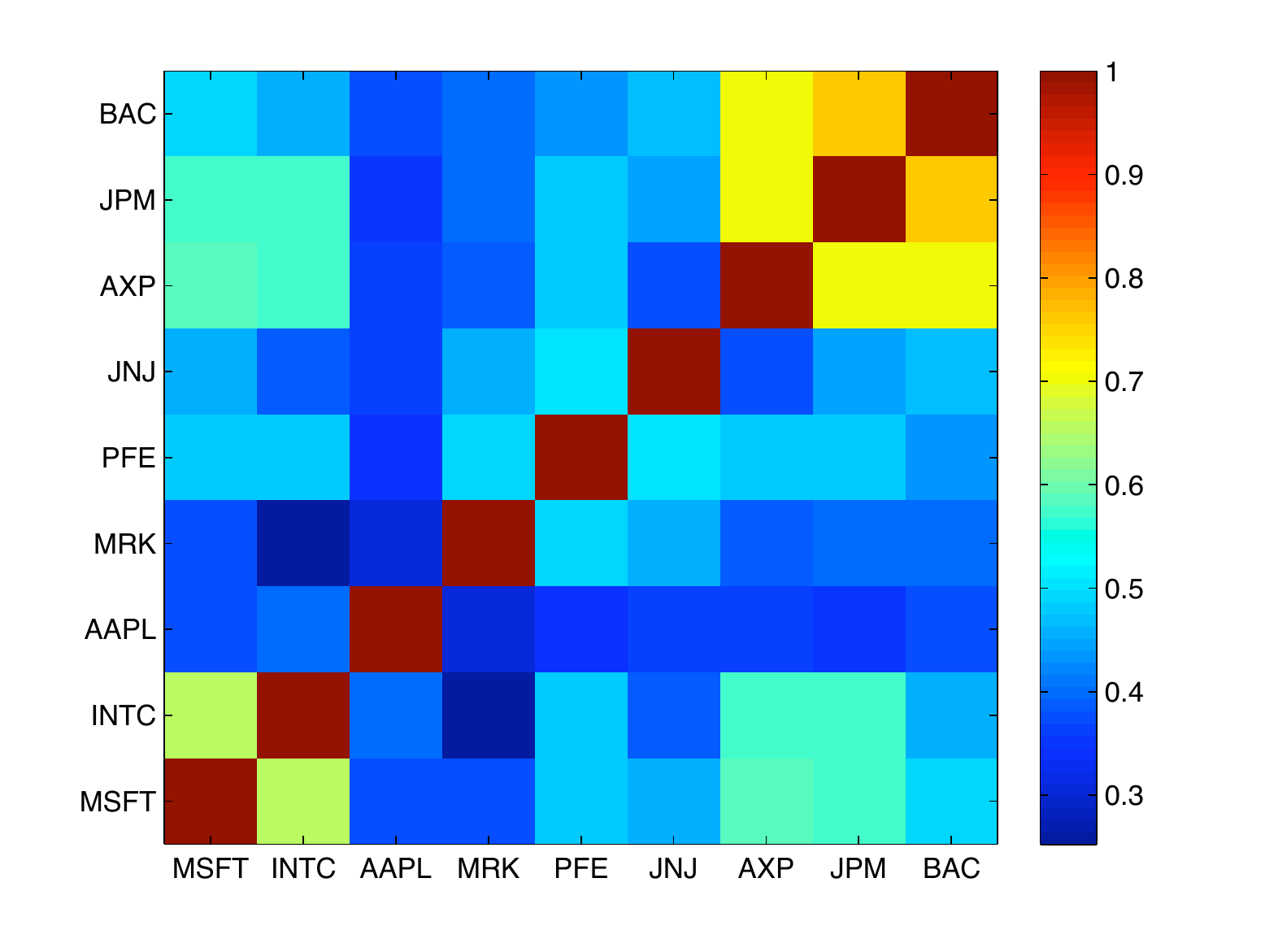}}
    \subfigure[]{\label{Fig_2_e}\includegraphics[scale=0.5]{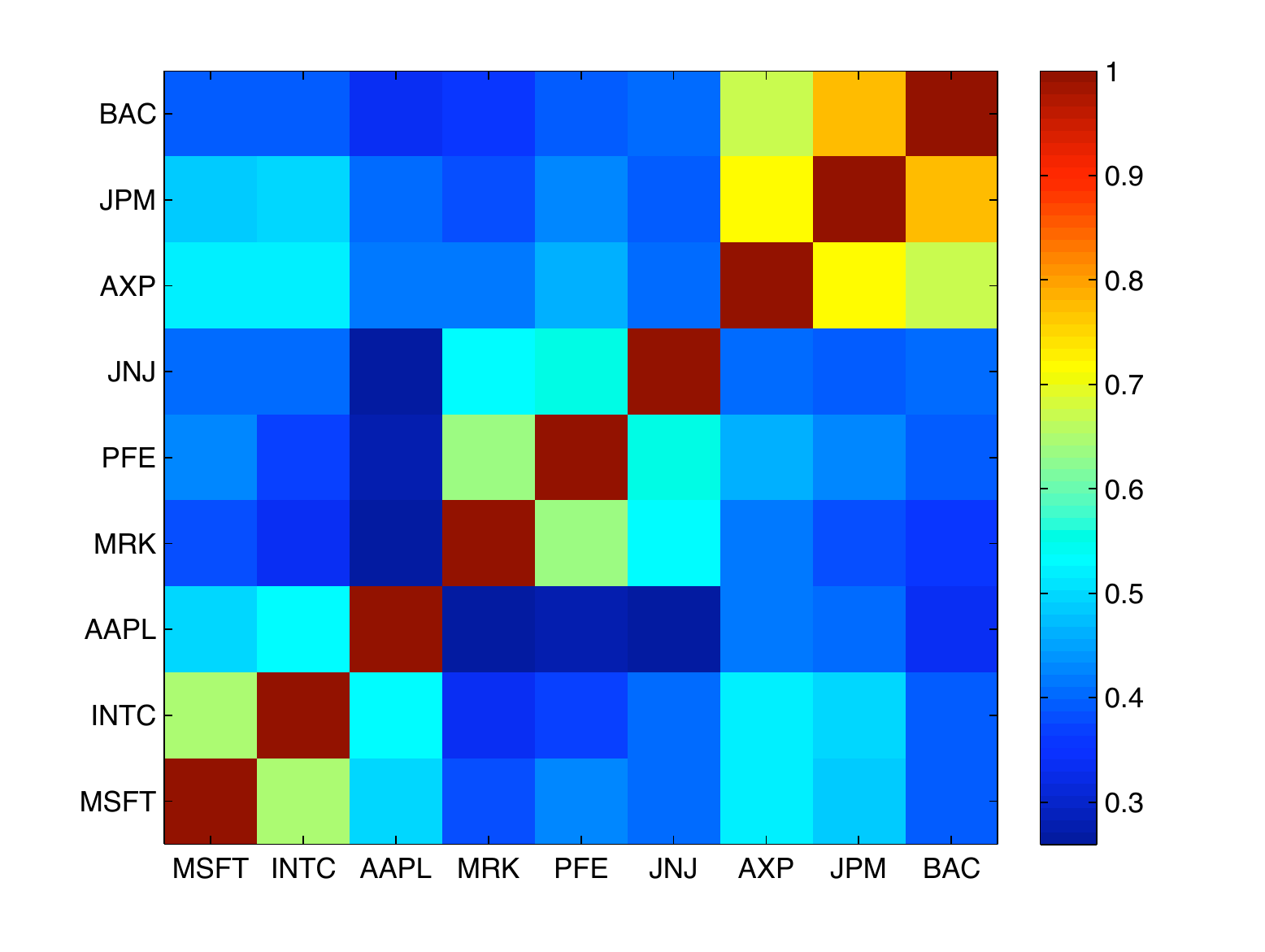}}
  \end{center}
  \caption{NASDAQ/NYSE experiment. Empirical first-order canonical correlation coefficients obtained by (a) MTCCA with exponential MT functions, (b) MTCCA with Gaussian MT-functions. (c) LCCA, (d) ICCA, and (e) KCCA. Note the three blocks of mutually high canonical correlations revealed by MTCCA and KCCA; MTCCA and KCCA better cluster companies in similar sectors: (MSFT, INTC, AAPL) - technology, (MRK, PFE, JNJ) - pharmaceuticals, (AXP, JPM, BAC) - financial.} 
  \label{Fig2Corr}
\end{figure}

Figs. \ref{Fig_2_a} and \ref{Fig_2_b} display the matrix of empirical first-order MT-canonical correlation coefficients for the exponential and Gaussian MT-functions, respectively. Figs. \ref{Fig_2_c}-\ref{Fig_2_e} show the matrix of empirical first-order canonical correlation coefficients obtained by LCCA, ICCA and KCCA, respectively. Note that MTCCA  and KCCA better cluster companies in similar sectors: (MSFT, INTC, AAPL) - technology, (MRK, PFE, JNJ) - pharmaceuticals, (AXP, JPM, BAC) - financial. In this example, the $p$-values associated with all empirical first-order canonical correlation coefficients were less than $0.01$. 

\begin{figure}[htbp!]
  \begin{center}
    \frame{{\subfigure{\label{GraphModel_3_MTCCA_EXP_1}\includegraphics[scale=0.25]{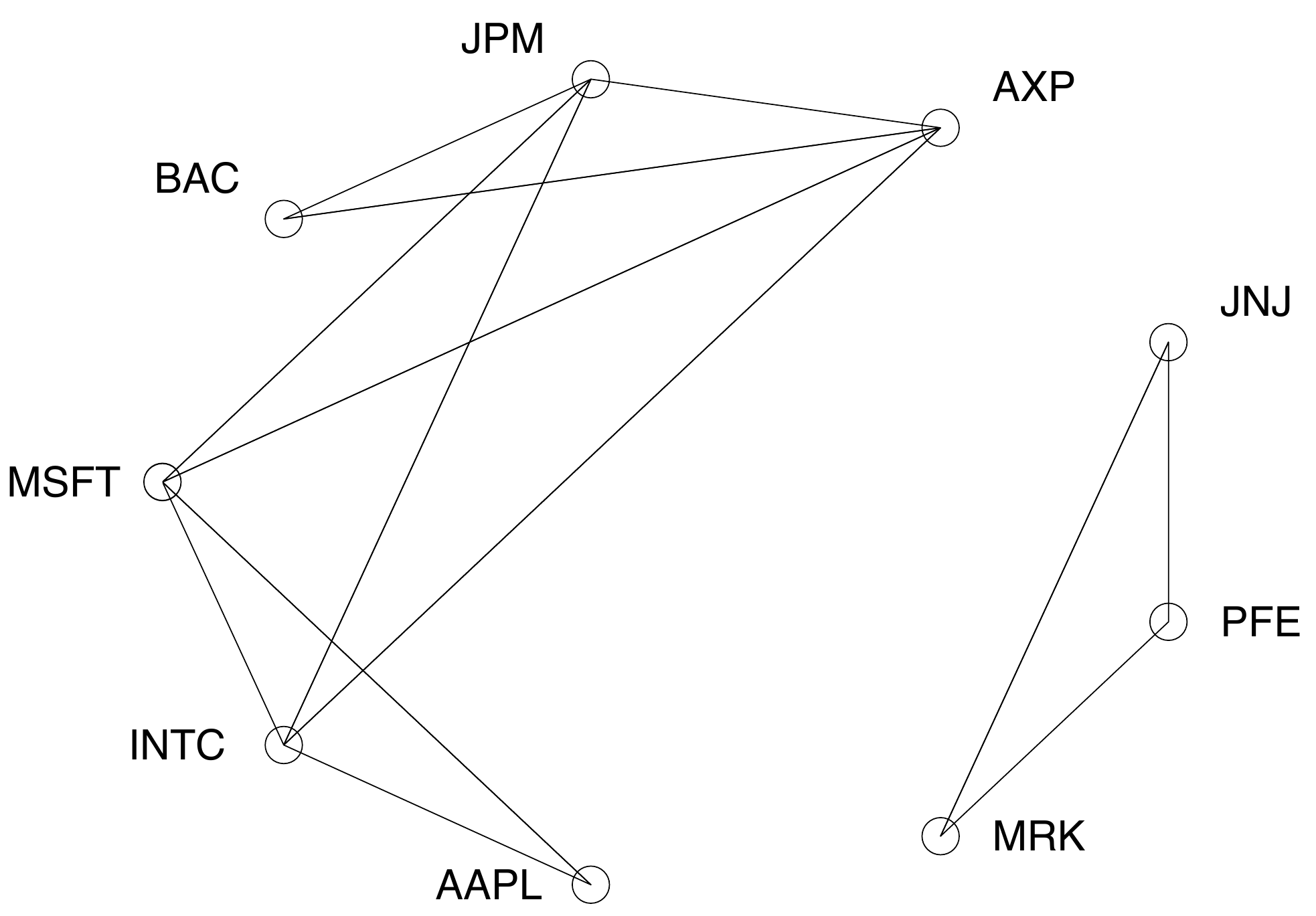}}}}
    \frame{{\subfigure{\label{GraphModel_3_MTCCA_EXP_2}\includegraphics[scale=0.25]{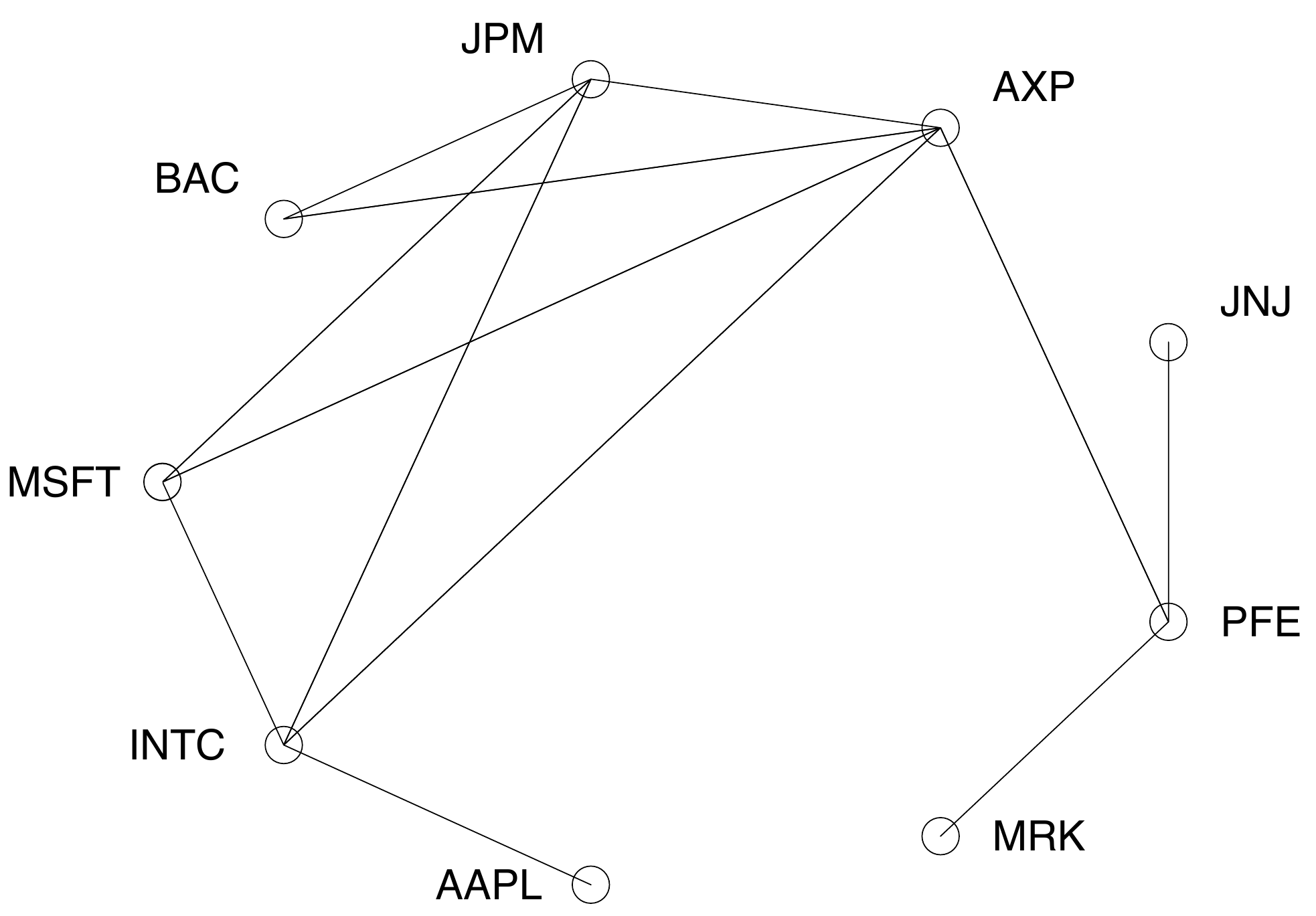}}}}
    \vspace{0.15cm}
    \frame{{\subfigure{\label{GraphModel_3_MTCCA_EXP_3}\includegraphics[scale=0.25]{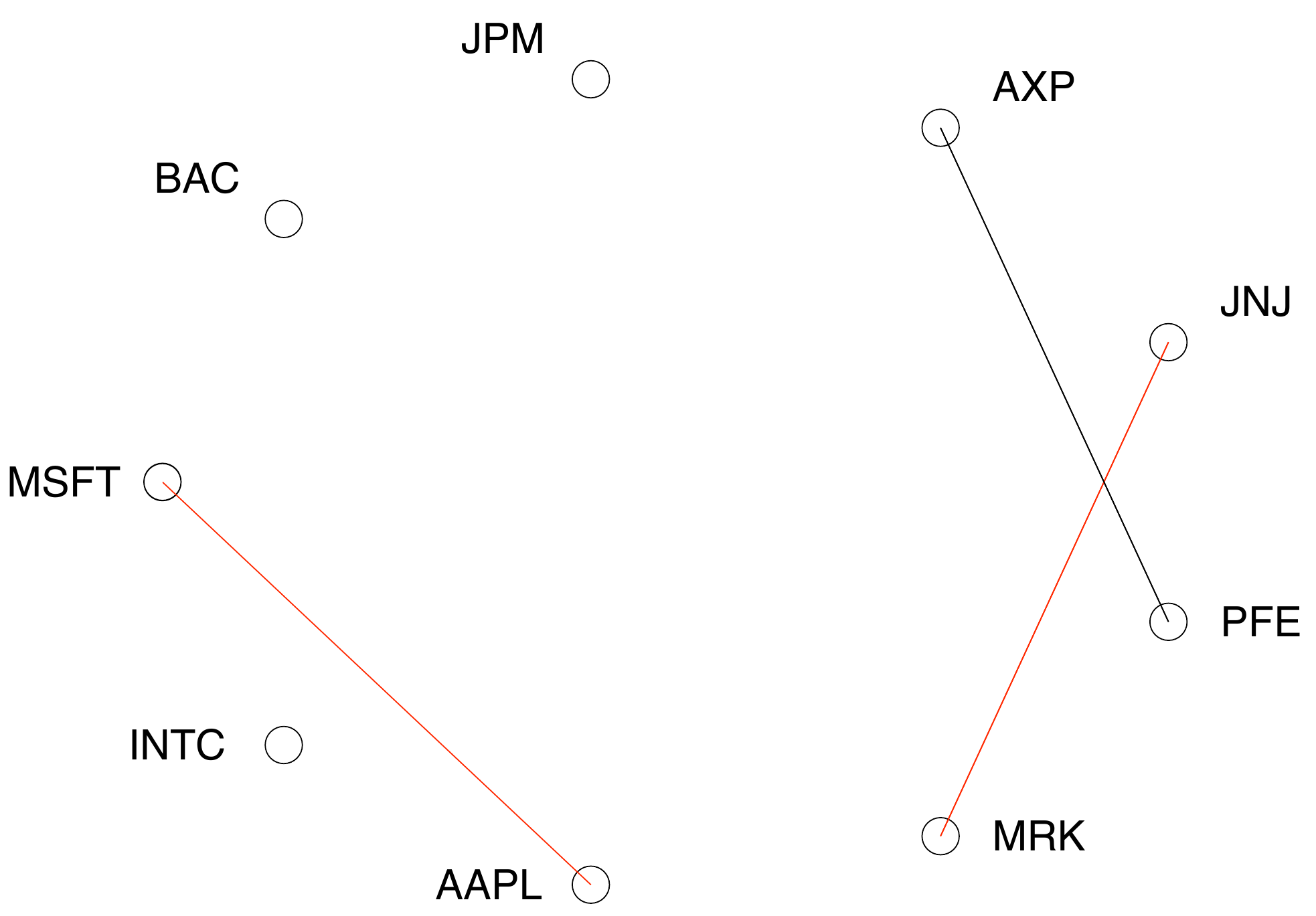}}}}
    \frame{{\subfigure{\label{GraphModel_3_MTCCA_EXP_4}\includegraphics[scale=0.25]{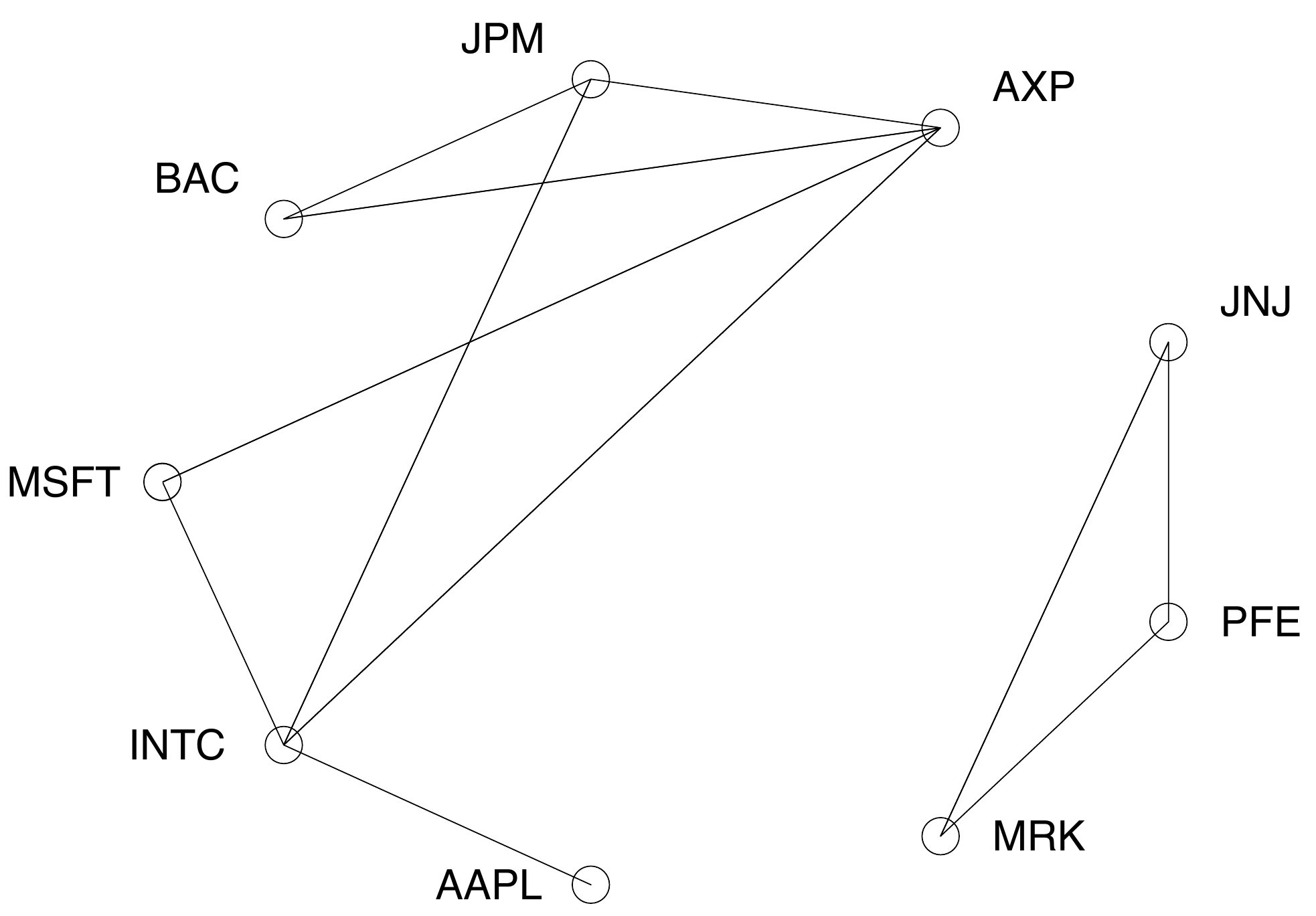}}}}
    \frame{{\subfigure{\label{GraphModel_3_MTCCA_EXP_5}\includegraphics[scale=0.25]{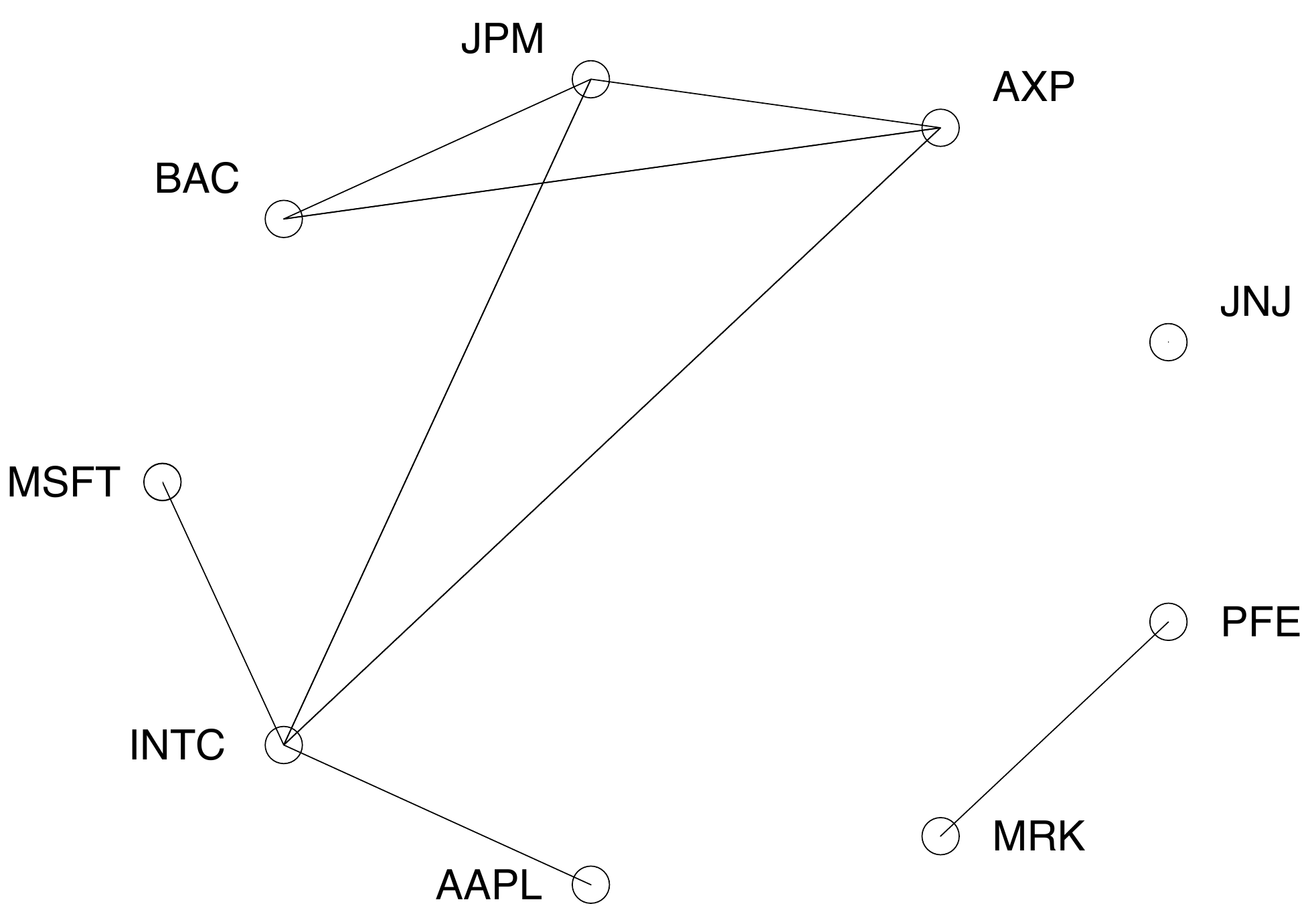}}}}
    \vspace{0.15cm}
    \frame{{\subfigure{\label{GraphModel_3_MTCCA_EXP_6}\includegraphics[scale=0.25]{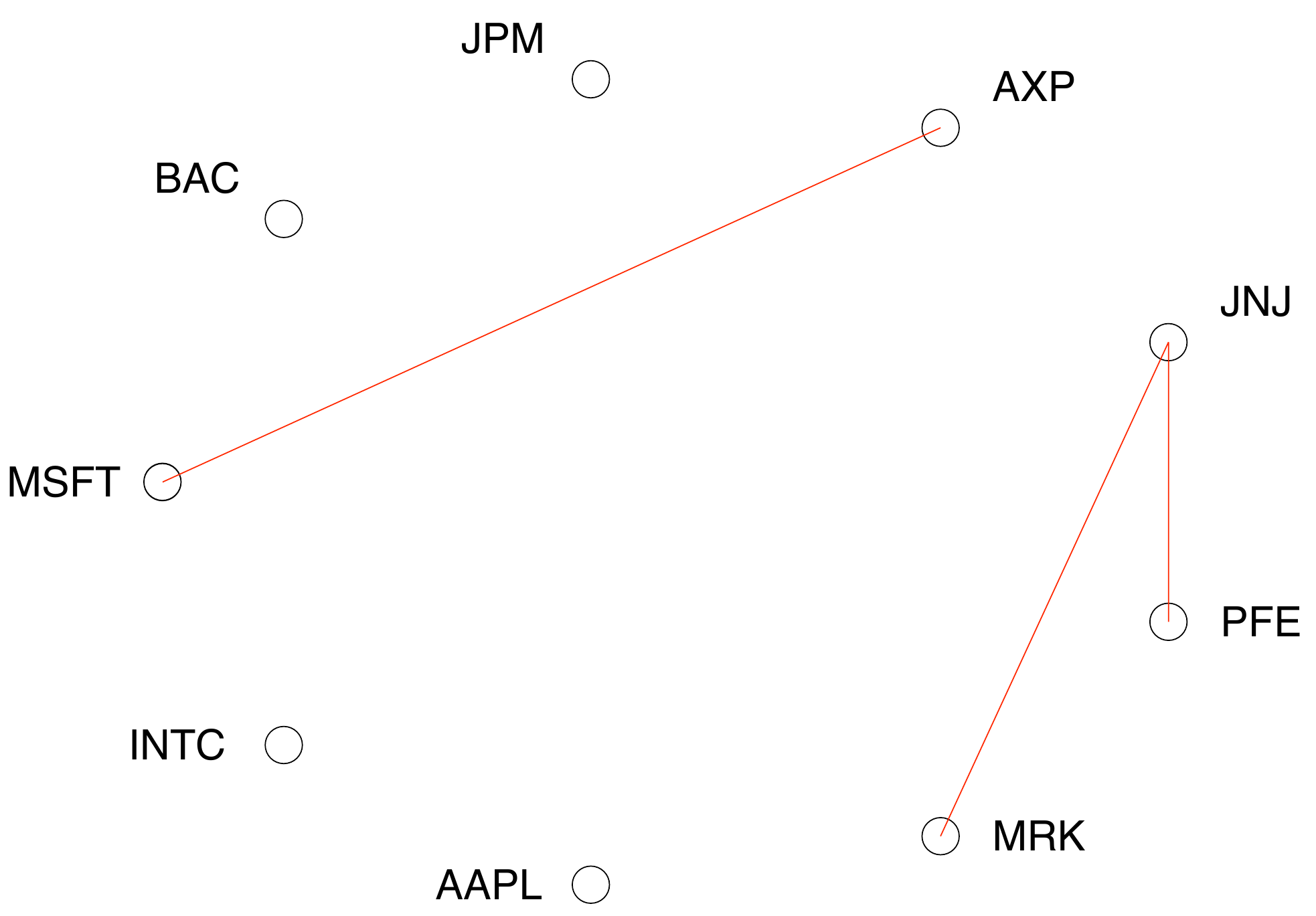}}}}
    \frame{{\subfigure{\label{GraphModel_3_MTCCA_EXP_7}\includegraphics[scale=0.25]{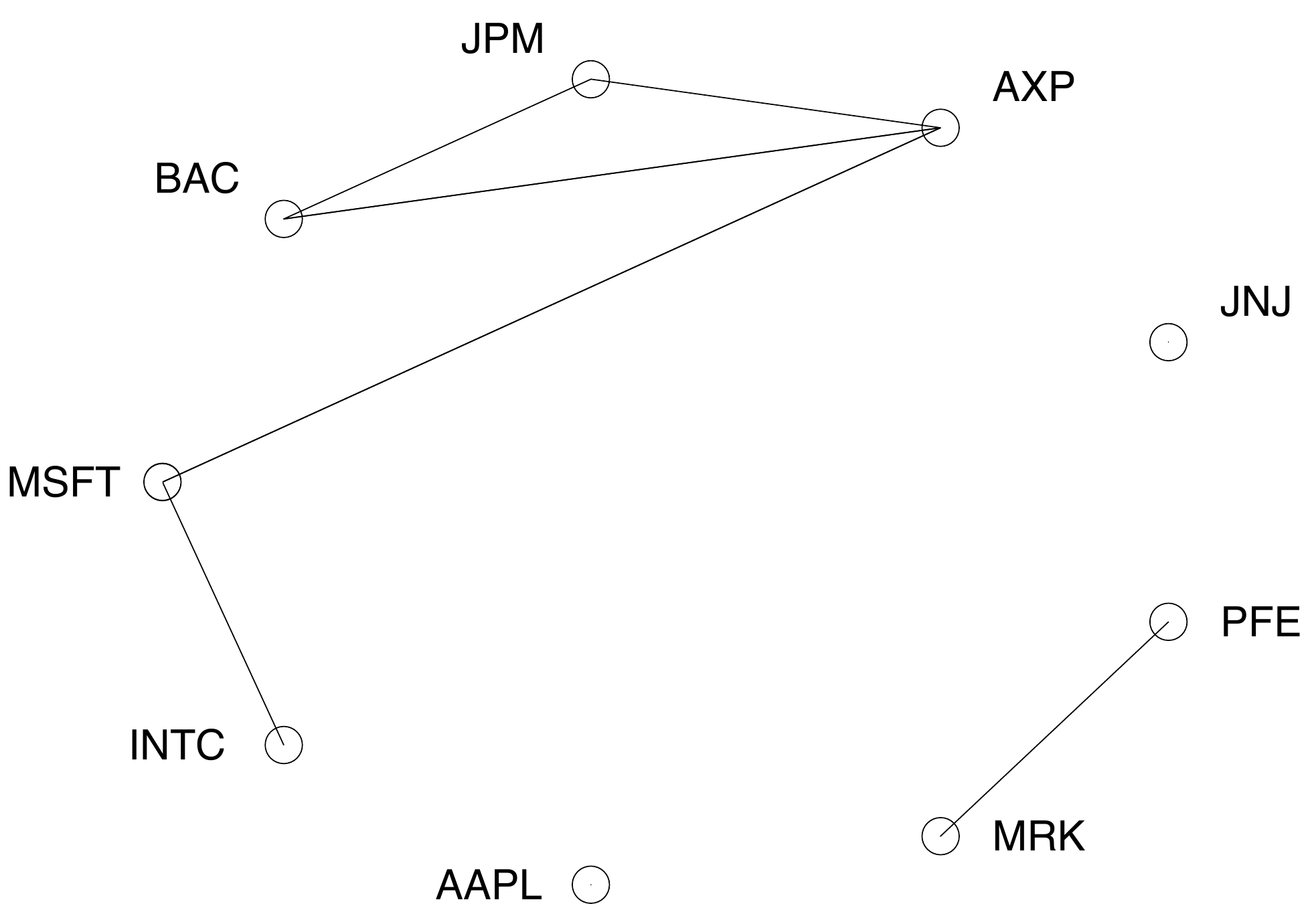}}}}
    \frame{{\subfigure{\label{GraphModel_3_MTCCA_EXP_8}\includegraphics[scale=0.25]{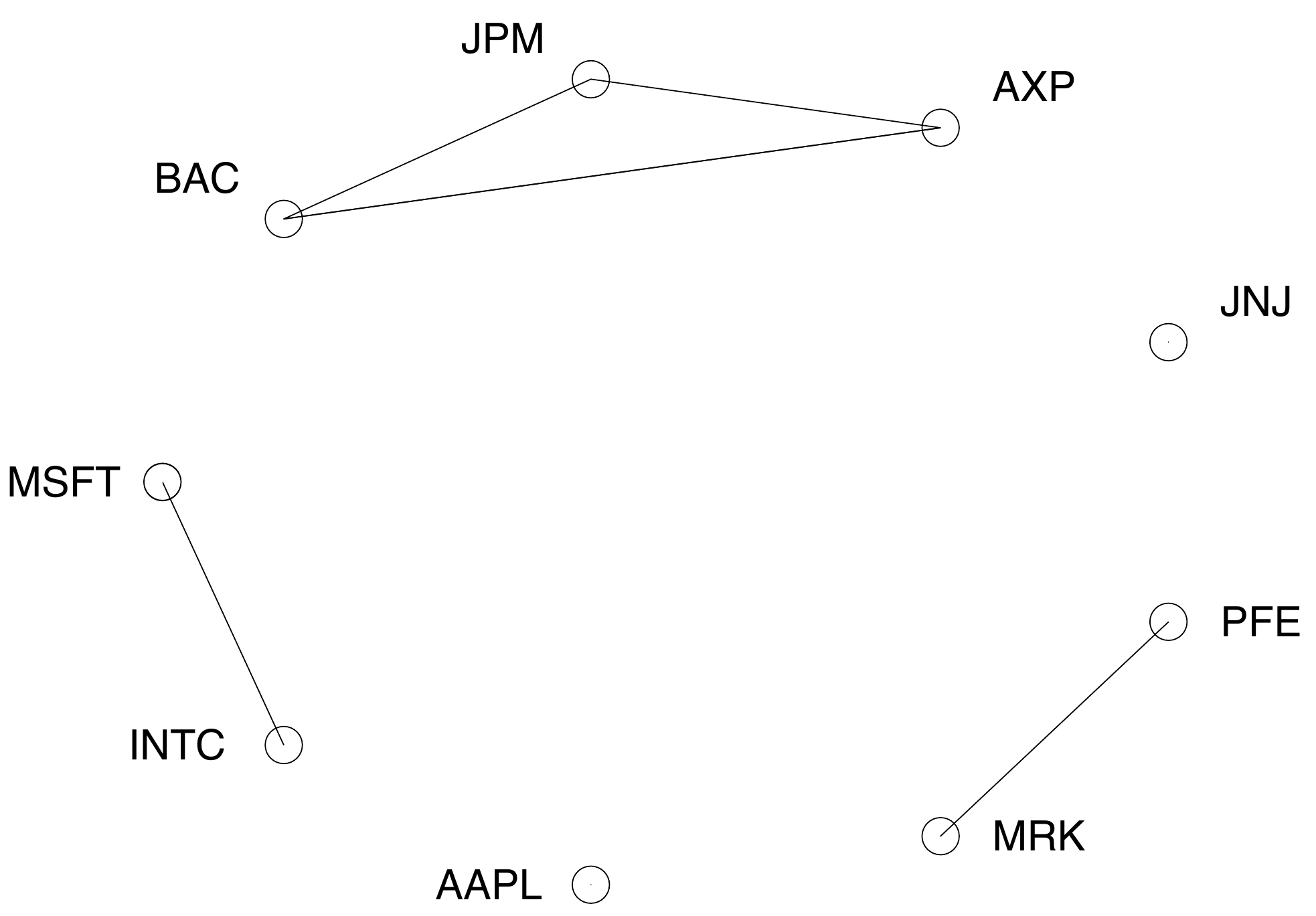}}}}
    \frame{{\subfigure{\label{GraphModel_3_MTCCA_EXP_9}\includegraphics[scale=0.25]{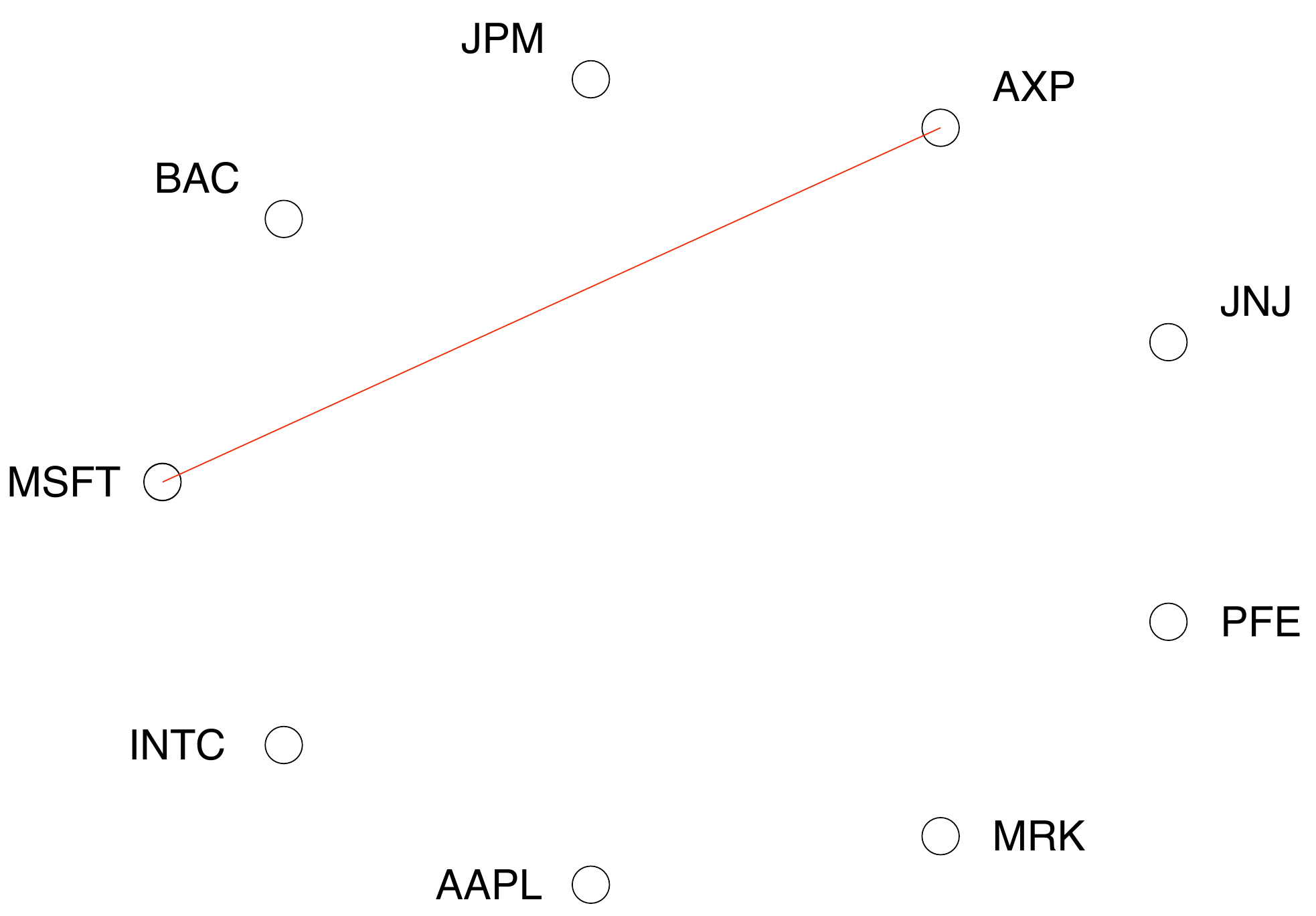}}}}
  \end{center}
  \caption{NASDAQ/NYSE experiment. {\bf{Left column}:} The graphical models selected by MTCCA with exponential MT-functions for $\lambda=0.5,0.55,0.58$. {\bf{Middle column}:} The closest graphs selected by LCCA. {\bf{Right column}:} The symmetric difference graphs: the red lines indicate edges found by MTCCA and not by LCCA, and vice-versa for the black lines. For these values of $\lambda$, exponential MTCCA detects more dependencies than LCCA: the MTCCA graph has more edges than the closest LCCA graph.}
\label{GraphModFinancial1}
\end{figure}

\begin{figure}[htbp!]
  \begin{center}
    \frame{{\subfigure{\label{GraphModel_3_MTCCA_EXP_1}\includegraphics[scale=0.25]{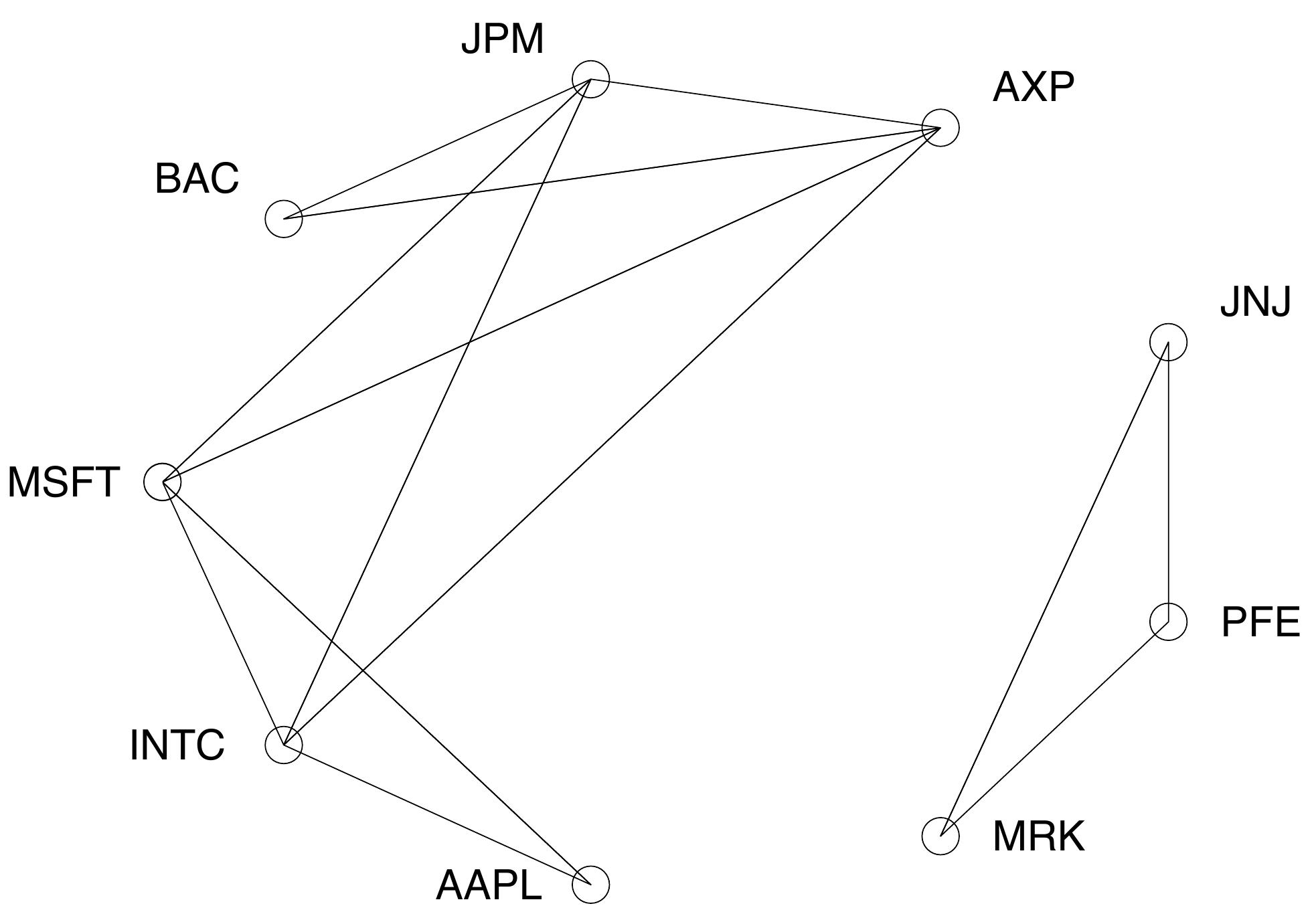}}}}
    \frame{{\subfigure{\label{GraphModel_3_MTCCA_EXP_2}\includegraphics[scale=0.25]{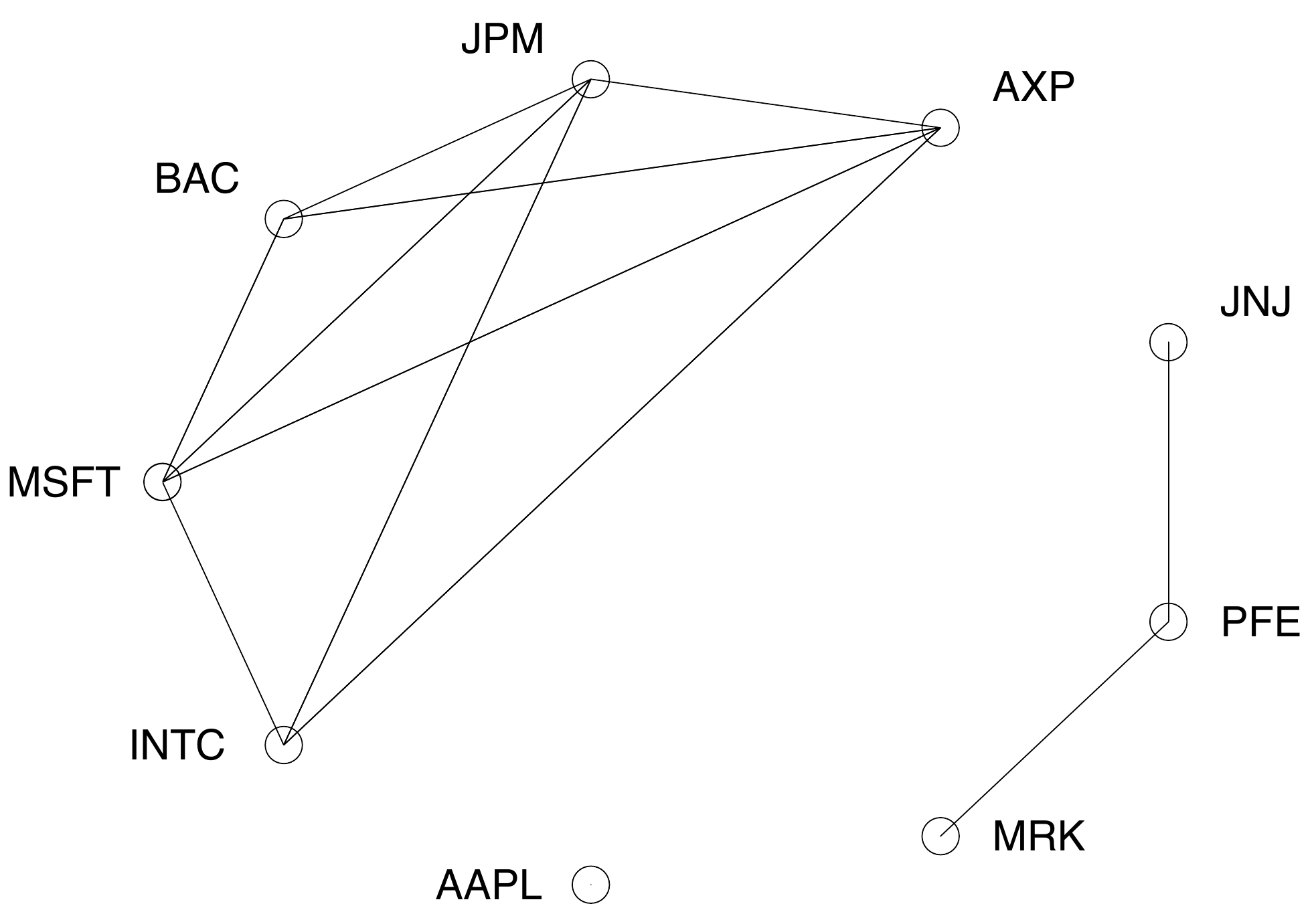}}}}
    \vspace{0.15cm}
    \frame{{\subfigure{\label{GraphModel_3_MTCCA_EXP_3}\includegraphics[scale=0.25]{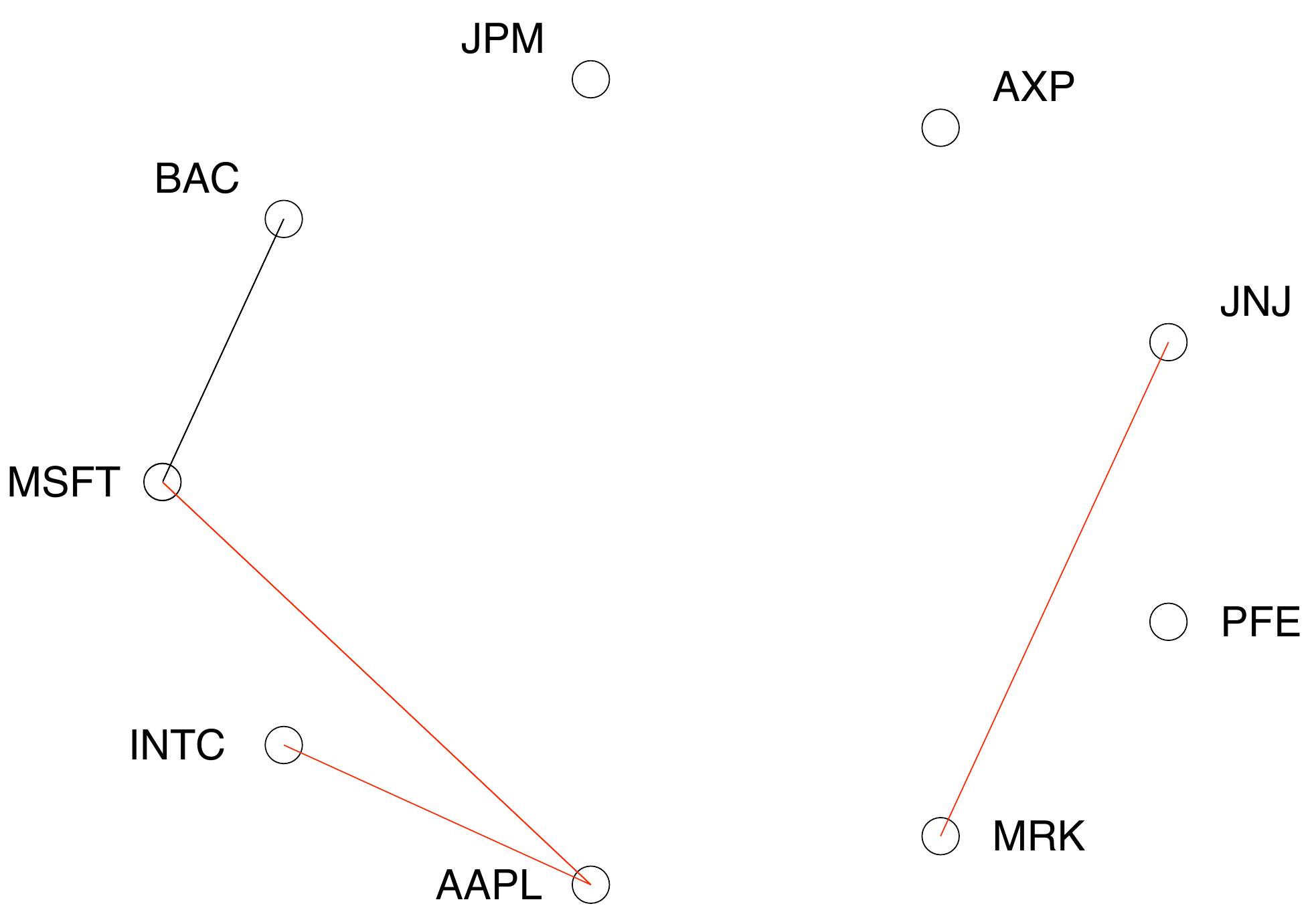}}}}
    \frame{{\subfigure{\label{GraphModel_3_MTCCA_EXP_4}\includegraphics[scale=0.25]{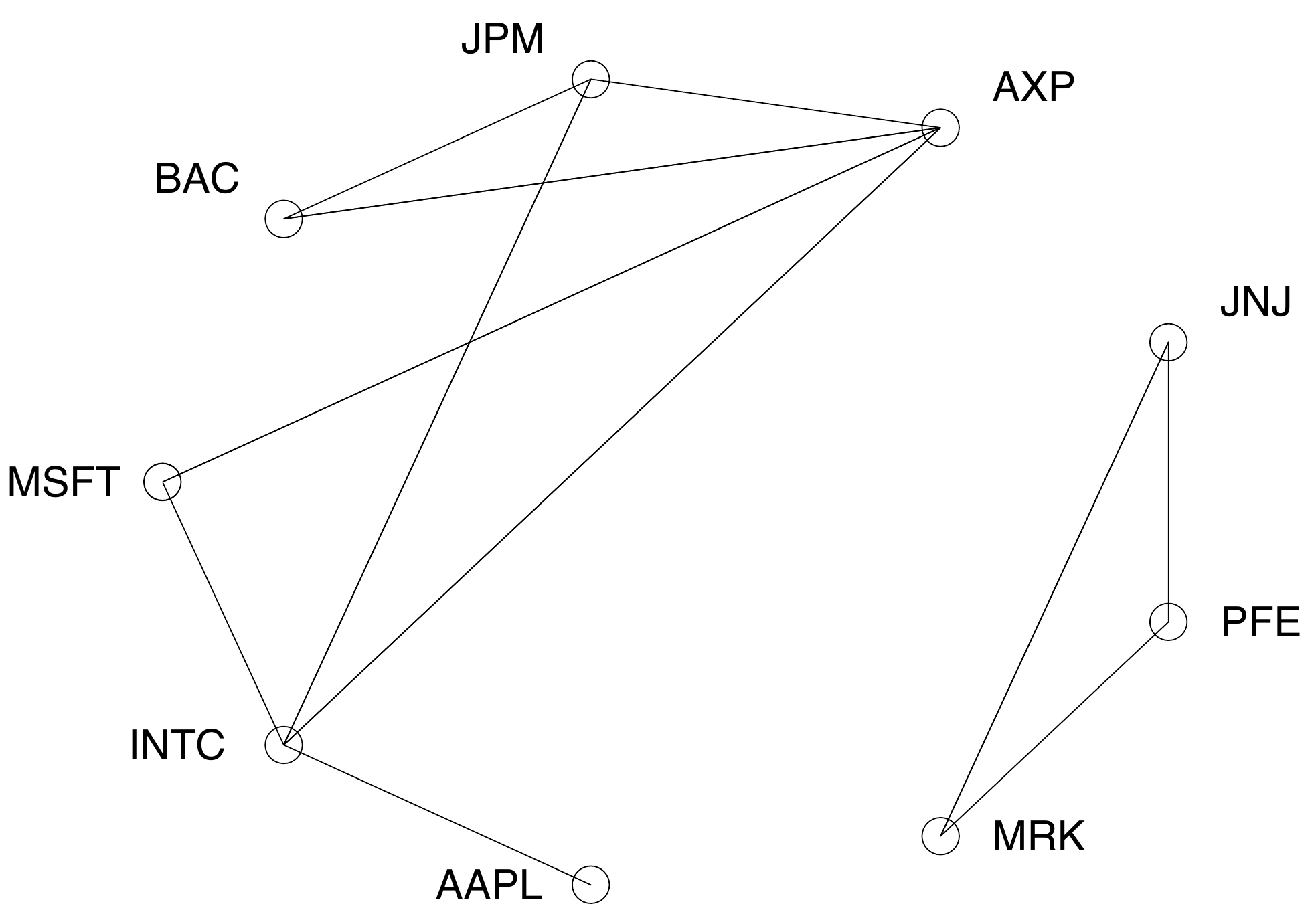}}}}
    \frame{{\subfigure{\label{GraphModel_3_MTCCA_EXP_5}\includegraphics[scale=0.25]{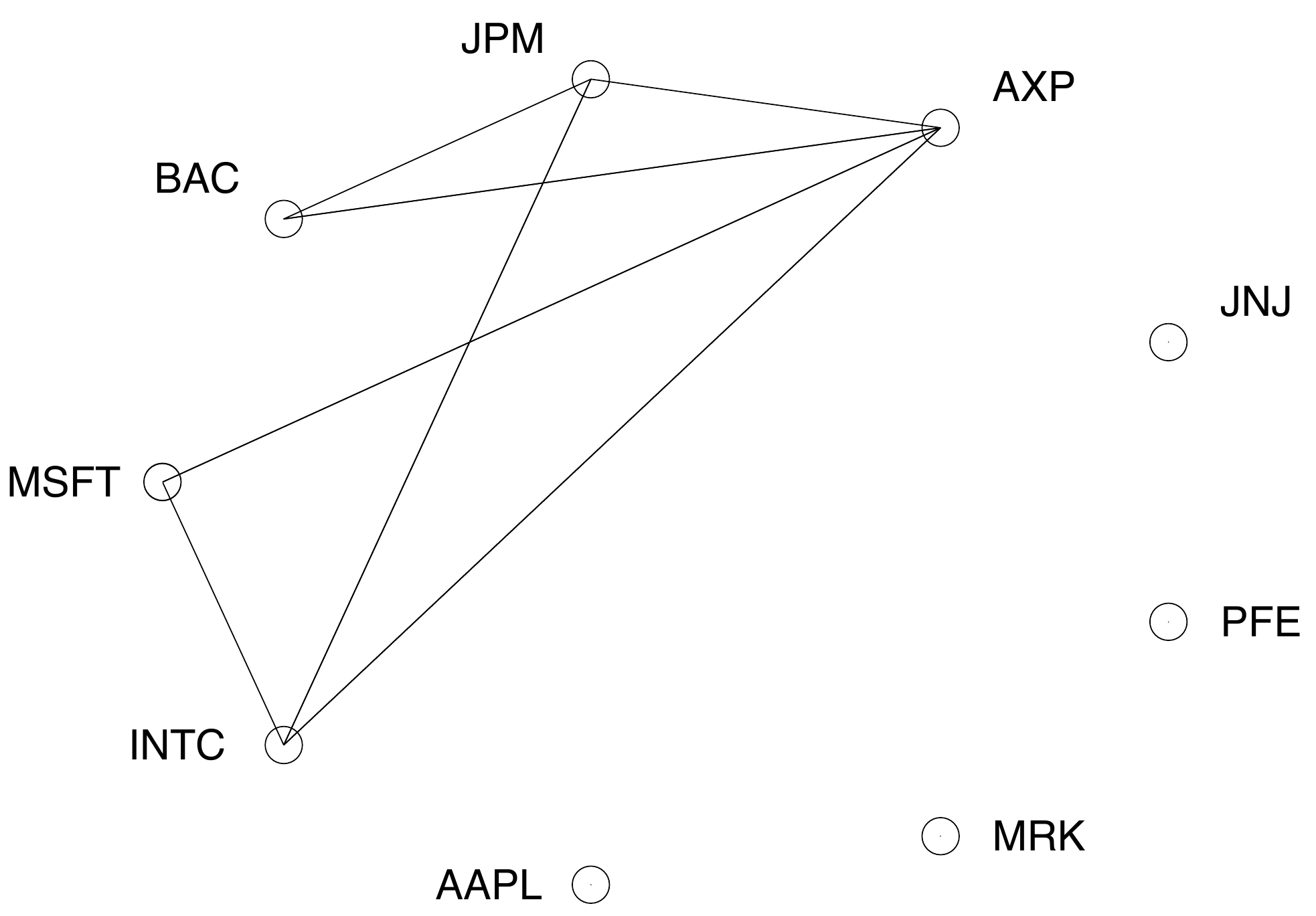}}}}
    \vspace{0.15cm}
    \frame{{\subfigure{\label{GraphModel_3_MTCCA_EXP_6}\includegraphics[scale=0.25]{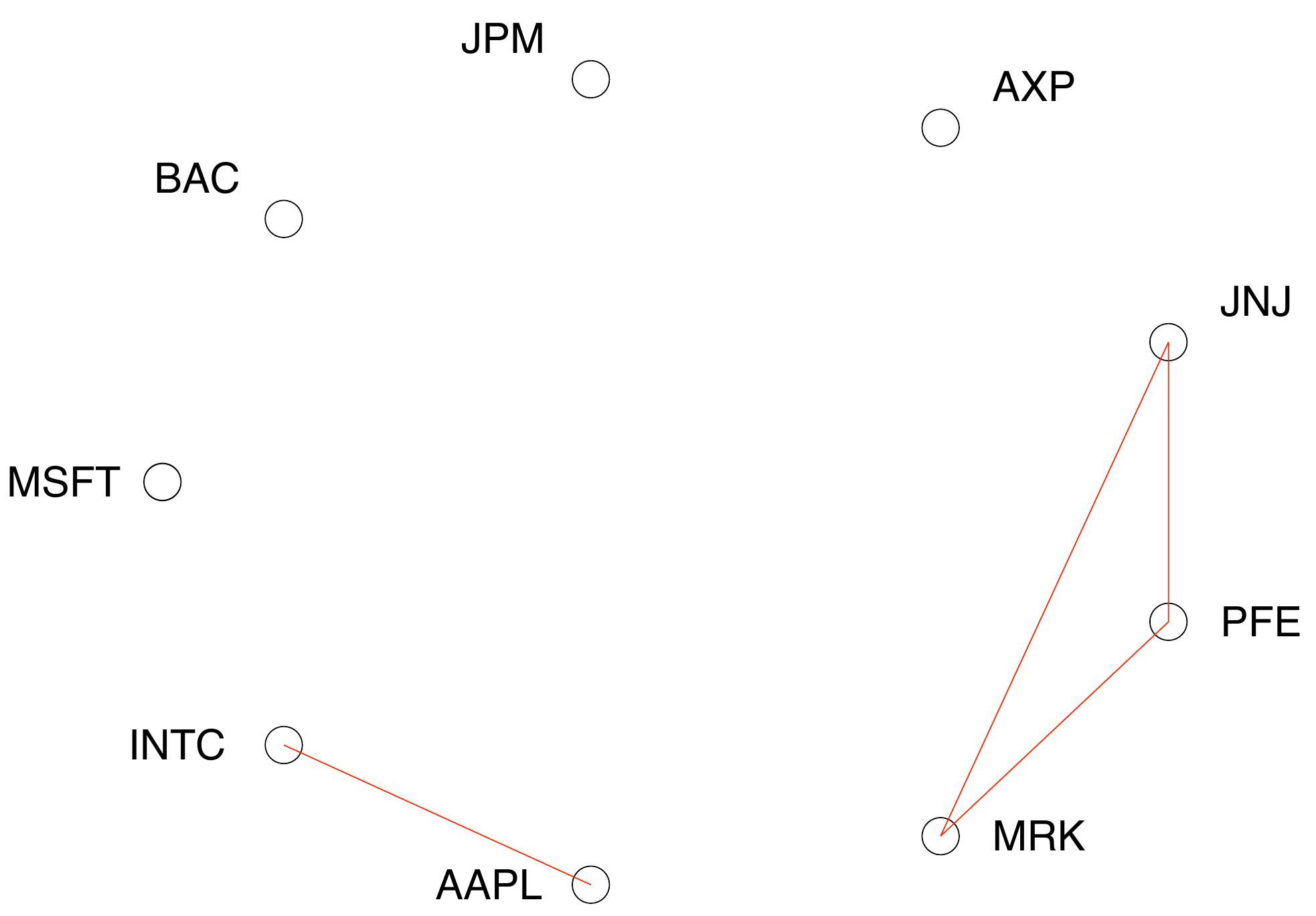}}}}
    \frame{{\subfigure{\label{GraphModel_3_MTCCA_EXP_7}\includegraphics[scale=0.25]{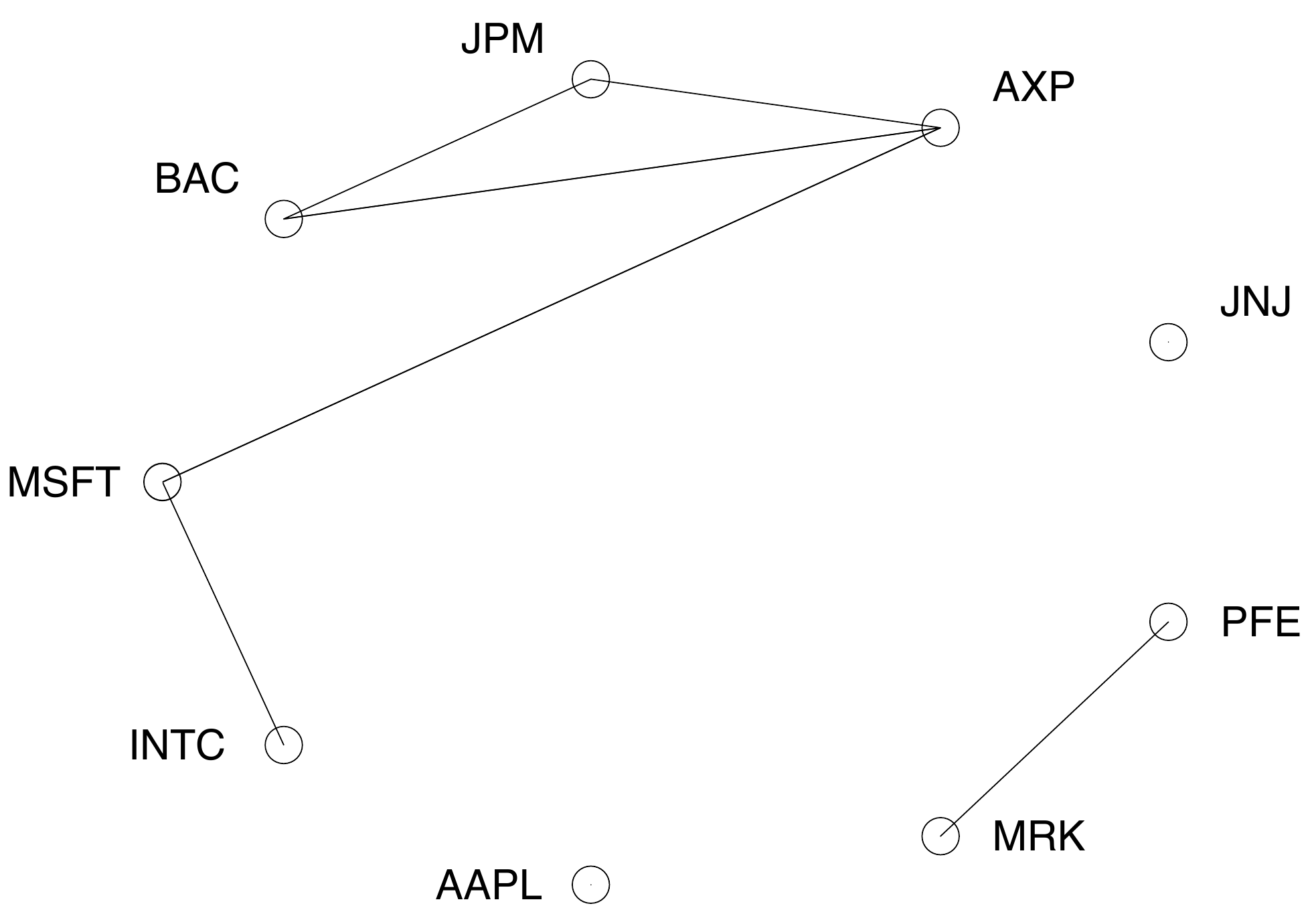}}}}
    \frame{{\subfigure{\label{GraphModel_3_MTCCA_EXP_8}\includegraphics[scale=0.25]{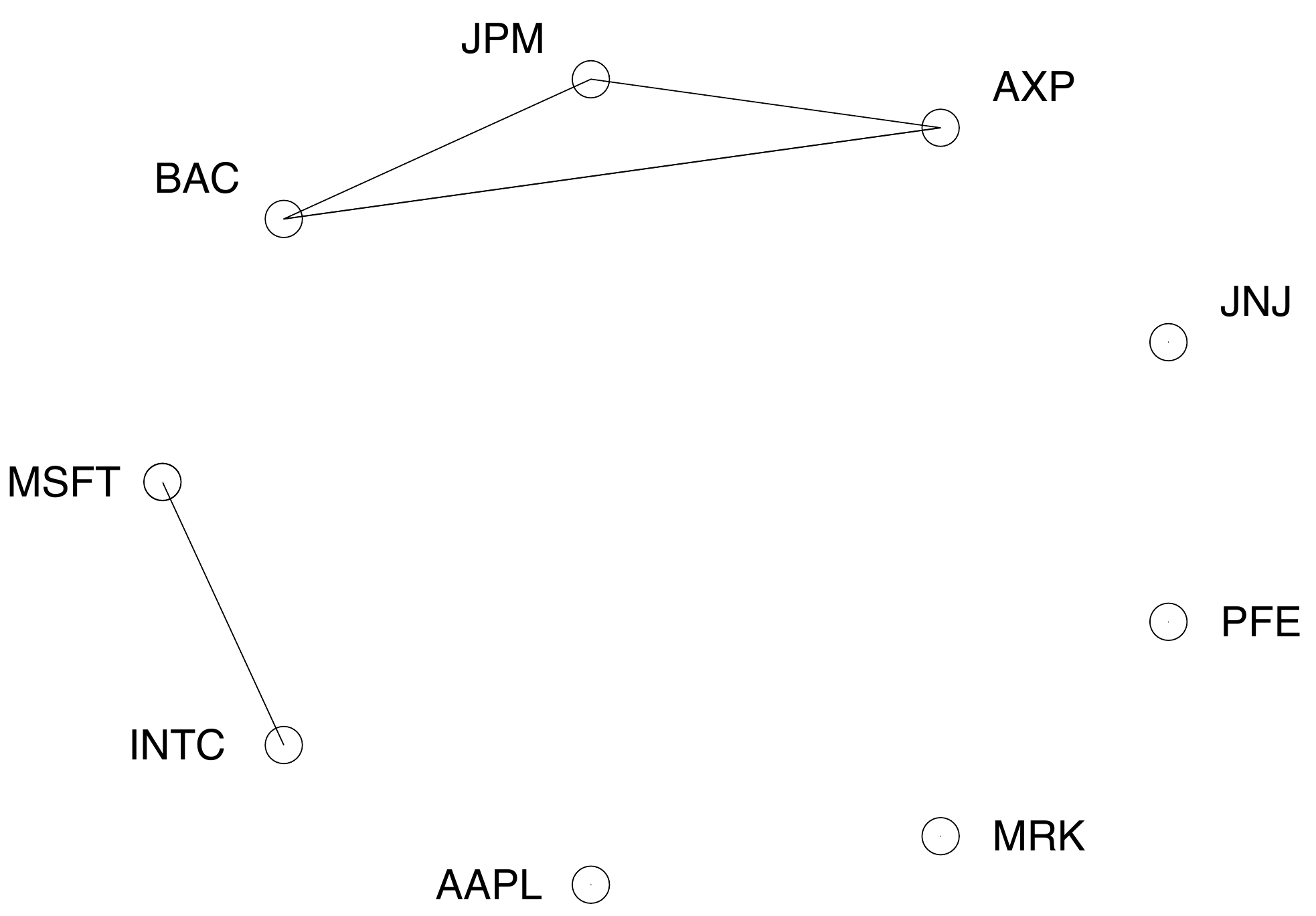}}}}
    \frame{{\subfigure{\label{GraphModel_3_MTCCA_EXP_9}\includegraphics[scale=0.25]{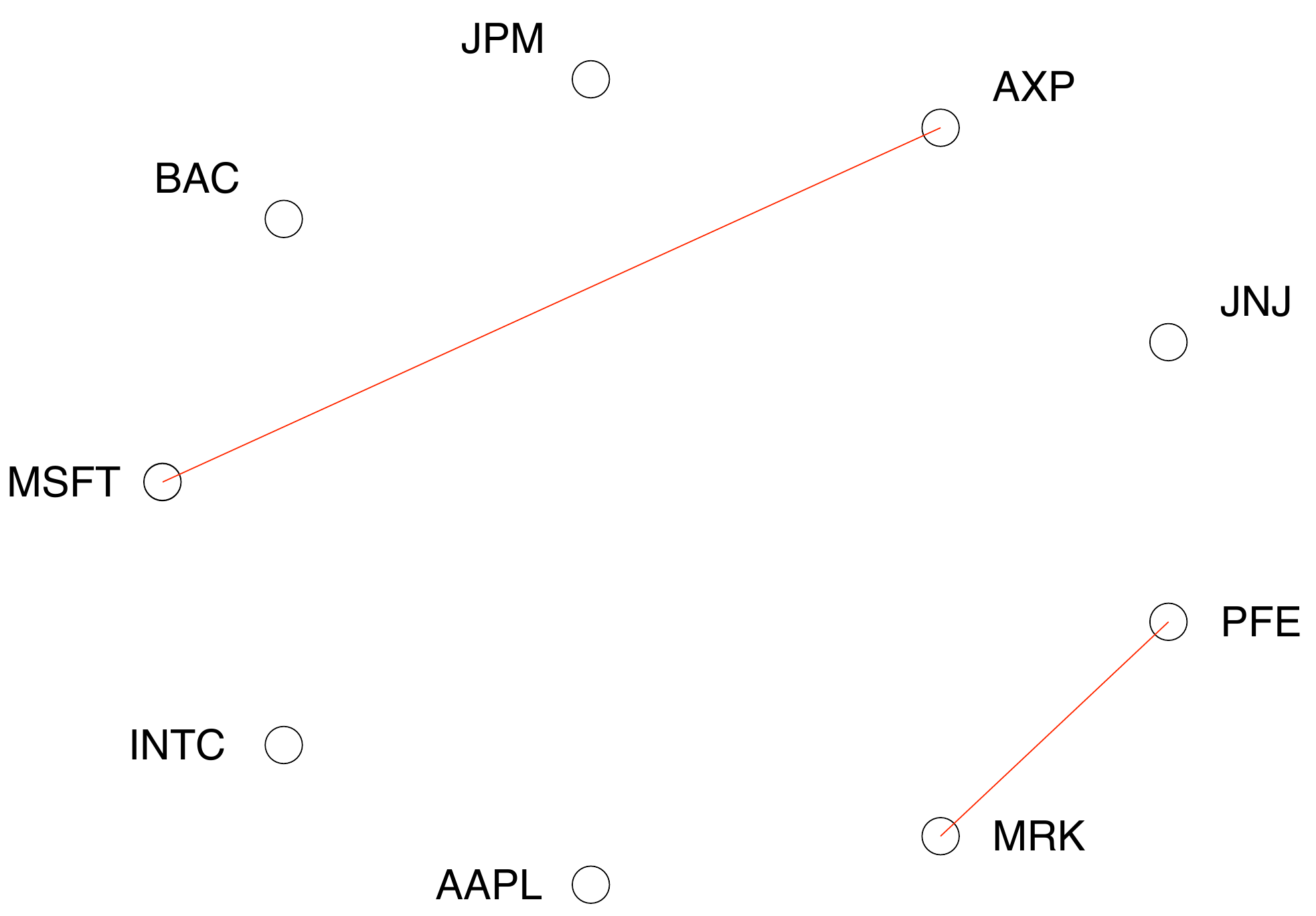}}}}
  \end{center}
  \caption{NASDAQ/NYSE experiment. {\bf{Left column}:} The graphical models selected by MTCCA with exponential MT-functions for $\lambda=0.5,0.55,0.58$. {\bf{Middle column}:} The closest graphs selected by ICCA. {\bf{Right column}:} The symmetric difference graphs: the red lines indicate edges found by MTCCA and not by ICCA, and vice-versa for the black lines. For these values of $\lambda$, exponential MTCCA detects more dependencies than ICCA: the MTCCA graph has more edges than the closest ICCA graph.}
\label{GraphModFinancial2}
\end{figure}

\begin{figure}[htbp!]
  \begin{center}
    \frame{{\subfigure{\label{GraphModel_3_MTCCA_EXP_1}\includegraphics[scale=0.25]{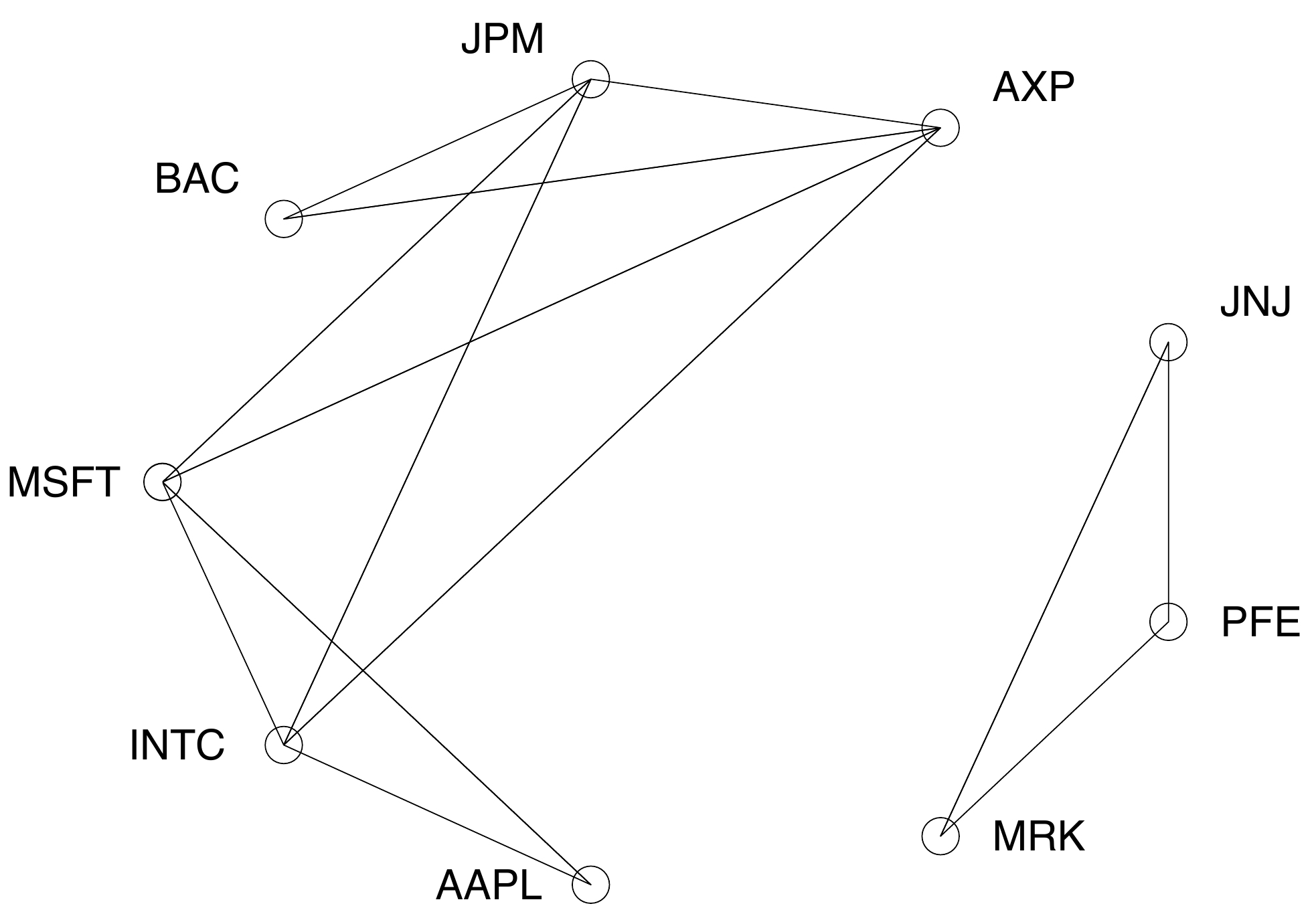}}}}
    \frame{{\subfigure{\label{GraphModel_3_MTCCA_EXP_2}\includegraphics[scale=0.25]{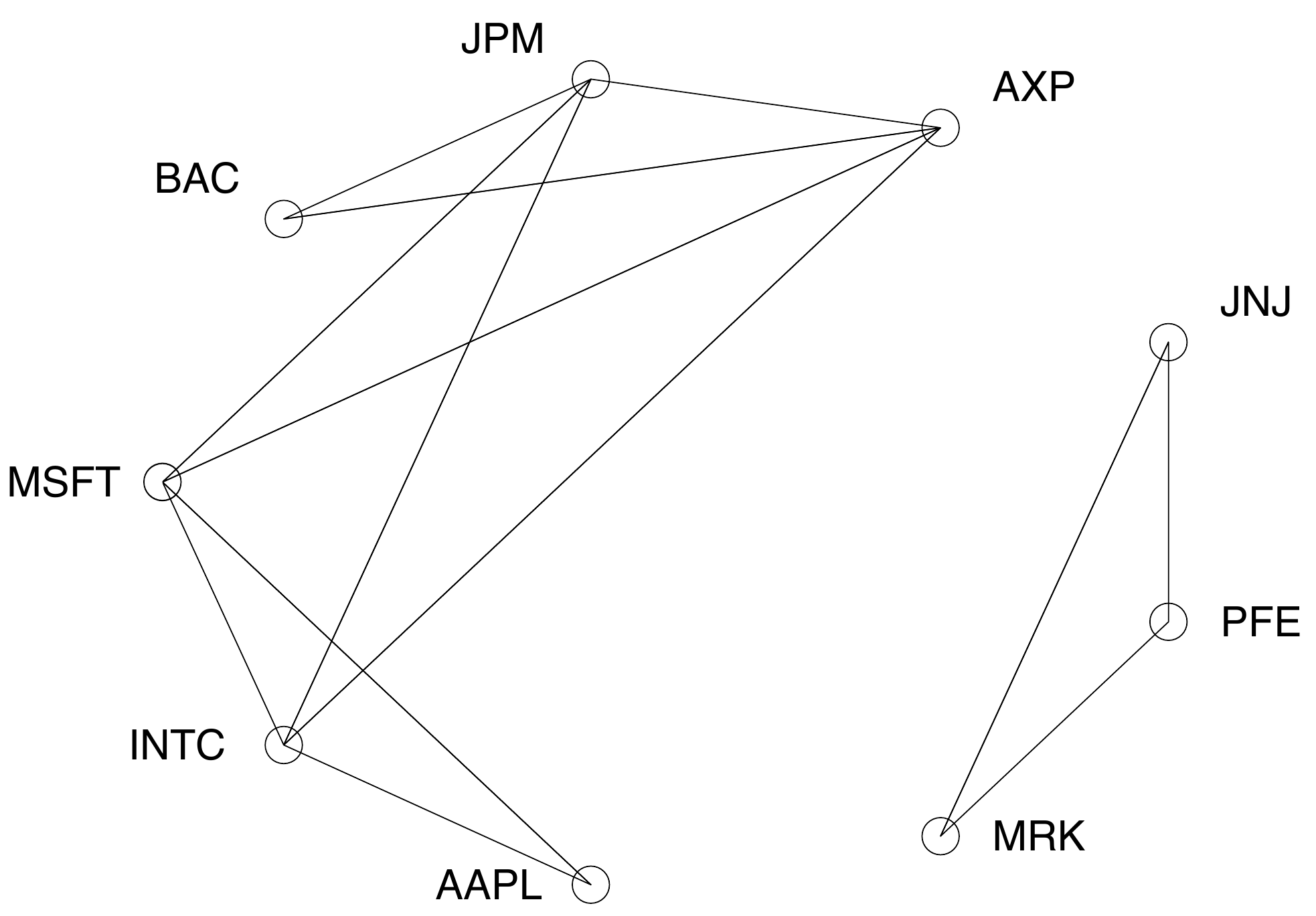}}}}
    \vspace{0.15cm}
    \frame{{\subfigure{\label{GraphModel_3_MTCCA_EXP_3}\includegraphics[scale=0.25]{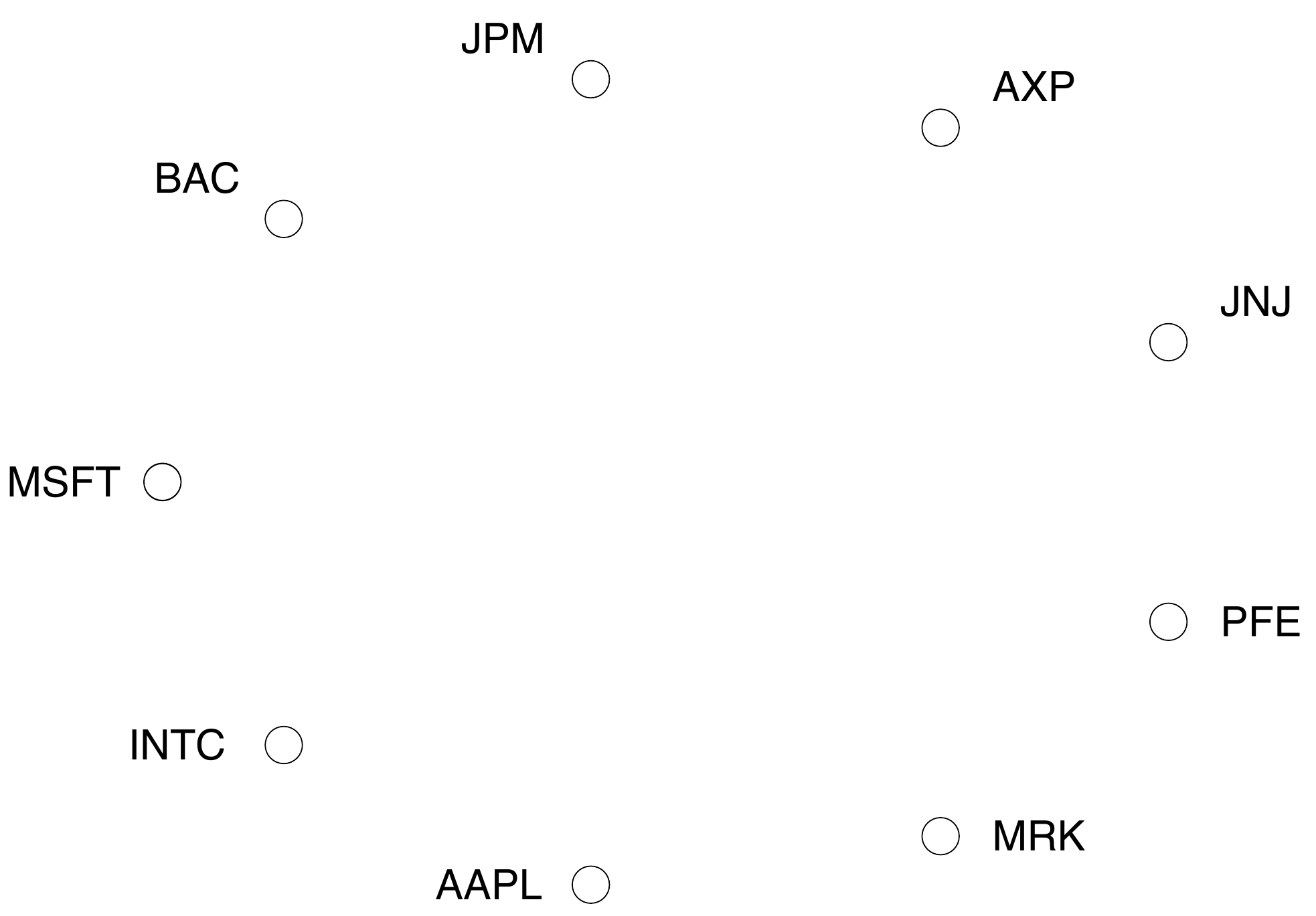}}}}
    \frame{{\subfigure{\label{GraphModel_3_MTCCA_EXP_4}\includegraphics[scale=0.25]{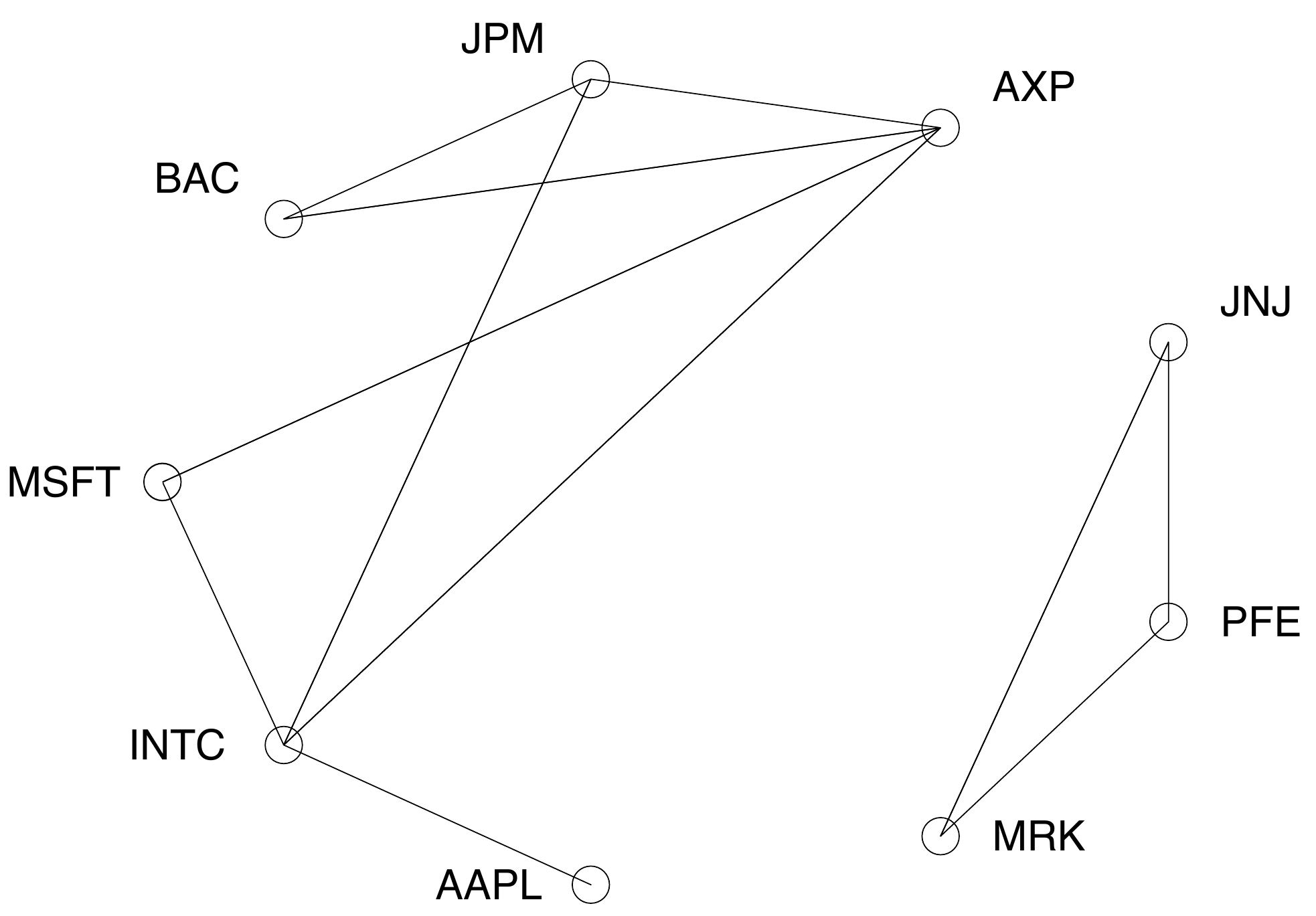}}}}
    \frame{{\subfigure{\label{GraphModel_3_MTCCA_EXP_5}\includegraphics[scale=0.25]{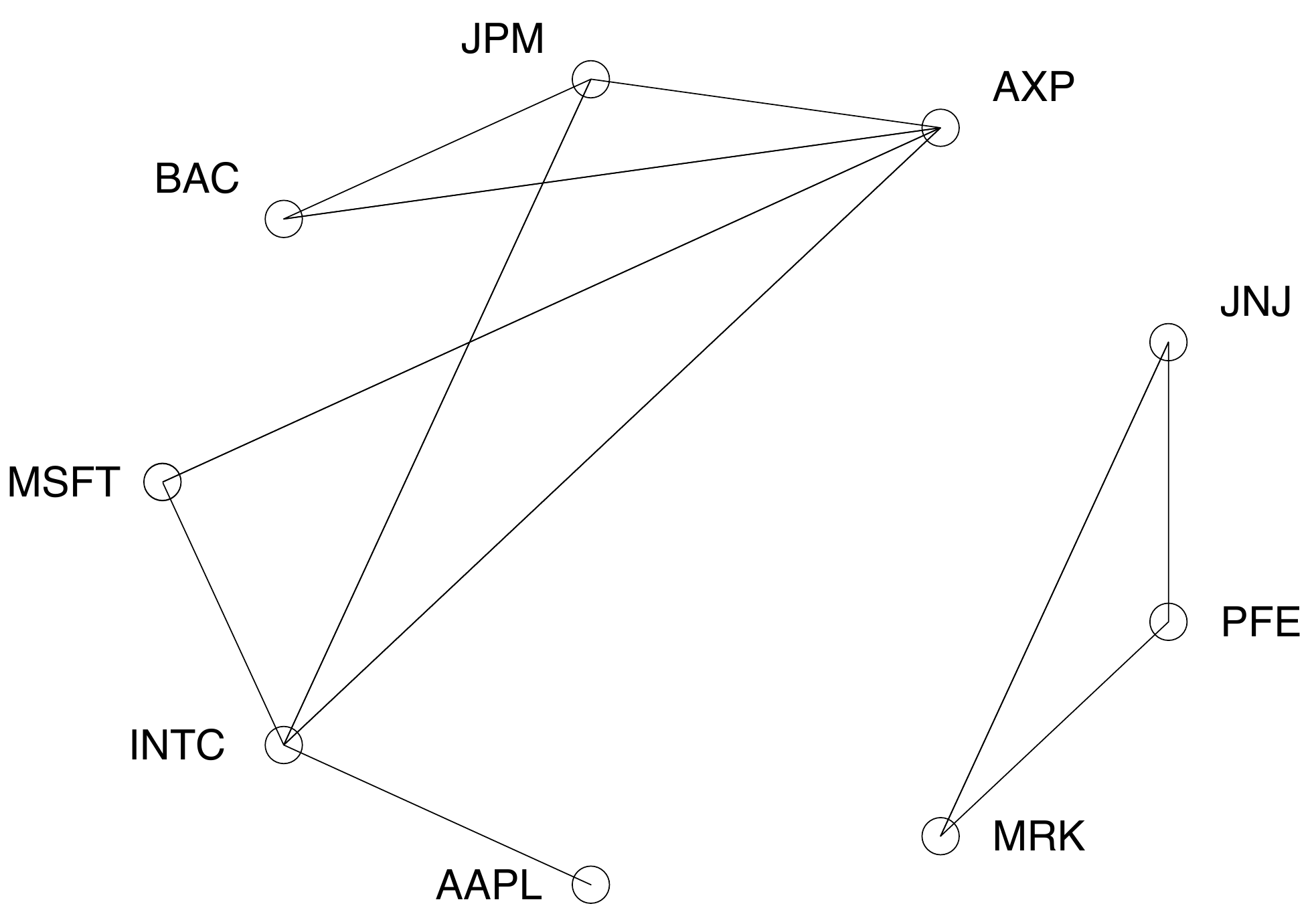}}}}
    \vspace{0.15cm}
    \frame{{\subfigure{\label{GraphModel_3_MTCCA_EXP_6}\includegraphics[scale=0.25]{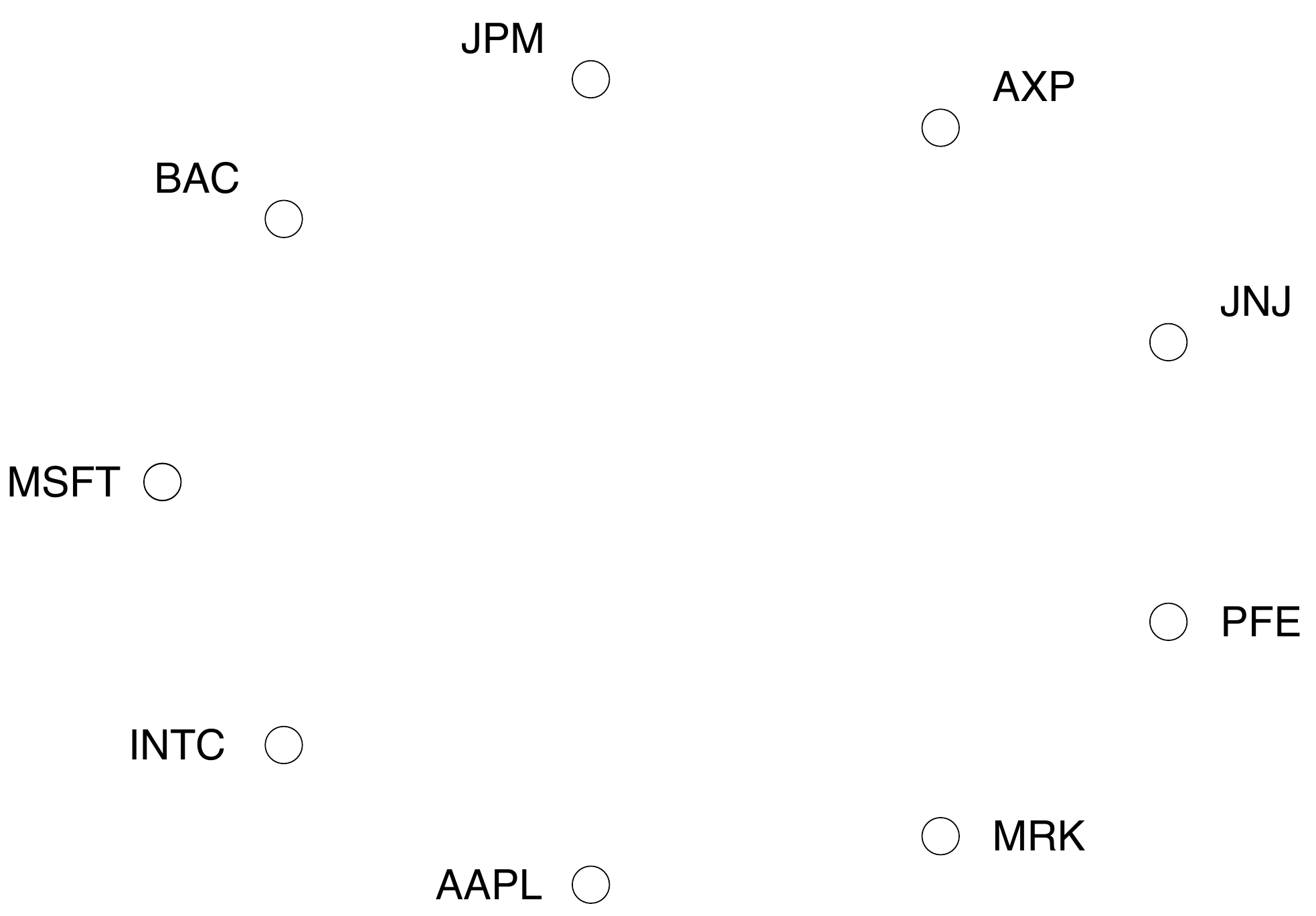}}}}
    \frame{{\subfigure{\label{GraphModel_3_MTCCA_EXP_7}\includegraphics[scale=0.25]{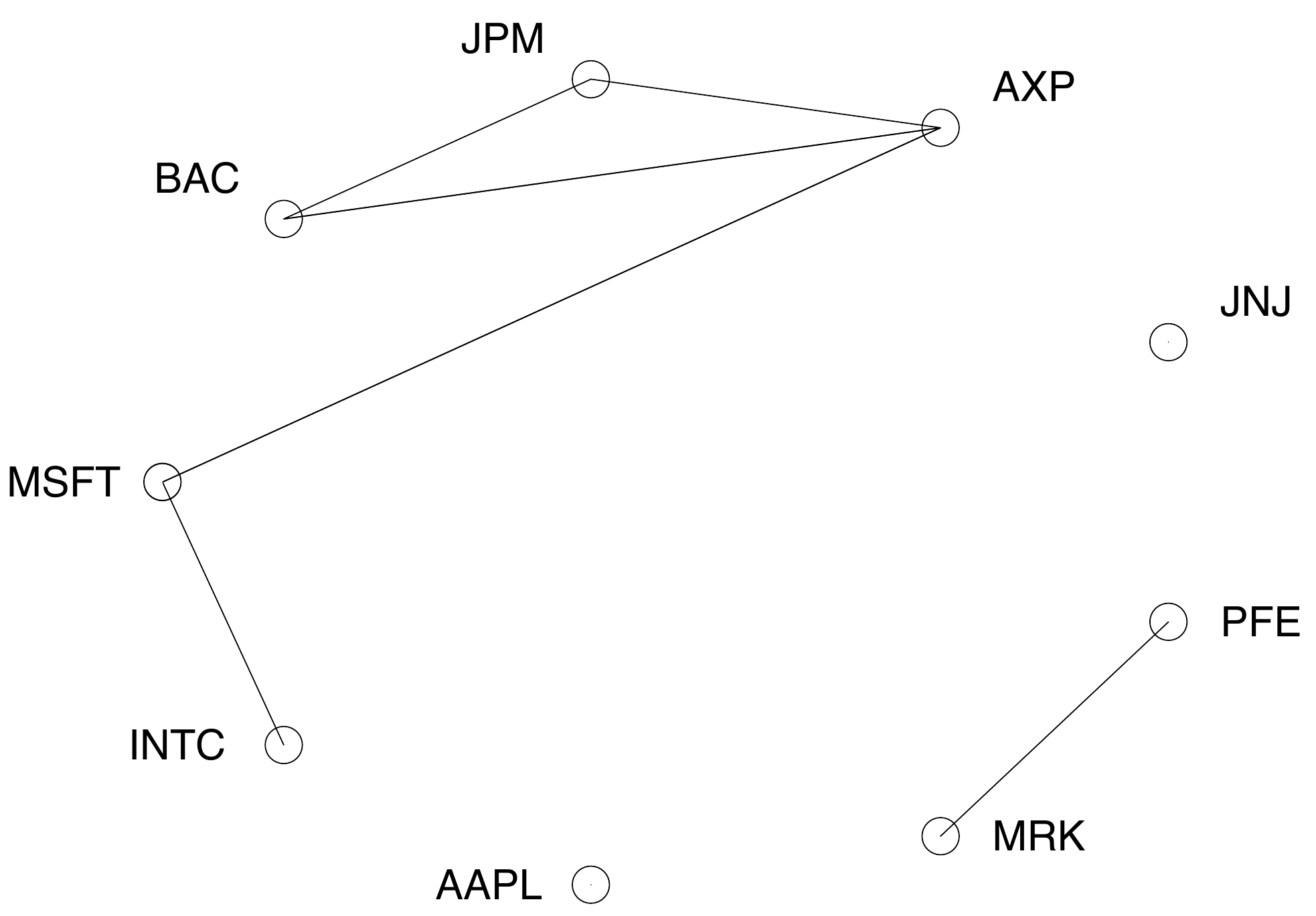}}}}
    \frame{{\subfigure{\label{GraphModel_3_MTCCA_EXP_8}\includegraphics[scale=0.25]{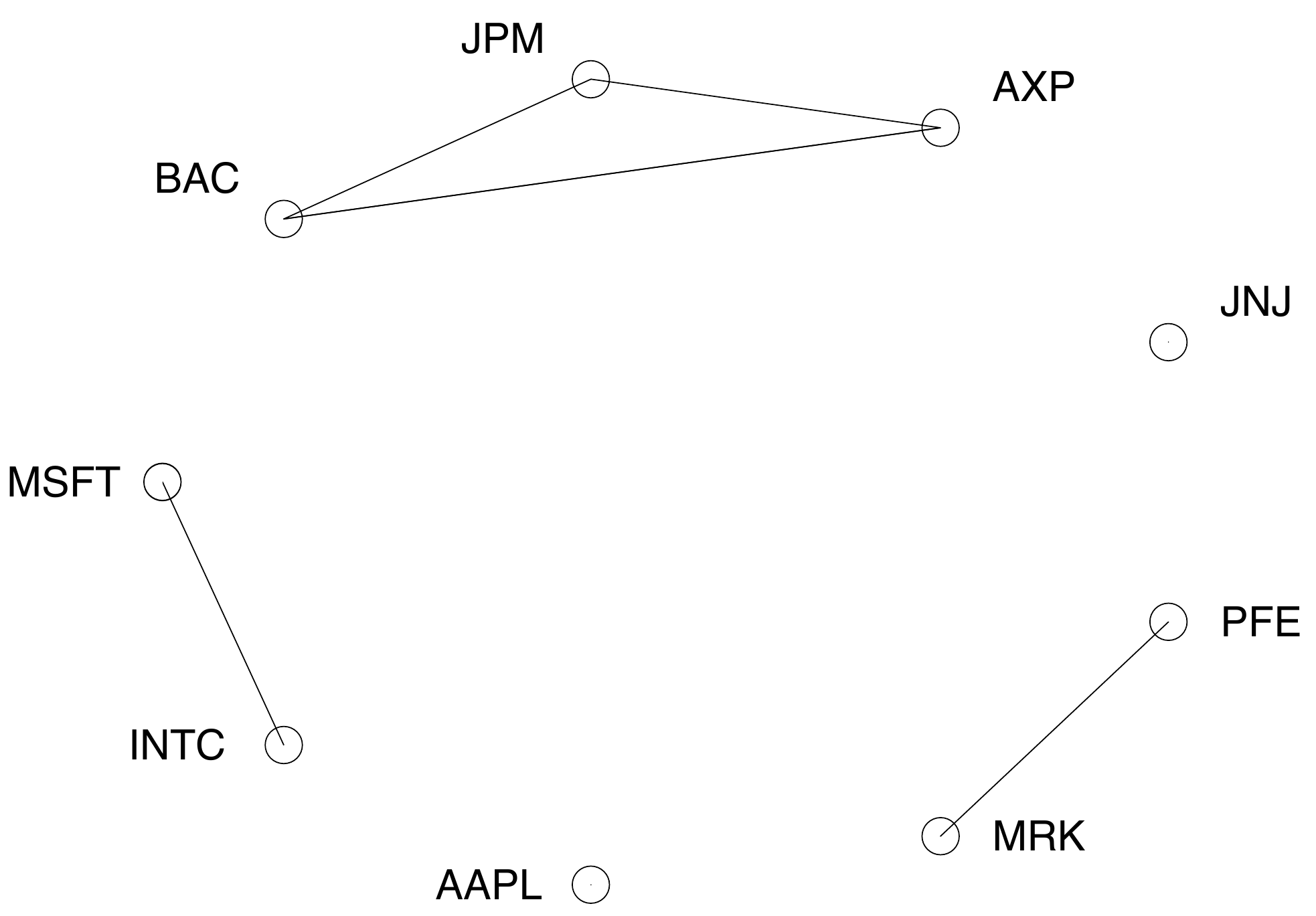}}}}
    \frame{{\subfigure{\label{GraphModel_3_MTCCA_EXP_9}\includegraphics[scale=0.25]{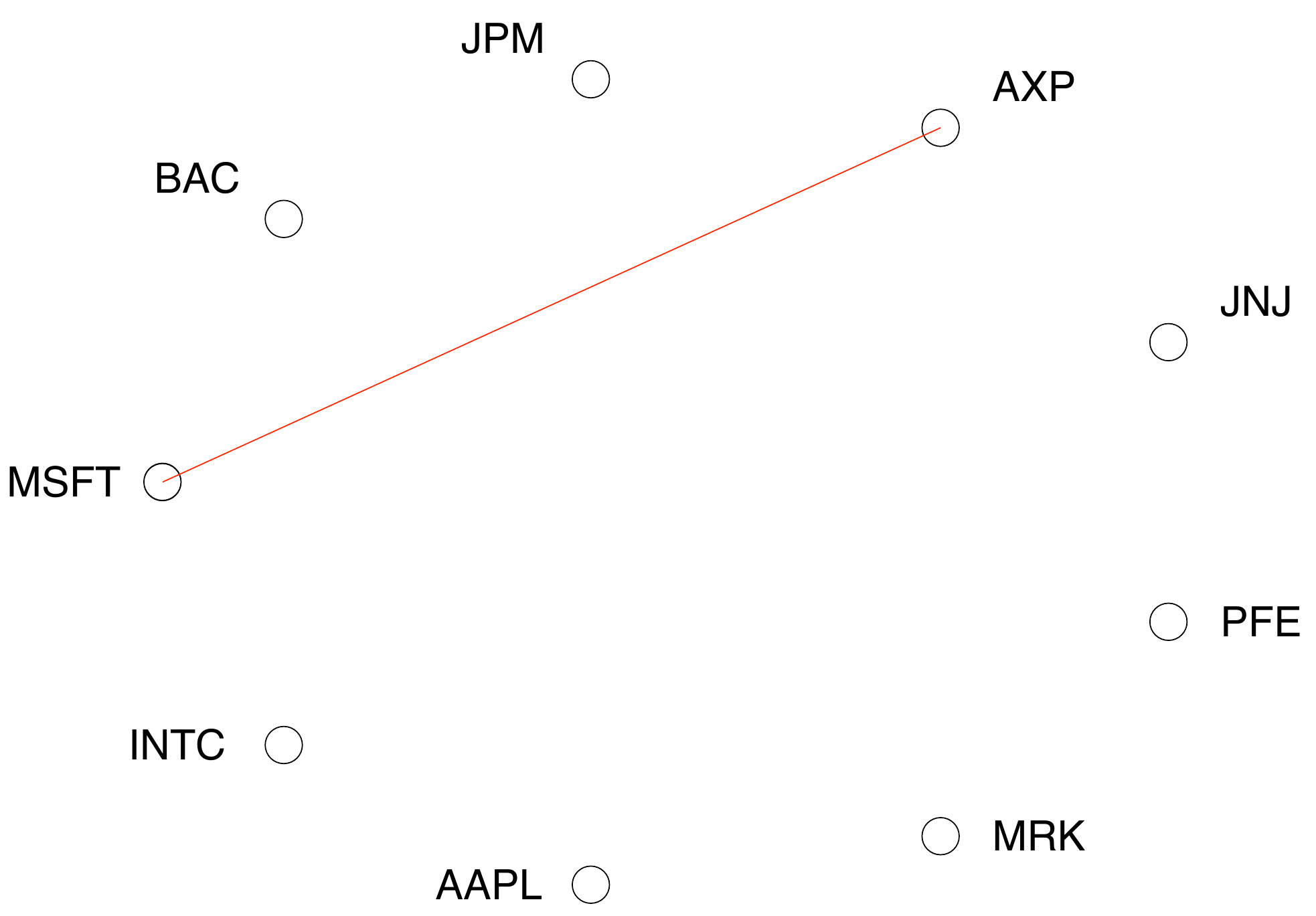}}}}
  \end{center}
  \caption{NASDAQ/NYSE experiment. {\bf{Left column}:} The graphical models selected by MTCCA with exponential MT-functions for $\lambda=0.5,0.55,0.58$. {\bf{Middle column}:} The closest graphs selected by KCCA. {\bf{Right column}:} The symmetric difference graphs: the red lines indicate edges found by MTCCA and not by KCCA, and vice-versa for the black lines. For $\lambda=0.58$ the MTCCA graph has one more edge than the closest KCCA graph.}
\label{GraphModFinancial3}
\end{figure}

\begin{figure}[htbp!]
  \begin{center}
    \frame{{\subfigure{\label{GraphModel_3_MTCCA_GAUSS_1}\includegraphics[scale=0.25]{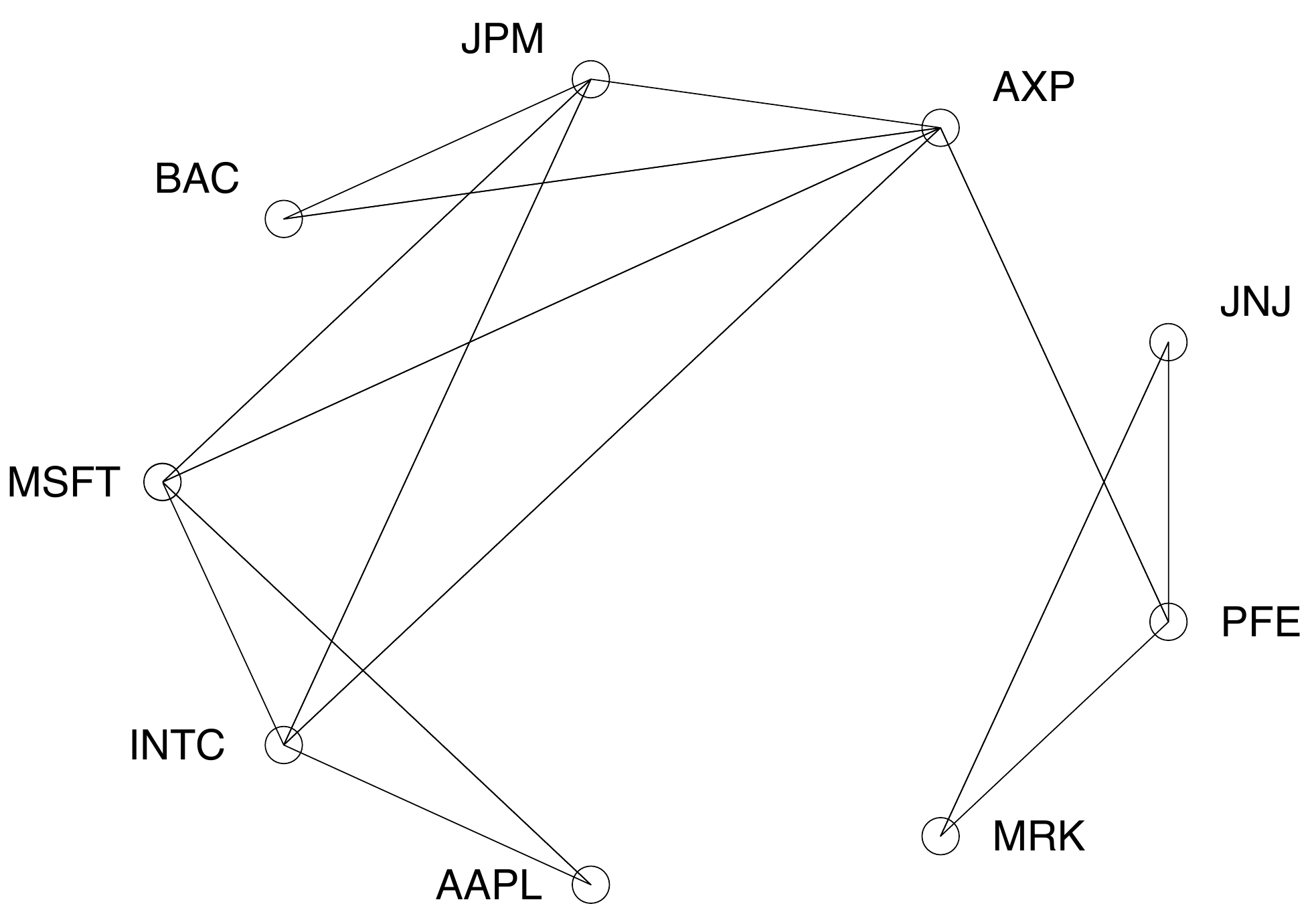}}}}
    \frame{{\subfigure{\label{GraphModel_3_MTCCA_GAUSS_2}\includegraphics[scale=0.25]{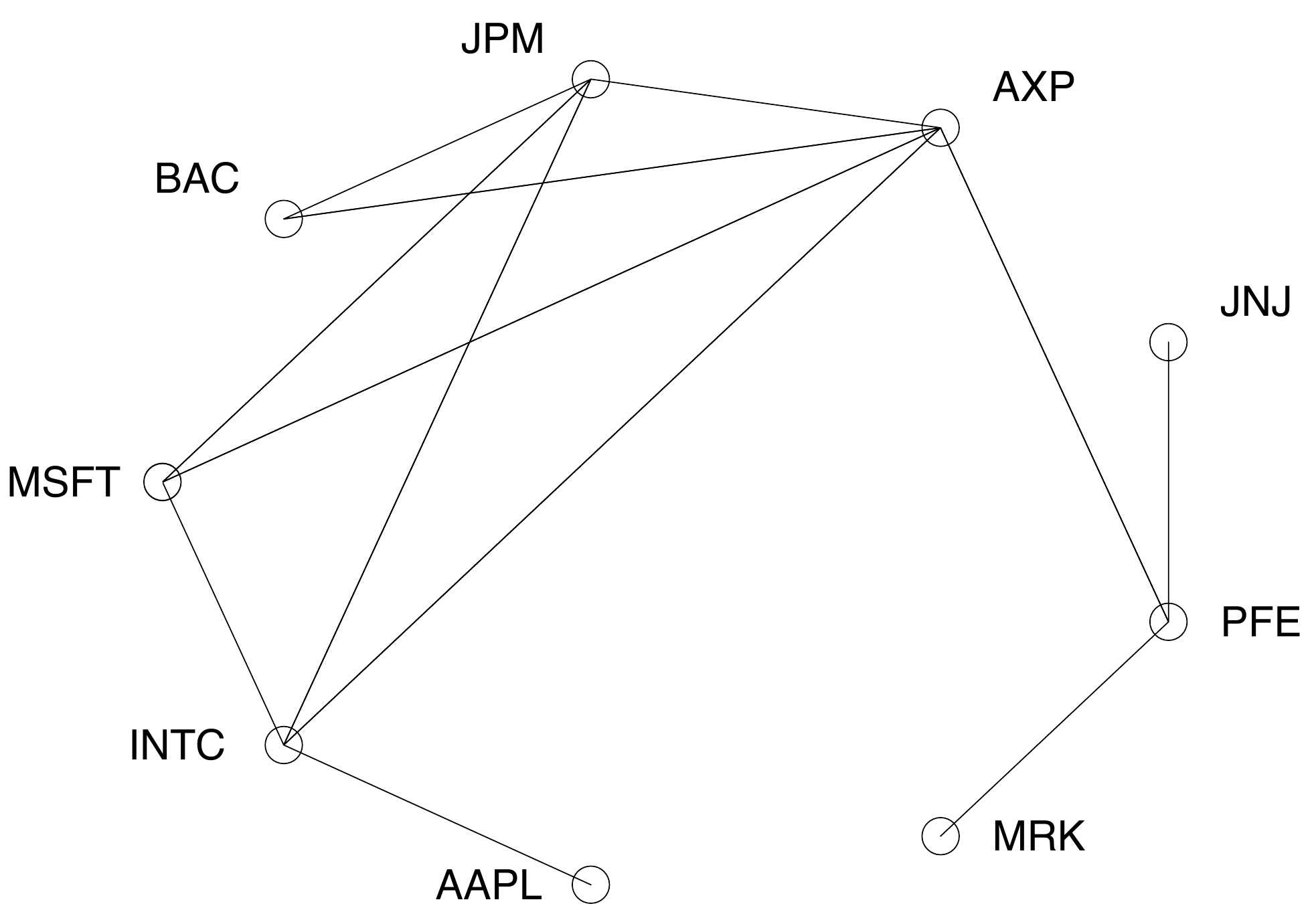}}}}
    \vspace{0.15cm}
    \frame{{\subfigure{\label{GraphModel_3_MTCCA_GAUSS_3}\includegraphics[scale=0.25]{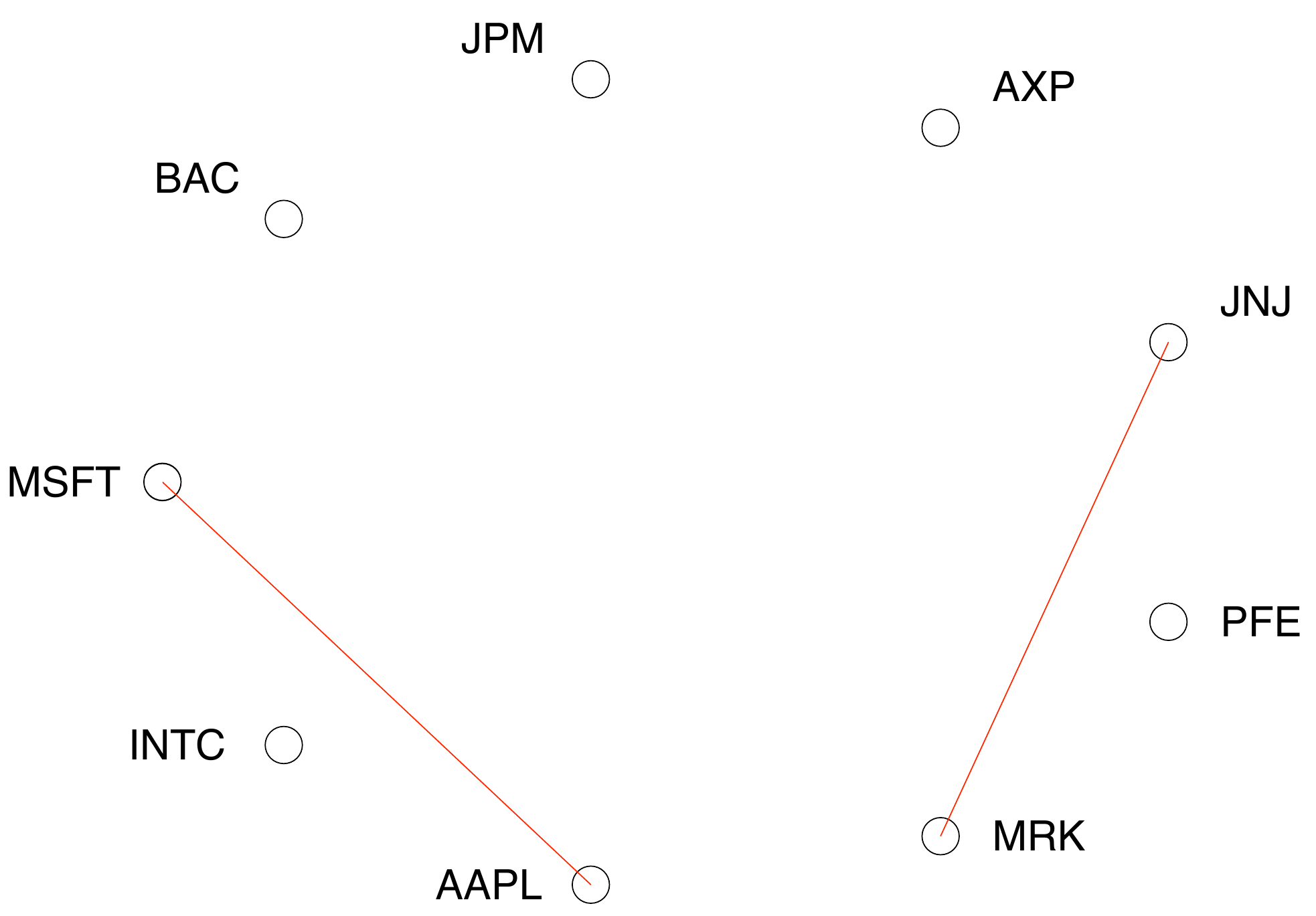}}}}
    \frame{{\subfigure{\label{GraphModel_3_MTCCA_GAUSS_4}\includegraphics[scale=0.25]{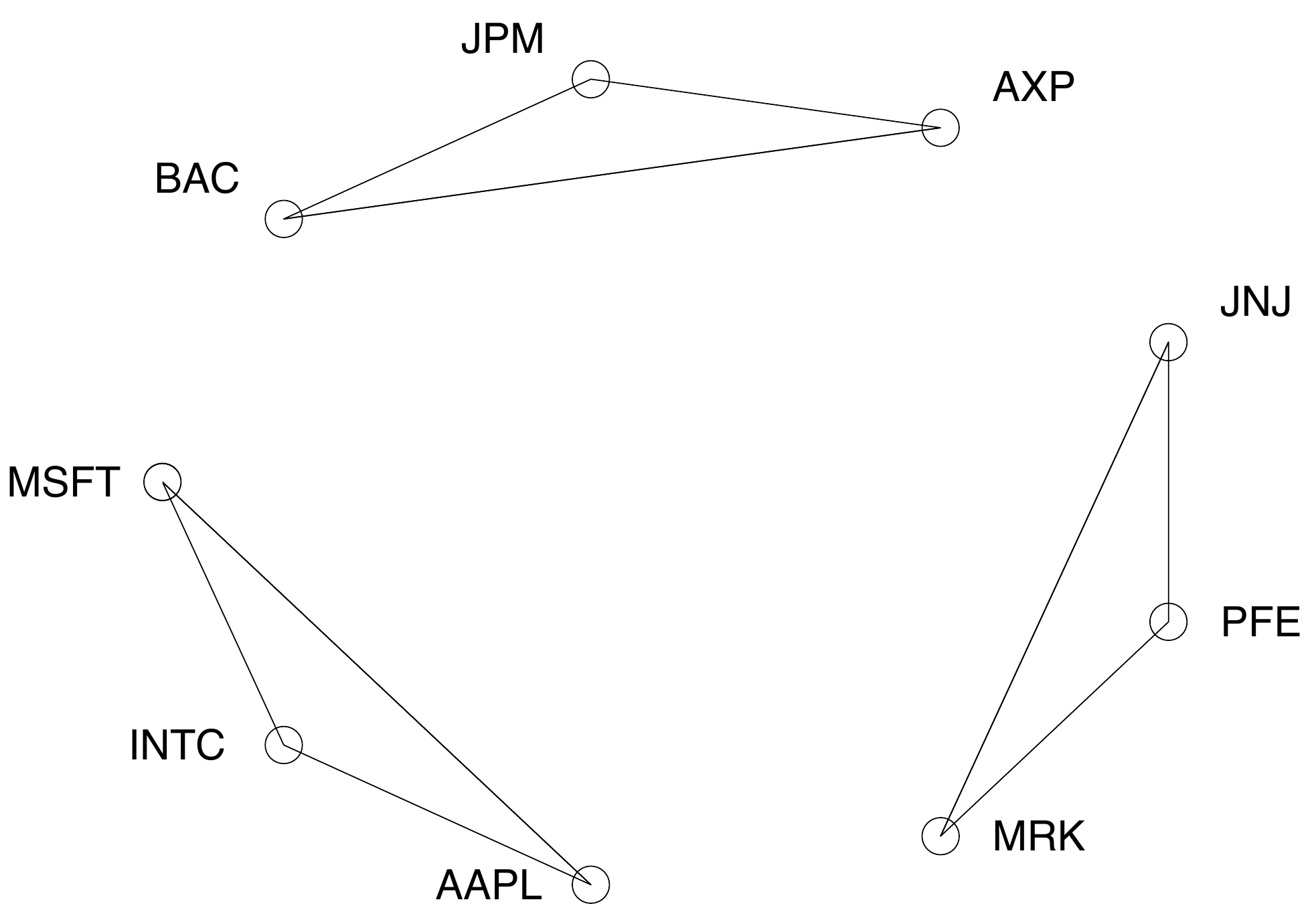}}}}
    \frame{{\subfigure{\label{GraphModel_3_MTCCA_GAUSS_5}\includegraphics[scale=0.25]{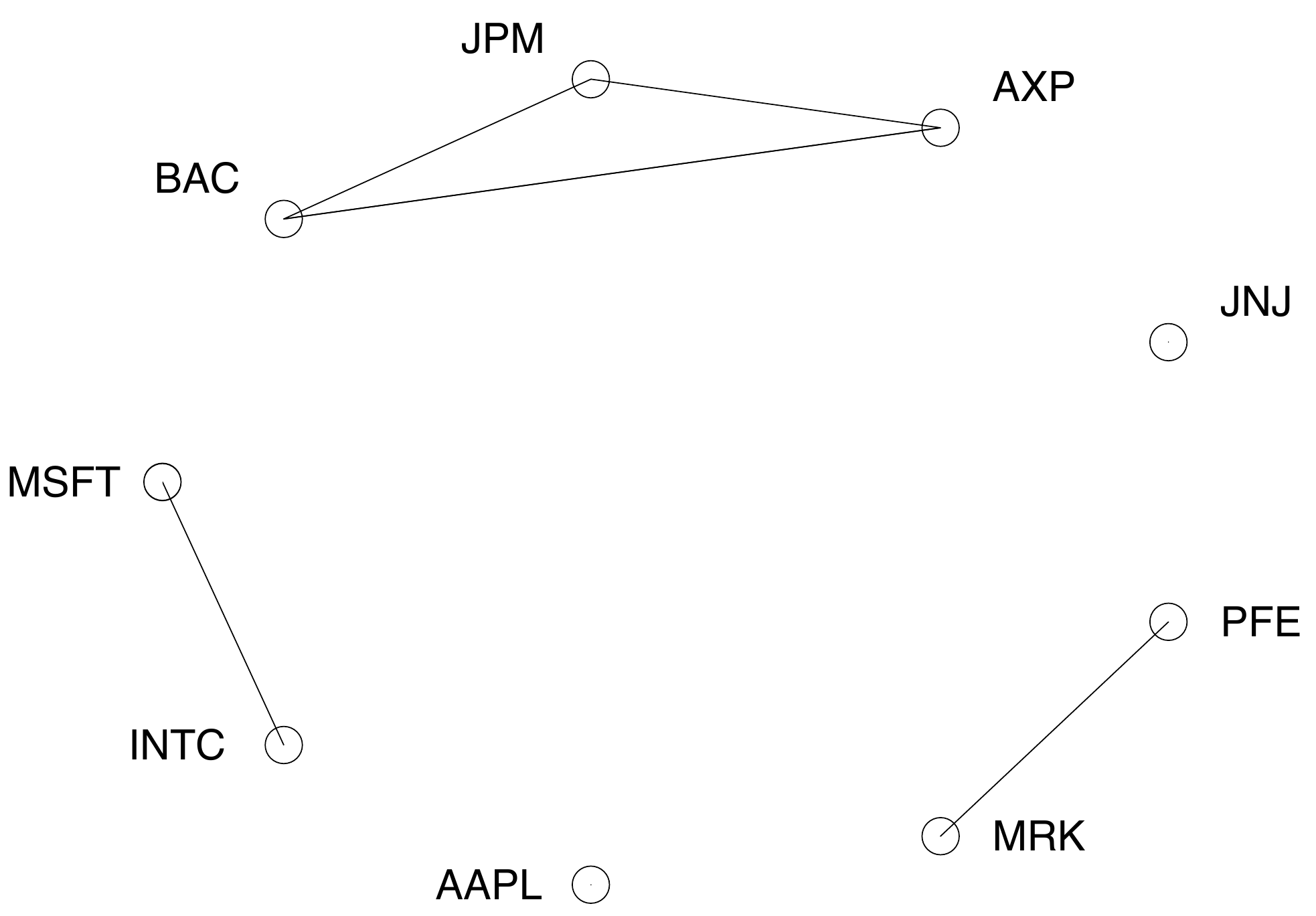}}}}
    \vspace{0.15cm}
    \frame{{\subfigure{\label{GraphModel_3_MTCCA_GAUSS_6}\includegraphics[scale=0.25]{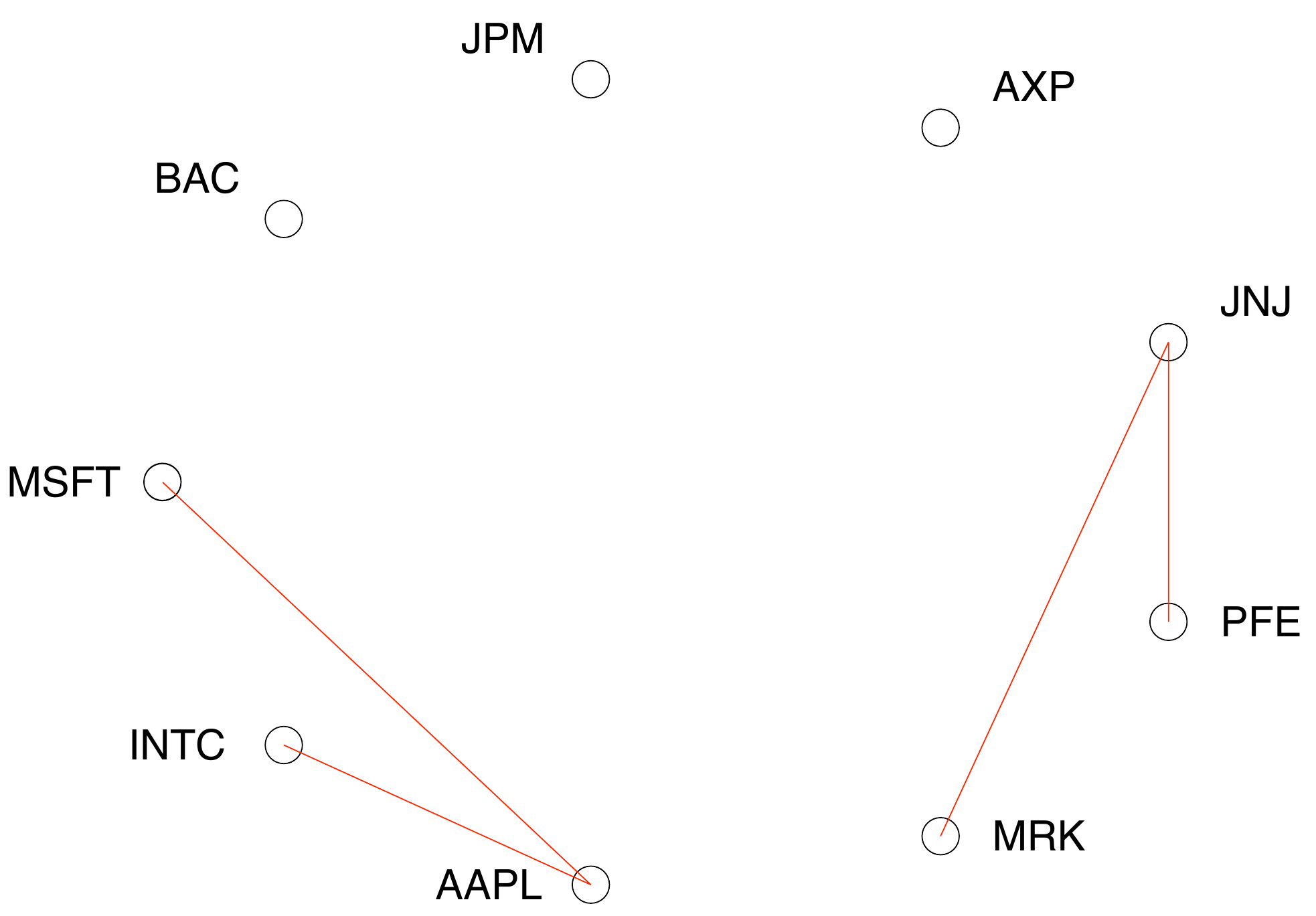}}}}
    \frame{{\subfigure{\label{GraphModel_3_MTCCA_GAUSS_7}\includegraphics[scale=0.25]{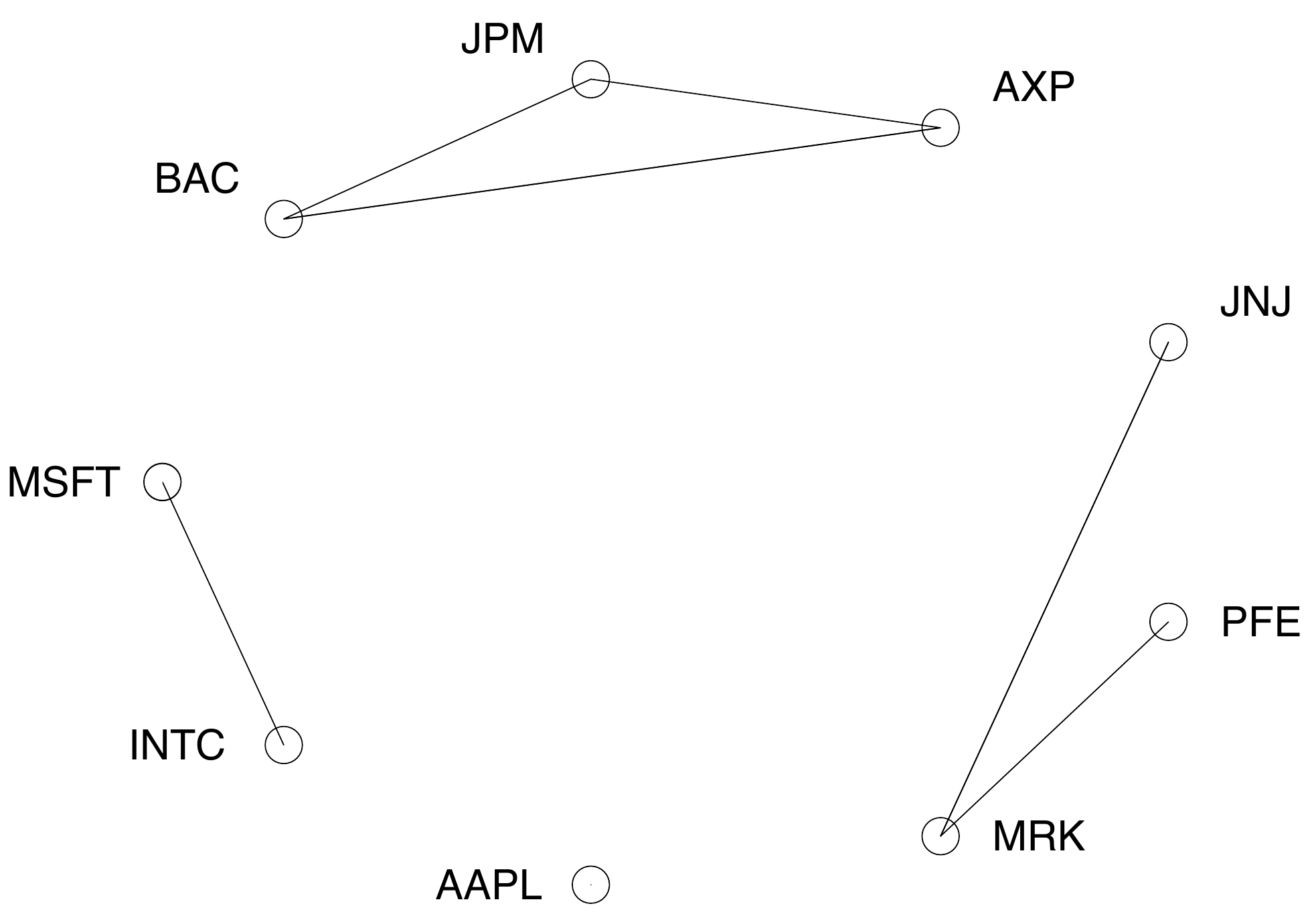}}}}
    \frame{{\subfigure{\label{GraphModel_3_MTCCA_GAUSS_8}\includegraphics[scale=0.25]{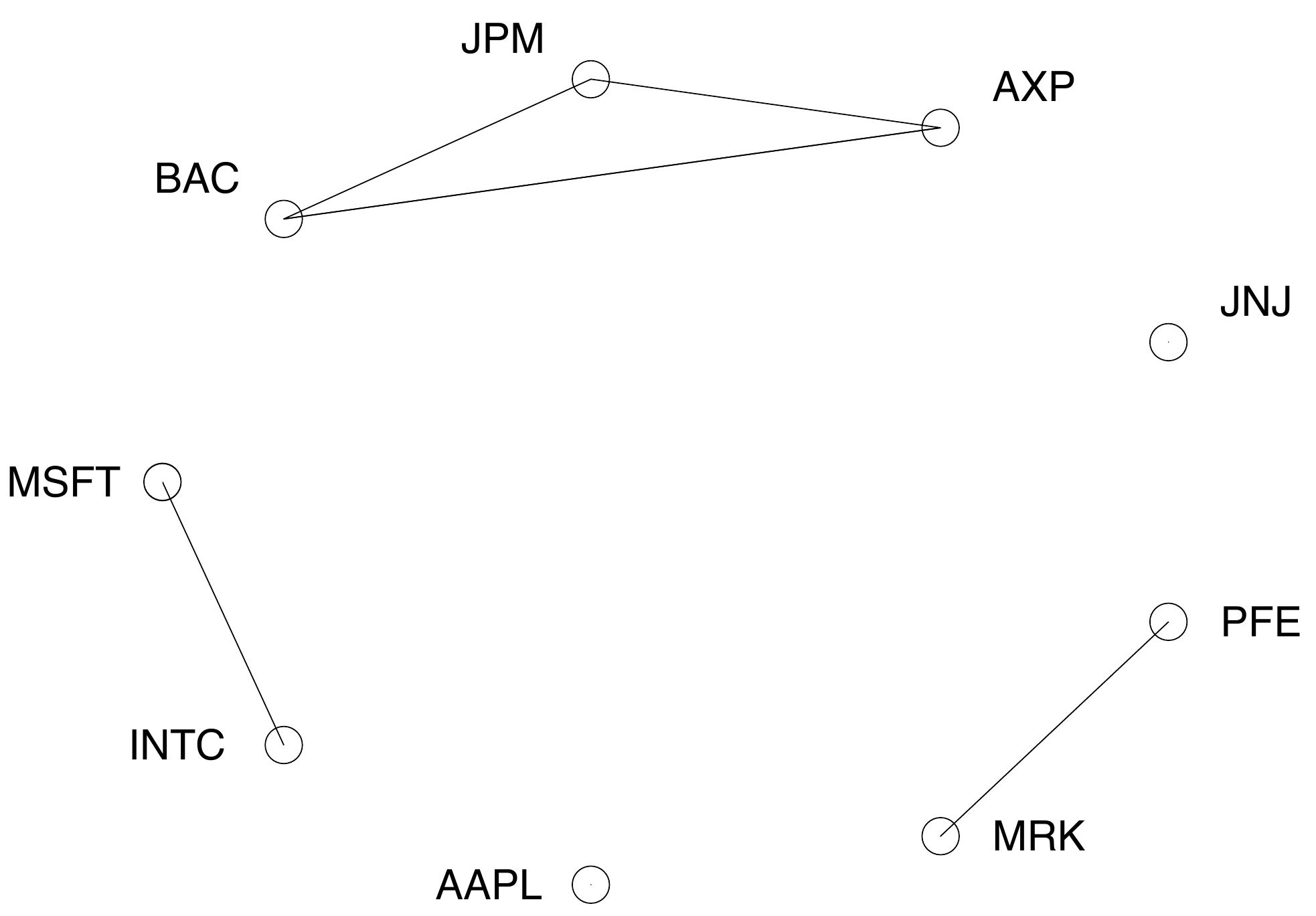}}}}
    \frame{{\subfigure{\label{GraphModel_3_MTCCA_GAUSS_9}\includegraphics[scale=0.25]{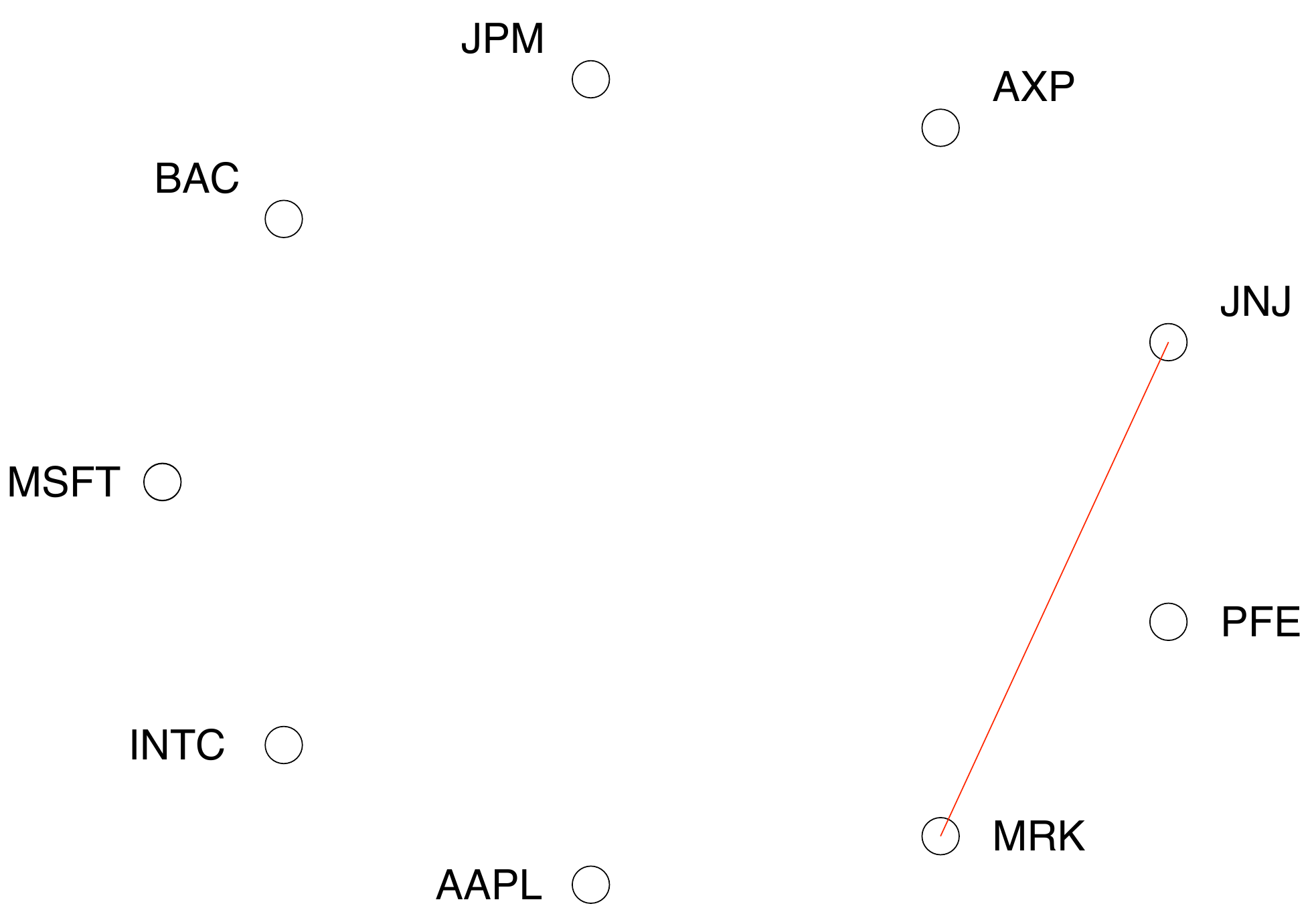}}}}
  \end{center}
  \caption{NASDAQ/NYSE experiment. {\bf{Left column}:} The graphical models selected by MTCCA with Gaussian MT-functions for $\lambda=0.5,0.55,0.58$. {\bf{Middle column}:} The closest graphs selected by LCCA. {\bf{Right column}:} The symmetric difference graphs: the red lines indicate edges found by MTCCA and not by LCCA, and vice-versa for the black lines. For these values of $\lambda$, Gaussian MTCCA detects more dependencies than LCCA: the MTCCA graph has more edges than the closest LCCA graph.}
\label{GraphModFinancial4}
\end{figure}

\begin{figure}[htbp!]
  \begin{center}
    \frame{{\subfigure{\label{GraphModel_3_MTCCA_GAUSS_1}\includegraphics[scale=0.25]{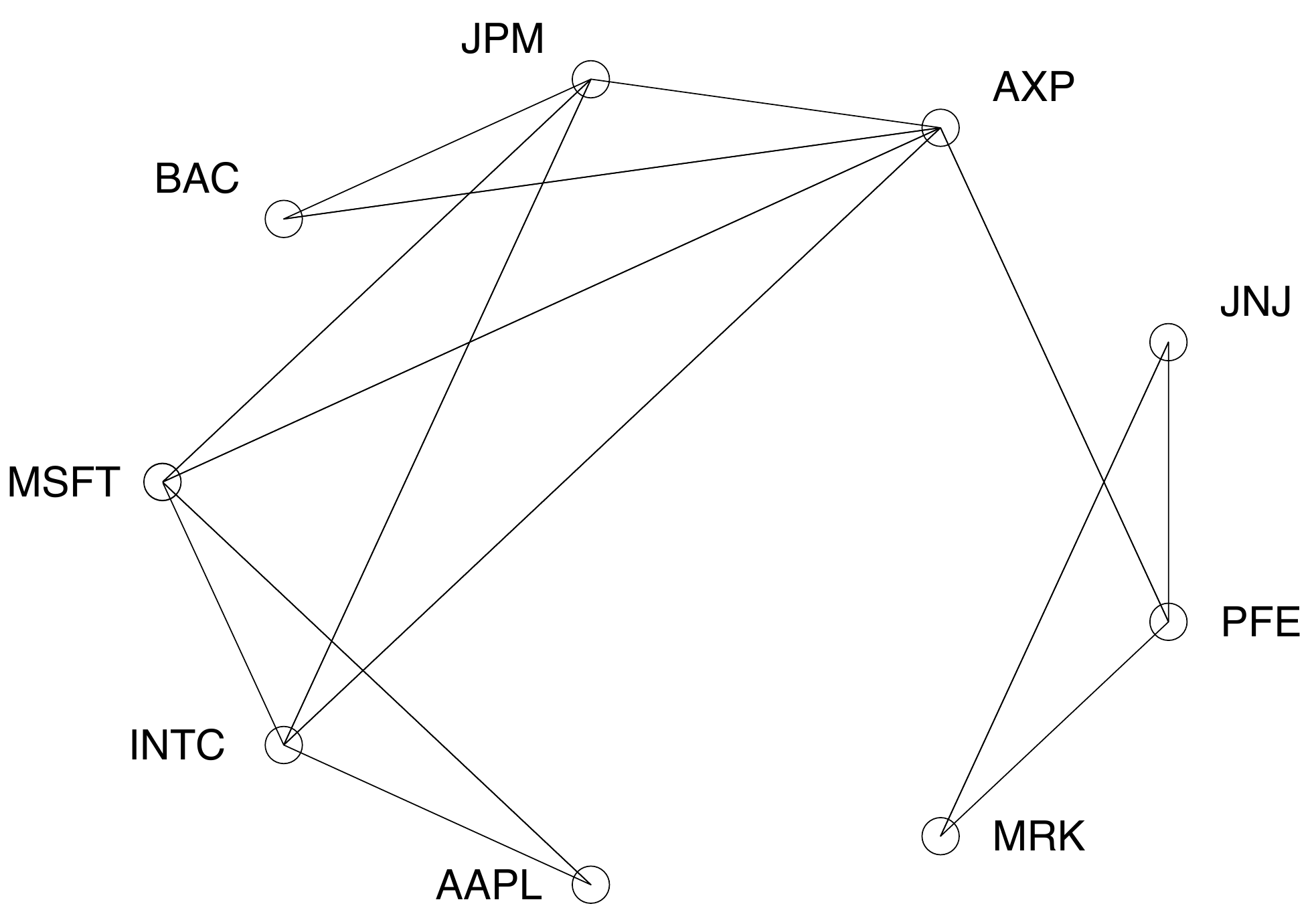}}}}
    \frame{{\subfigure{\label{GraphModel_3_MTCCA_GAUSS_2}\includegraphics[scale=0.25]{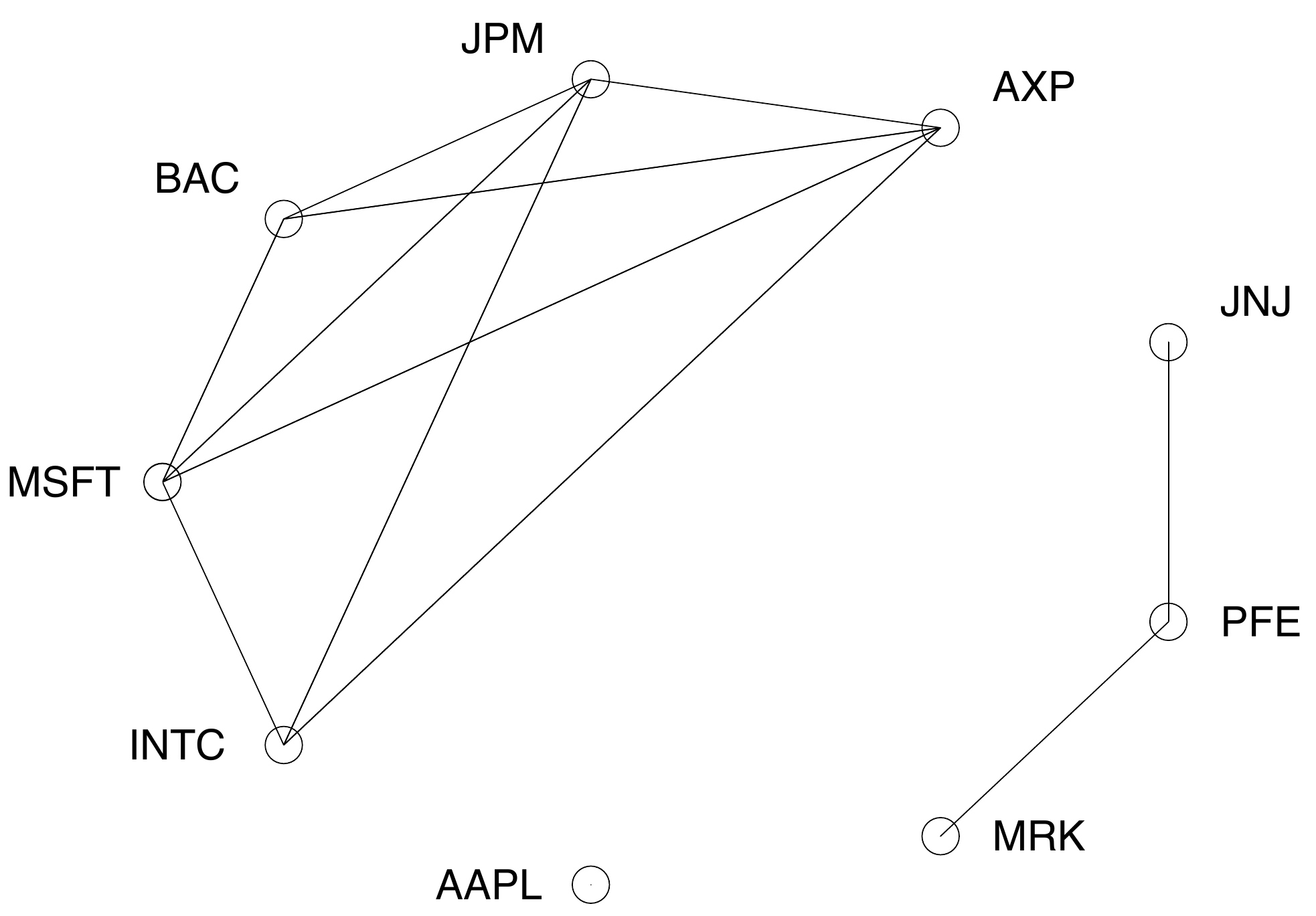}}}}
    \vspace{0.15cm}
    \frame{{\subfigure{\label{GraphModel_3_MTCCA_GAUSS_3}\includegraphics[scale=0.25]{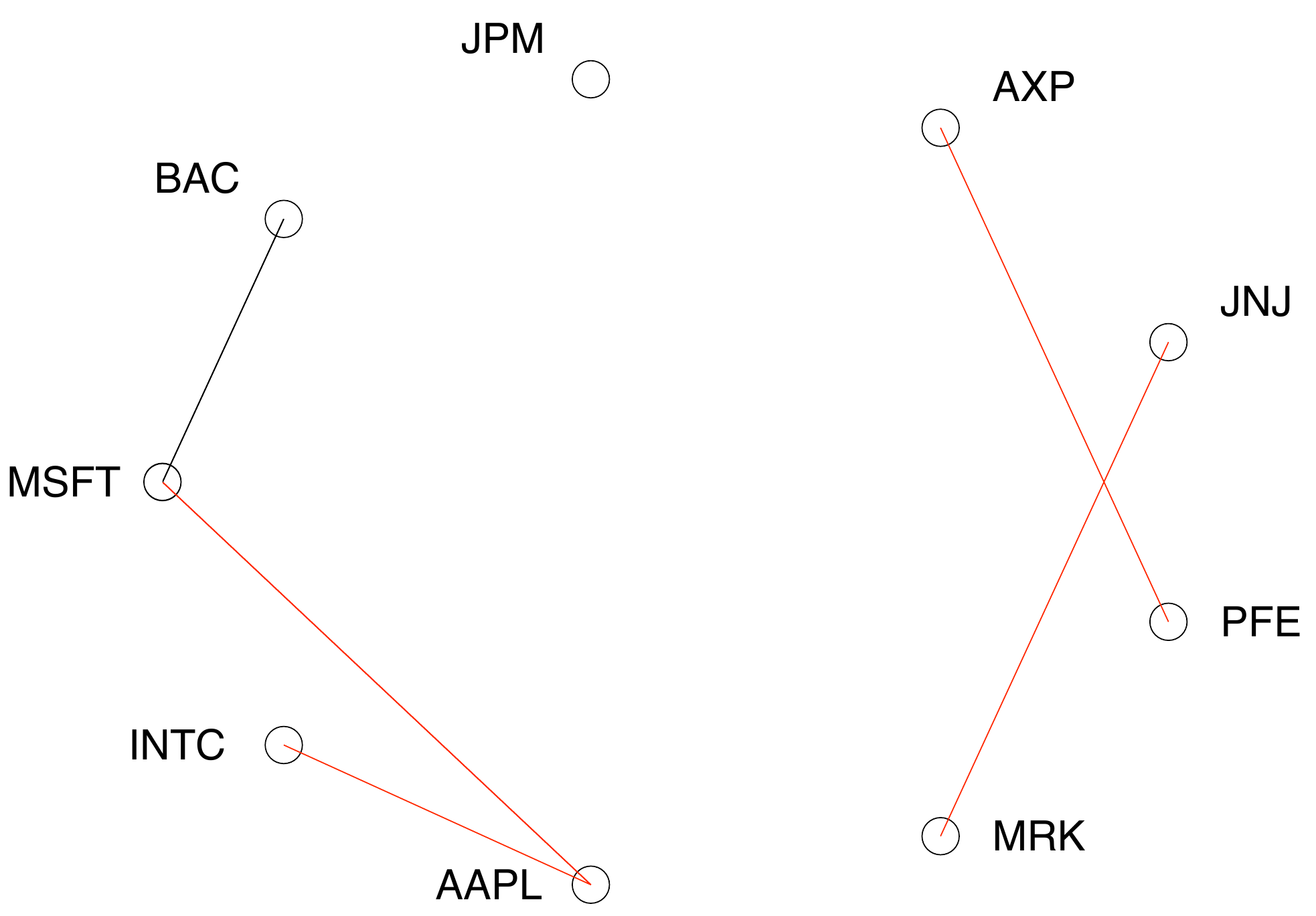}}}}
    \frame{{\subfigure{\label{GraphModel_3_MTCCA_GAUSS_4}\includegraphics[scale=0.25]{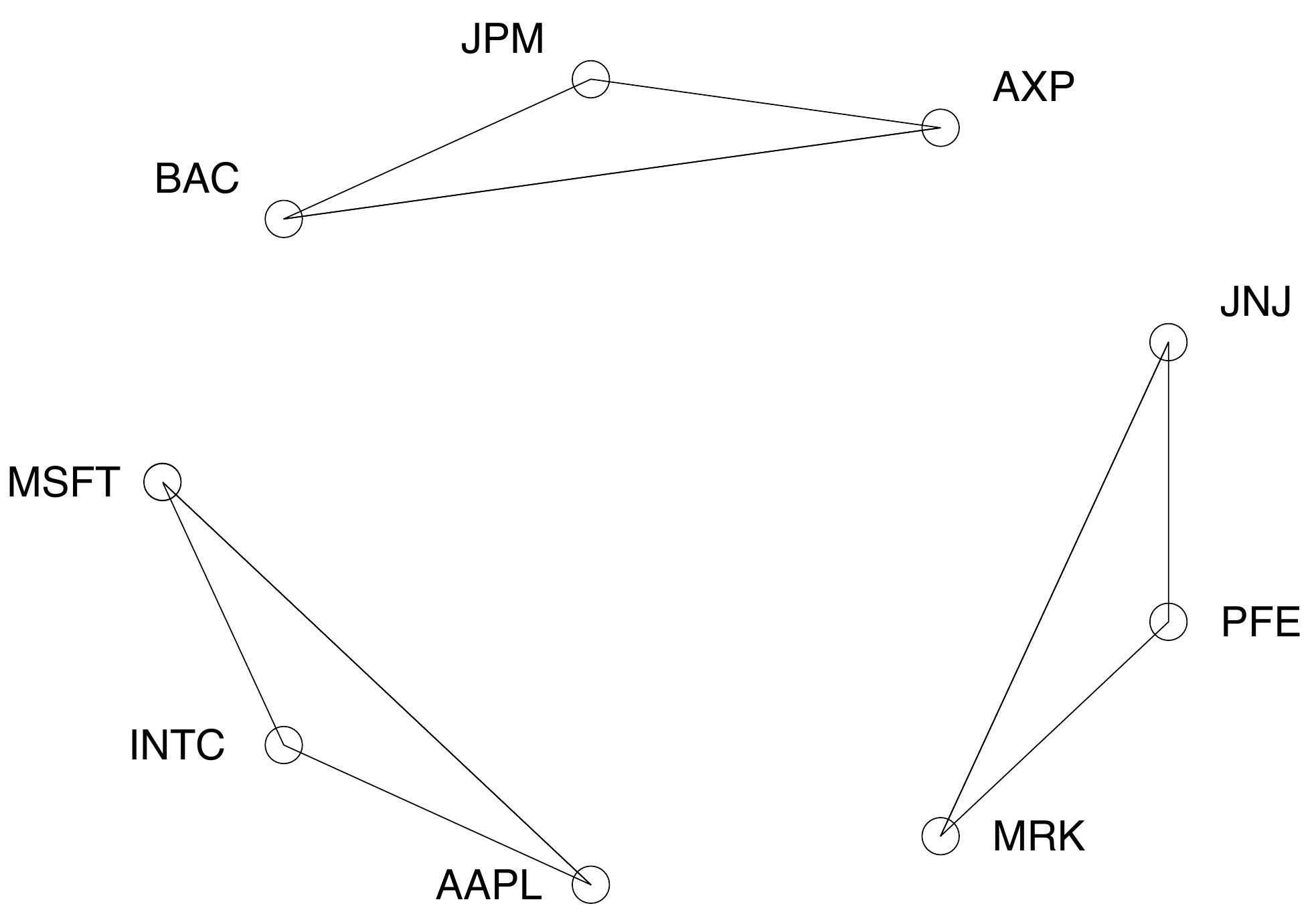}}}}
    \frame{{\subfigure{\label{GraphModel_3_MTCCA_GAUSS_5}\includegraphics[scale=0.25]{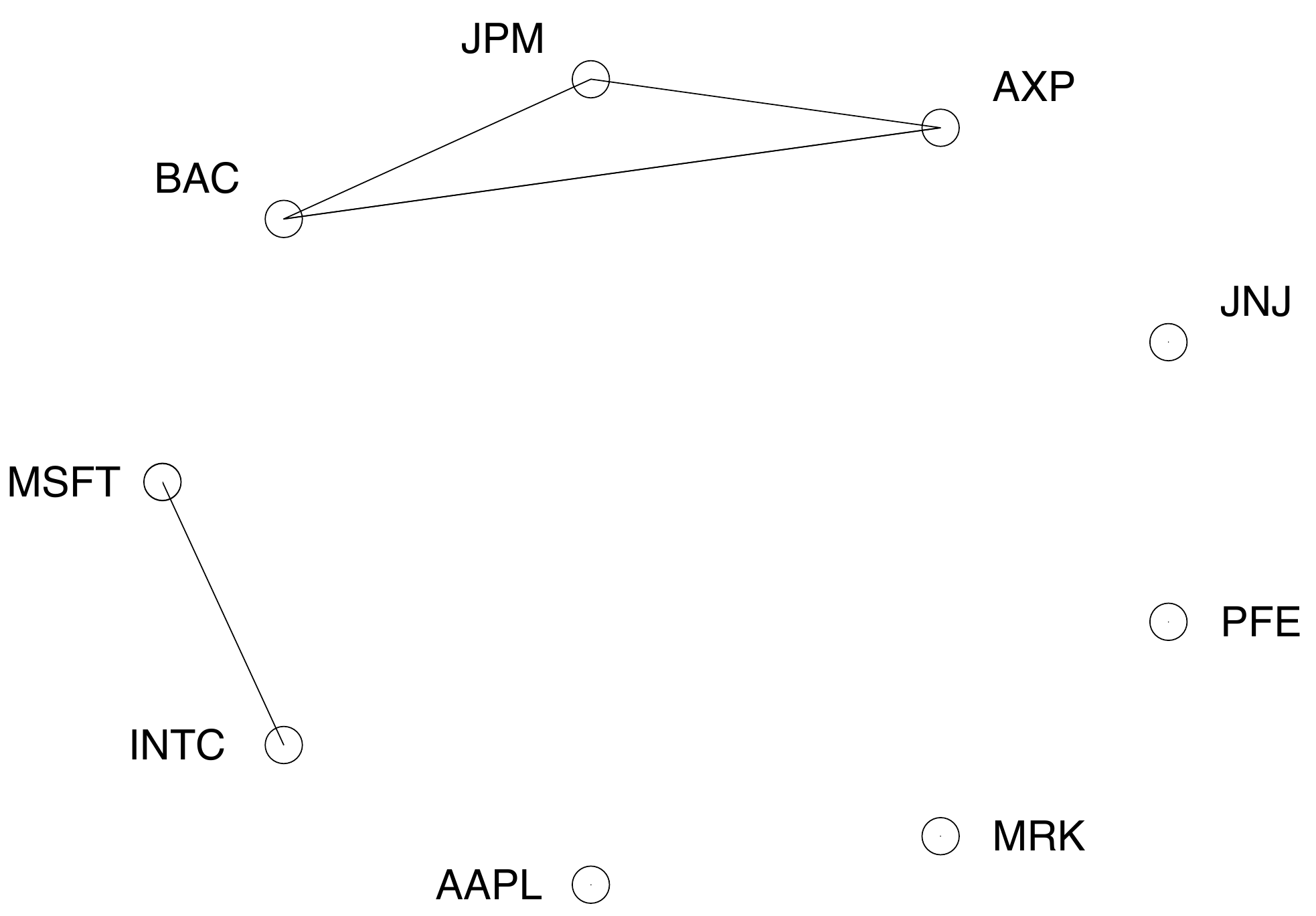}}}}
    \vspace{0.15cm}
    \frame{{\subfigure{\label{GraphModel_3_MTCCA_GAUSS_6}\includegraphics[scale=0.25]{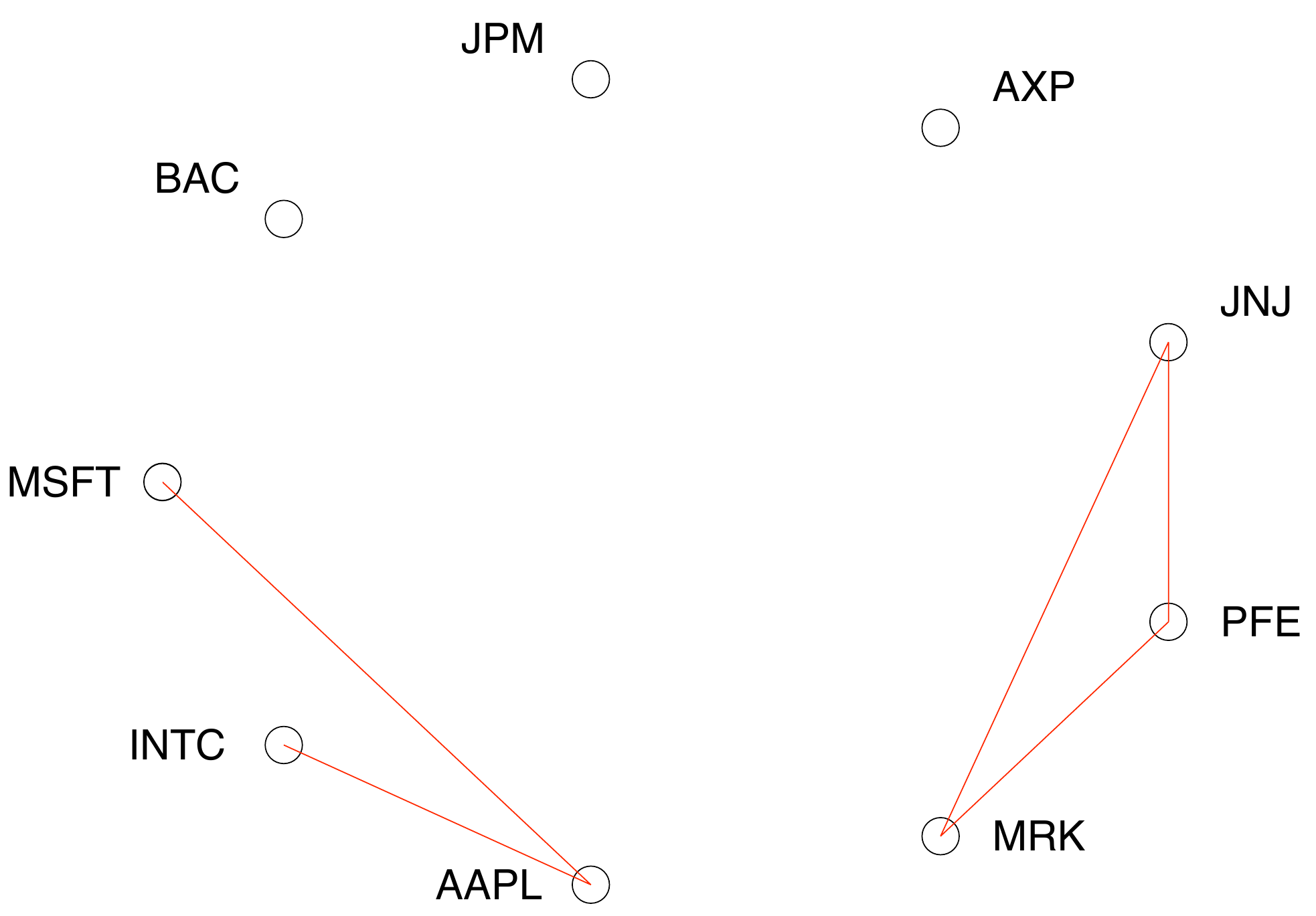}}}}
    \frame{{\subfigure{\label{GraphModel_3_MTCCA_GAUSS_7}\includegraphics[scale=0.25]{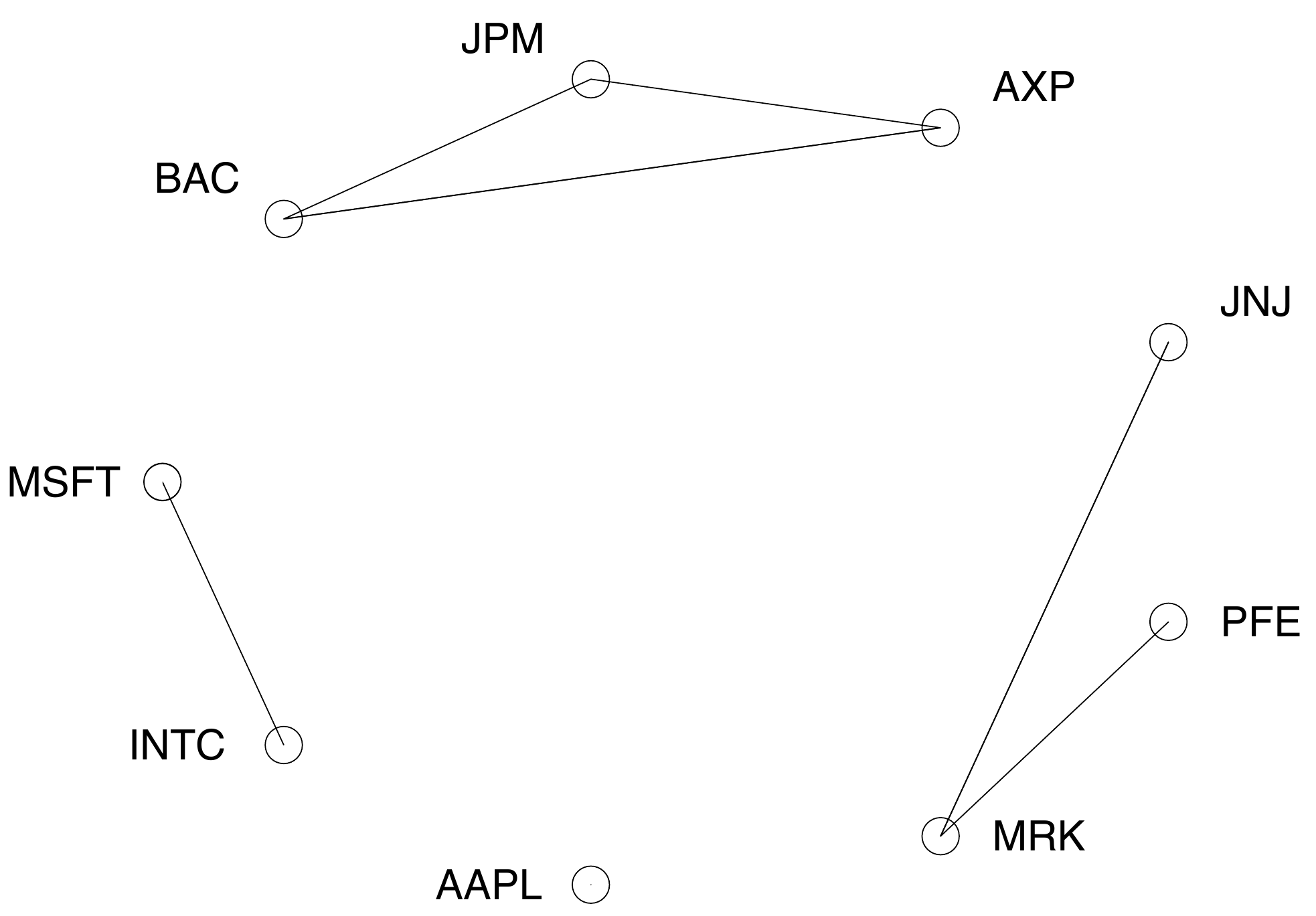}}}}
    \frame{{\subfigure{\label{GraphModel_3_MTCCA_GAUSS_8}\includegraphics[scale=0.25]{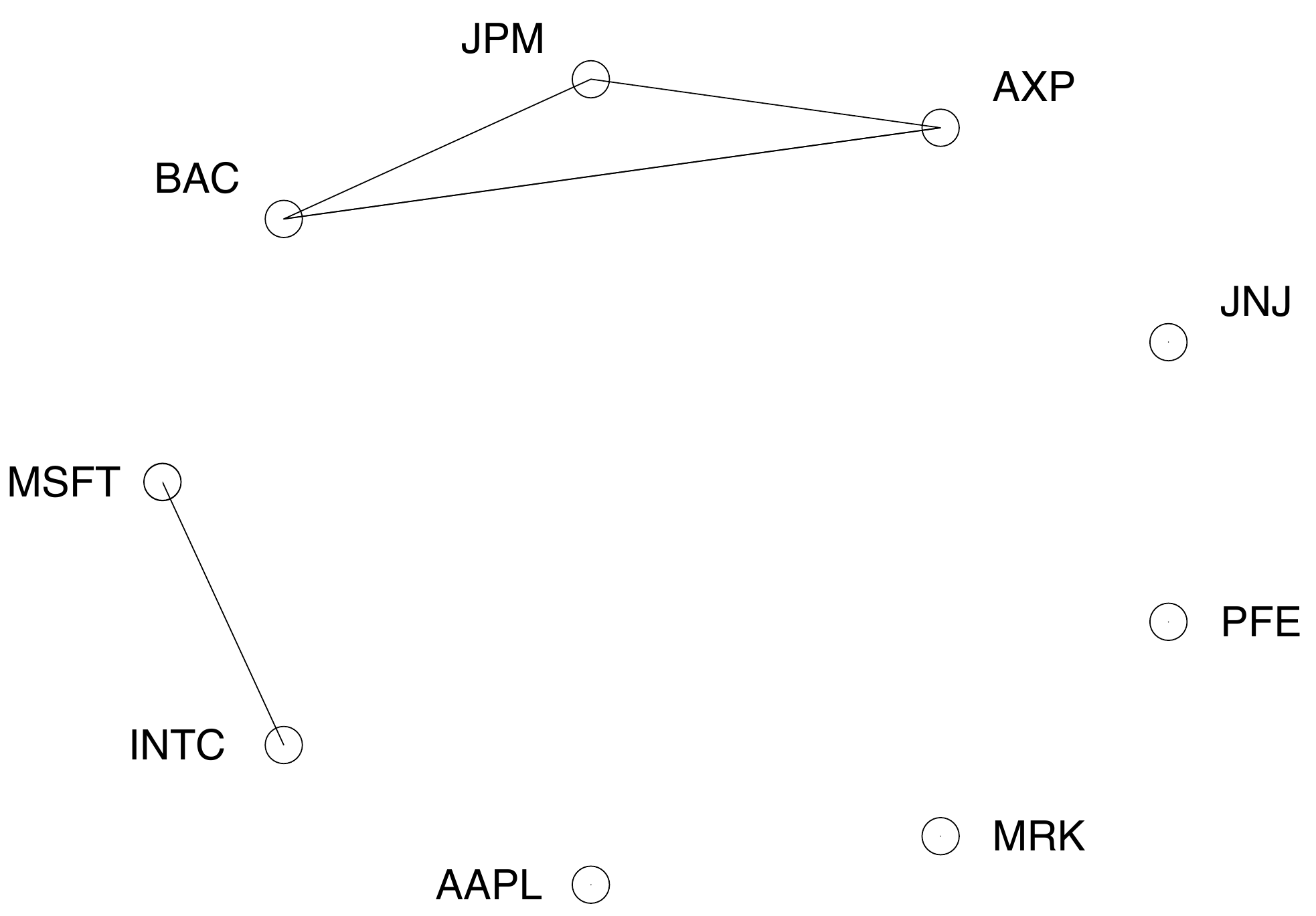}}}}
    \frame{{\subfigure{\label{GraphModel_3_MTCCA_GAUSS_9}\includegraphics[scale=0.25]{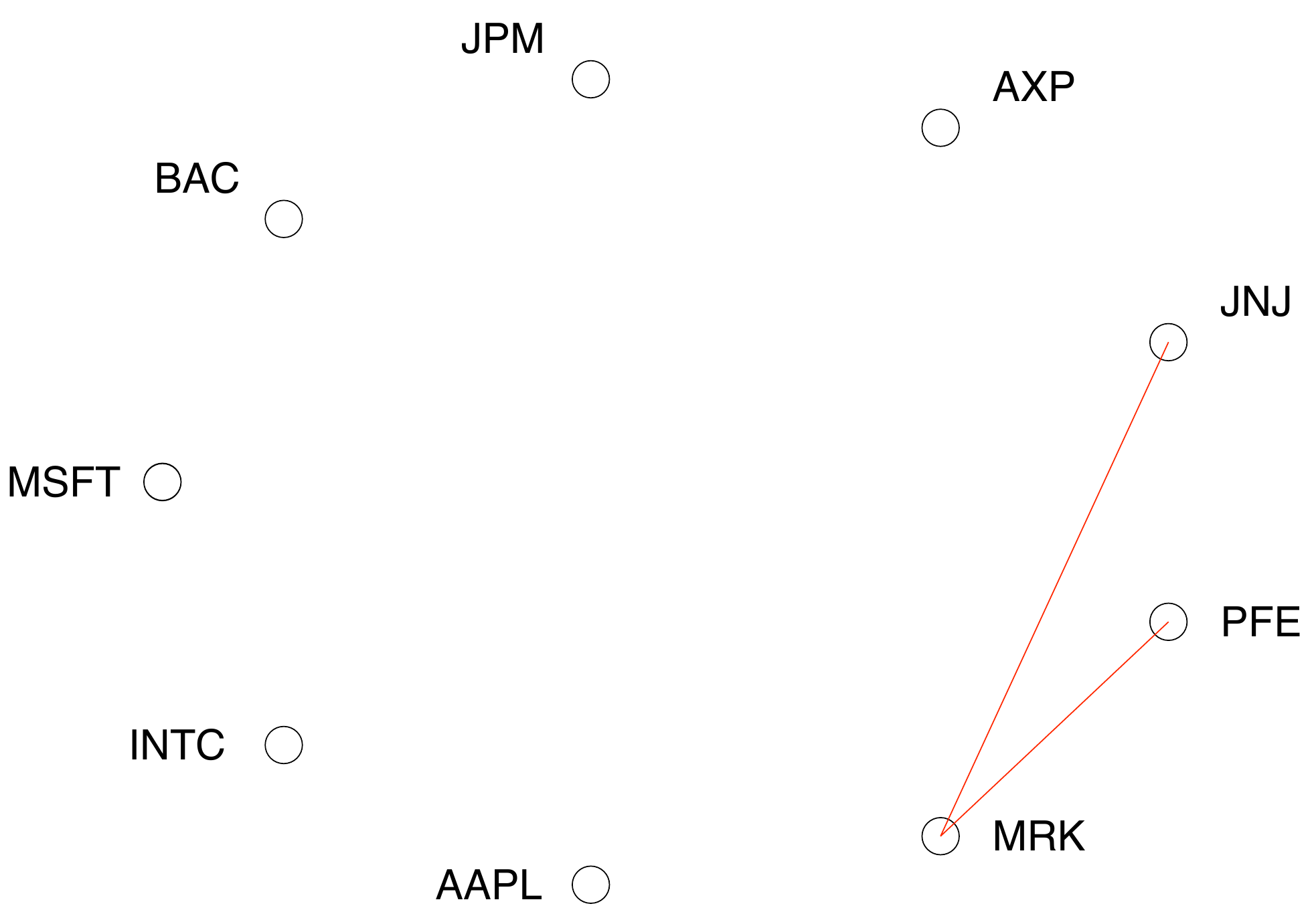}}}}
  \end{center}
  \caption{NASDAQ/NYSE experiment. {\bf{Left column}:} The graphical models selected by MTCCA with Gaussian MT-functions for $\lambda=0.5,0.55,0.58$. {\bf{Middle column}:} The closest graphs selected by ICCA. {\bf{Right column}:} The symmetric difference graphs: the red lines indicate edges found by MTCCA and not by ICCA, and vice-versa for the black lines. For these values of $\lambda$, Gaussian MTCCA detects more dependencies than ICCA: the MTCCA graph has more edges than the closest ICCA graph.}
\label{GraphModFinancial5}
\end{figure}

\begin{figure}[htbp!]
  \begin{center}
    \frame{{\subfigure{\label{GraphModel_3_MTCCA_GAUSS_1}\includegraphics[scale=0.25]{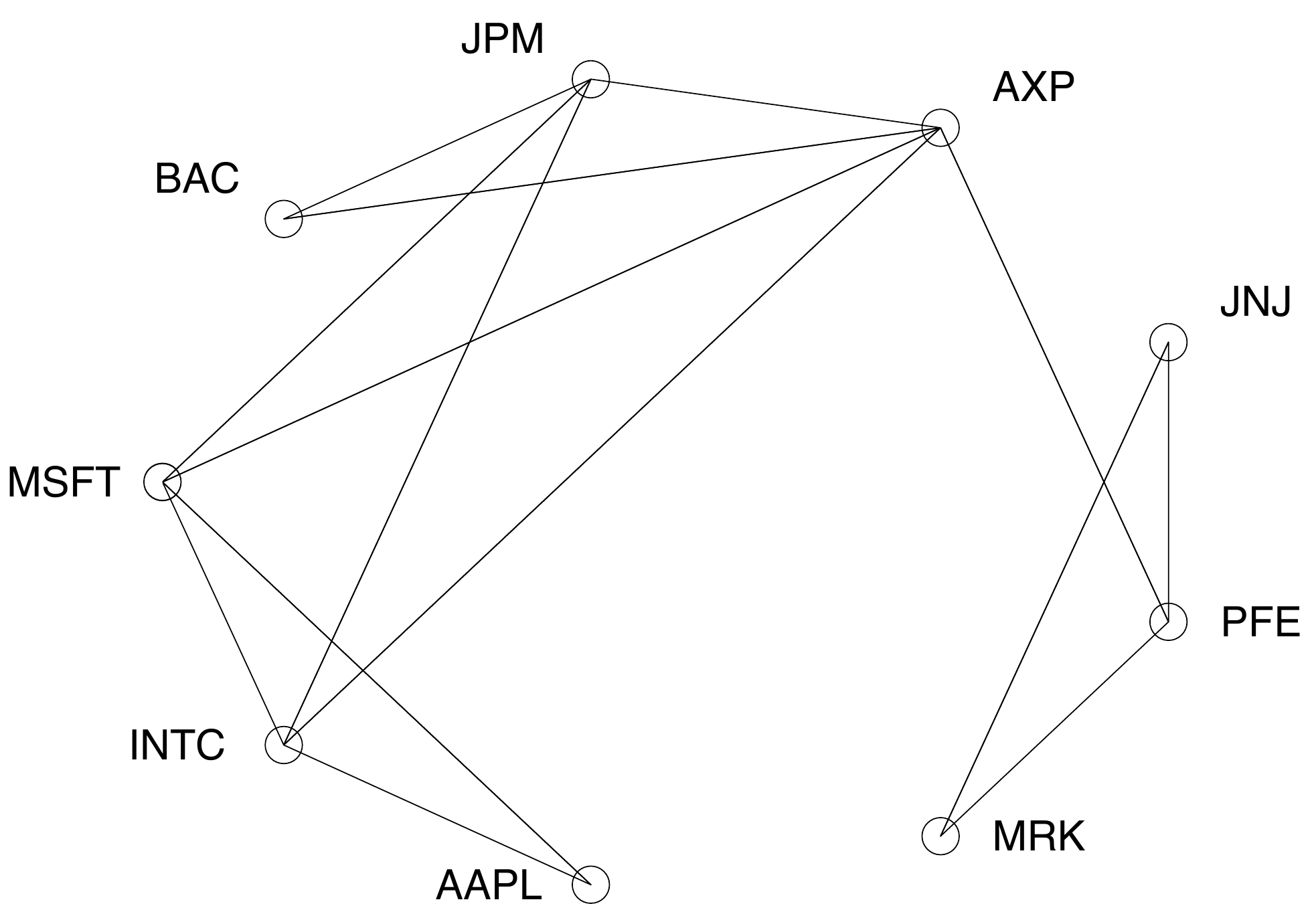}}}}
    \frame{{\subfigure{\label{GraphModel_3_MTCCA_GAUSS_2}\includegraphics[scale=0.25]{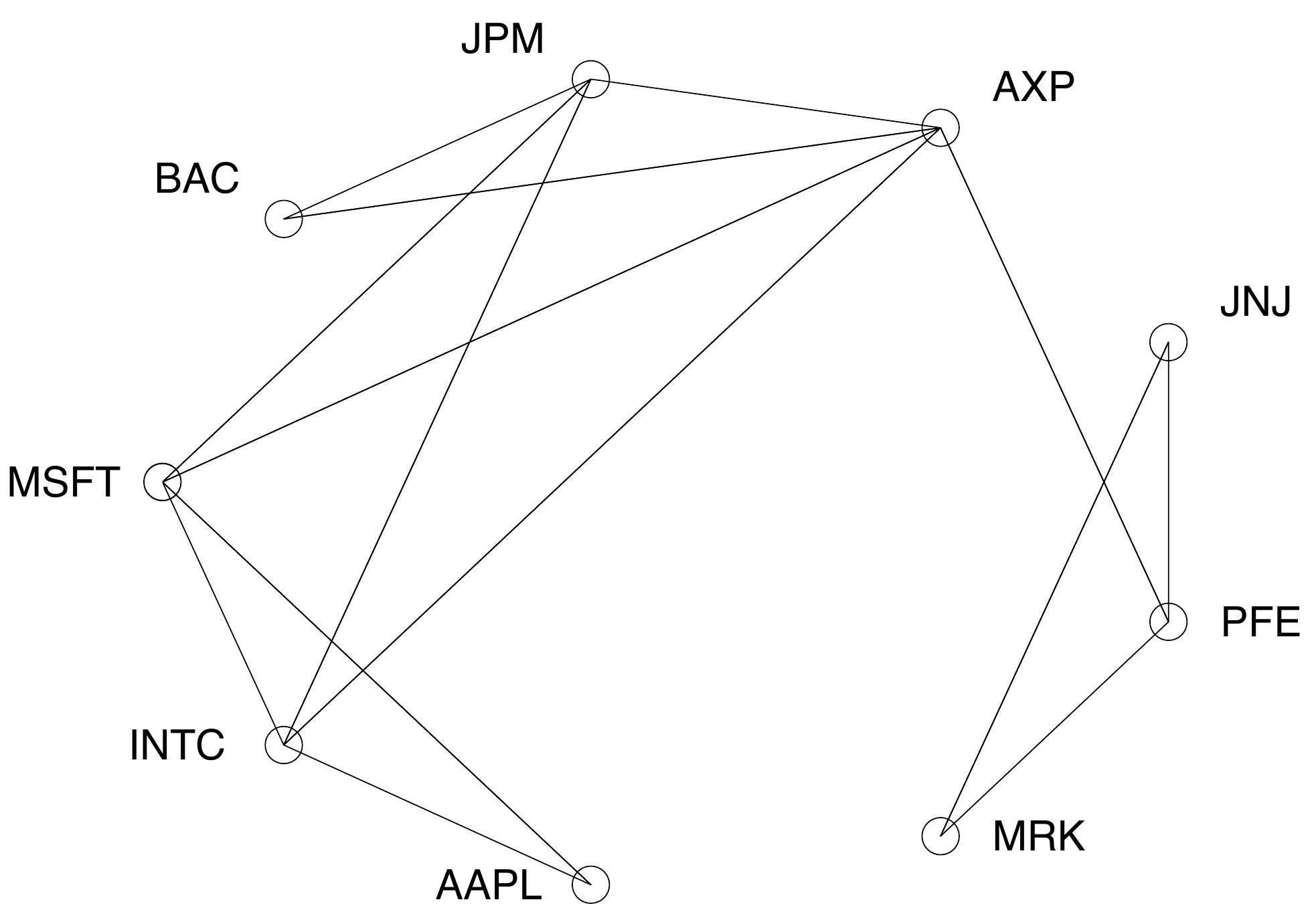}}}}
    \vspace{0.15cm}
    \frame{{\subfigure{\label{GraphModel_3_MTCCA_GAUSS_3}\includegraphics[scale=0.25]{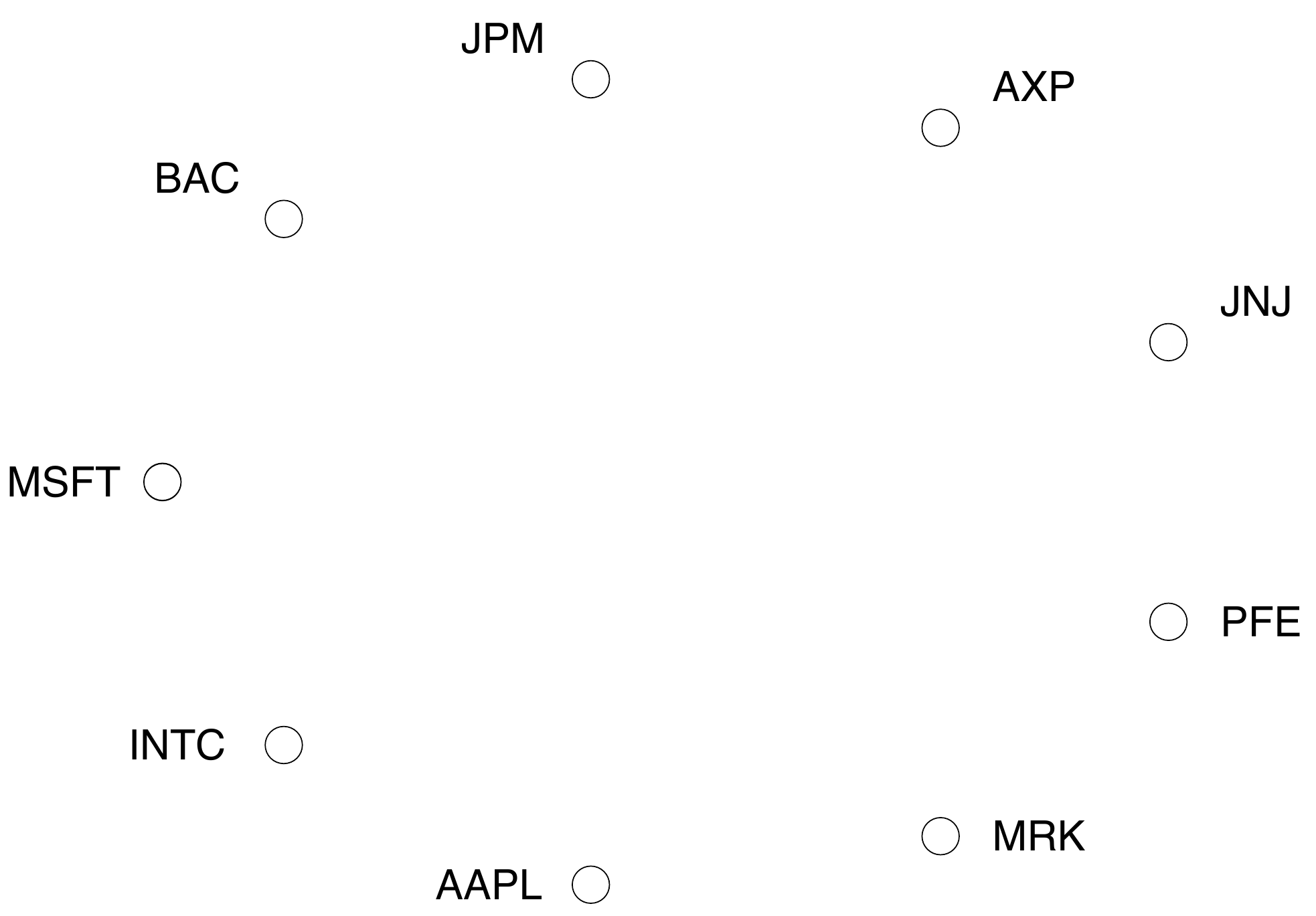}}}}
    \frame{{\subfigure{\label{GraphModel_3_MTCCA_GAUSS_4}\includegraphics[scale=0.25]{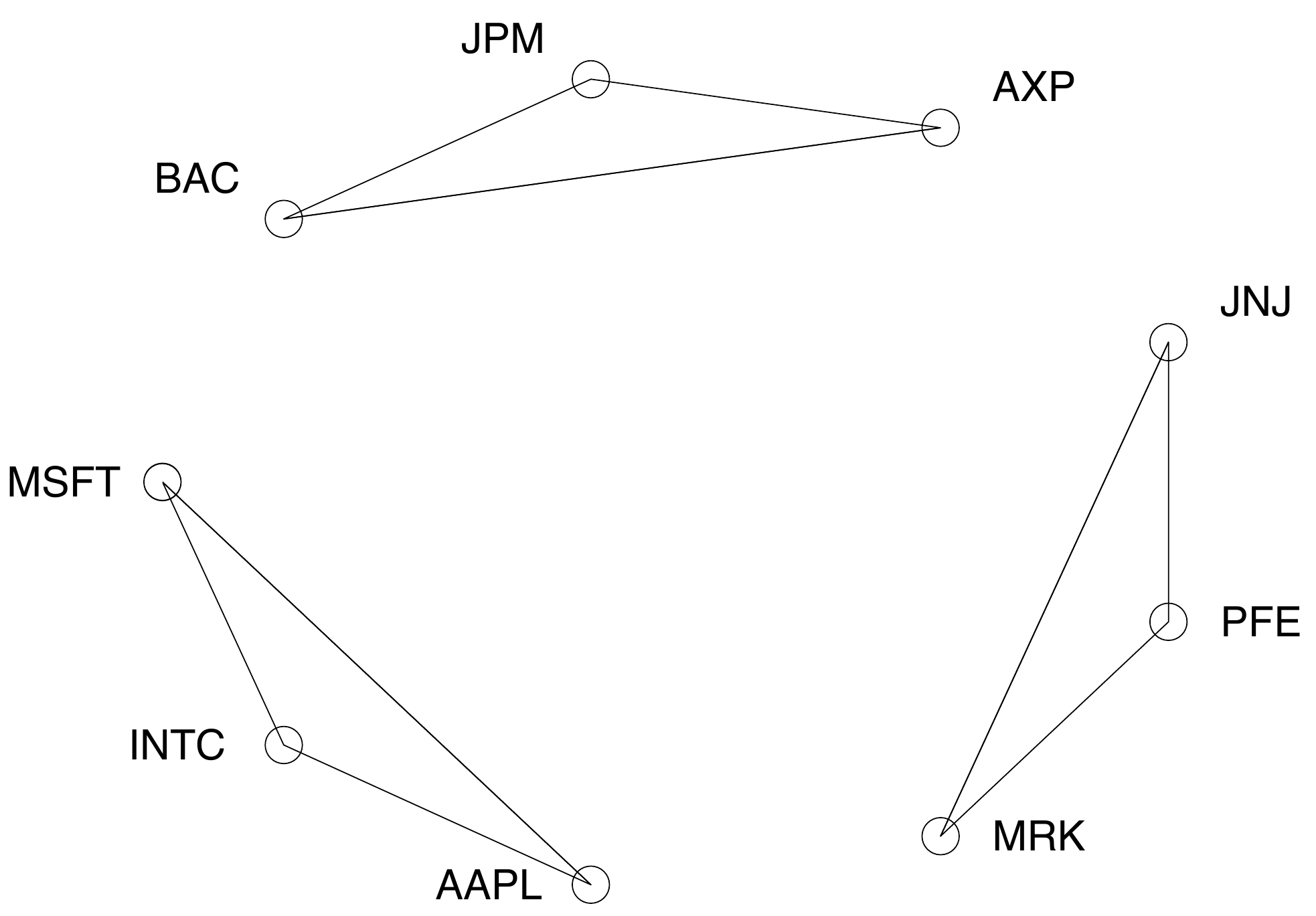}}}}
    \frame{{\subfigure{\label{GraphModel_3_MTCCA_GAUSS_5}\includegraphics[scale=0.25]{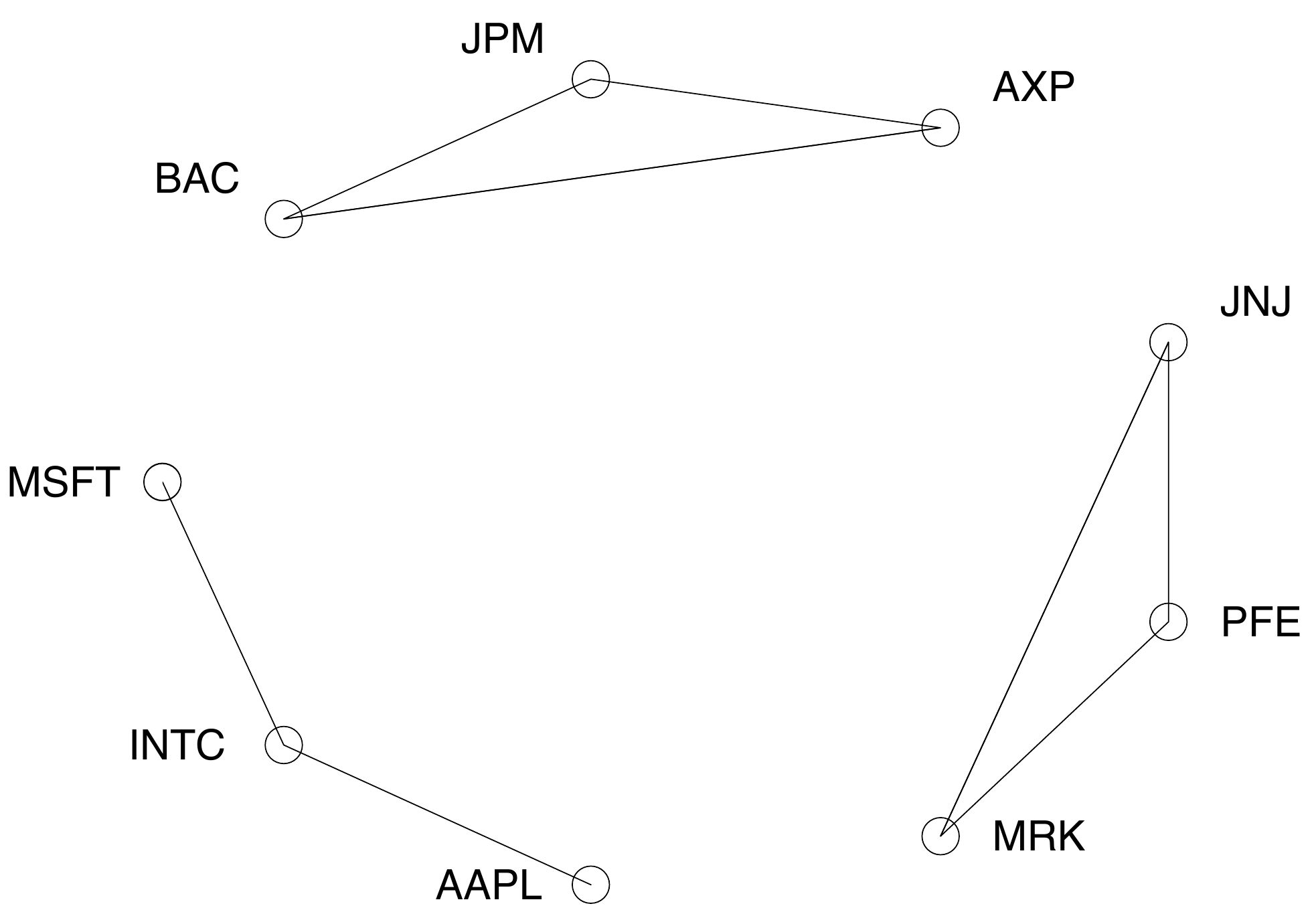}}}}
    \vspace{0.15cm}
    \frame{{\subfigure{\label{GraphModel_3_MTCCA_GAUSS_6}\includegraphics[scale=0.25]{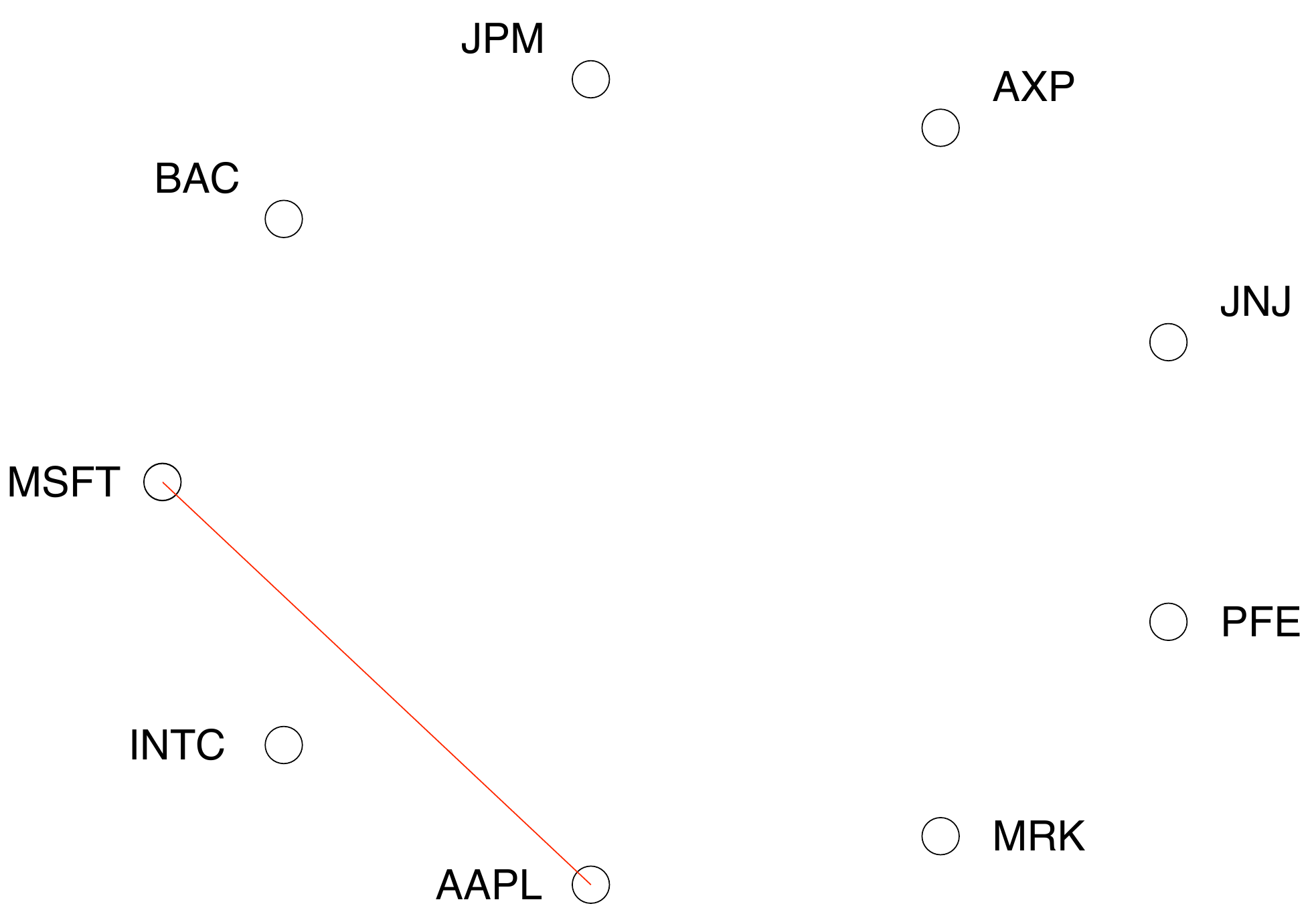}}}}
    \frame{{\subfigure{\label{GraphModel_3_MTCCA_GAUSS_7}\includegraphics[scale=0.25]{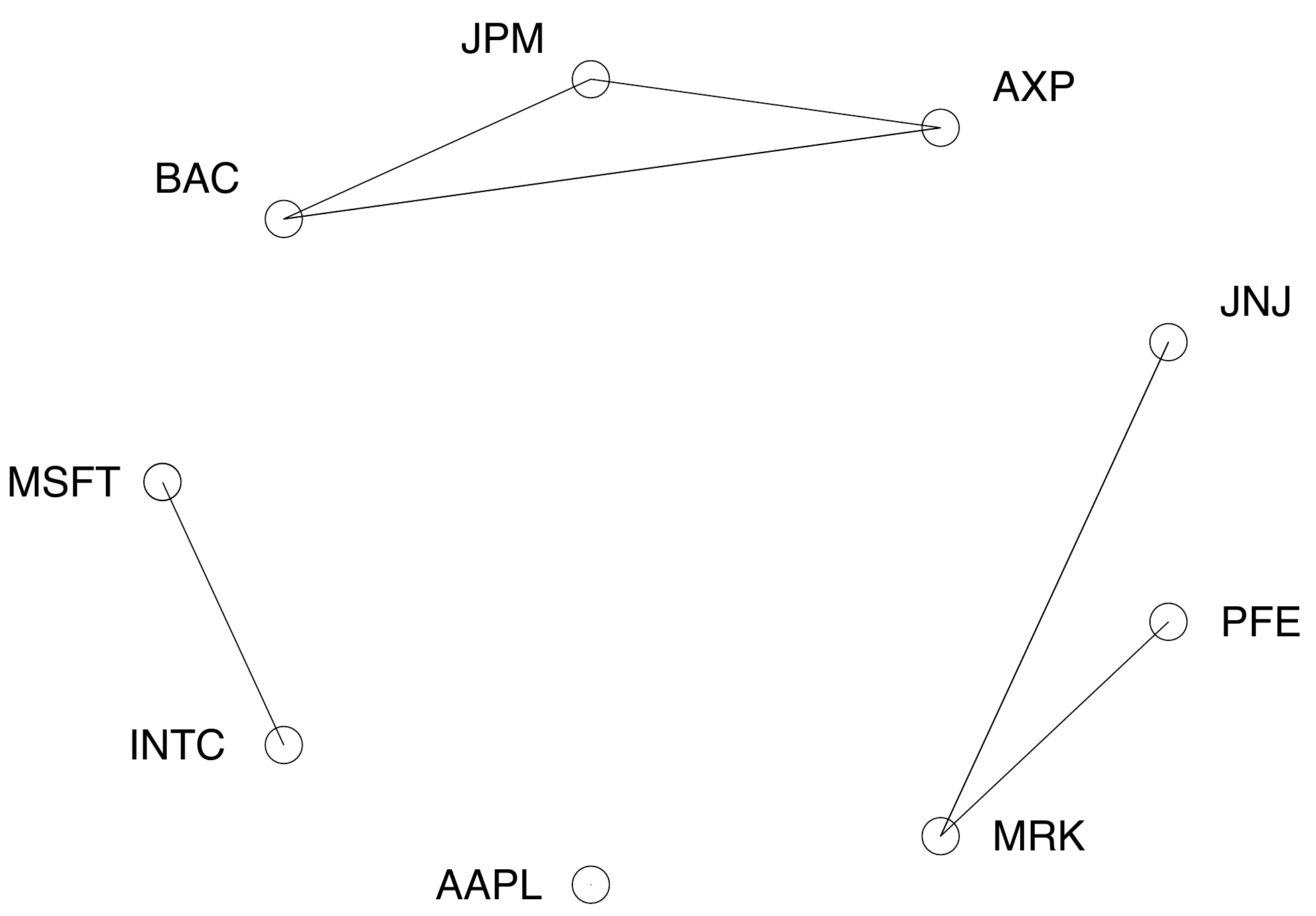}}}}
    \frame{{\subfigure{\label{GraphModel_3_MTCCA_GAUSS_8}\includegraphics[scale=0.25]{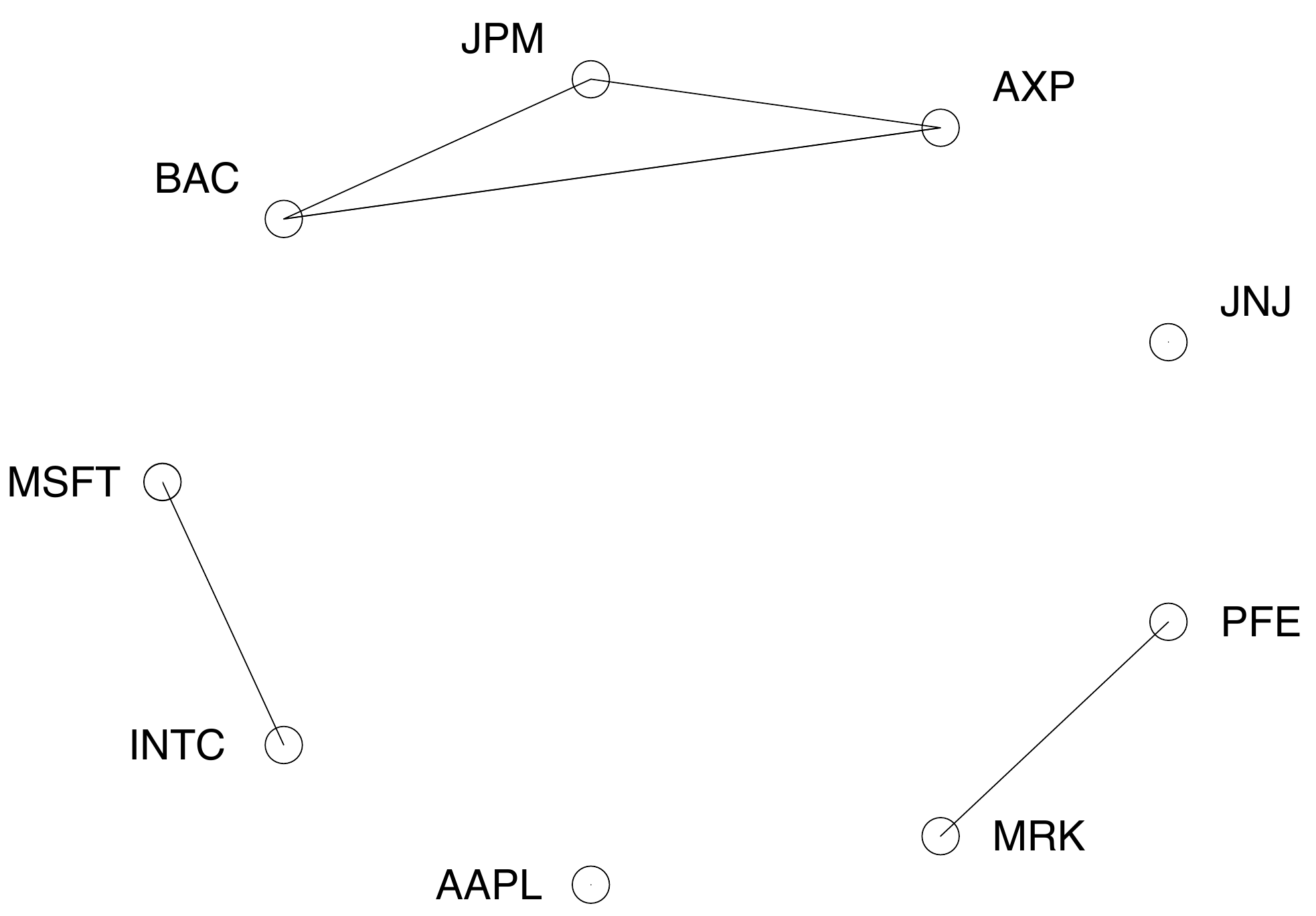}}}}
    \frame{{\subfigure{\label{GraphModel_3_MTCCA_GAUSS_9}\includegraphics[scale=0.25]{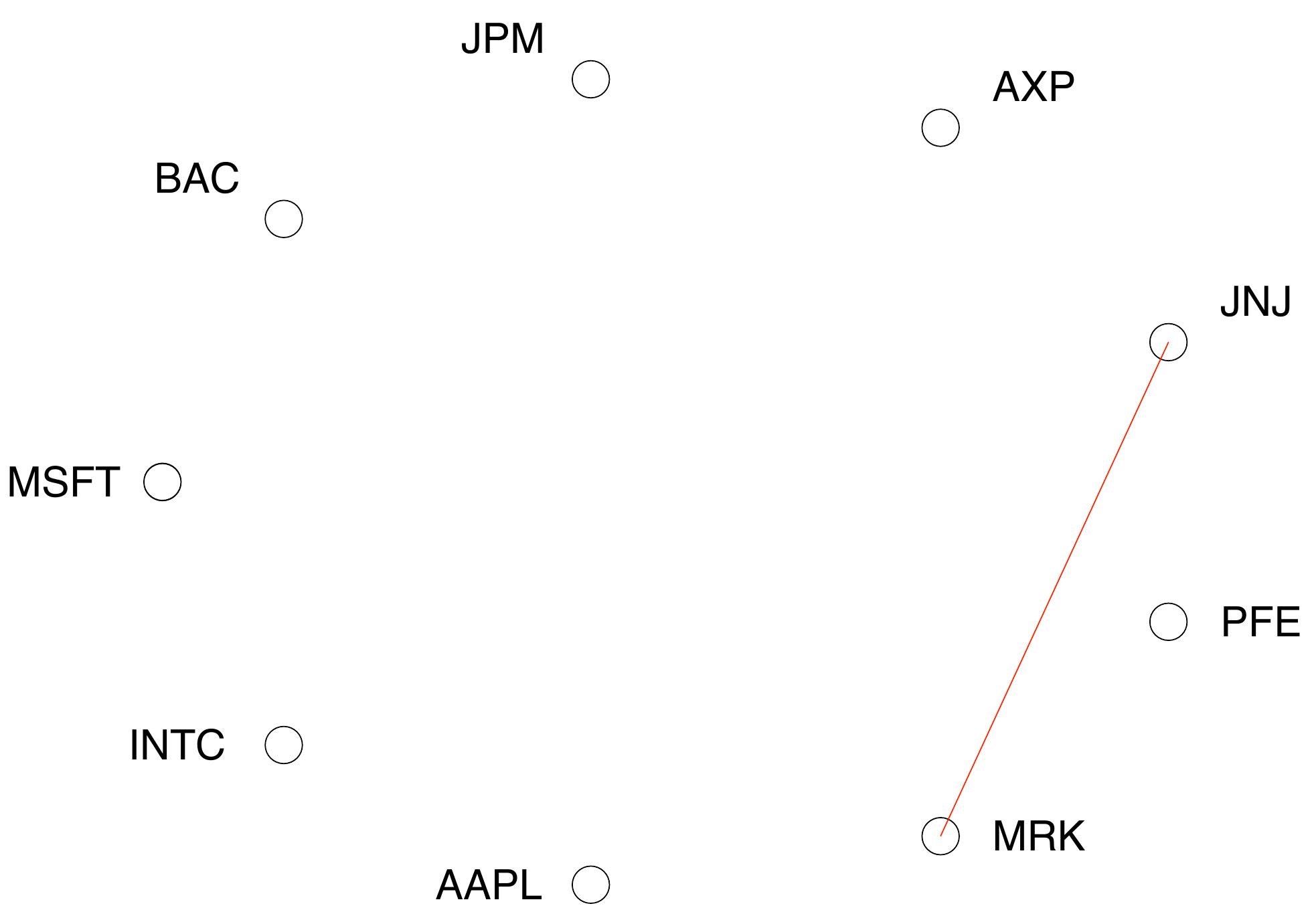}}}}
  \end{center}
  \caption{NASDAQ/NYSE experiment. {\bf{Left column}:} The graphical models selected by MTCCA with Gaussian MT-functions for $\lambda=0.5,0.55,0.58$. {\bf{Middle column}:} The closest graphs selected by KCCA. {\bf{Right column}:} The symmetric difference graphs: the red lines indicate edges found by MTCCA and not by KCCA, and vice-versa for the black lines. For $\lambda=0.55,0.58$ the MTCCA graph has more edges than the closest KCCA graph.}
\label{GraphModFinancial6}
\end{figure}

The empirical first-order canonical correlation coefficients were used for constructing graphical models in which the nodes represent the compared companies. The criterion for connecting a pair of nodes was set to empirical first-order canonical correlation coefficient greater than a threshold $\lambda$.  
In Figs. \ref{GraphModFinancial1}-\ref{GraphModFinancial3} the graphical models selected by MTCCA with exponential MT-functions are compared to LCCA, ICCA and KCCA, respectively. Similarly, in Figs. \ref{GraphModFinancial4}-\ref{GraphModFinancial6} the graphical models selected by MTCCA with Gaussian MT-functions are compared to LCCA, ICCA and KCCA, respectively. In the first column of each figure we show the graphs selected by MTCCA for $\lambda=0.5,0.55,0.58$. In the second column we show the corresponding graphs selected by the other compared method by scanning $\lambda$ over the interval $\left[0,1\right]$ and finding the graph with minimum edit distance \cite{EditDist}. The symmetric difference graphs are shown in the third column. The red lines in the symmetric difference graphs indicate edges found by MTCCA and not by the other compared method, and vice-versa for the black lines. Note that for all of the threshold parameters $\lambda$ investigated, the MTTCA graph shows equal or larger number of dependencies than the closest LCCA, ICCA and KCCA graphs. This result suggests that MTCCA has captured more dependencies than LCCA, ICCA and KCCA. While there is no ground truth validation, the fact that MTCCA clusters together companies in similar sectors (Banking, pharmaceuticals, and technology) provides anecdotal support for the power and applicability of MTCCA. 

\begin{figure}[htbp!]
\centerline{\psfig{figure=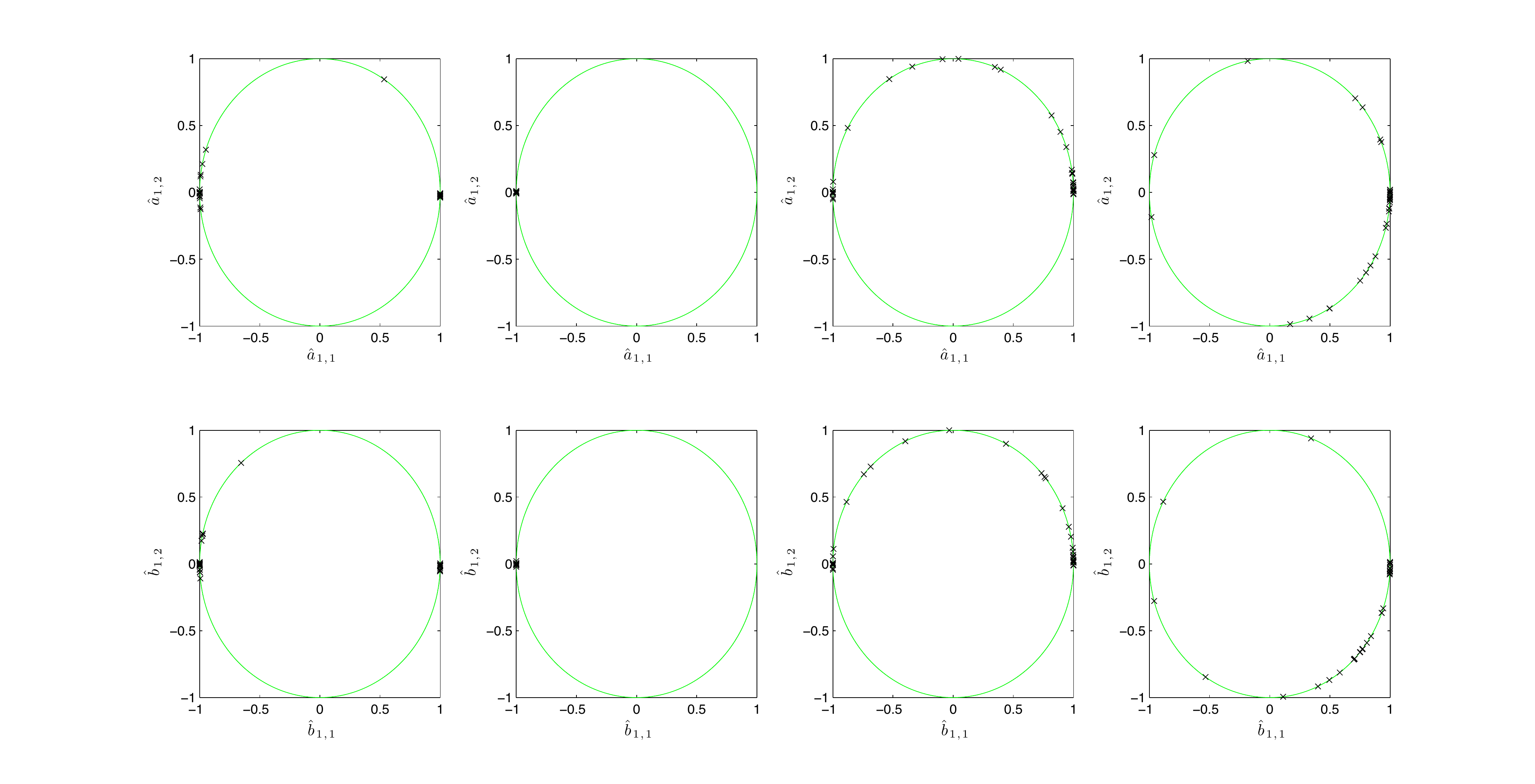,scale=0.5}}
\caption{NASDAQ/NYSE experiment. Distribution of the empirical MT, linear, and informational first-order canonical directions on the unit circle. Left to right ordering: {\textbf{First column}} - MTCCA with exponential MT-functions. {\textbf{Second column}} - MTCCA with Gaussian MT-functions. {\textbf{Third column}} - LCCA. {\textbf{Fourth column}} - ICCA. The estimated MT-canonical directions in first and second columns are much more concentrated than the linear and informational canonical directions in third and fourth columns, respectively. In particular, while linear and informational canonical directions appear to be equally sensitive to the daily log-returns and the daily log-volume ratios, MT-canonical directions are much more sensitive to the former as contrasted to the latter.}
\label{Fig3}
\end{figure}

Fig. \ref{Fig3} depicts the distribution of the empirical MT, linear, and informational first-order canonical directions. Let $\hat{\avec}_{1}=\left[\hat{a}_{1,1},\hat{a}_{1,2}\right]^{T}$ and $\hat{\bvec}_{1}=\left[\hat{b}_{1,1},\hat{b}_{1,2}\right]^{T}$ on the unit circle. Observe that in MTCCA (first and second columns) $\hat{a}_{1,2}$ and $\hat{b}_{1,2}$ are relatively small in comparison to  $\hat{a}_{1,1}$ and $\hat{b}_{1,1}$. One can conclude that, unlike LCCA and ICCA, MTCCA is zeroing in on the strong non-linear dependencies between the daily log-returns of these companies and is de-emphasizing the daily log-volume ratios. This analysis is not performed for KCCA since the empirical canonical directions obtained by KCCA do not correspond to the original coordinates of $\Xmat$ and $\Ymat$. 

We note that in this example the difference between MTCCA and ICCA may possibly arise from the sensitivity of fixed kernel density estimation, preformed in ICCA, to the heavy-tailed financial data \cite{Silverman}. 
\section{Conclusion}
\label{Disc}
In this paper, LCCA was generalized by applying a structured transform to the joint probability distribution of $\Xmat$ and $\Ymat$. By modifying the functions associated  with the transform, this generalization, called MTCCA, preserves independence and captures non-linear dependencies. Two classes of MTCCA were proposed based on specification of MT-functions in the exponential and Gaussian families, respectively. The proposed MTCCA approach was compared to LCCA, ICCA and KCCA for graphical model selection in simulated data having non-linear dependencies, and for measuring long-term associations between pairs of companies traded on the NASDAQ and NYSE stock markets. It is likely that there exist other classes of MT-functions that have a similar capability to accurately detect non-linear dependencies. 

In the paper we have shown that the Hessian of the joint cumulant generating function (\ref{OffHessxy}) is a special case of measure transformed covariance matrix with exponential MT-functions. Therefore, in similar to the generalization proposed in this paper, the techniques in \cite{Yeredor1}-\cite{Yeredor8}, which are based on Hessians of the cumulant generating function, may also be generalized by the measure-transformation framework.
\section{Acknowledgment} 
This research was supported in part by ARO grant W911NF-11-1-0391. 
\appendix
\subsection{Proof of Proposition \ref{Prop1}:}
\label{Prop1Proof}
\begin{enumerate}
\item
\textbf{Property \ref{P1}:}\\
Since $\varphi_{u,v}\left(\xvec,\yvec\right)$ is nonnegative, then by Corollary 2.3.6 in \cite{MeasureTheory} $\qxy$ is a measure on $\mathcal{S}_{\XCalsc\times\YCalsc}$. Furthermore, $\qxy\left(\XCal\times\YCal\right)=1$ so that $\qxy$ is a probability measure on $\mathcal{S}_{\XCalsc\times\YCalsc}$. 
\item
\textbf{Property \ref{P2}:}\\
Follows from definitions 4.1.1 and 4.1.3 in \cite{MeasureTheory}.
\item
\textbf{Property \ref{P3}}:\\
Let $\qx$ and $\qy$ denote the marginal probability measures of $\qxy$, defined on $\SCal_{\XCalsc}$ and $\SCal_{\YCalsc}$, respectively. Additionally, let $A_{x}$ and $A_{y}$ denote arbitrary sets in the $\sigma$-algebras $\SCal_{\XCalsc}$ and $\SCal_{\YCalsc}$, respectively.  Using (\ref{MeasureTransform}) and (\ref{VarPhiDef}), the assumed statistical independence of $\Xmat$ and $\Ymat$ under $\pxy$, and Tonelli's Theorem \cite{Folland}:
\begin{eqnarray}
\label{Th1PrEq2}
\qx\left(A_{x}\right)
&=&\int\limits_{A_{x}\times{\YCalsc}}d\qxy\left(\xvec,\yvec\right)
=\int\limits_{A_{x}\times{\YCalsc}}\frac{u\left(\xvec\right)v\left(\yvec\right)}{{\rm{E}}\left[u\left(\Xmat\right)v\left(\Ymat\right);\pxy\right]}d\pxy\left(\xvec,\yvec\right)
\\\nonumber
&=&
\int\limits_{A_{x}}\frac{u\left(\xvec\right)}{{\rm{E}}\left[u\left(\Xmat\right);\px\right]}d\px\left(\xvec\right)
\int\limits_{\YCalsc}\frac{v\left(\yvec\right)}{{\rm{E}}\left[v\left(\Ymat\right);\py\right]}d\py\left(\yvec\right)
=
\int\limits_{A_{x}}\frac{u\left(\xvec\right)}{{\rm{E}}\left[u\left(\Xmat\right);\px\right]}d\px\left(\xvec\right).
\end{eqnarray}
Similarly, it can be shown that $\qy\left(A_{y}\right)=\int\limits_{A_{y}}\frac{v\left(\yvec\right)}{{\rm{E}}\left[v\left(\Ymat\right);\py\right]}d\py\left(\yvec\right)$, and
\begin{equation}
\label{Th1PrEq4}
\qxy\left({A}_{x}\times{A}_{y}\right)
=
\int\limits_{A_{x}}\frac{u\left(\xvec\right)}{{\rm{E}}\left[u\left(\Xmat\right);\px\right]}d\px\left(\xvec\right)
\int\limits_{A_{y}}\frac{v\left(\xvec\right)}{{\rm{E}}\left[v\left(\Ymat\right);\px\right]}d\py\left(\yvec\right)
=\qx\left(A_{x}\right)\qy\left(A_{y}\right).
\end{equation}
Therefore, since $A_{x}$ and $A_{y}$ are arbitrary, $\Xmat$ and $\Ymat$ are statistically independent under the transformed probability measure $\qxy$.
\item
\textbf{Property \ref{P4}}:\\
According to the definition of $\varphi_{u,v}\left(\xvec,\yvec\right)$ in (\ref{VarPhiDef}), the strict positivity of $u\left(\xvec\right)$ and $v\left(\yvec\right)$, and Property \ref{P2}, we have that $\qxy$ is absolutely continuous w.r.t. $\pxy$ with strictly positive Radon-Nikodym derivative $\frac{d\qxy\left(\xvec,\yvec\right)}{d\pxy\left(\xvec,\yvec\right)}=\varphi_{u,v}\left(\xvec,\yvec\right)$. Therefore, by Proposition 4.1.2 in \cite{MeasureTheory} it is implied that $\pxy$ is absolutely continuous w.r.t. $\qxy$ with a strictly positive Radon-Nikodym derivative given by 
\begin{equation}
\label{RadNik2}
\frac{d\pxy\left(\xvec,\yvec\right)}{d\qxy\left(\xvec,\yvec\right)}=\frac{1}{\varphi_{u,v}\left(\xvec,\yvec\right)}.
\end{equation}

Hence, let $A_{x}$ and $A_{y}$ denote arbitrary sets in the $\sigma$-algebras $\SCal_{\XCalsc}$ and $\SCal_{\YCalsc}$, respectively.  Using (\ref{VarPhiDef}), (\ref{RadNik2}), the assumed statistical independence of $\Xmat$ and $\Ymat$ under $\qxy$, and Tonelli's Theorem \cite{Folland}:
\begin{eqnarray}
\label{Prop4Eq1}
\pxy\left(A_{x}\times{A}_{y}\right)&=&\int\limits_{A_{x}\times{A}_{y}}\frac{1}{\varphi_{u,v}\left(\xvec,\yvec\right)}d\qxy\left(\xvec,\yvec\right)\\\nonumber
&=&
{\rm{E}}\left[u\left(\Xmat\right)v\left(\Ymat\right);\pxy\right]\int\limits_{A_{x}}\frac{1}{u\left(\xvec\right)}d\qx\left(\xvec\right)\int\limits_{A_{y}}\frac{1}{v\left(\yvec\right)}d\qy\left(\yvec\right).
\end{eqnarray}
Similarly, it can be shown that 
\begin{eqnarray}
\label{Prop4Eq2}
\px\left(A_{x}\right)=\pxy\left(A_{x}\times\YCal\right)=
{\rm{E}}\left[u\left(\Xmat\right)v\left(\Ymat\right);\pxy\right]{\rm{E}}\left[\frac{1}{v\left(\Ymat\right)};\qy\right]
\int\limits_{A_{x}}\frac{1}{u\left(\xvec\right)}d\qx\left(\xvec\right)
\end{eqnarray}
and
\begin{eqnarray}
\label{Prop4Eq3} 
\py\left(A_{y}\right)=\pxy\left(\XCal\times{A}_{y}\right)=
{\rm{E}}\left[u\left(\Xmat\right)v\left(\Ymat\right);\pxy\right]{\rm{E}}\left[\frac{1}{u\left(\Xmat\right)};\qx\right]
\int\limits_{A_{y}}\frac{1}{v\left(\yvec\right)}d\qy\left(\yvec\right).
\end{eqnarray}
Now, using (\ref{ExpDef}), (\ref{VarPhiDef}), and (\ref{MeasureTransformRadNik}) we have that  
\begin{equation}
\label{Prop4Eq4}
{\rm{E}}\left[\frac{1}{u\left(\Xmat\right)};\qx\right]={\rm{E}}\left[\frac{1}{u\left(\Xmat\right)};\qxy\right]
={\rm{E}}\left[\frac{\varphi_{u,v}\left(\Xmat,\Ymat\right)}{u\left(\Xmat\right)};\pxy\right]
=\frac{{\rm{E}}\left[{v\left(\Ymat\right)};\py\right]}{{\rm{E}}\left[u\left(\Xmat\right)v\left(\Ymat\right);\pxy\right]},
\end{equation}
and similarly,
\begin{equation}
\label{Prop4Eq5}
{\rm{E}}\left[\frac{1}{v\left(\Ymat\right)};\qy\right]
=\frac{{\rm{E}}\left[{u\left(\Xmat\right)};\px\right]}{{\rm{E}}\left[u\left(\Xmat\right)v\left(\Ymat\right);\pxy\right]}.
\end{equation}
Additionally, by setting $A_{x}=\XCal$ and $A_{y}=\YCal$ in (\ref{Prop4Eq1}), followed by using (\ref{ExpDef}), (\ref{Prop4Eq4}), and (\ref{Prop4Eq5}) it is implied that
\begin{eqnarray}
\label{Prop4Eq7}
{\rm{E}}\left[u\left(\Xmat\right)v\left(\Ymat\right);\pxy\right]={\rm{E}}\left[u\left(\Xmat\right);\px\right]{\rm{E}}\left[v\left(\Ymat\right);\py\right].
\end{eqnarray}
Finally, substitution of (\ref{Prop4Eq7}) into (\ref{Prop4Eq1}), (\ref{Prop4Eq5}) into (\ref{Prop4Eq2}), and (\ref{Prop4Eq4}) into (\ref{Prop4Eq3}) yields
\begin{equation}
\label{Prop4Eq8}
\pxy\left(A_{x}\times{A}_{y}\right)
=
\int\limits_{A_{x}}\frac{{\rm{E}}\left[u\left(\Xmat\right);\px\right]}{u\left(\xvec\right)}d\qx\left(\xvec\right)
\int\limits_{A_{y}}\frac{{\rm{E}}\left[v\left(\Ymat\right);\py\right]}{v\left(\yvec\right)}d\qy\left(\yvec\right)
=
\px\left(A_{x}\right)\py\left(A_{y}\right),
\end{equation}
and therefore, since $A_{x}$ and $A_{y}$ are arbitrary, $\Xmat$ and $\Ymat$ are statistically independent under $\pxy$.
\qed
\end{enumerate}
\subsection{Proof of Theorem \ref{OffOriginIndepProp}:}
\label{ExpKerTh}
Using (\ref{OffHessxy}) and (\ref{MonGenFunc}) one can verify that if the condition in (\ref{IndCond}) is satisfied, then 
\begin{equation} 
\label{ExpKerThEq1}
{M}_{\Xmatsc\Ymatsc}\left(\svec,\tvec\right)={M}_{\Xmatsc}\left(\svec\right){M}_{\Ymatsc}\left(\tvec\right) \hspace{0.2cm}\forall\left(\svec,\tvec\right)\in{U}, \end{equation}
where ${M}_{\Xmatsc}\left(\cdot\right)$ and  ${M}_{\Ymatsc}\left(\cdot\right)$ are the marginal moment generating functions of $\Xmat$ and $\Ymat$, respectively. 
The joint moment generating function reduced to any open region containing the origin, within its region of convergence, uniquely determines the joint distribution \cite{Severini}, \cite{DasGupta} (this property stems from the analyticity of the joint moment generating function about the origin). Hence, by the relation above we have that $\Xmat$ and $\Ymat$ are statistically independent. Conversely, if $\Xmat$ and $\Ymat$ are statistically independent under $\pxy$, then  by Property \ref{P3} of Proposition \ref{Prop1} we have that $\bSigma^{\left(\uexp,\vexp\right)}_{\Xmatsc\Ymatsc}\left(\svec,\tvec\right)=\zerovec$ for all $\left(\svec,\tvec\right)\in{U}$.\qed
\subsection{Proof of Theorem \ref{GaussIndepProp}:}
\label{GaussKerTh}
Using (\ref{VarPhiDef}), (\ref{RxyMod_pxy}), and  (\ref{GaussKernel}) one can easily verify that  
\begin{equation}
\label{OffHessxy21}
\begin{gathered}
\bSigma^{\left(\uGausss,\vGausss\right)}_{\Xmatsc\Ymatsc}\left(\svec,\tvec\right)
=
\frac{{\rm{E}}\left[\Xmat\Ymat^{T}g\left(\Xmat\right)h\left(\Ymat\right)\exp\left(\frac{\svec^{T}\Xmat}{\sigma^{2}}
+\frac{\tvec^{T}\Ymat}{\tau^{2}}\right);\pxy\right]}
{{\rm{E}}\left[g\left(\Xmat\right)h\left(\Ymat\right)\exp\left(\frac{\svec^{T}\Xmat}{\sigma^{2}}
+\frac{\tvec^{T}\Ymat}{\tau^{2}}\right);\pxy\right]}-
\\
\frac{{\rm{E}}\left[\Xmat{g}\left(\Xmat\right)h\left(\Ymat\right)\exp\left(\frac{\svec^{T}\Xmat}{\sigma^{2}}
+\frac{\tvec^{T}\Ymat}{\tau^{2}}\right);\pxy\right]{\rm{E}}\left[\Ymat^{T}g\left(\Xmat\right)h\left(\Ymat\right)\exp\left(\frac{\svec^{T}\Xmat}{\sigma^{2}}
+\frac{\tvec^{T}\Ymat}{\tau^{2}}\right);\pxy\right]}
{{\rm{E}}^{2}\left[g\left(\Xmat\right)h\left(\Ymat\right)\exp\left(\frac{\svec^{T}\Xmat}{\sigma^{2}}
+\frac{\tvec^{T}\Ymat}{\tau^{2}}\right);\pxy\right]},
\end{gathered}
\end{equation}
where 
\begin{eqnarray}
\label{ghDef} 
g\left(\Xmat\right)\triangleq\exp\left(-\frac{\left\|\Xmat\right\|^{2}}{2\sigma^{2}}\right) &{\rm{and}}& h\left(\Ymat\right)\triangleq\exp\left(-\frac{\left\|\Ymat\right\|^{2}}{2\tau^{2}}\right).
\end{eqnarray}
Additionally, define 
\begin{equation}
\label{MXY22}
M^{\left(g,h\right)}_{\Xmatsc\Ymatsc}\left(\svec,\tvec\right)\triangleq{\rm{E}}\left[\exp\left(\svec^{T}\Xmat+\tvec^{T}\Ymat\right);Q^{\left(g,h\right)}_{\Xmatsc\Ymatsc}\right]
\end{equation}
as the joint moment generating function of $\Xmat$ and $\Ymat$ under the transformed probability measure $Q^{\left(g,h\right)}_{\Xmatsc\Ymatsc}$ associated with the MT-functions $g\left(\Xmat\right)$ and $h\left(\Ymat\right)$ in (\ref{ghDef}). Using (\ref{ExpDef}) and (\ref{MeasureTransformRadNik}) it can be shown that 
\begin{equation}
\label{MXY23}
M^{\left(g,h\right)}_{\Xmatsc\Ymatsc}\left(\svec,\tvec\right)={\rm{E}}\left[\exp\left(\svec^{T}\Xmat+\tvec^{T}\Ymat\right)\varphi_{g,h}\left(\Xmat,\Ymat\right);\pxy\right],\end{equation}
where $\varphi_{g,h}\left(\Xmat,\Ymat\right)$ is defined in (\ref{VarPhiDef}). Therefore, by (\ref{OffHessxy21}) and (\ref{MXY23}) we have that 
\begin{equation}
\label{OffHessxy2}
\bSigma^{\left(\uGausss,\vGausss\right)}_{\Xmatsc\Ymatsc}\left(\svec,\tvec\right)=\sigma^{2}\tau^{2}
\frac{\partial^{2}\log{M^{\left(g,h\right)}_{\Xmatsc\Ymatsc}}\left({\sigma^{-2}}\svec,{\tau^{-2}}\tvec\right)}{\partial\svec\partial\tvec^{T}}.
\end{equation}
Hence, if the condition in (\ref{IndCond2}) is satisfied, then by the properties of the joint moment generating function 
\cite{Severini}, \cite{DasGupta}, it is implied that $\Xmat$ and $\Ymat$ are statistically independent under $Q^{\left(g,h\right)}_{\Xmatsc\Ymatsc}$. 
Thus, since the MT-functions $g\left(\Xmat\right)$ and $h\left(\Ymat\right)$ are strictly positive, then by Property \ref{P4} of Proposition \ref{Prop1} we conclude that 
$\Xmat$ and $\Ymat$ are statistically independent under $\pxy$.
Conversely, if $\Xmat$ and $\Ymat$ are statistically independent under $\pxy$, then  by Property \ref{P3} of Proposition \ref{Prop1} we have that $\bSigma^{\left(\uGausss,\vGausss\right)}_{\Xmatsc\Ymatsc}\left(\svec,\tvec\right)=\zerovec$ for all $\left(\svec,\tvec\right)\in{U}$.\qed
\subsection{Proof of Proposition \ref{LowerBoundTh}:}
\label{LowerBoundProof}
Let $\evec^{\left(p\right)}_{i}$ denote a $p$-dimensional column vector, where $\left[\evec^{\left(p\right)}_{i}\right]_{k}=\delta_{i,k}$, and $\delta_{\left(\cdot,\cdot\right)}$ denotes the Kronecker delta function. It is easily verified that
\begin{equation}
\label{LowerBoundProofEq1}
\sum\limits_{i=1}^{p}\sum\limits_{j=1}^{q}\frac{\left(\evec^{\left(p\right)T}_{i}\bSigma^{\left(u,v\right)}_{\Xmatsc\Ymatsc}\left(\svec,\tvec\right)
\evec^{\left(q\right)}_{j}\right)^{2}}
{\evec^{\left(p\right)T}_{i}\bSigma^{\left(u,v\right)}_{\Xmatsc}\left(\svec,\tvec\right)
\evec^{\left(q\right)}_{j}\evec^{\left(p\right)T}_{i}\bSigma^{\left(u,v\right)}_{\Ymatsc}\left(\svec,\tvec\right)
\evec^{\left(q\right)}_{j}}
\leq{p}\cdot{q}\cdot\max_{\avec\neq\zerovec,\bvec\neq\zerovec}\frac{\left(\avec^{T}\bSigma^{\left(u,v\right)}_{\Xmatsc\Ymatsc}\left(\svec,\tvec\right)\bvec\right)^{2}}
{\avec^{T}\bSigma^{\left(u,v\right)}_{\Xmatsc}\left(\svec,\tvec\right)\avec
\bvec^{T}\bSigma^{\left(u,v\right)}_{\Ymatsc}\left(\svec,\tvec\right)\bvec}.
\end{equation}
Hence, by (\ref{LowerBoundProofEq1})
\begin{eqnarray}
\label{LowerBoundProofEq2}
&&\left(\frac{1}{p\cdot{q}}\sum\limits_{i=1}^{p}\sum\limits_{j=1}^{q}\frac{\left(\evec^{\left(p\right)T}_{i}\bSigma^{\left(u,v\right)}_{\Xmatsc\Ymatsc}\left(\svec,\tvec\right)
\evec^{\left(q\right)}_{j}\right)^{2}}
{\evec^{\left(p\right)T}_{i}\bSigma^{\left(u,v\right)}_{\Xmatsc}\left(\svec,\tvec\right)
\evec^{\left(q\right)}_{j}\evec^{\left(p\right)T}_{i}\bSigma^{\left(u,v\right)}_{\Ymatsc}\left(\svec,\tvec\right)
\evec^{\left(q\right)}_{j}}\right)^{1/2}
\\\nonumber
&\leq&
\left(\max\limits_{\avec\neq\zerovec,\bvec\neq\zerovec}\frac{\left(\avec^{T}\bSigma^{\left(u,v\right)}_{\Xmatsc\Ymatsc}\left(\svec,\tvec\right)\bvec\right)^{2}}
{\avec^{T}\bSigma^{\left(u,v\right)}_{\Xmatsc}\left(\svec,\tvec\right)\avec
\bvec^{T}\bSigma^{\left(u,v\right)}_{\Ymatsc}\left(\svec,\tvec\right)\bvec}\right)^{1/2}
\\\nonumber
&=&
\max\limits_{\avec\neq\zerovec,\bvec\neq\zerovec}\frac{\avec^{T}\bSigma^{\left(u,v\right)}_{\Xmatsc\Ymatsc}\left(\svec,\tvec\right)\bvec}
{\sqrt{\avec^{T}\bSigma^{\left(u,v\right)}_{\Xmatsc}\left(\svec,\tvec\right)\avec}
\sqrt{\bvec^{T}\bSigma^{\left(u,v\right)}_{\Ymatsc}\left(\svec,\tvec\right)\bvec}}
\\\nonumber
&=&
\max\limits_{\avec,\bvec}{\avec^{T}\bSigma^{\left(u,v\right)}_{\Xmatsc\Ymatsc}\left(\svec,\tvec\right)\bvec}\hspace{0.3cm}{\rm{s.t.}}\hspace{0.3cm}\avec^{T}\bSigma^{\left(u,v\right)}_{\Xmatsc}\avec=\bvec^{T}\bSigma^{\left(u,v\right)}_{\Ymatsc}\bvec=1,
\end{eqnarray}
where the last equality stems from the invariance of $\frac{\avec^{T}\bSigma^{\left(u,v\right)}_{\Xmatsc\Ymatsc}\left(\svec,\tvec\right)\bvec}
{\sqrt{\avec^{T}\bSigma^{\left(u,v\right)}_{\Xmatsc}\left(\svec,\tvec\right)\avec}
\sqrt{\bvec^{T}\bSigma^{\left(u,v\right)}_{\Ymatsc}\left(\svec,\tvec\right)\bvec}}$ to normalization of $\avec$ and $\bvec$. 
Therefore, according to (\ref{CCAOptProbMod}), (\ref{PsiDef}) and (\ref{LowerBoundProofEq2}), the relation in (\ref{LowerBoundEq}) is verified.\qed
\subsection{Proof of Theorem \ref{IndProp}:}
\label{IndPropProof}
If $\rho_{1}\left(\bSigma^{\left(u,v\right)}_{\Xmatsc}\left(\svec^{*},\tvec^{*}\right),\bSigma^{\left(u,v\right)}_{\Ymatsc}\left(\svec^{*},\tvec^{*}\right),\bSigma^{\left(u,v\right)}_{\Xmatsc\Ymatsc}\left(\svec^{*},\tvec^{*}\right)\right)=0$, then by (\ref{LowerBoundEq}) and the positivity of $\psi\left(\cdot,\cdot,\cdot\right)$ $$\psi\left(\bSigma^{\left(u,v\right)}_{\Xmatsc}\left(\svec^{*},\tvec^{*}\right),\bSigma^{\left(u,v\right)}_{\Ymatsc}\left(\svec^{*},\tvec^{*}\right),\bSigma^{\left(u,v\right)}_{\Xmatsc\Ymatsc}\left(\svec^{*},\tvec^{*}\right)\right)=0.$$ 
Therefore, since by (\ref{MaxProb}) $\left(\svec^{*},\tvec^{*}\right)$ are the maximizers of $\psi\left(\bSigma^{\left(u,v\right)}_{\Xmatsc}\left(\svec,\tvec\right),\bSigma^{\left(u,v\right)}_{\Ymatsc}\left(\svec,\tvec\right),\bSigma^{\left(u,v\right)}_{\Xmatsc\Ymatsc}\left(\svec,\tvec\right)\right)$ over $V$, which is a closed region in $\Rsp^{p}\times\Rsp^{q}$ containing the origin, we have that 
$$\psi\left(\bSigma^{\left(u,v\right)}_{\Xmatsc}\left(\svec,\tvec\right),\bSigma^{\left(u,v\right)}_{\Ymatsc}\left(\svec,\tvec\right),\bSigma^{\left(u,v\right)}_{\Xmatsc\Ymatsc}\left(\svec,\tvec\right)\right)=0\hspace{0.4cm}\forall\left(\svec,\tvec\right)\in{V}.$$
Hence, by the definition (\ref{PsiDef}) of $\psi\left(\cdot,\cdot,\cdot\right)$, $\bSigma^{\left(u,v\right)}_{\Xmatsc\Ymatsc}\left(\svec,\tvec\right)=\zerovec$ on the interior of $V$, which is an open region in $\Rsp^{p}\times\Rsp^{q}$ containing the origin. Thus, since the MT-functions $u\left(\cdot\right)$ and $v\left(\cdot\right)$ are chosen according to  (\ref{ExpKernel}) or (\ref{GaussKernel}), by Theorems \ref{OffOriginIndepProp} and \ref{GaussIndepProp} $\Xmat$ and $\Ymat$ must be statistically independent under $\pxy$.

Conversely, if $\Xmat$ and $\Ymat$ are statistically independent under $\pxy$, then by Property \ref{P3} of Proposition \ref{Prop1} we have that $\bSigma^{\left(u,v\right)}_{\Xmatsc\Ymatsc}\left(\svec,\tvec\right)=\zerovec$ for all $\left(\svec,\tvec\right)\in{V}$, and in particular for $\left(\svec^{*},\tvec^{*}\right)$. Therefore, by (\ref{CCAOptProbMod}), $\rho_{1}\left(\bSigma^{\left(u,v\right)}_{\Xmatsc}\left(\svec^{*},\tvec^{*}\right),\bSigma^{\left(u,v\right)}_{\Ymatsc}\left(\svec^{*},\tvec^{*}\right),\bSigma^{\left(u,v\right)}_{\Xmatsc\Ymatsc}\left(\svec^{*},\tvec^{*}\right)\right)=0$.\qed
\subsection{Proof of Proposition \ref{ConsistentEst}:}
\label{ConsistEstProof}
It suffices to show that if the conditions in (\ref{Cond1}) and (\ref{Cond2}) are satisfied, then $\hat{\bSigma}^{\left(u,v\right)}_{\Xmatsc\Ymatsc}\rightarrow{\bSigma}^{\left(u,v\right)}_{\Xmatsc\Ymatsc}$ almost surely as $N\rightarrow\infty$. Convergence proofs for $\hat{\bSigma}^{\left(u,v\right)}_{\Xmatsc}$ and $\hat{\bSigma}^{\left(u,v\right)}_{\Ymatsc}$ are very similar and therefore omitted. 

According to (\ref{Rxy_uv_Est})-(\ref{varphi_hat})
\begin{equation}
\label{Rxy_uv_Est_Lim}
\lim\limits_{N\rightarrow\infty}\hat{\bSigma}^{\left(u,v\right)}_{\Xmatsc\Ymatsc}=
\lim\limits_{N\rightarrow\infty}\frac{1}{N}\sum\limits_{n=1}^{N}\Xmat_{n}\Ymat^{T}_{n}\hat{\varphi}_{u,v}\left(\Xmat_{n},\Ymat_{n}\right)
-\lim\limits_{N\rightarrow\infty}\hat{\muvec}^{\left(u,v\right)}_{\Xmatsc}
\lim\limits_{N\rightarrow\infty}\hat{\muvec}^{\left(u,v\right)T}_{\Ymatsc},
\end{equation}
where
\begin{equation}
\label{XY_Lim}
\lim\limits_{N\rightarrow\infty}\frac{1}{N}\sum\limits_{n=1}^{N}\Xmat_{n}\Ymat^{T}_{n}\hat{\varphi}_{u,v}\left(\Xmat_{n},\Ymat_{n}\right)
=\frac{\lim\limits_{N\rightarrow\infty}\frac{1}{N}\sum\limits_{n=1}^{N}\Xmat_{n}\Ymat^{T}_{n}u\left(\Xmat_{n}\right)v\left(\Ymat_{n}\right)}
{\lim\limits_{N\rightarrow\infty}\frac{1}{N}\sum\limits_{n=1}^{N}u\left(\Xmat_{n}\right)v\left(\Ymat_{n}\right)},
\end{equation}
\begin{equation}
\label{mux_Lim}
\lim\limits_{N\rightarrow\infty}\hat{\muvec}^{\left(u,v\right)}_{\Xmatsc}=
\frac{\lim\limits_{N\rightarrow\infty}\frac{1}{N}\sum\limits_{n=1}^{N}\Xmat_{n}u\left(\Xmat_{n}\right)v\left(\Ymat_{n}\right)}
{\lim\limits_{N\rightarrow\infty}\frac{1}{N}\sum\limits_{n=1}^{N}u\left(\Xmat_{n}\right)v\left(\Ymat_{n}\right)},
\end{equation}
\begin{equation}
\label{muy_Lim} 
\lim\limits_{N\rightarrow\infty}\hat{\muvec}^{\left(u,v\right)}_{\Ymatsc}=
\frac{\lim\limits_{N\rightarrow\infty}\frac{1}{N}\sum\limits_{n=1}^{N}\Ymat_{n}u\left(\Xmat_{n}\right)v\left(\Ymat_{n}\right)}
{\lim\limits_{N\rightarrow\infty}\frac{1}{N}\sum\limits_{n=1}^{N}u\left(\Xmat_{n}\right)v\left(\Ymat_{n}\right)},
\end{equation}
and it is assumed that the denominator
\begin{equation}
\label{NonZeroAssumption}
{\lim\limits_{N\rightarrow\infty}\frac{1}{N}\sum\limits_{n=1}^{N}u\left(\Xmat_{n}\right)v\left(\Ymat_{n}\right)}\neq0\hspace{0.2cm}{\rm{a.s.}}
\end{equation}

In the following, the limits of the series in the r.h.s. of (\ref{XY_Lim})-(\ref{muy_Lim}) are obtained. Additionally, in Remark \ref{Remark3} below, we show that the assumption in (\ref{NonZeroAssumption}) is satisfied. Since $\left(\Xmat_{n},\Ymat_{n}\right)$, $n=1,\ldots,N$ is a sequence of i.i.d. samples of $\left(\Xmat,\Ymat\right)$, then the random matrices 
$\Xmat_{n}\Ymat^{T}_{n}u\left(\Xmat_{n}\right)v\left(\Ymat_{n}\right)$, $n=1,\ldots,N$, in the r.h.s. of (\ref{XY_Lim}), define a sequence of i.i.d. samples of $\Xmat\Ymat^{T}u\left(\Xmat\right)v\left(\Ymat\right)$. Moreover, if  ${\rm{E}}\left[X^{4}_{k};\px\right]<\infty$, for any $k=1,\ldots,p$, ${\rm{E}}\left[Y^{4}_{l};\py\right]<\infty$,  for any $l=1,\ldots,q$, 
${\rm{E}}\left[u^{4}\left(\Xmat\right);\px\right]<\infty$, and ${\rm{E}}\left[v^{4}\left(\Ymat\right);\py\right]<\infty$, then
\begin{eqnarray}
{\rm{E}}\left[\left|X_{k}Y_{l}u\left(\Xmat\right)v\left(\Ymat\right)\right|;\pxy\right]
&\leq&
\left({\rm{E}}\left[\left(X_{k}Y_{l}\right)^{2};\pxy\right]{\rm{E}}\left[\left(u\left(\Xmat\right)v\left(\Ymat\right)\right)^{2};\pxy\right]\right)^{\frac{1}{2}}
\\\nonumber
&\leq&
\left({\rm{E}}\left[X^{4}_{k};\px\right]{\rm{E}}\left[Y^{4}_{l};\py\right]{\rm{E}}\left[u^{4}\left(\Xmat\right);\px\right]{\rm{E}}\left[v^{4}\left(\Ymat\right);\py\right]\right)^{\frac{1}{4}}<\infty,
\end{eqnarray}
for any $k=1,\ldots,p$ and any $l=1,\ldots,q$, where the second and third semi-inequalities stem from the H\"older inequality for random variables \cite{MeasureTheory}.
Therefore, by Khinchine's strong law of large numbers (KSLLN) \cite{Folland}
\begin{equation}
\label{NomxyLim}
\lim\limits_{N\rightarrow\infty}\frac{1}{N}\sum\limits_{n=1}^{N}\Xmat_{n}\Ymat^{T}_{n}u\left(\Xmat_{n}\right)v\left(\Ymat_{n}\right)
={\rm{E}}\left[\Xmat\Ymat^{T}u\left(\Xmat\right)v\left(\Ymat\right);\pxy\right]\hspace{0.2cm}{\rm{a.s.}}
\end{equation}
Similarly, it can be shown that if the conditions in (\ref{Cond1}) and (\ref{Cond2}) are satisfied, then by the KSLLN 
\begin{equation}
\label{NomxLim}
\lim\limits_{N\rightarrow\infty}\frac{1}{N}\sum\limits_{n=1}^{N}\Xmat_{n}u\left(\Xmat_{n}\right)v\left(\Ymat_{n}\right)
={\rm{E}}\left[\Xmat{u}\left(\Xmat\right)v\left(\Ymat\right);\pxy\right]\hspace{0.2cm}{\rm{a.s.}},
\end{equation}
\begin{equation}
\label{NomyLim}
\lim\limits_{N\rightarrow\infty}\frac{1}{N}\sum\limits_{n=1}^{N}\Ymat_{n}u\left(\Xmat_{n}\right)v\left(\Ymat_{n}\right)
={\rm{E}}\left[\Ymat{u}\left(\Xmat\right)v\left(\Ymat\right);\pxy\right]\hspace{0.2cm}{\rm{a.s.}},
\end{equation}
and
\begin{equation}
\label{DenLim}
\lim\limits_{N\rightarrow\infty}\frac{1}{N}\sum\limits_{n=1}^{N}u\left(\Xmat_{n}\right)v\left(\Ymat_{n}\right)
={\rm{E}}\left[u\left(\Xmat\right)v\left(\Ymat\right);\pxy\right]\hspace{0.2cm}{\rm{a.s.}}
\end{equation}
\begin{Remark}
\label{Remark3}
By (\ref{DenLim}) and the assumption in (\ref{Assumption2}) the denominator in the r.h.s. of (\ref{XY_Lim})-(\ref{muy_Lim}) is non-zero almost surely.
\end{Remark}

Therefore, since the sequences in the l.h.s. of (\ref{XY_Lim})-(\ref{muy_Lim}) are obtained by continuous mappings of the elements of the sequences in their r.h.s., then by  (\ref{NomxyLim})-(\ref{DenLim}), and the Mann-Wald Theorem \cite{MannWald}
\begin{equation}
\label{XY_Lim_Closed}
\lim\limits_{N\rightarrow\infty}\frac{1}{N}\sum\limits_{n=1}^{N}\Xmat_{n}\Ymat^{T}_{n}\hat{\varphi}_{u,v}\left(\Xmat_{n},\Ymat_{n}\right)
=\frac{{\rm{E}}\left[\Xmat\Ymat^{T}u\left(\Xmat\right)v\left(\Ymat\right);\pxy\right]}{{\rm{E}}\left[u\left(\Xmat\right)v\left(\Ymat\right);\pxy\right]}
={\rm{E}}\left[\Xmat\Ymat^{T}\varphi_{u,v}\left(\Xmat,\Ymat\right);\pxy\right]\hspace{0.2cm}{\rm{a.s.}}
\end{equation}
\begin{equation}
\label{mux_Lim_Closed}
\lim\limits_{N\rightarrow\infty}\hat{\muvec}^{\left(u,v\right)}_{\Xmatsc}
=\frac{{\rm{E}}\left[\Xmat{u}\left(\Xmat\right)v\left(\Ymat\right);\pxy\right]}{{\rm{E}}\left[u\left(\Xmat\right)v\left(\Ymat\right);\pxy\right]}
={\rm{E}}\left[\Xmat\varphi_{u,v}\left(\Xmat,\Ymat\right);\pxy\right]\hspace{0.2cm}{\rm{a.s.}}
\end{equation}
and
\begin{equation}
\label{muy_Lim_Closed} 
\lim\limits_{N\rightarrow\infty}\hat{\muvec}^{\left(u,v\right)}_{\Ymatsc}
=\frac{{\rm{E}}\left[\Ymat^{T}u\left(\Xmat\right)v\left(\Ymat\right);\pxy\right]}{{\rm{E}}\left[u\left(\Xmat\right)v\left(\Ymat\right);\pxy\right]}
={\rm{E}}\left[\Ymat\varphi_{u,v}\left(\Xmat,\Ymat\right);\pxy\right]\hspace{0.2cm}{\rm{a.s.}},
\end{equation}
where the last equalities in (\ref{XY_Lim_Closed})-(\ref{muy_Lim_Closed}) follow from the definition of $\varphi_{u,v}\left(\Xmat,\Ymat\right)$ in (\ref{VarPhiDef}).

Thus, since the sequence in the l.h.s. of (\ref{Rxy_uv_Est_Lim}) is obtained by continuous mappings of the elements of the sequences in its r.h.s., then by (\ref{XY_Lim_Closed})-(\ref{muy_Lim_Closed}), the Mann-Wald Theorem, and (\ref{RxyMod_pxy}) it is concluded that $\hat{\bSigma}^{\left(u,v\right)}_{\Xmatsc\Ymatsc}\rightarrow{\bSigma}^{\left(u,v\right)}_{\Xmatsc\Ymatsc}$ a.s. as $N\rightarrow\infty$.\qed
\subsection{The empirical MTCCA procedure with the exponential and Gaussian MT-functions}
\label{EmpMTCCA}
Given $N$ i.i.d. samples of $\Xmat$ and $\Ymat$, the empirical MTCCA procedure with the exponential and Gaussian MT-functions was carried out via the following steps:
\begin{enumerate}
\item
Estimate the optimal MT-functions parameters in (\ref{ExpKernel})  and (\ref{GaussKernel}) according to 
\begin{equation}
\label{MaxProb2}
\left(\hat{\svec}^{*},\hat{\tvec}^{*}\right)=\arg\max\limits_{\left(\svecsc,\tvecsc\right)\in{V}}\psi\left(\hat{\bSigma}^{\left(u,v\right)}_{\Xmatsc}\left(\svec,\tvec\right),\hat{\bSigma}^{\left(u,v\right)}_{\Ymatsc}\left(\svec,\tvec\right),\hat{\bSigma}^{\left(u,v\right)}_{\Xmatsc\Ymatsc}\left(\svec,\tvec\right)\right),
\end{equation}
where $\psi\left(\cdot,\cdot,\cdot\right)$ is defined in (\ref{PsiDef}), and $\hat{\bSigma}^{\left(u,v\right)}_{\Xmatsc}\left(\svec,\tvec\right)$, $\hat{\bSigma}^{\left(u,v\right)}_{\Ymatsc}\left(\svec,\tvec\right)$, and $\hat{\bSigma}^{\left(u,v\right)}_{\Xmatsc\Ymatsc}\left(\svec,\tvec\right)$ are the estimates in (\ref{Rx_uv_Est})-(\ref{Rxy_uv_Est}) of the covariance matrices $\bSigma^{\left(u,v\right)}_{\Xmatsc}\left(\svec,\tvec\right)$, $\bSigma^{\left(u,v\right)}_{\Ymatsc}\left(\svec,\tvec\right)$, and $\bSigma^{\left(u,v\right)}_{\Xmatsc\Ymatsc}\left(\svec,\tvec\right)$, respectively. The maximization in (\ref{MaxProb2}) was carried out numerically using gradient ascent over the search region $V$, which was selected as follows: 
\begin{enumerate}
\item
For the exponential MT-functions, we chose 
\begin{equation}
\nonumber
V_{\rm{E}}=\left\{\svec\in\Rsp^{p},\tvec\in\Rsp^{q}:\hat{J}_{\Xmatsc\Ymatsc}\left(\svec,\tvec\right)\leq{D}\right\},
\end{equation}
where $D=\sqrt{2}$, and 
$\hat{J}_{\Xmatsc\Ymatsc}\left(\svec,\tvec\right)\triangleq
{1}+
\svec^{T}\hat{\muvec}_{\Xmatsc}+\tvec^{T}\hat{\muvec}_{\Ymatsc}+\frac{1}{2}\svec^{T}\hat{\Rmat}_{\Xmatsc}\svec
+\svec^{T}\hat{\Rmat}_{\Xmatsc\Ymatsc}\tvec+\frac{1}{2}\tvec^{T}\hat{\Rmat}_{\Ymatsc}\tvec$ is a quadratic empirical approximation of the joint moment generating function ${M}_{\Xmatsc\Ymatsc}\left(\svec,\tvec\right)$ in (\ref{MonGenFunc}). The vectors $\hat{\muvec}_{\Xmatsc}$ and $\hat{\muvec}_{\Ymatsc}$ denote the sample expectations of $\Xmat$ and $\Ymat$, respectively. The matrices $\hat{\Rmat}_{\Xmatsc}$, $\hat{\Rmat}_{\Ymatsc}$, and $\hat{\Rmat}_{\Xmatsc\Ymatsc}$ denote sample auto-correlation matrix of $\Xmat$, the sample auto-correlation matrix of $\Ymat$,  and their cross-correlation matrix, respectively. Since $D=\sqrt{2}$ and $\hat{J}_{\Xmatsc\Ymatsc}\left(\svec,\tvec\right)$ is quadratic and takes a unit value at the origin, then $V_{\rm{E}}$ defines a closed region in $\Rsp^{p}\times\Rsp^{q}$ containing the origin. 
\item
For the Gaussian MT-functions, the search region was set to 
\begin{eqnarray}
\nonumber
V_{\rm{G}}=\{\svec\in\Rsp^{p},\tvec\in\Rsp^{q}:\hat{\nu}\left(X_{k},5\right)\leq{s}_{k}\leq\hat{\nu}\left(X_{k},95\right),\hat{\nu}\left(Y_{l},5\right)\leq{t}_{l}\leq\hat{\nu}\left(Y_{l},95\right),\\\nonumber{k}=1,\ldots,p,l=1,\ldots,q\},
\end{eqnarray}
where $s_{k}$ and $t_{l}$ are the $k$-th and $l$-th entries of $\svec$ and $\tvec$, respectively, and $\hat{\nu}\left(X,\alpha\right)$ is the empirical $\alpha$-th percentile of the random variable $X$. One can notice that $V_{\rm{G}}$ defines a closed rectangle in $\Rsp^{p}\times\Rsp^{q}$. In the considered examples it was verified that $V_{\rm{G}}$ contains the origin. We note that in case where $V_{\rm{G}}$ does not contain the origin, one can always subtract the expectations of $\Xmat$ and $\Ymat$ and perform MTCCA on $\Xmat^{\prime}=\Xmat-{\rm{E}}\left[\Xmat;\px\right]$ and $\Ymat^{\prime}=\Ymat-{\rm{E}}\left[\Ymat;\py\right]$. 
\end{enumerate}
\item
Obtain estimates of the MT-canonical correlation coefficients, $$\hat{\rho}_{k}\triangleq\rho_{k}\left({\hat{\bSigma}^{\left(u,v\right)}_{\Xmatsc}\left(\hat{\svec}^{*},\hat{\tvec}^{*}\right)},{\hat{\bSigma}^{\left(u,v\right)}_{\Ymatsc}\left(\hat{\svec}^{*},\hat{\tvec}^{*}\right)},{\hat{\bSigma}^{\left(u,v\right)}_{\Xmatsc\Ymatsc}\left(\hat{\svec}^{*},\hat{\tvec}^{*}\right)}\right),\hspace{0.4cm}k=1,\ldots,{r},$$ and estimates of the MT-canonical directions,
$$\left(\hat{\avec}_{k},\hat{\bvec}_{k}\right),\hspace{0.4cm}k=1,\ldots,{r},$$
by solving the following GEVD equation
\begin{equation}
\label{GEVD_PMTCCA_2} 
\left[\begin{array}{cc}{\zerovec} & {\hat{\bSigma}^{\left(u,v\right)}_{\Xmatsc\Ymatsc}\left(\hat{\svec}^{*},\hat{\tvec}^{*}\right)} \\ {\hat{\bSigma}^{\left(u,v\right)T}_{\Xmatsc\Ymatsc}\left(\hat{\svec}^{*},\hat{\tvec}^{*}\right)} & {\zerovec}\end{array}\right]\left[\begin{array}{c}{\avec} \\{\bvec}\end{array}\right]=\rho\left[\begin{array}{cc}{\hat{\bSigma}^{\left(u,v\right)}_{\Xmatsc}\left(\hat{\svec}^{*},\hat{\tvec}^{*}\right)} & {\zerovec} \\ {\zerovec} & {\hat{\bSigma}^{\left(u,v\right)}_{\Ymatsc}\left(\hat{\svec}^{*},\hat{\tvec}^{*}\right)}\end{array}\right]\left[\begin{array}{c}{\avec} \\{\bvec}\end{array}\right],
\end{equation}
where $\rho=\hat{\rho}_{k}$ is the $k$-th largest generalized eigenvalue of the pencil in (\ref{GEVD_PMTCCA_2}), and $\left[{\avec}^{T},{\bvec}^{T}\right]^{T}=\left[\hat{\avec}^{T}_{k},\hat{\bvec}^{T}_{k}\right]^{T}$ is its corresponding generalized eigenvector.
\end{enumerate}
In all considered examples the width parameters $\sigma$ and $\tau$ of the Gaussian MT-functions (\ref{GaussKernel}) were set to 
$\sigma=\frac{1}{p}\sum\limits_{k=1}^{p}\hat{\sigma}\left(X_{k}\right)$ and $\tau=\frac{1}{q}\sum\limits_{l=1}^{q}\hat{\sigma}\left(Y_{l}\right)$, where $\hat{\sigma}\left(X\right)$ denotes the empirical standard deviation the random variable $X$.
\subsection{Testing the statistical significance of the empirical canonical correlation coefficients}
\label{PVal}
Let $\Xmat^{N}\triangleq\left\{\Xmat_{n}\right\}_{n=1}^{N}$ and $\Ymat^{N}\triangleq\left\{\Ymat_{n}\right\}_{n=1}^{N}$ denote sequences of $N$ i.i.d. samples of $\Xmat$ and $\Ymat$, respectively. Additionally, let $\hat{\rho}_{k}\left(\Xmat^{N},\Ymat^{N}\right)$ denote the empirical $k$-th order canonical correlation coefficient based on $\Xmat^{N}$ and $\Ymat^{N}$. A bootstrap based procedure for testing the statistical significance of the empirical $k$-th order canonical correlation coefficient is specified below:
\begin{enumerate}
\item Repeat the following procedure for $M$ times (with index $m=1,\ldots,M$):
\begin{enumerate}
\item 
Generate a randomly permuted version of the sequence $\Ymat^{N}$, denoted by $\Ymat^{N}_{m}$.
\item
Compute the statistic $\theta_{m}=\hat{\rho}_{k}\left(\Xmat^{N},\Ymat^{N}_{m}\right)$. 
\end{enumerate}
\item
Construct an empirical cumulative distribution function from the sample statistics $\theta_{m}$, $m=1,\ldots,M$, as
\begin{equation}\nonumber
F_{\Theta}\left(\theta\right)={\rm{Pr}}\left(\Theta\leq\theta\right)=\frac{1}{M}\sum\limits_{m=1}^{M}{\mathbbm{1}}_{x\geq0}\left(x=\theta-\theta_{m}\right),
\end{equation}
where $\mathbbm{1}$ is an indicator random variable on its argument $x$.
\item
Compute the $p$-value 
\begin{equation}\nonumber
p_{0}=1-F_{\Theta}\left(\theta_{0}\right),
\end{equation}
where $\theta_{0}=\hat{\rho}_{k}\left(\Xmat^{N},\Ymat^{N}\right)$ is the true detection statistic.
\item
If $p_{0}<\alpha$, then we have that $\hat{\rho}_{k}\left(\Xmat^{N},\Ymat^{N}\right)$ is significant at level $\alpha$, leading to rejection of the null-hypothesis of no dependence between $\Xmat$ and $\Ymat$.
\end{enumerate}
In all considered examples, the number of permutations $M$ and the significance level $\alpha$ were set to $1000$ and $0.01$, respectively.


\begin{thebibliography}{1}
\footnotesize

\bibitem{Hotelling} H. Hotelling, ``Relations between two sets of variates,'' {\em Biometrika,} vol. 28, pp. 321-377, 1936.

\bibitem{Pearson} K. Pearson, ``Mathematical contributions to the theory of evolution. III. Regression, heredity and panmixia,'' {\em Philos. Trans. Royal Soc. London Ser. A,} vol. 187, pp. 253-318, 1896.

\bibitem{BSSCCA} Y.O. Li, T. Adal\i, W. Wang, and V.D. Calhoun, ``Joint Blind Source Separation by Multiset Canonical Correlation Analysis,'' {\em IEEE Trans. Signal Processing,} vol. 57, pp. 3918-3929, 2009.

\bibitem{ImProcCCA} T.K. Kim, J. Klitter, and R. Cipolla, ``Discriminative learning and recognition of image set classes using canonical correlations,'' {\em IEEE Trans. Pattern Analysis and Machine Intelligence,} vol. 29, pp. 1005-1018, 2007.

\bibitem{DOA_CCA_1} G. Huang, L. Yang, and Z. He, ``Canonical correlation analysis using for DOA estimation of multiple audio sources,'' {\em Computational Intelligence and Bioinspired Systems,} vol. 29, pp. 228-302, 2005.

\bibitem{DOA_CCA_2} X. Wang, H. Ge, and I. P. Kirsteins,  ``Direction-of-arrival estimation using distributed arrays: A canonical coordinates perspective with limited array size and sample support,'' {\em Proc. of the ICASSP 2010,} pp. 2622-2625, 2010.

\bibitem{CCA_Med_Im} N. M. Correa, T. Adal\i, and Y. O. Li, ``Canonical correlation analysis for data fusion and group inferences,'' {\em IEEE Signal Processing Magazine,} vol. 27, pp. 39-50, 2010.

\bibitem{AV_SYNCH_1} E. Kidron, and Y. Y. Schechner, ``Pixels that sound,'' {\em Proc. of CVPR 2005,} vol. 1, pp. 88-95, 2005.

\bibitem{AV_SYNCH_2} J. S. Lee, and T. Ebrahimi, ``Audio-visual synchronization recovery in multimedia content,'' {\em Proc. of the ICASSP 2011,} pp. 2280 - 2283, 2011.

\bibitem{CCA_Sharf} A. Pezeshki, M.R. Azimi-Sadjadi, and L.L. Scharf, ``Undersea target classification using canonical correlation analysis,'' {\em IEEE Journal of Oceanic Engineering,} vol. 32, pp. 948-955, 2007.

\bibitem{MICCA} X. Yin, ``Canonical correlation analysis based on information theory,'' {\em Journal of Multivariate Analysis,} vol. 91, pp. 161-176, 2004.

\bibitem{InfTheory} T. M. Cover, and J. A. Thomas, {\em Elements of information theory,} John Wiley and Sons, 2006.

\bibitem{ConvOpt} S. Boyd, {\em Convex Optimization,} Cambridge University Press, 2004.

\bibitem{Akaho} S. Akaho, ``A kernel method for canonical correlation analysis,'' {\em Proc. of the IMPS 2001,}.

\bibitem{Melzer} T. Melzer, M. Reiter, and H. Bischof, ``Nonlinear feature extraction using generalized canonical correlation analysis,'' {\em Proc. of the ICANN 2001,} pp. 353 - 2283, 2001.

\bibitem{Bach} F. R. Bach, and M. I. Jordan, ``Kernel independent component analysis,'' {\em Journal of Machine
Learning Research,} vol. 3, pp. 1-48, 2002.

\bibitem{Yoshihiro} Y. Yamanishi, J. P. Vert, A. Nakaya, and M. Kanehisa, ``Extraction of
correlated gene clusters from multiple genomic data by generalized kernel canonical correlation
analysis,'' {\em Bioinformatics,} vol. 19, pp. 323i-330i, 2003.

\bibitem{Hardoon} D. R. Hardoon, J. Shawe-Taylor, and O. Friman, ``KCCA for fMRI analysis,'' {\em Proc. of Medical Image Understanding and Analysis 2004}.

\bibitem{Suetani} H. Suetani, Y. Iba, and K. Aihara, ``Detecting hidden synchronization of chaotic
dynamical systems: A kernel-based approach,'' {\em Journal of Physics A: Mathematical and General,} vol. 39, pp. 10723-10742, 2006.

\bibitem{StatDict} F. H. C. Marriott, {\em A dictionary of statistical terms,} Longman Scientific \& Technical, 1990.

\bibitem{Anderson} T. W. Anderson, {\em An introduction to multivariate statistical analysis,} John Wiley and Sons, 2003.

\bibitem{Yeredor1} A. Yeredor, ``Blind Source Separation via the Second Characteristic Function,'' {\em Signal Processing,} vol. 80, pp. 897-902, 2000.


\bibitem{Yeredor3} A. Yeredor, ``Blind Channel Estimation Using First and Second Derivatives of the Characteristic Function,'' {\em IEEE Signal Processing Letters,} vol. 9, pp. 100-103, 2002.

\bibitem{Yeredor4} E. Edinger, and A. Yeredor, ``Blind MIMO identification Using the Second Characteristic Function,'' {\em IEEE Trans. on Signal Processing,} vol. 53, pp. 4067-4079, 2005.

\bibitem{Yeredor5} A. Yeredor, ``MUSIC using Off-Origin Hessians of the Second Characteristic Function,'' {\em Proc. of SAM 2006,} 2006.

\bibitem{Yeredor6} A. Slapak, and A. Yeredor, ``Weighting for More": Enhancing Characteristic-Function Based ICA with Asymptotically Optimal Weighting,'' {\em Signal Processing,} vol. 91, pp. 2016-2027, 2011.

\bibitem{Yeredor7} A. Yeredor, ``Yule-Walker equations applied to Hessians of the characteristic function for improved AR estimation,'' {\em Proc. of the ICASSP 2007}.

\bibitem{Yeredor8} A. Yeredor, ``Substituting the cumulants in the super-exponential blind equalization algorithm,'' {\em Proc. of the ICASSP 2008}.

\bibitem{Severini} T. A. Severini, {\em Elements of distribution theory}. Cambridge University Press, 2005.

\bibitem{DasGupta} A. DasGupta, {\em Fundamentals of Probability: A First Course}. Springer Verlag, 2010.

\bibitem{AdCal} P. M. Fitzpatrick, {\em Advanced calculus}. American Mathematical Society, 2006.

\bibitem{MeasureTheory} K. B. Athreya and S. N. Lahiri, {\em Measure theory and probability theory,} Springer-Verlag, 2006.

\bibitem{Folland} G. B. Folland, {\em Real Analysis}. John Wiley and Sons, 1984.

\bibitem{MannWald} H. B. Mann, and A. Wald, ``On stochastic limit and order relationships,'' {\em Ann. Math. Stat.,} vol. 14, pp. 217-226, 1943.

\bibitem{WRDS} Wharton Research Data Services, {\em https://wrds-web.wharton.upenn.edu/wrds}.

\bibitem{EditDist} X. Gao, B. Xiao, D. Tao, and X. Li, ``A survey of graph edit distance,'' {\em Pattern Analysis Applications,} vol. 13, pp. 113-129, 2010.

\bibitem{Silverman} B. W. Silverman, {\em Density estimation for statistics and data analysis,} Chapman \& Hall/CRC, 1986.

\end{thebibliography}
\end{document}